\documentclass[12pt]{article}
\usepackage{amsmath,amssymb,epsfig}
\makeatletter \@addtoreset{equation}{section} \makeatother
\renewcommand{\theequation}{\thesection.\arabic{equation}}
\newcommand{\ba}{\begin{array}}
\newcommand{\ea}{\end{array}}
\newcommand{\beq}{\begin{equation}}
\newcommand{\eeq}{\end{equation}}
\newcommand{\bea}{\begin{eqnarray}}
\newcommand{\eea}{\end{eqnarray}}
\def\bce{\begin{center}}
\def\ece{\end{center}}

\def\nonu{\nonumber}

\def\be{\beta}

\newcommand{\tr}{\mbox{Tr}}

\def\eps6{{\displaystyle \mathop{\epsilon}^{6}}{}}

\def\nab6{{\displaystyle \mathop{\nabla}^{6}}{}}
\def\0{{\sst{(0)}}}
\def\1{{\sst{(1)}}}
\def\2{{\sst{(2)}}}
\def\3{{\sst{(3)}}}
\def\4{{\sst{(4)}}}
\def\5{{\sst{(5)}}}
\def\6{{\sst{(6)}}}
\def\7{{\sst{(7)}}}
\def\8{{\sst{(8)}}}

\def\ba{\begin{array}}
\def\ea{\end{array}}
\def\beq{\begin{equation}}
\def\eeq{\end{equation}}
\def\be{\begin{equation}}
\def\ee{\end{equation}}

\def\tr{\mathop{\rm tr}}

\def\eps{\epsilon}

\def\ba{\begin{array}}
\def\ea{\end{array}}
\def\beq{\begin{equation}}
\def\eeq{\end{equation}}
\def\be{\begin{equation}}
\def\ee{\end{equation}}

\def\tr{\mathop{\rm tr}}

\def\eps{\epsilon}

\newcommand{\bean}{\begin{eqnarray*}}
\newcommand{\eean}{\end{eqnarray*}}

\begin{document}
\thispagestyle{empty} \addtocounter{page}{-1}
\begin{flushright}
%KIAS-P07034 \\
%CALT-68-nnnn \\
%{\tt arXiv:yymm.nnnn}\\
\end{flushright}

\vspace*{1.3cm}

\centerline{ \Large \bf  Meta-Stable Brane Configurations with
Multiple NS5-Branes}
%\vspace{.3cm} 
%\centerline{ \Large \bf by Adding an Orientifold-Plane to Giveon-Kutasov } 
\vspace*{1.5cm}
\centerline{{\bf Changhyun Ahn} 
%and {\bf Yutaka Ookouchi $^{2}$}
} 
\vspace*{1.0cm} 
\centerline{\it 
Department of Physics, Kyungpook National University, Taegu
702-701, Korea} 
%\centerline{\it $^{2}$ California Institute of 
%Technology, Pasadena, CA91125, USA }
\vspace*{0.8cm} 
\centerline{\tt ahn@knu.ac.kr} 
%\qquad
%yutaka@caltech.edu} 
\vskip2cm

\centerline{\bf Abstract}
\vspace*{0.5cm}

Starting from an ${\cal N}=1$ supersymmetric electric gauge theory 
with the multiple product gauge group  
and the bifundamentals, we apply Seiberg dual 
to each gauge group, obtain the ${\cal N}=1$
supersymmetric 
dual magnetic gauge theories with dual matters including the
gauge singlets. Then we 
describe the intersecting brane configurations, 
where there are NS-branes and
D4-branes(and anti D4-branes), of
type 
IIA string theory corresponding to the meta-stable nonsupersymmetric 
vacua of this gauge theory.

We also discuss the case where the orientifold 4-planes are added into
the above brane configuration.
Next, by adding an orientifold 6-plane, we apply to 
an ${\cal N}=1$ supersymmetric electric gauge theory 
with the multiple product gauge group(where a single 
symplectic or orthogonal gauge group is present)  
and the bifundamentals. 
Finally, we describe the other cases where the orientifold 6-plane
intersects with NS-brane.

\baselineskip=18pt
\newpage
\renewcommand{\theequation}
{\arabic{section}\mbox{.}\arabic{equation}}

%%%%%%%%%%%%%%%%%%%%%%%%%%%%%%%%%%%%%%%%%%%%%%%%%%%%%%%%%%%%%%%%%%%%%%
%%%%%%%%%%%%%%%%%%%%%%%%%%%%%%%%%%%%%%%%%%%%%%%%%%%%%%%%%%%%%%%%%%%%%%
\section{Introduction}
%%%%%%%%%%%%%%%%%%%%%%%%%%%%%%%%%%%%%%%%%%%%%%%%%%%%%%%%%%%%%%%%%%%%%%
%%%%%%%%%%%%%%%%%%%%%%%%%%%%%%%%%%%%%%%%%%%%%%%%%%%%%%%%%%%%%%%%%%%%%%

It is known that 
the dynamical supersymmetry breaking in meta-stable vacua \cite{ISS} 
arises 
in the standard ${\cal N}=1$ SQCD with massive fundamental 
flavors(for recent developments
on supersymmetry breaking, see also the review paper \cite{IS}).
Due to the extra mass term
for quarks in the magnetic superpotential, 
not all F-term equations can be satisfied and the
supersymmetry is broken.
The meta-stable brane configurations of type IIA string theory
corresponding to ${\cal N}=1$ SQCD with massive  fundamental flavors 
have been found in \cite{OO1,FGU,BGHSS}(for 
the brane dynamics and supersymmetric gauge theory, 
see also the review paper \cite{GK98}). 
There is  also the meta-stable brane configuration  \cite{Ahn06} 
corresponding to the gauge theory 
where there exists an extra adjoint matter as well as fundamentals.
 
Giveon and Kutasov \cite{GK} have found 
the type IIA brane configuration consisting of 
three NS-branes, D4-branes and anti-D4-branes($\overline{D4}$-branes).
The meta-stable vacua of \cite{ISS} occur in some
region of parameter space when the D4-branes and
$\overline{D4}$-branes can decay and the geometric misalignment of
flavor D4-branes arises. 
The mass term in the magnetic superpotential 
corresponds to the relative displacement of two NS5'-branes 
along the (45) directions and the dual quarks can be represented by the 
bifundamentals of product gauge group. 

Adding an orientifold 4-plane(O4-plane) only to the brane configuration 
of \cite{GK} implies that the gauge group is a product of a symplectic
group and an orthogonal group and the geometric misalignment of flavor 
D4-branes \cite{Ahn07-5} leads to the brane configuration of \cite{Ahn06-1}. 
The ${\cal N}=1$ product gauge group theory \cite{ILS,BIWW} 
is realized by three
NS-branes, two kinds of D4-branes, and two kinds of 
D6-branes \cite{BH}.  
When an orientifold 6-plane(O6-plane) is added into the brane configuration of 
\cite{GK}, the gauge group is a product of two unitary groups 
with extra matters as well as bifundamentals and  
the type IIA brane configuration consists of 
five NS-branes, 
D4-branes, $\overline{D4}$-branes and O6-plane.
Similarly,  the geometric misalignment of flavor 
D4-branes \cite{Ahn07-5}  leads to the brane configurations of \cite{Ahn07} or 
\cite{Ahn07-1} depending on the O6-plane charge. 

One can generalize the work of \cite{GK} further by adding more NS-branes to
that brane configuration.
For the ${\cal N}=1$ triple product group gauge theory, the
supersymmetric electric brane configuration 
consists of  four
NS-branes, three kinds of D4-branes, and three kinds of 
D6-branes \cite{BH,AT97}.
Then the triple product gauge group theory with bifundamentals only
can be realized by four NS-branes, three kinds of 
D4-branes and $\overline{D4}$-branes \cite{Ahn07-6}.
The meta-stable vacua of \cite{Ahn07-3} occur in some
region of parameter space when the D4-branes and
$\overline{D4}$-branes can decay and the geometric misalignment of
flavor D4-branes arises.

Now adding an orientifold 4-plane only to the brane configuration 
of \cite{Ahn07-6} leads to the fact that the gauge group is a triple 
product of a symplectic
group and an orthogonal group alternatively 
and the geometric misalignment of flavor 
D4-branes \cite{Ahn07-6} reduces to the brane 
configuration of \cite{Ahn07-2}. 
One can add an orientifold 6-plane into the brane configuration of 
\cite{BH,AT97} together with extra outer NS-branes. 
Then one of 
the type IIA brane configurations consists of 
six NS-branes, 
D4-branes, $\overline{D4}$-branes and O6-plane.
The geometric misalignment of flavor 
D4-branes \cite{Ahn07-6} leads to the brane configurations of \cite{Ahn07-3}. 
Different O6-plane charge will give rise to other brane configuration. 
When there are seven NS-branes, 
D4-branes, $\overline{D4}$-branes and O6-plane, 
the geometric misalignment of flavor 
D4-branes \cite{Ahn07-7} gives rise to the brane configurations 
\cite{Ahn07-4}. 

Recently, 
from the ${\cal N}=1$ triple product group gauge theory with
fundamentals and bifundamentals, the
meta-stable brane configurations 
consisting of  four
NS-branes, three kinds of D4-branes, and three kinds of 
D6-branes are found \cite{Ahn07-8} and its generalization to multiple
product gauge groups also is discussed in \cite{Ahn07-8,Ahn07-9}.

In this paper, along the line of \cite{GK,Ahn07-5,Ahn07-6,Ahn07-7}, 
we  present the intersecting 
brane configurations of type IIA string
theory corresponding to 
the new meta-stable nonsupersymmetric meta-stable vacua in four dimensional ${\cal
N}=1$ multiple product  gauge theory with matters.
For the ${\cal N}=1$ multiple product gauge group theory with 
bifundamentals, the
supersymmetric electric brane configuration in type IIA string theory 
consists of  $(n+1)$
NS-branes, and $n$ kinds of D4-branes(and anti D4-branes) and 
the gauge group is a product of $n$ unitary groups 
with bifundamentals. 
For a given supersymmetric electric gauge theory which does not have
any superpotential, 
one takes both the mass deformation for bifundamentals and its 
Seiberg dual. Then we construct the meta-stable 
brane configurations in type IIA string theory.    
Adding the O4-plane or O6-plane to these brane configurations are 
described.

In section 2,
we describe the type IIA brane configuration corresponding
to the electric theory based on the ${\cal N}=1$ gauge theory with 
gauge group $\prod_{i=1}^{n} 
SU(N_{c,i})$  
and bifundamentals, 
and deform this theory by adding the mass term
for the bifundamentals for each gauge group. 
Then we construct the Seiberg dual magnetic theories for each gauge
group with corresponding dual
matters as well as additional gauge singlets. 
After that, the
nonsupersymmetric brane configurations are  found by recombination and
splitting for the flavor D4-branes.

In section 3, 
we consider the type IIA brane configuration corresponding
to the electric theory based on the ${\cal N}=1$ $[ \prod_{i=1}^{n-2} 
Sp(N_{c,i}) \times
SO(2N_{c,i+1})] \times Sp(N_{c,n})$ or 
 $[ \prod_{i=1}^{n-1} Sp(N_{c,i}) \times
SO(2N_{c,i+1})]$
gauge theory 
and bifundamentals, 
and deform this theory by adding the mass terms
for the bifundamentals for each gauge group. 
We present the Seiberg dual magnetic theories for each gauge
group with corresponding dual
matters as well as additional gauge singlets. 
The
nonsupersymmetric brane configurations are  found.
When the orientifold 4-plane  charge is reversed, it is
straightforward to proceed similarly.

In section 4,
we study the type IIA brane configuration corresponding
to the electric theory based on the ${\cal N}=1$ $Sp(N_{c,1}) \times
\prod_{i=2}^{n} SU(N_{c,i})$ 
gauge theory 
with bifundamentals, 
and deform this theory by adding the mass term
for the quarks for each gauge group.
Explicitly we construct the Seiberg dual magnetic theories for each gauge
group factor with corresponding dual
matters as well as extra gauge singlets and the
nonsupersymmetric brane configurations are  found.

In section 5,
we explain the type IIA brane configuration corresponding
to the electric theory based on the ${\cal N}=1$ $SO(N_{c,1}) \times
\prod_{i=2}^{n} SU(N_{c,i}) $ 
gauge theory 
with bifundamentals, 
and deform this theory by adding the mass term
for the quarks for each gauge group.
After that we describe the Seiberg dual magnetic theories for each gauge
group factor with corresponding dual
matters as well as extra gauge singlets. 
The
nonsupersymmetric brane configurations are  found from the magnetic
brane configurations. 

In section 6,
we describe the type IIA brane configuration corresponding
to the electric theory based on the ${\cal N}=1$ $\prod_{i=1}^{n} SU(N_{c,i})$ 
gauge theory 
with bifundamentals, a symmetric flavor and a conjugate
symmetric flavor for $SU(N_{c,1})$, 
and deform this theory by adding the mass term
for the quarks for each gauge group. 
Then we construct the Seiberg dual magnetic theories for each gauge
group with corresponding dual
matters as well as additional gauge singlets. 
After that, the
nonsupersymmetric brane configurations are  found by recombination and
splitting for the flavor D4-branes.

In section 7,
we consider the type IIA brane configuration corresponding
to the electric theory based on the ${\cal N}=1$ $\prod_{i=1}^{n} SU(N_{c,i})$ 
gauge theory 
with bifundamentals, eight-fundamentals, 
an antisymmetric flavor and a conjugate
symmetric flavor for $SU(N_{c,1})$, 
and deform this theory by adding the mass terms
for the quarks for each gauge group. 
We present the Seiberg dual magnetic theories for each gauge
group with corresponding dual
matters as well as additional gauge singlets. 
The
nonsupersymmetric brane configurations are  found.

Finally, in section 8, 
we summarize what we have found in this paper.

There exist some related works \cite{BMV}-\cite{FU} 
on the meta-stable vacua in different
directions.

%%%%%%%%%%%%%%%%%%%%%%%%%%%%%%%%%%%%%%%%%%%%%%%%%%%%%%%%%%%%%%%%%%%%%
%%%%%%%%%%%%%%%%%%%%%%%%%%%%%%%%%%%%%%%%%%%%%%%%%%%%%%%%%%%%%%%%%%%%%
\section{Meta-stable brane configurations with $(n+1)$ NS-branes}
%section 2
%%%%%%%%%%%%%%%%%%%%%%%%%%%%%%%%%%%%%%%%%%%%%%%%%%%%%%%%%%%%%%%%%%%%
%%%%%%%%%%%%%%%%%%%%%%%%%%%%%%%%%%%%%%%%%%%%%%%%%%%%%%%%%%%%%%%%%%%%%

The type IIA brane configuration  \cite{BH,AT97} corresponding to 
${\cal N}=1$ supersymmetric electric gauge theory(see also
\cite{Ahn07-8})  with
gauge group
\bea
SU(N_{c,1}) \times \cdots
\times SU(N_{c,n})
\nonu
\eea
and with 
the $(n-1)$ bifundametals $F_i$ charged under 
$({\bf 1_1, \cdots, 1, \Box_i, \overline{\Box}_{i+1}, 1, \cdots,  1_n})$
and their
complex conjugate fields $\widetilde{F}_i$ 
charged $({\bf 1_1, \cdots, 1, \overline{\Box}_i, \Box_{i+1}, 1, 
\cdots, 1_n})$ where $i=1, 2, \cdots, (n-1)$
can be described by 
the $NS5_1'$-brane(012389), 
the  
$NS5_2$-brane(012345), $\cdots$, the $NS5_{n+1}$-brane for odd number
of gauge groups(or 
the $NS5_{n+1}'$-brane for even number of gauge groups),
$N_{c,1}$-, $N_{c,2}$-,  $\cdots$, and $N_{c,n}$-color D4-branes(01236). 
See the Figure 1 for the details on the brane configuration.

%%%%%%%%%%%%%%%%%%%%%%%%%%%%%%%%%%%%%%%%%%%%%%%%%%%%%%%%%%%%%%%%%%%%%
%%%%%%%%%%%%%%%%%%%%%%%%%%%%%%%%%%%%%%%%%%%%%%%%%%%%%%%%%%%%%%%%%%%%%%
\begin{figure}[ht]
   \epsfxsize=4.0in 
\centerline{\epsffile{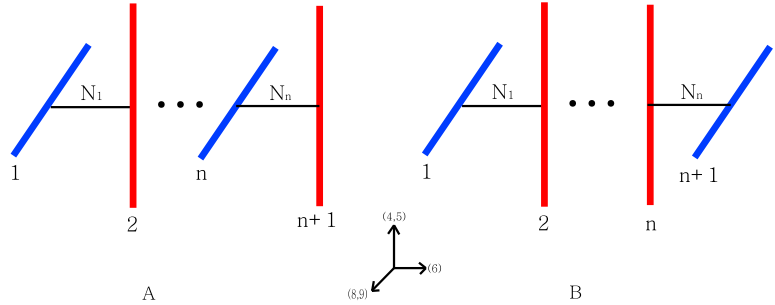}}
   \caption[FIG. \arabic{figure}.]{ 
The 
 ${\cal N}=1$ supersymmetric 
electric brane configuration for the gauge group $\prod_{i=1}^{n}
 SU(N_{c,i})$ 
and  bifundamentals $F_i$ and $\widetilde{F}_i$  with vanishing mass
for the bifundamentals when the number of gauge groups factor 
$n$ is odd(1A) and even(1B). The number of D4-branes $N_{c,i}$ is 
denoted by
$N_i$ in this Figure for simplicity 
and the NS5'-branes are labelled by $1, 3, \cdots$, and 
$n[n+1]$ while
the NS5-branes are labelled by $2, 4, \cdots$, 
and $(n+1)[n]$ in 1A[1B]. 
%We do not draw the middle region of this
% chain denoted by $\cdots$ for simplicity. 
}
\end{figure}
%%%%%%%%%%%%%%%%%%%%%%%%%%%%%%%%%%%%%%%%%%%%%%%%%%%%%%%%%%%%%%%%%%%%%
%Figure 1A and 1B
%%%%%%%%%%%%%%%%%%%%%%%%%%%%%%%%%%%%%%%%%%%%%%%%%%%%%%%%%%%%%%%%%%%%%

Let us place the $NS5_1'$-brane at the origin $x^6=0$
and denote the $x^6$ 
coordinates for 
the  
$NS5_2$-brane, $\cdots$, the $NS5_{n+1}$-brane for odd $n$(or 
the $NS5_{n+1}'$-brane for even $n$)
are given by $x^6=y_1, y_1+y_2, \cdots, \sum_{j=1}^{n-1} y_j + y_n$
respectively.
The $N_{c,1}$ D4-branes 
are suspended between the 
$NS5_1'$-brane and the $NS5_2$-brane, 
the $N_{c,2}$ D4-branes 
are suspending between the 
$NS5_2$-brane and the $NS5_3'$-brane, $\cdots$ and 
the $N_{c,n}$ D4-branes  
are suspended between the $NS5_n'$-brane and the $NS5_{n+1}$-brane for
odd $n$(or 
between the $NS5_n$-brane and the $NS5_{n+1}'$-brane for even $n$).
The fields $F_i$ and $\widetilde{F}_i$  correspond to 4-4 strings connecting 
the $N_{c,i}$-color D4-branes with $N_{c,i+1}$-color D4-branes.
We draw this ${\cal N}=1$ supersymmetric 
electric brane configuration in Figure 1A(1B) 
when $n$ is odd(even) for the vanishing mass
for the fields $F_i$ and $\widetilde{F}_i$. 
The gauge couplings of $SU(N_{c,1})$, $SU(N_{c,2})$, $\cdots$, 
and $ SU(N_{c,n})$
are given by a string coupling constant $g_s$, a string scale $\ell_s$ 
and the $x^6$ coordinates $y_i$ for $n$ NS-branes through
\bea
g_1^2 =\frac{g_s \ell_s}{y_1}, \qquad
g_2^2 = \frac{g_s \ell_s}{y_2}, \qquad \cdots \qquad, \qquad
g_{n}^2=\frac{g_s \ell_s}{y_n}.
\nonu
\eea

Let us deform the theory by Figure 1A.
Displacing the two NS5'-branes, $NS5_{i}'$-brane and
$NS5_{i+2}'$-brane, 
relative each other in the 
\bea
v \equiv
x^4 + i x^5
\nonu
\eea 
direction, characterized by $(\Delta x)_{i+1}$, 
corresponds to turning on a quadratic
mass-deformed superpotential
for the fields $F_i$ and $\widetilde{F}_i$ as follows:
\bea
W_{elec} = m_{i+1} F_i \widetilde{F}_i (\equiv m_{i+1} \Phi_{i+1}),
\qquad
\mbox{when $i$ is odd}
\label{mass}
\eea
where 
the $i$-th gauge group indices in $F_i$ and $\widetilde{F}_i$ 
are contracted, each $(i+1)$-th gauge group index in them is encoded in 
$\Phi_{i+1}$ and the mass $m_{i+1}$ is given by
\bea
m_{i+1} 
= 
\frac{(\Delta x)_{i+1}}{\ell_s^2}.
\label{m}
\eea

The gauge-singlet $\Phi_{i+1}$ for the $i$-th  gauge group which was in the 
adjoint representation for the $(i+1)$-th  gauge group 
can be represented by
\bea
{(\bf 1_1, \cdots, 1_i, (N_{c,i+1}-N_{c,i+2})^2-1, 1_{i+2}, \cdots, 1_n)  
\oplus (1_1, \cdots, 1_n)}
\nonu
\eea 
under the  gauge group where the gauge group is broken from
$SU(N_{c,i+1})$ 
to $SU(N_{c,i+1}-N_{c,i+2})$. 
Then the $\Phi_{i+1}$ is a $(N_{c,i+1}-N_{c,i+2}) \times 
(N_{c,i+1}-N_{c,i+2})$ matrix.
The $NS5_{i+2}'$-brane together with $(N_{c,i+1}-N_{c,i+2})$-color D4-branes 
is moving to the $+v$ direction  for
fixed other branes during this mass deformation. 
In other words, the $N_{c, i+2}$ D4-branes among $N_{c,i+1}$ D4-branes 
are not participating in 
the mass deformation.
Then the $x^5$ coordinate($\equiv x$) 
of $NS5_i'$-brane is equal to
zero
while the $x^5$ coordinate of $NS5_{i+2}'$-brane is given by 
$(\Delta x)_{i+1}$.
Giving an expectation value to the meson field $\Phi_{i+1}$
corresponds to recombination of $N_{c,i}$- and $N_{c,i+1}$- color 
D4-branes, which will become $N_{c,i}$- or $N_{c,i+1}$-color D4-branes
in Figure 1A such that they are suspended between 
the $NS5_i'$-brane and the $NS5_{i+2}'$-brane 
and pushing them into the 
\bea
w \equiv x^8 + i
x^9
\nonu
\eea
direction. We assume that the number of colors satisfies
$
N_{c,i+1} \geq N_{c,i}-N_{c,i-1} \geq N_{c,i+2}$.
Now 
we draw this brane configuration in Figure 2A for nonvanishing mass
for the fields $F_i$ and $\widetilde{F}_i$. 

%%%%%%%%%%%%%%%%%%%%%%%%%%%%%%%%%%%%%%%%%%%%%%%%%%%%%%%%%%%%%%%%%%%%%%
%%%%%%%%%%%%%%%%%%%%%%%%%%%%%%%%%%%%%%%%%%%%%%%%%%%%%%%%%%%%%%%%%%%%%%
\begin{figure}[ht]
   \epsfxsize=4.0in 
\centerline{\epsffile{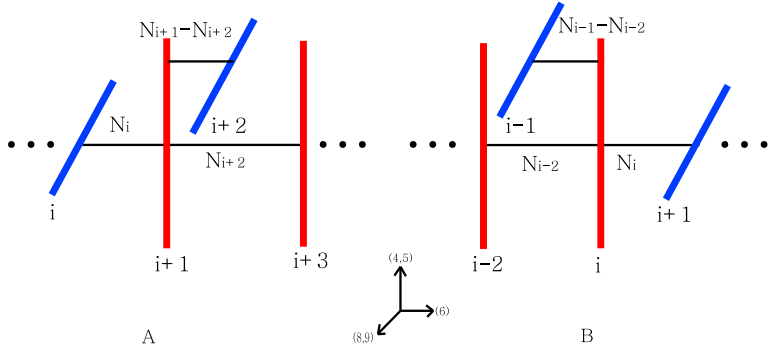}}
   \caption[FIG. \arabic{figure}.]{ 
The 
 ${\cal N}=1$ supersymmetric 
electric brane configuration for the gauge group $\prod_{i=1}^{n}
 SU(N_{c,i})$ 
and  bifundamentals  
with nonvanishing mass
for the bifundamentals  $F_i(F_{i-1})$ and
 $\widetilde{F}_i(\widetilde{F}_{i-1})$ when the number of gauge groups factor 
$n$ is odd(2A) and even(2B).  
The $N_{c,i+1}$ D4-branes in 1A are decomposed into 
$(N_{c,i+1}-N_{c,i+2})$ D4-branes which are moving to $+v$ direction
 with $(\Delta x)_{i+1}$ coordinate in 2A 
and $N_{c,i+2}$ D4-branes which are recombined with those D4-branes
connecting between $NS5_{i+2}'$-brane and $NS5_{i+3}$-brane in 2A.
%The $N_{c,i-1}$ D4-branes in 1B are decomposed into 
%$(N_{c,i-1}-N_{c,i-2})$ D4-branes which are moving to $+v$ 
%direction with $(\Delta x)_{i-1}$ coordinate in 2B 
%and $N_{c,i-2}$ D4-branes which are recombined with those D4-branes
%connecting between $NS5_{i-2}$-brane and $NS5_{i-1}'$-brane in 2B.
}
\end{figure}
%%%%%%%%%%%%%%%%%%%%%%%%%%%%%%%%%%%%%%%%%%%%%%%%%%%%%%%%%%%%%%%%%%%%%
%Figure 2A and 2B
%%%%%%%%%%%%%%%%%%%%%%%%%%%%%%%%%%%%%%%%%%%%%%%%%%%%%%%%%%%%%%%%%%%%%

Let us deform the theory by Figure 1B.
Displacing the two NS5'-branes, the $NS_{i-1}'$-brane and the 
$NS_{i+1}'$-brane, 
relative each other in the 
$v$ 
direction, charaterized by $(\Delta x)_{i-1}$, 
corresponds to turning on a quadratic
mass-deformed superpotential
for the fields $F_{i-1}$ and $\widetilde{F}_{i-1}$ as follows:
\bea
W_{elec} = m_{i-1} F_{i-1} \widetilde{F}_{i-1} (\equiv m_{i-1}
\Phi_{i-1}),
\qquad
\mbox{when $i$ is even}
\label{masseven}
\eea
where 
the $i$-th gauge group indices in $F_{i-1}$ and $\widetilde{F}_{i-1}$ 
are contracted, each $(i-1)$-th gauge group index in them is encoded in 
$\Phi_{i-1}$ and the mass $m_{i-1}$ is given by
\bea
m_{i-1} 
= 
\frac{(\Delta x)_{i-1}}{\ell_s^2}.
\label{m1even}
\eea

The gauge-singlet $\Phi_{i-1}$ for the $i$-th  gauge group which was  in the 
adjoint representation for the $(i-1)$-th  gauge group 
is in the  representation 
\bea
{(\bf 1_1, \cdots, 1_{i-2}, (N_{c,i-1}-N_{c,i-2})^2-1, 1_{i}, \cdots,
  1_n)  
\oplus (1_1, \cdots, 1_n)}
\nonu
\eea 
under the gauge group where the gauge group is broken from
$SU(N_{c,i-1})$ 
to $SU(N_{c,i-1}-N_{c,i-2})$. 
Then the $\Phi_{i-1}$ is a $(N_{c,i-1}-N_{c,i-2}) \times 
(N_{c,i-1}-N_{c,i-2})$ matrix.
The $NS5_{i-1}'$-brane together with $(N_{c,i-1}-N_{c,i-2})$-color D4-branes 
is moving to the $+v$ direction  for
fixed other branes during this mass deformation. 
In other words, the $N_{c, i-2}$ D4-branes among $N_{c,i-1}$ D4-branes 
are not participating in 
the mass deformation.
Then the $x^5$ coordinate($\equiv x$) 
of $NS5_{i+1}'$-brane is equal to
zero
while the $x^5$ coordinate of $NS5_{i-1}'$-brane is given by 
$(\Delta x)_{i-1}$.
Giving an expectation value to the meson field $\Phi_{i-1}$
corresponds to recombination of $N_{c,i-1}$- and $N_{c,i}$- color 
D4-branes, which will become $N_{c,i-1}$- or $N_{c,i}$-color D4-branes
in Figure 1B such that they are suspended between 
the $NS5_{i-1}'$-brane and the $NS5_{i+1}'$-brane 
and pushing them into the 
$w$ direction. We assume that the number of colors satisfies
$
N_{c,i-1} \geq N_{c,i}-N_{c,i+1} \geq N_{c,i-2}$.
Now 
we draw this brane configuration in Figure 2B for nonvanishing mass
for the fields $F_{i-1}$ and $\widetilde{F}_{i-1}$. 

%Next we describe five different magnetic dual theories by taking each
%corresponding mass deformation.

%%%%%%%%%%%%%%%%%%%%%%%%%%%%%%%%%%%%%%%%%%%%%%%%%%%%%%%%%%%%%%%%%%%%%%%%%
%%%%%%%%%%%%%%%%%%%%%%%%%%%%%%%%%%%%%%%%%%%%%%%%%%%%%%%%%%%%%%%%%%%%%%%%%
\subsection{${\cal N}=1$ 
$SU(N_{c,1}) \times \cdots \times 
SU(\widetilde{N}_{c,i}) \times \cdots \times SU(N_{c,n})$ magnetic theory}
%%%%%%%%%%%%%%%%%%%%%%%%%%%%%%%%%%%%%%%%%%%%%%%%%%%%%%%%%%%%%%%%%%%%%%%%
%%%%%%%%%%%%%%%%%%%%%%%%%%%%%%%%%%%%%%%%%%%%%%%%%%%%%%%%%%%%%%%%%%%%%%%%

Let us first consider the Seiberg dual for the middle gauge group
factor.
There are two magnetic duals depending on whether the gauge group factor
occurs at odd chain or even chain.

%%%%%%%%%%%%%%%%%%%%%%%%%%%%%%%%%%%%%%%%%%%%%%%%%%%%%%%%%%%%%
%%%%%%%%%%%%%%%%%%%%%%%%%%%%%%%%%%%%%%%%%%%%%%%%%%%%%%%%%%%%%
\subsubsection{When the dual gauge group occurs at odd chain} 
%%%%%%%%%%%%%%%%%%%%%%%%%%%%%%%%%%%%%%%%%%%%%%%%%%%%%%%%%%%%%
%%%%%%%%%%%%%%%%%%%%%%%%%%%%%%%%%%%%%%%%%%%%%%%%%%%%%%%%%%%%%

Starting from Figure 1A, moving the $NS5_{i+2}'$-brane 
with $(N_{c,i+1}-N_{c,i+2})$
D4-branes 
to the $+v$ direction leading to Figure 2A, 
and interchanging the $NS5_{i}'$-brane and the $NS5_{i+1}$-brane,
one obtains the Figure 3A.

%%%%%%%%%%%%%%%%%%%%%%%%%%%%%%%%%%%%%%%%%%%%%%%%%%%%%%%%%%%%%%%%%%%%%
%%%%%%%%%%%%%%%%%%%%%%%%%%%%%%%%%%%%%%%%%%%%%%%%%%%%%%%%%%%%%%%%%%%%%%
\begin{figure}[ht]
   \epsfxsize=4.0in 
\centerline{\epsffile{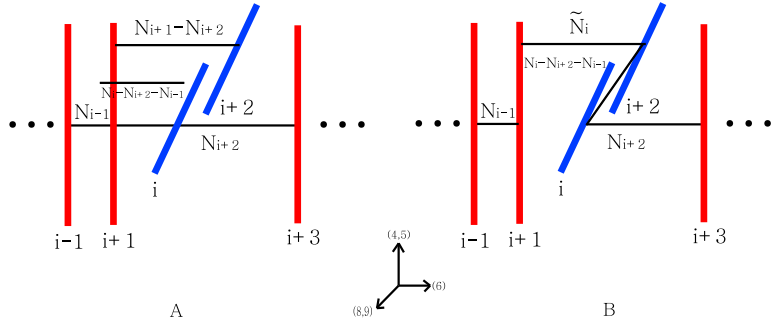}}
   \caption[FIG. \arabic{figure}.]{ 
The 
 ${\cal N}=1$ magnetic brane configuration for the gauge group 
containing $SU(\widetilde{N}_{c,i}=N_{c,i-1}+N_{c,i+1}-N_{c,i})$ where
 $i$
is odd, 
corresponding to Figure 2A with D4-
and $\overline{D4}$-branes(3A) and with 
a misalignment between D4-branes(3B) when the two NS5'-branes are close to
each other. 
The number of tilted D4-branes in 3B can be written as
$N_{c,i}-N_{c,i-1}-N_{c,i+2}
 =N_{c,i+1}-N_{c,i+2}-\widetilde{N}_{c,i}$.
%The $x$ coordinate of $NS5_{i+2}'$-brane is given by $(\Delta
% x)_{i+1}$.
}
\end{figure}
%%%%%%%%%%%%%%%%%%%%%%%%%%%%%%%%%%%%%%%%%%%%%%%%%%%%%%%%%%%%%%%%%%%%%
%Figure 3A and 3B
%%%%%%%%%%%%%%%%%%%%%%%%%%%%%%%%%%%%%%%%%%%%%%%%%%%%%%%%%%%%%%%%%%%%%

Before arriving at the Figure 3A, there exists an intermediate 
step where the $(N_{c,i+1}+N_{c,i-1}-N_{c,i})$ D4-branes are 
connecting between the 
$NS5_{i+1}$-brane and the  $NS5_{i}'$-brane,  
$(N_{c,i+1}-N_{c,i+2})$ D4-branes connecting between the  
$NS5_{i}'$-brane and   
$NS5_{i+2}'$-brane, and $N_{c,i+2}$ D4-branes between the 
$NS5_{i}'$-brane and
the $NS5_{i+3}$-brane. By introducing $-N_{c,i+2}$ D4-branes and $-N_{c,i+2}$ 
anti-D4-branes  between the  $NS5_{i+1}$-brane and   
$NS5_{i}'$-brane, reconnecting the former with  
the $N_{c,i+1}$ D4-branes connecting between  
$NS5_{i+1}$-brane 
and the $NS5_{i}'$-brane (therefore $N_{c,i+1}-N_{c,i+2}$ D4-branes)
and moving those combined
$(N_{c,i+1}-N_{c,i+2})$ 
D4-branes
to $+v$-direction, 
one gets the final Figure 3A where we are left with 
$(N_{c,i}-N_{c,i+2}-N_{c,i-1})$ 
anti-D4-branes between the $NS5_{i+1}$-brane and   
$NS5_{i}'$-brane.

The dual gauge group from Seiberg dual relation is given by 
\bea
SU(N_{c,1}) \times \cdots \times SU(N_{c,i-1}) \times 
SU(\widetilde{N}_{c,i}) 
\times SU(N_{c,i+1}) \times \cdots \times SU(N_{c,n})
\label{dual}
\eea
with 
$\widetilde{N}_{c,i} \equiv N_{c,i+1}+N_{c,i-1}-N_{c,i}$
where the matter contents are   
the bifundamentals $f_i$ in 
 $({\bf 1_1, \cdots, 1, \Box_i, \overline{\Box}_{i+1}, 1, \cdots, 1_n})$,
$\cdots$, and $\widetilde{f}_i$ in the representation 
$({\bf 1_1, \cdots, 1, \overline{\Box}_i, \Box_{i+1}, 1,
\cdots, 1_n})$ in
addition to $(n-2)$ bifundamentals $F_j$ and $\widetilde{F}_j$, 
 $j=1,2, \cdots, i-1, i+1,
\cdots, n$ and
the gauge singlet $\Phi_{i+1}$
for the $i$-th dual gauge group in the 
adjoint representation for the $(i+1)$-th dual gauge group, 
i.e.,  
$
{(\bf 1_1, \cdots, 1_i, (N_{c,i+1}-N_{c,i+2})^2-1, 1_{i+2}, \cdots, 1_n)  
}
$ plus a singlet
under the 
dual gauge group (\ref{dual}) where the gauge group is broken from
$SU(N_{c,i+1})$ 
to $SU(N_{c,i+1}-N_{c,i+2})$.

When two NS5'-branes in Figure 3A are close to each other, then 
it leads to Figure 3B by realizing that the number of $(N_{c,i+1}-N_{c,i+2})$
D4-branes connecting between $NS5_{i+1}$-brane and $NS5_{i+2}'$-brane can
be rewritten as $(N_{c,i}-N_{c,i+2}-N_{c,i-1})$ plus $\widetilde{N}_{c,i}$.
If we ignore all the D4-branes and NS-branes located at the left hand
side of $NS_{i+1}$-brane and at the right hand side of $NS_i'$-brane  
from Figure 3, then
the brane configuration becomes the one in \cite{GK}.
The Figure 2 of \cite{Ahn07-6} is contained in the Figure 3. In
particular, the brane configuration from the $NS5_{i+1}$-brane to 
the $NS5_{i+3}$-brane is exactly same as the one of \cite{Ahn07-6}.

The cubic superpotential with the mass term (\ref{mass}) with (\ref{m}) in the dual
theory \footnote{In general, there are also 
the extra terms in the superpotential
$\Phi' f_{i-1} f_{i} + \Phi'' \widetilde{f}_{i-1} \widetilde{f}_i + \Phi_{i-1} f_{i-1} 
\widetilde{f}_{i-1}$ where we define 
$\Phi' \equiv F_i F_{i-1}$ and 
$\Phi'' \equiv \widetilde{F}_{i} \widetilde{F}_{i-1}$, coming from 
different bifundamentals. However, the F- term conditions,
$\Phi' f_i + \Phi_{i-1} \widetilde{f}_{i-1}=0=\Phi''
\widetilde{f}_i + \Phi_{i-1} f_{i-1}$ lead to 
$<\Phi'>=<\Phi''>=<f_{i-1}>=<\widetilde{f}_{i-1}>=0$. 
Therefore, these extra terms do not
contribute to the one loop computation up to quadratic order.} is given by
\bea
W_{dual} = \Phi_{i+1} f_{i} \widetilde{f}_{i}  + m_{i+1} \tr \Phi_{i+1}.
\label{W}
\eea
Here the magnetic fields $f_i$ and $\widetilde{f}_i$  
correspond to 4-4 strings connecting 
the $\widetilde{N}_{c,i}$-color D4-branes(that are 
connecting between the $NS5_{i+1}$-brane
and the $NS5_{i+2}'$-brane in Figure 3B) with $N_{c,i+1}$-flavor 
D4-branes(that are 
a combination of three different D4-branes in Figure 3B).
Among these $N_{c,i+1}$-flavor D4-branes, only the strings ending on
the upper $(N_{c,i+1}-N_{c,i}+N_{c,i-1})$ D4-branes and 
on the tilted middle $(N_{c,i}-N_{c,i+2}-N_{c,i-1})$ 
D4-branes in Figure 3B enter the cubic superpotential term. 
Although the $(N_{c,i+1}-N_{c,i+2})$ D4-branes in Figure 3A cannot move any
directions for fixed other branes,
the tilted $(N_{c,i}-N_{c,i+2}-N_{c,i-1})$-flavor D4-branes 
can move $w$ direction in Figure 3B.
The remaining upper $\widetilde{N}_{c,i}$ D4-branes are 
fixed also and cannot 
move any direction. Note that 
there is a decomposition 
\bea
N_{c,i+1}-N_{c,i+2}=(N_{c,i}-N_{c,i-1}-N_{c,i+2}) +\widetilde{N}_{c,i}.
\nonu
\eea

The brane configuration for zero mass for the bifundamental 
$F_i$ and $\widetilde{F}_i$,
which has only a cubic superpotential from (\ref{W}),
can be obtained from Figure 3A by moving
the upper $NS5_{i+2}'$-brane together with $(N_{c,i+1}-N_{c,i+2})$ 
color D4-branes 
into the origin $v=0$.
Then the number of dual colors for D4-branes 
becomes  
$\widetilde{N}_{c,i}$ between $NS5_{i+1}$-brane and $NS5_{i}'$-brane 
and
$N_{c,i+1}$ between two NS5'-branes
as well
as $N_{c,i-1}$ D4-branes between $NS5_{i-1}$-brane and $NS5_{i+1}$-brane.
Or starting from Figure 1A and moving the $NS5_{i+1}$-brane to 
the left all the
way past the $NS5_{i}'$-brane,
one also obtains the corresponding magnetic brane configuration
for massless case.

The brane configuration in Figure 3A is stable as long as the
distance $(\Delta x)_{i+1}$ between the upper NS5'-brane and 
the lower NS5'-brane is large, as
in \cite{GK}. If they are close to each other, then this brane
configuration is unstable to decay and leads to 
the brane configuration in Figure
3B.
One can regard these brane configurations as particular states in the
magnetic gauge theory with the gauge group (\ref{dual}) and
superpotential (\ref{W}).
The   $(N_{c,i+1}-N_{c,i+2}-\widetilde{N}_{c,i})$ flavor D4-branes of 
straight brane configuration
of
Figure 3B  bend due to the fact that there exists an attractive
gravitational interaction
between those flavor D4-branes and $NS5_{i+1}$-brane from the DBI action, by
following the procedure of \cite{GK}, as long as the distance $y_{i+2}$
corresponding to the $NS5_{i+3}$-brane 
goes to the infinity because the presence of an extra $NS5_{i+3}$-brane does
not affect the DBI action. 
For the finite and small $y_{i+2}$, the careful analysis for DBI action is
needed in
order to obtain the bending curve connecting  two NS5'-branes.  

When the upper NS5'-brane(or $NS5_{i+2}'$-brane) 
is replaced by coincident $(N_{c,i+1}-N_{c,i+2})$ 
D6-branes and 
the $NS5_{i+3}$ is rotated by an angle $\frac{\pi}{2}$ in the $(v,w)$
plane in Figure 3B, this brane configuration reduces to the one 
found in \cite{Ahn07-8} where the gauge group was given by 
$ \cdots \times 
SU(n_{c,i-1}) \times SU(n_{f,i}+n_{c,i+1}+n_{c,i-1}-n_{c,i}) \times
SU(n_{c,i+1}) 
\times \cdots $ 
with $n_{f,i}$ multiplets,  $\widetilde{n}_{f,i}$ multiplets, 
bifundamentals, and gauge 
singlets. 
In particular, the Figure 3B of \cite{Ahn07-8} with vanishing flavors
$Q$ and $Q''$
is contained in
this modified Figure 3B running from the $NS5_{i-1}$-brane to 
the $NS5_{i+3}$-brane. 
Then the present number $(N_{c,i+1}-N_{c,i+2})$ corresponds to the $n_{f,i}$, the
number $N_{c,i}$ corresponds to $n_{c,i}$,
 the
number $N_{c,i-1}$ corresponds to $n_{c,i-1}$,
 and 
the number $N_{c,i+2}$ corresponds to the $n_{c,i+1}$.
Note that the number of D4-branes touching $NS5_{i+2}'$-brane in Figure 3B
is equal to $(N_{c,i+1}-N_{c,i+2})$.

The quantum corrections can be understood for small $(\Delta x)_{i+1}$ by 
using the low energy field theory on the branes.
The low energy dynamics of the magnetic brane configuration 
can be described by the ${\cal N}=1$ supersymmetric gauge theory
with gauge group (\ref{dual})
and the gauge couplings for the three gauge group factors are
given by
\bea
g_{i-1,mag}^2  = \frac{g_s \ell_s}{(y_{i}+y_{i-1})}, \qquad 
g_{i,mag}^2 = \frac{g_s \ell_s}{y_{i}}, \qquad
g_{i+1,mag}^2  = \frac{g_s \ell_s}{(y_{i}+y_{i+1})}.
\nonu
\eea
The dual gauge theory has  a gauge singlet $\Phi_{i+1}$ and 
bifundamentals $f_i, \widetilde{f}_i, F_j$, and $\widetilde{F}_j$ 
under the dual gauge
group (\ref{dual}) and the superpotential 
corresponding to Figures 3A and 3B is given by 
\bea
W_{dual} = h \Phi_{i+1} f_i \widetilde{f}_i - h \mu_{i+1}^2 \tr  \Phi_{i+1}, 
\qquad h^2 = g_{i+1,
  mag}^2,
\qquad \mu_{i+1}^2 = -\frac{(\Delta x)_{i+1}}{ 2\pi g_s \ell_s^3}.
\nonu
\eea
Then $ f_i \widetilde{f}_i$ is a $\widetilde{N}_{c,i} \times \widetilde{N}_{c,i}$ 
matrix where the $(i+1)$-th gauge group indices for $f_i$ and $\widetilde{f}_i$ 
are contracted with those
of $\Phi_{i+1}$ while $\Phi_{i+1}$ is a 
$(N_{c,i+1}-N_{c,i+2}) \times (N_{c,i+1}-N_{c,i+2})$ matrix.
Although the field $f_i$ itself is an antifundamental in the $(i+1)$-th gauge
group
which is a different  
representation for the usual standard quark
coming from D6-branes,
the product $f_i \widetilde{f}_i$ has the same representation for the 
product of quarks
and moreover, 
the $(i+1)$-th gauge group indices for the field $\Phi_{i+1}$ play the
role of the flavor indices, as in comparison with the brane
configuration in the presence of D6-branes before.

Therefore, the F-term equation, the derivative $W_{dual}$ with respect to the
meson field $\Phi_{i+1}$ cannot be satisfied if the $(N_{c,i+1}-N_{c,i+2})$ exceeds
$\widetilde{N}_{c,i}$.
So the supersymmetry is broken.   
That is, 
there exist three equations from F-term conditions:
$
f_i^a \widetilde{f}_{i,b} -\mu_{i+1}^2 \delta^a_b =0$, and 
$\Phi_{i+1} f_i =0=\widetilde{f}_i \Phi_{i+1}$.
Then the solutions for these
are given by 
\bea
<f_i>   & = & 
\left(
\begin{array}{c}
\mu_{i+1}  {\bf 1}_{\widetilde{N}_{c,i}}  \\
0
\end{array}
\right), 
\qquad
<\widetilde{f}_i>   = 
\left(
\begin{array}{cc}
\mu_{i+1}  {\bf 1}_{\widetilde{N}_{c,i}} & 0  \\
\end{array}
\right), \nonu \\ 
<\Phi_{i+1}> & = &
 \left(
\begin{array}{cc}
0  & 0  \\
0 & M_{i+1}  {\bf 1}_{(N_{c,i+1}-N_{c,i+2}-\widetilde{N}_{c,i})} 
\end{array}
\right) 
\label{point}
\eea
where the zero of $<f_i>$ is a $
(N_{c,i+1}-N_{c,i+2}-\widetilde{N}_{c,i}) \times \widetilde{N}_{c,i}$ 
matrix, the zero of $<\widetilde{f}_i>$ is a
$\widetilde{N}_{c,i} \times (N_{c,i+1}-N_{c,i+2}-\widetilde{N}_{c,i})
$
 matrix and 
the zeros of $<\Phi_{i+1}>$ are $\widetilde{N}_{c,i} \times \widetilde{N}_{c,i}$,
$\widetilde{N}_{c,i} \times (N_{c,i+1}-N_{c,i+2}-\widetilde{N}_{c,i})$, 
and $(N_{c,i+1}-N_{c,i+2}-\widetilde{N}_{c,i}) \times
\widetilde{N}_{c,i}$ matrices.
Then one can expand these fields around on a point (\ref{point}) 
and arrives at the relevant superpotential
up to quadratic order in the fluctuation. 
At one loop, the effective potential $V_{eff}^{(1)}$ for $M_{i+1}$
leads to the positive value for $m_{M_{i+1}}^2$ implying that these
vacua are stable.

%%%%%%%%%%%%%%%%%%%%%%%%%%%%%%%%%%%%%%%%%%%%%%%%%%%%%%%%%%%%%%
%%%%%%%%%%%%%%%%%%%%%%%%%%%%%%%%%%%%%%%%%%%%%%%%%%%%%%%%%%%%%%
\subsubsection{When the dual gauge group occurs at even chain}
%%%%%%%%%%%%%%%%%%%%%%%%%%%%%%%%%%%%%%%%%%%%%%%%%%%%%%%%%%%%%%
%%%%%%%%%%%%%%%%%%%%%%%%%%%%%%%%%%%%%%%%%%%%%%%%%%%%%%%%%%%%%%

Let us discuss the other case for
the Seiberg dual of the middle gauge group
factor.
Starting from Figure 1B, moving the $NS5_{i-1}'$-brane 
with $(N_{c,i-1}-N_{c,i-2})$
D4-branes 
to the $+v$ direction leading to Figure 2B, 
and interchanging the $NS5_{i}$-brane and the $NS5_{i+1}'$-brane,
one obtains the Figure 4A.

%%%%%%%%%%%%%%%%%%%%%%%%%%%%%%%%%%%%%%%%%%%%%%%%%%%%%%%%%%%%%%%%%%%%%%
%%%%%%%%%%%%%%%%%%%%%%%%%%%%%%%%%%%%%%%%%%%%%%%%%%%%%%%%%%%%%%%%%%%%%%
\begin{figure}[ht]
   \epsfxsize=4.0in 
\centerline{\epsffile{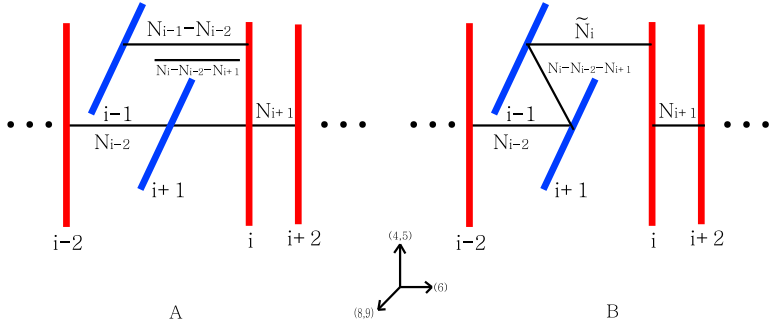}}
   \caption[FIG. \arabic{figure}.]{ 
The 
 ${\cal N}=1$ magnetic brane configuration for the gauge group 
containing $SU(\widetilde{N}_{c,i}=N_{c,i-1}+N_{c,i+1}-N_{c,i})$
where $i$ is even, 
corresponding to Figure 2B with D4-
and $\overline{D4}$-branes(4A) and with 
a misalignment between D4-branes(4B) when the two NS5'-branes are close to
each other. 
The number of tilted D4-branes in 4B can be written as
$N_{c,i}-N_{c,i+1}-N_{c,i-2}
 =(N_{c,i-1}-N_{c,i-2})-\widetilde{N}_{c,i}$.
%The $x$ coordinate of $NS5_{i-1}'$-brane is given by $(\Delta
% x)_{i-1}$.
}
\end{figure}
%%%%%%%%%%%%%%%%%%%%%%%%%%%%%%%%%%%%%%%%%%%%%%%%%%%%%%%%%%%%%%%%%%%%%
%Figure 4A and 4B 
%%%%%%%%%%%%%%%%%%%%%%%%%%%%%%%%%%%%%%%%%%%%%%%%%%%%%%%%%%%%%%%%%%%%%%

Before arriving at the Figure 4A, there exists an intermediate 
step where the $(N_{c,i+1}+N_{c,i-1}-N_{c,i})$ D4-branes are 
connecting between the 
$NS5_{i+1}'$-brane and the  $NS5_{i}$-brane,  
$(N_{c,i-1}-N_{c,i-2})$ D4-branes connecting between the  $NS5_{i-1}'$-brane and   
$NS5_{i+1}'$-brane, and $N_{c,i-2}$ D4-branes between the $NS5_{i-2}$-brane and
the $NS5_{i+1}'$-brane. By introducing $-N_{c,i-2}$ D4-branes and $-N_{c,i-2}$ 
anti-D4-branes  between the  $NS5_{i+1}'$-brane and   
$NS5_{i}$-brane, reconnecting the former with  
the $N_{c,i-1}$ D4-branes connecting between  
$NS5_{i+1}'$-brane 
and the $NS5_{i}$-brane (therefore $(N_{c,i-1}-N_{c,i-2})$ D4-branes)
and moving those combined
$(N_{c,i-1}-N_{c,i-2})$ 
D4-branes
to $+v$-direction, 
one gets the final Figure 4A where we are left with 
$(N_{c,i}-N_{c,i-2}-N_{c,i+1})$ 
anti-D4-branes between the $NS5_{i+1}'$-brane and   
$NS5_{i}$-brane.

The dual gauge group which is the same as previous expression 
is given by 
\bea
\cdots \times SU(N_{c,i-1}) \times 
SU(\widetilde{N}_{c,i} \equiv N_{c,i+1}+N_{c,i-1}-N_{c,i}) 
\times SU(N_{c,i+1}) \times \cdots
\label{dual1}
\eea
where the matter contents are   
the bifundamentals $f_{i-1}$ in 
 $({\bf 1_1, \cdots, 1, \Box_{i-1}, \overline{\Box}_{i}, 1, \cdots, 1_n})$,
$\cdots$, and $\widetilde{f}_{i-1}$ in the representation 
$({\bf 1_1, \cdots, 1, \overline{\Box}_{i-1}, \Box_{i}, 1,
\cdots, 1_n})$ in
addition to $(n-2)$ bifundamentals $F_j$ and $\widetilde{F}_j$ with  
 $j=1,2, \cdots, i-2, i,
\cdots, (n-1)$ and
the gauge singlet $\Phi_{i-1}$
for the $i$-th dual gauge group which was in the 
adjoint representation for the $(i-1)$-th dual gauge group 
is in the representation
$
{(\bf 1_1, \cdots, 1_{i-2}, (N_{c,i-1}-N_{c,i-2})^2-1, 1_{i}, \cdots, 1_n)  
}
$ plus a singlet
under the 
dual gauge group (\ref{dual1}) where the gauge group is broken from
$SU(N_{c,i-1})$ 
to $SU(N_{c,i-1}-N_{c,i-2})$.

When two NS5'-branes in Figure 4A are close to each other, then 
it leads to Figure 4B by realizing that the number of $(N_{c,i-1}-N_{c,i-2})$
D4-branes connecting between $NS5_{i-1}'$-brane and $NS5_{i}$-brane can
be rewritten as $(N_{c,i}-N_{c,i-2}-N_{c,i+1})$ plus $\widetilde{N}_{c,i}$.
If we ignore all the D4-branes and NS-branes located at the left hand
side of $NS5_{i+1}'$-brane and at the right hand side of $NS5_i$-brane  
from Figure 4, then
the brane configuration becomes the one in \cite{GK}.
The Figure 4 of \cite{Ahn07-6} is contained in the Figure 4. In
particular, the brane configuration from the $NS5_{i-1}'$-brane to 
the $NS5_{i+2}$-brane is exactly same as the one of \cite{Ahn07-6}.

The cubic superpotential with the mass term (\ref{masseven}) with
(\ref{m1even}) 
in the dual
theory 
\footnote{In general, there are also 
the extra terms in the superpotential
$\Phi_{i+1} f_{i} \widetilde{f}_{i} + 
\Phi'' \widetilde{f}_{i-1} \widetilde{f}_i + \Phi' f_{i-1} 
f_{i}$ where we define 
$\Phi' \equiv F_i F_{i-1}$ and 
$\Phi'' \equiv \widetilde{F}_{i} \widetilde{F}_{i-1}$, coming from 
different bifundamentals. However, the F- term conditions,
$\Phi_{i+1} \widetilde{f}_i + \Phi' f_{i-1}=0=\Phi_{i+1}
f_i + \Phi'' \widetilde{f}_{i-1}$ lead to 
$<\Phi'>=<\Phi''>=<f_i>=<\widetilde{f}_i>=0$. 
In this case also these extra terms do not
contribute to the one loop computation up to quadratic order.}
is given by
\bea
W_{dual} = \Phi_{i-1} f_{i-1} \widetilde{f}_{i-1}  + m_{i-1} \tr \Phi_{i-1}.
\label{Wdual1}
\eea
Here the magnetic fields $f_{i-1}$ and $\widetilde{f}_{i-1}$  
correspond to 4-4 strings connecting 
the $\widetilde{N}_{c,i}$-color D4-branes(that are 
connecting between the $NS5_{i-1}'$-brane
and the $NS5_{i}$-brane in Figure 4B) with $N_{c,i-1}$-flavor 
D4-branes(that are 
a combination of three different D4-branes in Figure 4B).
Among these $N_{c,i-1}$-flavor D4-branes, only the strings ending on
the upper $(N_{c,i+1}-N_{c,i}+N_{c,i-1})$ D4-branes and 
on the tilted middle $(N_{c,i}-N_{c,i-2}-N_{c,i+1})$ 
D4-branes in Figure 4B enter the cubic superpotential term. 
Although the $(N_{c,i-1}-N_{c,i-2})$ D4-branes in Figure 4A for fixed
other branes cannot move any
directions,
the tilted $(N_{c,i}-N_{c,i-2}-N_{c,i+1})$-flavor D4-branes 
can move $w$ direction in Figure 4B.
The remaining upper $\widetilde{N}_{c,i}$ D4-branes are fixed also and cannot 
move any direction. Note that 
there is a decomposition 
\bea
N_{c,i-1}-N_{c,i-2}=(N_{c,i}-N_{c,i+1}-N_{c,i-2}) +\widetilde{N}_{c,i}.
\nonu
\eea

The brane configuration for zero mass for the bifundamental $F_{i-1}$
and 
$\widetilde{F}_{i-1}$,
which has only a cubic superpotential (\ref{Wdual1}),
can be obtained from Figure 4A by moving
the upper $NS5_{i-1}'$-brane together with $(N_{c,i-1}-N_{c,i-2})$ color D4-branes 
into the origin $v=0$.
Then the number of dual colors for D4-branes 
becomes  
$\widetilde{N}_{c,i}$ between $NS5_{i+1}'$-brane and $NS5_{i}$-brane 
and
$N_{c,i-1}$ between two NS5'-branes
as well
as $N_{c,i+1}$ D4-branes between $NS5_{i}$-brane and $NS5_{i+2}$-brane.
Or starting from Figure 1B and moving the $NS5_{i+1}'$-brane to the left all the
way past the $NS5_{i}$-brane,
one also obtains the corresponding magnetic brane configuration
for massless case.

The brane configuration in Figure 4A is stable as long as the
distance $(\Delta x)_{i-1}$ between the upper NS5'-brane and 
the lower NS5'-brane is large, as
in \cite{GK}. If they are close to each other, then this brane
configuration is unstable to decay and leads to 
the brane configuration in Figure
4B.
One can regard these brane configurations as particular states in the
magnetic gauge theory with the gauge group (\ref{dual1}) and
superpotential (\ref{Wdual1}).
The   $(N_{c,i}-N_{c,i+1}-N_{c,i-2})$ flavor D4-branes of 
straight brane configuration
of
Figure 4B  bend due to the fact that there exists an attractive
gravitational interaction
between those flavor D4-branes and $NS5_{i}$-brane from the DBI action, by
following the procedure of \cite{GK}, as long as the distance
$y_{i+1}$
corresponding to the $NS5_{i+2}$-brane
goes to $\infty$ because the presence of an extra $NS5_{i+2}$-brane does
not affect the DBI action. 
For the finite and small $y_{i+1}$, the careful analysis for DBI action is
needed in
order to obtain the bending curve connecting  two NS5'-branes.  

When the upper NS5'-brane(or $NS5_{i-1}'$-brane) 
is replaced by coincident $(N_{c,i-1}-N_{c,i-2})$ 
D6-branes in Figure 4B, this brane configuration reduces to the one 
found in \cite{Ahn07-8} where the gauge group was given by 
$ \cdots \times SU(n_{c,i-1}) \times SU(n_{f,i}+n_{c,i+1}+n_{c,i-1}-n_{c,i}) 
\times SU(n_{c,i+1}) \times \cdots $ 
with $n_{f,i}$ multiplets,  $\widetilde{n}_{f,i}$ multiplets, 
bifundamentals and gauge 
singlets. 
Then the present number $(N_{c,i-1}-N_{c,i-2})$ corresponds to the $n_{f,i}$, the
number $N_{c,i}$ corresponds to $n_{c,i}$,
 the
number $N_{c,i+1}$ corresponds to $n_{c,i+1}$ 
and 
the number $N_{c,i-2}$ corresponds to the $n_{c,i-1}$.
Note that the number of D4-branes touching $NS5_{i-1}'$-brane in Figure 4B
is equal to $(N_{c,i-1}-N_{c,i-2})$.
In particular, the Figure 5B of \cite{Ahn07-8} with vanishing flavors
$Q'$ and $Q''$
is contained in
this modified Figure 4B running from the $NS5_{i-1}'$-brane to 
the $NS5_{i+3}'$-brane. 

The quantum corrections can be understood for small $(\Delta x)_{i-1}$ by 
using the low energy field theory on the branes.
The low energy dynamics of the magnetic brane configuration 
can be described by the ${\cal N}=1$ supersymmetric gauge theory
with gauge group (\ref{dual1})
and the gauge couplings for the three gauge group factors are
given by
\bea
g_{i-1,mag}^2  = \frac{g_s \ell_s}{(y_{i}+y_{i-1})}, \qquad 
g_{i,mag}^2 = \frac{g_s \ell_s}{y_{i}}, \qquad
g_{i+1,mag}^2  = \frac{g_s \ell_s}{(y_{i}+y_{i+1})}.
\nonu
\eea
The dual gauge theory has  a meson field $\Phi_{i-1}$  and 
bifundamentals $f_{i-1}, \widetilde{f}_{i-1}, F_j$, and
$\widetilde{F}_j$  
and the superpotential 
corresponding to Figures 4A and 4B is given by 
\bea
W_{dual} = h \Phi_{i-1} f_{i-1} \widetilde{f}_{i-1} - h \mu_{i-1}^2
\tr \Phi_{i-1}, 
\qquad h^2 = g_{i-1,
  mag}^2,
\qquad \mu_{i-1}^2 = -\frac{(\Delta x)_{i-1}}{ 2\pi g_s \ell_s^3}.
\nonu
\eea
Then $ f_{i-1} \widetilde{f}_{i-1}$ is a $\widetilde{N}_{c,i} \times \widetilde{N}_{c,i}$ 
matrix where the $(i-1)$-th gauge group indices for $f_{i-1}$ and $\widetilde{f}_{i-1}$ 
are contracted with those
of $\Phi_{i-1}$ while the $\Phi_{i-1}$ is a 
$(N_{c,i-1}-N_{c,i-2}) \times (N_{c,i-1}-N_{c,i-2})$ matrix.
Although the field $f_{i-1}$ itself is an antifundamental in the $i$-th gauge
group
which is a different  
representation for the usual standard quark
coming from D6-branes,
the product $f_{i-1} \widetilde{f}_{i-1}$ has the same representation for the 
product of quarks
and moreover, 
the $(i-1)$-th gauge group indices for the field $\Phi_{i-1}$ play the
role of the flavor indices, as in comparison with the brane
configuration in the presence of D6-branes before.

Therefore, the F-term equation, the derivative $W_{dual}$ with respect to the
meson field $\Phi_{i-1}$ cannot be satisfied if the $(N_{c,i-1}-N_{c,i-2})$ exceeds
$\widetilde{N}_{c,i}$.
So the supersymmetry is broken.   
That is, 
there exist three equations from F-term conditions:
$
f_{i-1}^a \widetilde{f}_{i-1,b} -\mu_{i-1}^2 \delta^a_b =0$, and 
$\Phi_{i-1} f_{i-1} =0=\widetilde{f}_{i-1} \Phi_{i-1}$.
Then the solutions for these
are given by 
\bea
<f_{i-1}>   & = & 
\left(
\begin{array}{c}
\mu_{i-1}  {\bf 1}_{\widetilde{N}_{c,i}}  \\
0
\end{array}
\right), 
\qquad
<\widetilde{f}_{i-1}>   = 
\left(
\begin{array}{cc}
\mu_{i-1}  {\bf 1}_{\widetilde{N}_{c,i}} & 0  \\
\end{array}
\right), \nonu \\ 
<\Phi_{i-1}>  & = &
 \left(
\begin{array}{cc}
0  & 0  \\
0 & M_{i-1}  {\bf 1}_{(N_{c,i-1}-N_{c,i-2}-\widetilde{N}_{c,i})} 
\end{array}
\right) 
\label{point1}
\eea
where the zero of $<f_{i-1}>$ is a $
(N_{c,i-1}-N_{c,i-2}-\widetilde{N}_{c,i}) \times \widetilde{N}_{c,i}$ 
matrix, the zero of $<\widetilde{f}_{i-1}>$ is a
$\widetilde{N}_{c,i} \times (N_{c,i-1}-N_{c,i-2}-\widetilde{N}_{c,i}) $ matrix and 
the zeros of $<\Phi_{i-1}>$ are $\widetilde{N}_{c,i} \times \widetilde{N}_{c,i}$,
$\widetilde{N}_{c,i} \times (N_{c,i-1}-N_{c,i-2}-\widetilde{N}_{c,i})$, 
and $(N_{c,i-1}-N_{c,i-2}-\widetilde{N}_{c,i}) \times
\widetilde{N}_{c,i}$ matrices.
Then one can expand these fields around on a point (\ref{point1}) and 
arrives at the relevant superpotential
up to quadratic order in the fluctuation. 
At one loop, the effective potential $V_{eff}^{(1)}$ for $M_{i-1}$
leads to the positive value for $m_{M_{i-1}}^2$ implying that these
vacua are stable.

%%%%%%%%%%%%%%%%%%%%%%%%%%%%%%%%%%%%%%%%%%%%%%%%%%%%%%%%%%%%%%%%%%%%%%
%%%%%%%%%%%%%%%%%%%%%%%%%%%%%%%%%%%%%%%%%%%%%%%%%%%%%%%%%%%%%%%%%%%%%%%
\subsection{${\cal N}=1$ 
$SU(N_{c,1}) \times \cdots \times SU(\widetilde{N}_{c,n})$ magnetic theory}
%%%%%%%%%%%%%%%%%%%%%%%%%%%%%%%%%%%%%%%%%%%%%%%%%%%%%%%%%%%%%%%%%%%%%%%
%%%%%%%%%%%%%%%%%%%%%%%%%%%%%%%%%%%%%%%%%%%%%%%%%%%%%%%%%%%%%%%%%%%%%%%

Let us consider the Seiberg dual for the last gauge group
factor.
There are two magnetic duals depending on whether the gauge group factor
occurs at odd chain or even chain.

%%%%%%%%%%%%%%%%%%%%%%%%%%%%%%%%%%%%%%%%%%%%%%%%%%%%%%%%%%%%%
%%%%%%%%%%%%%%%%%%%%%%%%%%%%%%%%%%%%%%%%%%%%%%%%%%%%%%%%%%%%%
\subsubsection{When the dual gauge group occurs at odd chain} 
%%%%%%%%%%%%%%%%%%%%%%%%%%%%%%%%%%%%%%%%%%%%%%%%%%%%%%%%%%%%%
%%%%%%%%%%%%%%%%%%%%%%%%%%%%%%%%%%%%%%%%%%%%%%%%%%%%%%%%%%%%%

Let us consider other magnetic theory for the same electric theory in
previous section.
By applying the Seiberg dual to the $SU(N_{c,n})$ factor  and 
interchanging the $NS5_n'$-brane and the $NS5_{n+1}$-brane,
one obtains the Figure 5A.
Before arriving at the Figure 5A, there exists an intermediate 
step where 
$N_{c,n-1}$ D4-branes between
$NS5_{n-1}'$-brane and the $NS5_{n+1}$-brane,
the $N_{c,n-2}$ D4-branes are connecting between the 
$NS5_{n-2}'$-brane and the  $NS5_{n-1}'$-brane, and  
$(N_{c,n-1}-N_{c,n})$ D4-branes are connecting between the  $NS5_{n+1}$-brane and   
$NS5_n'$-brane. 
By rotating $NS5_{n-1}$-brane by an angle $\frac{\pi}{2}$ which will
become $NS5_{n-1}'$-brane, 
moving it with the $(N_{c,n-1}-N_{c, n-2})$ D4-branes 
to $+v$ direction where we introduce $(N_{c,n}-N_{c,n-2})$ D4-branes and
$(N_{c,n}-N_{c,n-2})$ anti D4-branes between the $NS5_{n+1}$-brane and the 
$NS5_n'$-brane, 
one gets the final Figure 5A where we are left with 
$(N_{c,n}-N_{c,n-2})$ 
anti-D4-branes between the $NS5_{n+1}$-brane and   
the $NS5_n'$-brane.

%%%%%%%%%%%%%%%%%%%%%%%%%%%%%%%%%%%%%%%%%%%%%%%%%%%%%%%%%%%%%%%%%%%%%%
%%%%%%%%%%%%%%%%%%%%%%%%%%%%%%%%%%%%%%%%%%%%%%%%%%%%%%%%%%%%%%%%%%%%%%
\begin{figure}[ht]
   \epsfxsize=4.0in 
\centerline{\epsffile{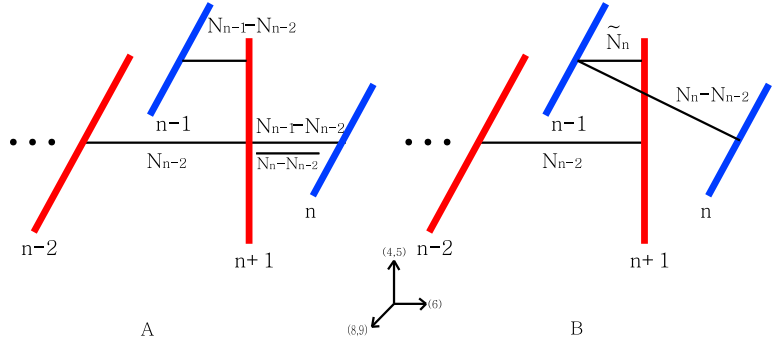}}
   \caption[FIG. \arabic{figure}.]{ 
The 
 ${\cal N}=1$ magnetic brane configuration for the gauge group 
containing $SU(\widetilde{N}_{c,n}=N_{c,n-1}-N_{c,n})$, where $n$ is
 odd,  
with D4-
and $\overline{D4}$-branes(5A) and with 
a misalignment between D4-branes(5B) when the two NS5'-branes are close to
each other. 
The number of tilted D4-branes in 5B can be written as
$N_{c,n}-N_{c,n-2} =(N_{c,n-1}-N_{c,n-2})-\widetilde{N}_{c,n}$.
%The $x$ coordinate of $NS5_{n-1}'$-brane is given by $(\Delta
% x)_{n-1}$.
}
\end{figure}
%%%%%%%%%%%%%%%%%%%%%%%%%%%%%%%%%%%%%%%%%%%%%%%%%%%%%%%%%%%%%%%%%%%%%
%Figure 5A and 5B
%%%%%%%%%%%%%%%%%%%%%%%%%%%%%%%%%%%%%%%%%%%%%%%%%%%%%%%%%%%%%%%%%%%%%%

The dual gauge group is given by
\bea
SU(N_{c,1}) \times \cdots \times 
SU(N_{c,n-1}) \times SU(\widetilde{N}_{c,n} \equiv N_{c,n-1}-N_{c,n})
\label{evendual}
\eea
and the matter contents are the field $f_{n-1}$ 
 charged under
$( {\bf 1, \cdots, 1_{n-2}, \Box_{n-1}, \overline{\Box}_{n} })$ 
 and their conjugates 
$\widetilde{f}_{n-1}$ 
$( {\bf 1, \cdots, 1_{n-2}, \overline{\Box}_{n-1}, \Box_{n} })$ 
under the dual gauge group (\ref{evendual})
and  
the gauge-singlet $\Phi_{n-1}$ which is in the 
adjoint representation for the $(n-1)$-th gauge group, 
in other words,   
$ ({ \bf   1_1, 
\cdots, 1_{n-2},  (N_{c,n-1}-N_{c,n-2})^2-1,1_n})  \oplus  ({\bf 1_1,
 \cdots, 1_n})$ under the
dual gauge group where the gauge group is broken from
$SU(N_{c,n-1})$ 
to $SU(N_{c,n-1}-N_{c,n-2})$.
Then the $\Phi_{n-1}$ is a $(N_{c,n-1}-N_{c,n-2}) \times
 (N_{c,n-1}-N_{c,n-2})$ 
matrix.
Only $(N_{c,n-1}-N_{c,n-2})$ D4-branes can participate in the mass deformation.

When two NS5'-branes in Figure 5A are close to each other, then 
it leads to Figure 5B
 by realizing that the number of $(N_{c,n-1}-N_{c,n-2})$
D4-branes connecting between $NS5_{n-1}'$-brane and $NS5_{n+1}$-brane can
be rewritten as $(N_{c,n}-N_{c,n-2})$ plus $\widetilde{N}_{c,n}$.
The Figure 5 of \cite{Ahn07-6} is contained in the Figure 5. In
particular, the brane configuration from the $NS5_{n-2}'$-brane to 
the $NS5_n'$-brane is exactly same as the one of \cite{Ahn07-6}.

The cubic superpotential with the mass term
is given by
\bea
W_{dual} = \Phi_{n-1} f_{n-1} \widetilde{f}_{n-1} + m_{n-1} \tr \Phi_{n-1}
\label{ssuper1}
\eea
where we define $\Phi_{n-1}$ as $\Phi_{n-1} \equiv F_{n-1} \widetilde{F}_{n-1}$ and 
the $n$-th gauge group indices in $F_{n-1}$ and $\widetilde{F}_{n-1}$ 
are contracted, each $(n-1)$-th gauge group index in them is encoded in 
$\Phi_{n-1}$. 
Here the magnetic fields $f_{n-1}$ and $\widetilde{f}_{n-1}$  
correspond to 4-4 strings connecting 
the $\widetilde{N}_{c,n}$-color D4-branes(that are 
connecting between the $NS5_{n-1}'$-brane
and the $NS5_{n+1}$-brane in Figure 5B) with $N_{c,n-1}$-flavor 
D4-branes.
Among these $N_{c,n-1}$-flavor D4-branes, only the strings ending on
the upper $(N_{c,n-1}-N_{c,n})$ D4-branes and 
on the tilted $(N_{c,n}-N_{c,n-2})$ 
D4-branes in Figure 5B enter the cubic superpotential term. 
Although the $(N_{c,n-1}-N_{c,n-2})$ D4-branes in Figure 5A cannot move any
directions for fixed other branes,
the tilted $(N_{c,n}-N_{c,n-2})$-flavor D4-branes can move $w$ direction.
The remaining upper $\widetilde{N}_{c,n}$ D4-branes are fixed also and cannot 
move any direction. 
Note that 
there is a decomposition 
\bea
(N_{c,n-1}-N_{c,n-2})=(N_{c,n}-N_{c,n-2})+\widetilde{N}_{c,n}.
\nonu
\eea 

The brane configuration in Figure 5A is stable as long as the
distance $(\Delta x)_{n-1}$ between the upper NS5'-brane and 
the lower NS5'-brane(or $NS5_n'$-brane) 
is large. If they are close to each other, then this brane
configuration is unstable to decay to 
the brane configuration in Figure
5B.
One can regard these brane configurations as particular states in the
magnetic gauge theory with the gauge group (\ref{evendual}) and
superpotential (\ref{ssuper1}).
The   $(N_{c,n-1}-N_{c,n-2}-\widetilde{N}_{c,n})$ flavor D4-branes of 
straight brane configuration
of
Figure 5B  bend since there exists an attractive
gravitational interaction
between those flavor D4-branes and NS5-brane from the DBI action. 
As mentioned in \cite{Ahn07-5},
the two NS5'-branes are located at different side of $NS5_{n+1}$-brane in
Figure 5B and the DBI action computation for this bending curve
should be taken into account. 

The brane configuration for zero mass for the bifundamental,
which has only a cubic superpotential (\ref{ssuper1}),
can be obtained from Figure 5A by moving
the upper  NS5'-brane together with $(N_{c,n-1}-N_{c,n-2})$ color D4-branes 
into the origin $v=0$.
Then the number of dual colors for D4-branes 
becomes $N_{c,n-1}$ between the $NS5_{n-1}'$-brane and the $NS5_{n+1}$-brane, 
$N_{c,n-2}$ between the $NS5_{n-2}'$-brane and the  $NS5_{n-1}'$-brane
and $\widetilde{N}_{c,n}$ 
between $NS5_{n+1}$-brane and $NS5_n'$-brane.
Or starting from Figure 1B and moving the 
$NS5_n'$-brane to the right all the
way past the $NS5_{n+1}'$-brane,
one also obtains the corresponding magnetic brane configuration
for massless case.

The low energy dynamics of the magnetic brane configuration 
can be described by the ${\cal N}=1$ supersymmetric gauge theory
with gauge group
and the gauge couplings for the three gauge group factors are
given by
\bea
g_{n-2,mag}^2  = \frac{g_s \ell_s}{y_{n-2}}, \qquad 
g_{n-1,mag}^2 = \frac{g_s \ell_s}{(y_n+y_{n-1})}, \qquad
g_{n,mag}^2  = \frac{g_s \ell_s}{y_{n}}.
\nonu
\eea
The dual gauge theory has  a meson field $\Phi_{n-1}$  and 
bifundamentals $f_{n-1}, \widetilde{f}_{n-1}, F_j$ and
$\widetilde{F}_j$ 
under the dual gauge
group (\ref{evendual}) and the superpotential (\ref{ssuper1}) 
corresponding to Figures 5A and 5B is given by 
\bea
W_{dual} = h \Phi_{n-1} f_{n-1} \widetilde{f}_{n-1} - 
h \mu_{n-1}^2 \tr \Phi_{n-1}, \qquad h^2 = g_{n-1,
  mag}^2,
\qquad \mu_{n-1}^2 = -\frac{(\Delta x)_{n-1}}{ 2\pi g_s \ell_s^3}.
\nonu
\eea
Then $ f_{n-1} \widetilde{f}_{n-1}$ is a $\widetilde{N}_{c,n} \times 
\widetilde{N}_{c,n}$ 
matrix where the $(n-1)$-th gauge group indices for $f_{n-1}$ and 
$\widetilde{f}_{n-1}$ 
are contracted with those
of $\Phi_{n-1}$ while the $\Phi_{n-1}$ is a 
$(N_{c,n-1}-N_{c,n-2}) \times (N_{c,n-1}-N_{c,n-2})$ matrix.
The product $f_{n-1} \widetilde{f}_{n-1}$ has the same representation for the 
product of quarks
and moreover, 
the $(n-1)$-th gauge group indices for the field $\Phi_{n-1}$ play the
role of the flavor indices.

Therefore, the F-term equation, the derivative $W_{dual}$ with respect to the
meson field $\Phi_{n-1}$ cannot be satisfied if the $(N_{c,n-1}-N_{c,n-2})$ exceeds
$\widetilde{N}_{c,n}$.
So the supersymmetry is broken.   
That is, 
there exist three equations from F-term conditions:
$
f_{n-1}^a \widetilde{f}_{n-1,b} -
\mu_{n-1}^2 \delta^a_b =0$ and $ \Phi_{n-1} f_{n-1} =0=
\widetilde{f}_{n-1} \Phi_{n-1}$.
Then the solutions for these
are given by 
\bea
<f_{n-1}>   & = & 
\left(
\begin{array}{c}
\mu_{n-1}  {\bf 1}_{\widetilde{N}_{c,n}}  \\
0
\end{array}
\right), \qquad
<\widetilde{f}_{n-1}>   = 
\left(
\begin{array}{cc}
\mu_{n-1}  {\bf 1}_{\widetilde{N}_{c,n}} & 0  \\
\end{array}
\right), \nonu \\
<\Phi_{n-1}>  & = &
 \left(
\begin{array}{cc}
0  & 0  \\
0 & M_{n-1}  {\bf 1}_{(N_{c,n-1}-N_{c,n-2})-\widetilde{N}_{c,n}} 
\end{array}
\right) 
\label{sol}
\eea
where the zero of $<f_{n-1}>$ is a $
(N_{c,n-1}-N_{c,n-2}-\widetilde{N}_{c,n}) \times \widetilde{N}_{c,n}$ 
matrix, the zero of $<\widetilde{f}_{n-1}>$ is a
$\widetilde{N}_{c,n} \times (N_{c,n-1}-N_{c,n-2}-\widetilde{N}_{c,n}) $ matrix and 
the zeros of $<\Phi_{n-1}>$ are $\widetilde{N}_{c,n} \times \widetilde{N}_{c,n}$,
$\widetilde{N}_{c,n} \times 
(N_{c,n-1}-N_{c,n-2}-\widetilde{N}_{c,n})$ and $(N_{c,n-1}-N_{c,n-2}-
\widetilde{N}_{c,n}) \times
\widetilde{N}_{c,n}$ 
matrices.
Then one can expand these fields around on a point (\ref{sol})
and one arrives at the relevant superpotential
up to quadratic order in the fluctuation. 
At one loop, the effective potential $V_{eff}^{(1)}$ for $M_{n-1}$
leads to the positive value for $m_{M_{n-1}}^2$ implying that these
vacua are stable.

%%%%%%%%%%%%%%%%%%%%%%%%%%%%%%%%%%%%%%%%%%%%%%%%%%%%%%%%%%%%%%
%%%%%%%%%%%%%%%%%%%%%%%%%%%%%%%%%%%%%%%%%%%%%%%%%%%%%%%%%%%%%%
\subsubsection{When the dual gauge group occurs at even chain}
%%%%%%%%%%%%%%%%%%%%%%%%%%%%%%%%%%%%%%%%%%%%%%%%%%%%%%%%%%%%%%
%%%%%%%%%%%%%%%%%%%%%%%%%%%%%%%%%%%%%%%%%%%%%%%%%%%%%%%%%%%%%%

Let us discuss the other case for
the Seiberg dual of the last gauge group
factor.
Starting from Figure 1A, moving the $NS5_{n-1}'$-brane 
with $(N_{c,n-1}-N_{c,n-2})$
D4-branes 
to the $+v$ direction leading to Figure 2B, 
and interchanging the $NS5_{n}$-brane and the $NS5_{n+1}'$-brane,
one obtains the Figure 6A.

%%%%%%%%%%%%%%%%%%%%%%%%%%%%%%%%%%%%%%%%%%%%%%%%%%%%%%%%%%%%%%%%%%%%%
%%%%%%%%%%%%%%%%%%%%%%%%%%%%%%%%%%%%%%%%%%%%%%%%%%%%%%%%%%%%%%%%%%%%%%
\begin{figure}[ht]
   \epsfxsize=4.0in 
\centerline{\epsffile{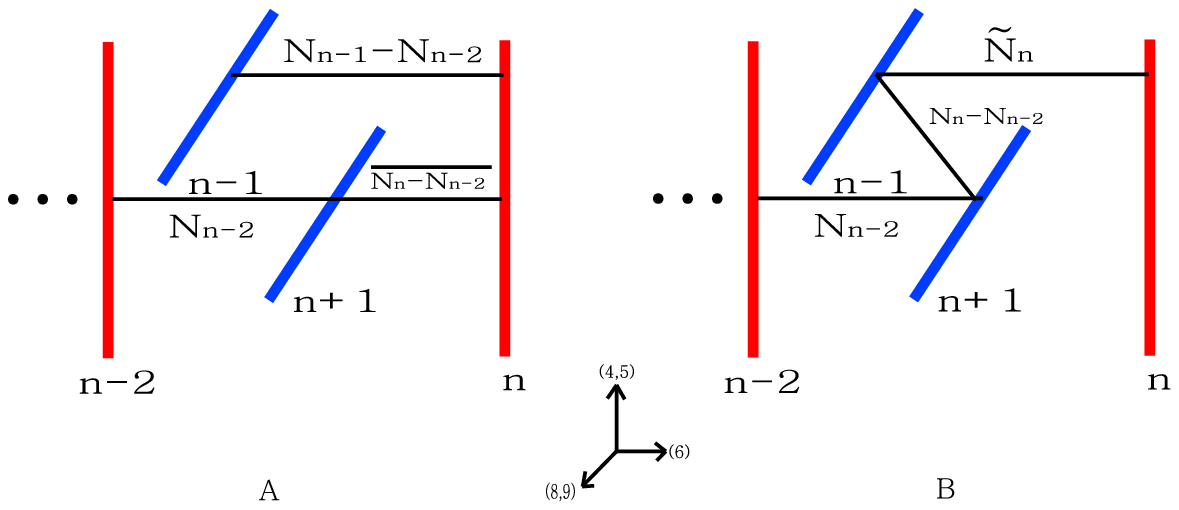}}
   \caption[FIG. \arabic{figure}.]{ 
The 
 ${\cal N}=1$ magnetic brane configuration for the gauge group 
containing $SU(\widetilde{N}_{c,n}=N_{c,n-1}-N_{c,n})$, where $n$ is even, 
with D4-
and $\overline{D4}$-branes(6A) and with 
a misalignment between D4-branes(6B) when the two NS5'-branes are close to
each other. 
The number of tilted D4-branes in 6B can be written as
$N_{c,n}-N_{c,n-2} =(N_{c,n-1}-N_{c,n-2})-\widetilde{N}_{c,n}$.
%The $x$ coordinate of $NS5_{n-1}'$-brane is given by $(\Delta
% x)_{n-1}$.
}
\end{figure}
%%%%%%%%%%%%%%%%%%%%%%%%%%%%%%%%%%%%%%%%%%%%%%%%%%%%%%%%%%%%%%%%%%%%%
%Figure 6A and 6B
%%%%%%%%%%%%%%%%%%%%%%%%%%%%%%%%%%%%%%%%%%%%%%%%%%%%%%%%%%%%%%%%%%%%%

Before arriving at the Figure 6A, there exists an intermediate 
step where the $(N_{c,n-1}-N_{c,n})$ D4-branes are 
connecting between the 
$NS5_{n+1}'$-brane and the  $NS5_{n}$-brane,  
$(N_{c,n-1}-N_{c,n-2})$ D4-branes connecting between the  $NS5_{n-1}'$-brane and   
$NS5_{n+1}'$-brane, and $N_{c,n-2}$ D4-branes between the $NS5_{n-2}$-brane and
the $NS5_{n+1}'$-brane. By introducing $-N_{c,n-2}$ D4-branes and $-N_{c,n-2}$ 
anti-D4-branes  between the  $NS5_{n+1}'$-brane and   
$NS5_{n}$-brane, reconnecting the former with  
the $N_{c,n-1}$ D4-branes connecting between  
$NS5_{n+1}'$-brane 
and the $NS5_{n}$-brane (therefore $N_{c,n-1}-N_{c,n-2}$ D4-branes)
and moving those combined
$(N_{c,n-1}-N_{c,n-2})$ 
D4-branes
to $+v$-direction, 
one gets the final Figure 6A where we are left with 
$(N_{c,n}-N_{c,n-2})$ 
anti-D4-branes between the $NS5_{n+1}'$-brane and   
$NS5_{n}$-brane.

The dual gauge group that is the same as before is given by 
\bea
SU(N_{c,1}) \times \cdots \times SU(N_{c,n-1}) \times 
SU(\widetilde{N}_{c,n} \equiv N_{c,n-1}-N_{c,n}) 
\label{dual22}
\eea
where 
the matter contents are   
the bifundamentals $f_{n-1}$ in 
 $({\bf 1_1, \cdots, 1, \Box_{n-1}, \overline{\Box}_{n}, })$,
and $\widetilde{f}_{n-1}$ in the representation 
$({\bf 1_1, \cdots, 1, \overline{\Box}_{n-1}, \Box_{n}})$ in
addition to $(n-2)$ bifundamentals $F_j$ and $\widetilde{F}_j$, 
 $j=1,2,
\cdots, n-2$ and
the gauge singlet $\Phi_{n-1}$
for the $n$-th dual gauge group in the 
adjoint representation for the $(n-1)$-th dual gauge group, 
i.e.,  
$
{(\bf 1_1, \cdots, 1_{n-2}, (N_{c,n-1}-N_{c,n-2})^2-1, 1_n)  
}
$ plus a singlet
under the 
dual gauge group (\ref{dual22}) where the gauge group is broken from
$SU(N_{c,n-1})$ 
to $SU(N_{c,n-1}-N_{c,n-2})$.

When two NS5'-branes in Figure 6A are close to each other, then 
it leads to Figure 6B by realizing that the number of $(N_{c,n-1}-N_{c,n-2})$
D4-branes connecting between $NS5_{n-1}'$-brane and $NS5_{n}$-brane can
be rewritten as $(N_{c,n}-N_{c,n-2})$ plus $\widetilde{N}_{c,n}$.
The Figure 6 looks similar to the previous Figure 4.
If we ignore all the D4-branes and NS-branes located at the left hand
side of $NS_{n+1}'$-brane   
from Figure 6, then
the brane configuration becomes the one in \cite{GK}.

The cubic superpotential with the mass term  in the dual
theory is given by
\bea
W_{dual} = \Phi_{n-1} f_{n-1} \widetilde{f}_{n-1}  + m_{n-1} \tr \Phi_{n-1}.
\label{Wdual2}
\eea
Here the magnetic fields $f_{n-1}$ and $\widetilde{f}_{n-1}$  
correspond to 4-4 strings connecting 
the $\widetilde{N}_{c,n}$-color D4-branes(that are 
connecting between the $NS5_{n-1}'$-brane
and the $NS5_{n}$-brane in Figure 6B) with $N_{c,n-1}$-flavor 
D4-branes(that are 
a combination of three different D4-branes in Figure 6B).
Among these $N_{c,n-1}$-flavor D4-branes, only the strings ending on
the upper $(N_{c,n-1}-N_{c,n})$ D4-branes and 
on the tilted $(N_{c,n}-N_{c,n-2})$ 
D4-branes in Figure 6B enter the cubic superpotential term (\ref{Wdual2}). 
Although the $(N_{c,n-1}-N_{c,n-2})$ D4-branes in Figure 6A cannot move any
directions for fixed other branes,
the tilted $(N_{c,n}-N_{c,n-2})$-flavor D4-branes 
can move $w$ direction in Figure 6B.
The remaining upper $\widetilde{N}_{c,n}$ D4-branes are fixed also and cannot 
move any direction. Note that 
there is a decomposition 
\bea
N_{c,n-1}-N_{c,n-2}=(N_{c,n}-N_{c,n-2}) +\widetilde{N}_{c,n}.
\nonu
\eea

The brane configuration for zero mass for the bifundamental $F_{n-1}$
and 
$\widetilde{F}_{n-1}$,
which has only a cubic superpotential (\ref{Wdual2}),
can be obtained from Figure 6A by moving
the upper $NS5_{n-1}'$-brane together with 
$(N_{c,n-1}-N_{c,n-2})$ color D4-branes 
into the origin $v=0$.
Then the number of dual colors for D4-branes 
becomes  
$\widetilde{N}_{c,n}$ between $NS5_{n+1}'$-brane and $NS5_{n}$-brane 
and
$N_{c,n-1}$ between two NS5'-branes
as well
as $N_{c,n-2}$ D4-branes between $NS5_{n-2}$-brane and $NS5_{n-1}'$-brane.
Or starting from Figure 1B and moving the $NS5_{n+1}'$-brane to the left all the
way past the $NS5_{n}$-brane,
one also obtains the corresponding magnetic brane configuration
for massless case.

The brane configuration in Figure 6A is stable as long as the
distance $(\Delta x)_{n-1}$ between the upper NS5'-brane and 
the lower NS5'-brane is large, as
in \cite{GK}. If they are close to each other, then this brane
configuration is unstable to decay and leads to 
the brane configuration in Figure
6B.
One can regard these brane configurations as particular states in the
magnetic gauge theory with the gauge group (\ref{dual22}) and
superpotential (\ref{Wdual2}).
The   $(N_{c,n}-N_{c,n-2})$ flavor D4-branes of 
straight brane configuration
of
Figure 6B  bend due to the fact that there exists an attractive
gravitational interaction
between those flavor D4-branes and $NS5_{n}$-brane from the DBI action, by
following the procedure of \cite{GK}, 
as long as the distance $y_{n-3}$
goes to $\infty$ because the presence of an extra $NS5_{n-2}$-brane does
not affect the DBI action. 
For the finite and small $y_{n-3}$, the careful analysis for DBI action is
needed in
order to obtain the bending curve connecting  two NS5'-branes.  

When the upper NS5'-brane(or $NS5_{n-1}'$-brane) 
is replaced by coincident $(N_{c,n-1}-N_{c,n-2})$ 
D6-branes and 
the $NS5_{n-2}$ is rotated by an angle $\frac{\pi}{2}$ in the $(v,w)$
plane in Figure 6B, this brane configuration reduces to the one 
found in \cite{Ahn07-8} where the gauge group was given by 
$ SU(n_{c,1}) \times \cdots \times SU(n_{c,n-2}) 
\times SU(n_{c,n-1}) \times SU(n_{f,n}+n_{c,n-1}-n_{c,n})  $ 
with $n_{f,n}$ multiplets,  $\widetilde{n}_{f,n}$ multiplets, 
bifundamentals and gauge 
singlets. 
Then the present number $(N_{c,n-1}-N_{c,n-2})$ corresponds to the $n_{f,n}$, the
number $N_{c,n}$ corresponds to $n_{c,n}$ and 
the number $N_{c,n-2}$ corresponds to the $n_{c,n-1}$.
Note that the number of D4-branes touching $NS5_{n-1}'$-brane in Figure 6B
is equal to $(N_{c,n-1}-N_{c,n-2})$.
In particular, the Figure 2B of \cite{Ahn07-8} with vanishing flavors
$Q$ and $Q'$
is contained in
this modified Figure 6B running from the $NS5_{n-2}$-brane to 
the $NS5_{n}$-brane. 

The quantum corrections can be understood for small $(\Delta x)_{n-1}$ by 
using the low energy field theory on the branes.
The low energy dynamics of the magnetic brane configuration 
can be described by the ${\cal N}=1$ supersymmetric gauge theory
with gauge group (\ref{dual22})
and the gauge couplings for the three gauge group factors are
given by
\bea
g_{n-2,mag}^2  = \frac{g_s \ell_s}{y_{n-2}}, \qquad
g_{n-1,mag}^2  = \frac{g_s \ell_s}{(y_{n}+y_{n-1})}, \qquad 
g_{n,mag}^2 = \frac{g_s \ell_s}{y_{n}}.
\nonu
\eea
The dual gauge theory has  a gauge singlet $\Phi_{n-1}$  and 
bifundamentals $f_{n-1}, \widetilde{f}_{n-1}, F_j$, and 
$\widetilde{F}_j$  and the superpotential (\ref{Wdual2}) 
corresponding to Figures 6A and 6B is given by 
\bea
W_{dual} = h \Phi_{n-1} f_{n-1} \widetilde{f}_{n-1} - h \mu_{n-1}^2
\tr \Phi_{n-1}, 
\qquad h^2 = g_{n-1,
  mag}^2,
\qquad \mu_{n-1}^2 = -\frac{(\Delta x)_{n-1}}{ 2\pi g_s \ell_s^3}.
\nonu
\eea
Then $ f_{n-1} \widetilde{f}_{n-1}$ is 
a $\widetilde{N}_{c,n} \times \widetilde{N}_{c,n}$ 
matrix where the $(n-1)$-th gauge group indices for $f_{n-1}$ and 
$\widetilde{f}_{n-1}$ 
are contracted with those
of $\Phi_{n-1}$ while $\Phi_{n-1}$ is a 
$(N_{c,n-1}-N_{c,n-2}) \times (N_{c,n-1}-N_{c,n-2})$ matrix.
Although the field $f_{n-1}$ itself is an antifundamental in the $n$-th gauge
group
which is a different  
representation for the usual standard quark
coming from D6-branes,
the product $f_{n-1} \widetilde{f}_{n-1}$ has the same representation for the 
product of quarks
and moreover, 
the $(n-1)$-th gauge group indices for the field $\Phi_{n-1}$ play the
role of the flavor indices, as in comparison with the brane
configuration in the presence of D6-branes before.

Therefore, the F-term equation, the derivative $W_{dual}$ with respect to the
meson field $\Phi_{n-1}$ cannot be satisfied if the $(N_{c,n-1}-N_{c,n-2})$ exceeds
$\widetilde{N}_{c,n}$.
So the supersymmetry is broken.   
That is, 
there exist three equations from F-term conditions:
$
f_{n-1}^a \widetilde{f}_{n-1,b} -\mu_{n-1}^2 \delta^a_b =0$, and 
$\Phi_{n-1} f_{n-1} =0=\widetilde{f}_{n-1} \Phi_{n-1}$.
Then the solutions for these
are given by 
\bea
<f_{n-1}>   & = & 
\left(
\begin{array}{c}
\mu_{n-1}  {\bf 1}_{\widetilde{N}_{c,n}}  \\
0
\end{array}
\right), \qquad 
<\widetilde{f}_{n-1}>   = 
\left(
\begin{array}{cc}
\mu_{n-1}  {\bf 1}_{\widetilde{N}_{c,n}} & 0  \\
\end{array}
\right), \nonu \\
<\Phi_{n-1}> & = &
 \left(
\begin{array}{cc}
0  & 0  \\
0 & M_{n-1}  {\bf 1}_{(N_{c,n-1}-N_{c,n-2}-\widetilde{N}_{c,n})} 
\end{array}
\right) 
\label{point22}
\eea
where the zero of $<f_{n-1}>$ is a $
(N_{c,n-1}-N_{c,n-2}-\widetilde{N}_{c,n}) \times \widetilde{N}_{c,n}$ 
matrix, the zero of $<\widetilde{f}_{n-1}>$ is a
$\widetilde{N}_{c,n} \times (N_{c,n-1}-N_{c,n-2}-\widetilde{N}_{c,n}) $ matrix and 
the zeros of $<\Phi_{n-1}>$ are $\widetilde{N}_{c,n} \times \widetilde{N}_{c,n}$,
$\widetilde{N}_{c,n} \times (N_{c,n-1}-N_{c,n-2}-\widetilde{N}_{c,n})$, 
and $(N_{c,n-1}-N_{c,n-2}-\widetilde{N}_{c,n}) \times
\widetilde{N}_{c,n}$ matrices.
Then one can expand these fields around on a point (\ref{point22})
and one arrives at the relevant superpotential
up to quadratic order in the fluctuation. 
At one loop, the effective potential $V_{eff}^{(1)}$ for $M_{n-1}$
leads to the positive value for $m_{M_{n-1}}^2$ implying that these
vacua are stable.

%%%%%%%%%%%%%%%%%%%%%%%%%%%%%%%%%%%%%%%%%%%%%%%%%%%%%%%%%%%%%%%%%%%%%%%
%%%%%%%%%%%%%%%%%%%%%%%%%%%%%%%%%%%%%%%%%%%%%%%%%%%%%%%%%%%%%%%%%%%%%%%
\subsection{
${\cal N}=1$ 
$SU(\widetilde{N}_{c,1}) \times \cdots \times SU(N_{c,n})$ magnetic theory}
%%%%%%%%%%%%%%%%%%%%%%%%%%%%%%%%%%%%%%%%%%%%%%%%%%%%%%%%%%%%%%%%%%%%%%%%
%%%%%%%%%%%%%%%%%%%%%%%%%%%%%%%%%%%%%%%%%%%%%%%%%%%%%%%%%%%%%%%%%%%%%%%%

Now we turn to the last case.
Let us consider the Seiberg dual for the first gauge group
factor.
Starting from Figure 1A, moving the $NS5_{3}'$-brane 
with $(N_{c,2}-N_{c,3})$
D4-branes 
to the $+v$ direction leading to Figure 2A, 
and interchanging the $NS5_{1}'$-brane and the $NS5_{2}$-brane,
one obtains the Figure 7A.

%%%%%%%%%%%%%%%%%%%%%%%%%%%%%%%%%%%%%%%%%%%%%%%%%%%%%%%%%%%%%%%%%%%%%
%%%%%%%%%%%%%%%%%%%%%%%%%%%%%%%%%%%%%%%%%%%%%%%%%%%%%%%%%%%%%%%%%%%%%%
\begin{figure}[ht]
   \epsfxsize=4.0in 
\centerline{\epsffile{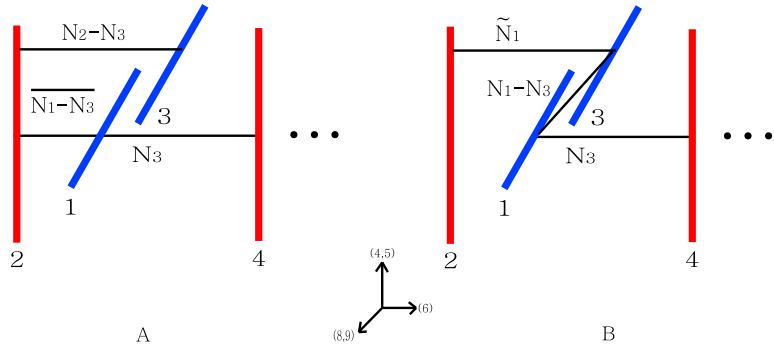}}
   \caption[FIG. \arabic{figure}.]{ 
The 
 ${\cal N}=1$ magnetic brane configuration for the gauge group 
containing $SU(\widetilde{N}_{c,1}=N_{c,2}-N_{c,1})$ 
with D4-
and $\overline{D4}$-branes(7A) and with 
a misalignment between D4-branes(7B) when the two NS5'-branes are close to
each other. 
The number of tilted D4-branes in 7B can be written as
$N_{c,1}-N_{c,3} =(N_{c,2}-N_{c,3})-\widetilde{N}_{c,1}$.
%The $x$ coordinate of $NS5_{3}'$-brane is given by $(\Delta x)_{2}$.
}
\end{figure}
%%%%%%%%%%%%%%%%%%%%%%%%%%%%%%%%%%%%%%%%%%%%%%%%%%%%%%%%%%%%%%%%%%%%%
%Figure 7A and 7B
%%%%%%%%%%%%%%%%%%%%%%%%%%%%%%%%%%%%%%%%%%%%%%%%%%%%%%%%%%%%%%%%%%%%%

Before arriving at the Figure 7A, there exists an intermediate 
step where the $(N_{c,2}-N_{c,1})$ D4-branes are 
connecting between the 
$NS5_{2}$-brane and the  $NS5_{1}'$-brane,  
$(N_{c,2}-N_{c,3})$ D4-branes connecting between the  $NS5_{1}'$-brane and   
$NS5_{3}'$-brane, and $N_{c,3}$ D4-branes between the $NS5_{1}'$-brane and
the $NS5_{4}$-brane. By introducing $-N_{c,3}$ D4-branes and $-N_{c,3}$ 
anti-D4-branes  between the  $NS5_{2}$-brane and   
$NS5_{1}'$-brane, reconnecting the former with  
the $N_{c,2}$ D4-branes connecting between  
$NS5_{2}$-brane 
and the $NS5_{1}'$-brane (therefore $(N_{c,2}-N_{c,3})$ D4-branes)
and moving those combined
$(N_{c,2}-N_{c,3})$ 
D4-branes
to $+v$-direction, 
one gets the final Figure 7A where we are left with 
$(N_{c,1}-N_{c,3})$ 
anti-D4-branes between the $NS5_{2}$-brane and   
$NS5_{1}'$-brane.

The dual gauge group is given by 
\bea
SU(\widetilde{N}_{c,1} \equiv N_{c,2}-N_{c,1}) \times 
SU(N_{c,2}) \times \cdots \times SU(N_{c,n})
\label{dualdual11}
\eea
where the matter contents are   
the bifundamentals $f_1$ in 
 $({\bf \Box_1, \overline{\Box}_{2}, \cdots, 1_n})$,
and $\widetilde{f}_1$ in the representation 
$({\bf \overline{\Box}_1, \Box_{2}, 1,
\cdots, 1_n})$ in
addition to $(n-2)$ bifundamentals $F_j$ and $\widetilde{F}_j$, 
 $j= 2, 3,
\cdots, n$ and
the gauge singlet $\Phi_{2}$
for the first dual gauge group in the 
adjoint representation for the second dual gauge group, 
i.e.,  
$
{(\bf 1_1, (N_{c,2}-N_{c,3})^2-1, 1_{3}, \cdots, 1_n)  
}
$ plus a singlet
under the 
dual gauge group where the gauge group is broken from
$SU(N_{c,2})$ 
to $SU(N_{c,2}-N_{c,3})$.

When two NS5'-branes in Figure 7A are close to each other, then 
it leads to Figure 7B by realizing that the number of $(N_{c,2}-N_{c,3})$
D4-branes connecting between $NS5_{2}$-brane and $NS5_{3}'$-brane can
be rewritten as $(N_{c,1}-N_{c,3})$ plus $\widetilde{N}_{c,1}$.
The Figure 7 looks similar to Figure 3.
If we ignore all the D4-branes and NS-branes at the right hand side of 
$NS_1'$-brane  
from Figure 7, then
the brane configuration becomes the one in \cite{GK}.

The cubic superpotential with the mass term (\ref{mass}) in the dual
theory is given by
\bea
W_{dual} = \Phi_{2} f_{1} \widetilde{f}_{1}  + m_{2} \tr \Phi_{2}.
\label{Wdual333}
\eea
Here the magnetic fields $f_1$ and $\widetilde{f}_1$  
correspond to 4-4 strings connecting 
the $\widetilde{N}_{c,1}$-color D4-branes(that are 
connecting between the $NS5_{2}$-brane
and the $NS5_{3}'$-brane in Figure 7B) with $N_{c,2}$-flavor 
D4-branes(that are 
a combination of three different D4-branes in Figure 7B).
Among these $N_{c,2}$-flavor D4-branes, only the strings ending on
the upper $(N_{c,2}-N_{c,1})$ D4-branes and 
on the tilted  $(N_{c,1}-N_{c,3})$ 
D4-branes in Figure 7B enter the cubic superpotential term (\ref{Wdual333}). 
Although the $(N_{c,2}-N_{c,3})$ D4-branes for fixed other branes 
in Figure 7A cannot move any
directions,
the tilted $(N_{c,1}-N_{c,3})$-flavor D4-branes 
can move $w$ direction in Figure 7B.
The remaining upper $\widetilde{N}_{c,1}$ D4-branes are fixed also and cannot 
move any direction. Note that 
there is a decomposition 
\bea
N_{c,2}-N_{c,3}=(N_{c,1}-N_{c,3}) +\widetilde{N}_{c,1}.
\nonu
\eea

The brane configuration for zero mass for the bifundamental $F_1$ and 
$\widetilde{F}_1$,
which has only a cubic superpotential (\ref{dualdual11}),
can be obtained from Figure 7A by moving
the upper $NS5_{3}'$-brane together with 
$(N_{c,2}-N_{c,3})$ color D4-branes 
into the origin $v=0$.
Then the number of dual colors for D4-branes 
becomes  
$\widetilde{N}_{c,1}$ between $NS5_{2}$-brane and $NS5_{1}'$-brane 
and
$N_{c,2}$ between two NS5'-branes
as well
as $N_{c,3}$ D4-branes between $NS5_{3}'$-brane and $NS5_{4}$-brane.
Or starting from Figure 1A and moving the $NS5_{2}$-brane to the left all the
way past the $NS5_{1}'$-brane,
one also obtains the corresponding magnetic brane configuration
for massless case.

The brane configuration in Figure 7A is stable as long as the
distance $(\Delta x)_{2}$ between the upper NS5'-brane and 
the lower NS5'-brane is large, as
in \cite{GK}. If they are close to each other, then this brane
configuration is unstable to decay and leads to 
the brane configuration in Figure
3B.
One can regard these brane configurations as particular states in the
magnetic gauge theory with the gauge group  (\ref{dualdual11}) and
superpotential (\ref{Wdual333}).
The   $(N_{c,2}-N_{c,3}-\widetilde{N}_{c,1})$ flavor D4-branes of 
straight brane configuration
of
Figure 7B  bend due to the fact that there exists an attractive
gravitational interaction
between those flavor D4-branes and $NS5_{2}$-brane from the DBI action, by
following the procedure of \cite{GK}, as long as the distance $y_{3}$
goes to $\infty$ because the presence of an extra $NS5_{4}$-brane does
not affect the DBI action. 
For the finite and small $y_{3}$, the careful analysis for DBI action is
needed in
order to obtain the bending curve connecting  two NS5'-branes.  

When the upper NS5'-brane(or $NS5_{3}'$-brane) 
is replaced by coincident $(N_{c,2}-N_{c,3})$ 
D6-branes and 
the $NS5_{4}$ is rotated by an angle $\frac{\pi}{2}$ in the $(v,w)$
plane in Figure 7B, this brane configuration reduces to the one 
found in \cite{Ahn07-8} where the gauge group was given by 
$ SU(n_{f,1}+n_{c,2}-n_{c,1}) \times
SU(n_{c,2}) 
\times \cdots $ 
with $n_{f,i}$ multiplets,  $\widetilde{n}_{f,i}$ multiplets,
bifundamentals and gauge 
singlets. 
Then the present number $(N_{c,2}-N_{c,3})$ corresponds to the $n_{f,1}$, the
number $N_{c,1}$ corresponds to $n_{c,1}$,
 the
number $N_{c,3}$ corresponds to $n_{c,2}$.
Note that the number of D4-branes touching $NS5_{3}'$-brane in Figure 7B
is equal to $(N_{c,2}-N_{c,3})$.
In particular, the Figure 2B of \cite{Ahn07-8} with vanishing flavors
$Q$ and $Q'$
is contained in
this modified Figure 7B running from the $NS5_{2}$-brane to 
the $NS5_{4}$-brane. 

The quantum corrections can be understood for small $(\Delta x)_{2}$ by 
using the low energy field theory on the branes.
The low energy dynamics of the magnetic brane configuration 
can be described by the ${\cal N}=1$ supersymmetric gauge theory
with gauge group
and the gauge couplings for the three gauge group factors are
given by
\bea
g_{1,mag}^2  = \frac{g_s \ell_s}{y_1}, \qquad 
g_{2,mag}^2 = \frac{g_s \ell_s}{(y_{2}+y_1)}, \qquad
g_{3,mag}^2  = \frac{g_s \ell_s}{y_3}.
\nonu
\eea
The dual gauge theory has  a gauge singlet $\Phi_{2}$  and 
bifundamentals $f_1, \widetilde{f}_1, F_j$, and $\widetilde{F}_j$ 
under the dual gauge
group and the superpotential 
corresponding to Figures 7A and 7B is given by 
\bea
W_{dual} = h \Phi_{2} f_1 \widetilde{f}_1 - h \mu_2^2 \tr \Phi_{2}, 
\qquad h^2 = g_{2,
  mag}^2,
\qquad \mu_2^2 = -\frac{(\Delta x)_{2}}{ 2\pi g_s \ell_s^3}.
\nonu
\eea
Then $ f_1 \widetilde{f}_1$ is a $\widetilde{N}_{c,1} \times \widetilde{N}_{c,1}$ 
matrix where the second gauge group indices for $f_1$ and $\widetilde{f}_1$ 
are contracted with those
of $\Phi_{2}$ while $\Phi_2$ is a 
$(N_{c,2}-N_{c,3}) \times (N_{c,2}-N_{c,3})$ matrix.
Although the field $f_1$ itself is an antifundamental in the second gauge
group
which is a different  
representation for the usual standard quark
coming from D6-branes,
the product $f_1 \widetilde{f}_1$ has the same representation for the 
product of quarks
and moreover, 
the second gauge group indices for the field $\Phi_{2}$ play the
role of the flavor indices, as in comparison with the brane
configuration in the presence of D6-branes before.

Therefore, the F-term equation, the derivative $W_{dual}$ with respect to the
meson field $\Phi_{2}$ cannot be satisfied if the $(N_{c,2}-N_{c,3})$ exceeds
$\widetilde{N}_{c,1}$.
So the supersymmetry is broken.   
That is, 
there exist three equations from F-term conditions:
$
f_1^a \widetilde{f}_{1,b} -\mu_1^2 \delta^a_b =0$,   and 
$\Phi_{2} f_1 =0=\widetilde{f}_1 \Phi_{2}$.
Then the solutions for these
are given by 
\bea
<f_1>   = 
\left(
\begin{array}{c}
\mu_2  {\bf 1}_{\widetilde{N}_{c,1}}  \\
0
\end{array}
\right), 
<\widetilde{f}_1>   = 
\left(
\begin{array}{cc}
\mu_2  {\bf 1}_{\widetilde{N}_{c,1}} & 0  \\
\end{array}
\right), 
<\Phi_{2}> =
 \left(
\begin{array}{cc}
0  & 0  \\
0 & M_{2}  {\bf 1}_{(N_{c,2}-N_{c,3}-\widetilde{N}_{c,1})} 
\end{array}
\right) 
\label{point333}
\eea
where the zero of $<f_1>$ is a $
(N_{c,2}-N_{c,3}-\widetilde{N}_{c,1}) \times \widetilde{N}_{c,1}$ 
matrix, the zero of $<\widetilde{f}_1>$ is a
$\widetilde{N}_{c,1} \times (N_{c,2}-N_{c,3}-\widetilde{N}_{c,1}) $ matrix and 
the zeros of $<\Phi_{2}>$ are $\widetilde{N}_{c,1} \times \widetilde{N}_{c,1}$,
$\widetilde{N}_{c,1} \times (N_{c,2}-N_{c,3}-\widetilde{N}_{c,1})$, 
and $(N_{c,2}-N_{c,3}-\widetilde{N}_{c,1}) \times
\widetilde{N}_{c,1}$ matrices.
Then one can expand these fields around on a point (\ref{point333})
and one arrives at the relevant superpotential
up to quadratic order in the fluctuation. 
At one loop, the effective potential $V_{eff}^{(1)}$ for $M_{2}$
leads to the positive value for $m_{M_{2}}^2$ implying that these
vacua are stable.

%%%%%%%%%%%%%%%%%%%%%%%%%%%%%%%%%%%%%%%%%%%%%%%%%%%%%%%%%%%%%%%%%%%%%
%%%%%%%%%%%%%%%%%%%%%%%%%%%%%%%%%%%%%%%%%%%%%%%%%%%%%%%%%%%%%%%%%%%%%
\section{Meta-stable brane configurations  with $(n+1)$ NS-branes plus
 O4-plane}
%section3%%%%%%%%%%%%%%%%%%%%%%%%%%%%%%%%%%%%%%%%%%%%%%%%%%%%%%%%%%%%
%%%%%%%%%%%%%%%%%%%%%%%%%%%%%%%%%%%%%%%%%%%%%%%%%%%%%%%%%%%%%%%%%%%%%

The type IIA brane configuration   \cite{AOT97} corresponding to 
${\cal N}=1$ supersymmetric electric gauge theory(see also
\cite{Ahn07-8})  with
gauge group
\bea
& & Sp(N_{c,1}) \times 
SO(2N_{c,2}) \times \cdots \times SO(2N_{c,n-1}) \times Sp(N_{c,n}) 
\;\;\; \mbox{for odd}\;\; n, 
\nonu \\
& & Sp(N_{c,1}) \times 
SO(2N_{c,2}) \times \cdots \times Sp(N_{c,n-1}) \times SO(2N_{c,n})
\;\;\; \mbox{for even} \;\; n, \nonu
\eea
and with 
the $(n-1)$ bifundametals $F_i$ charged under 
$({\bf 1_1, \cdots, 1, \Box_i, \Box_{i+1}, 1, \cdots,  1_n})$
where $i=1, 2, \cdots, (n-1)$
can be described by 
the $NS5_1'$-brane, 
the  
$NS5_2$-brane, $\cdots$, the $NS5_{n+1}$-brane for odd number
of gauge groups(or 
the $NS5_{n+1}'$-brane for even number of gauge groups),
$2N_{c,1}$-, $2N_{c,2}$-,  $\cdots$, and $2N_{c,n}$-color D4-branes. 
There exists the $O4^{\pm}$-planes(01236) which act as 
$(x^4, x^5, x^7, x^8, x^9) \rightarrow (-x^4, -x^5, -x^7, 
-x^8, -x^9)$ and they have RR charge $\pm 1$.
See the Figure 8 for the details on the brane configuration.

%%%%%%%%%%%%%%%%%%%%%%%%%%%%%%%%%%%%%%%%%%%%%%%%%%%%%%%%%%%%%%%%%%%%%
%%%%%%%%%%%%%%%%%%%%%%%%%%%%%%%%%%%%%%%%%%%%%%%%%%%%%%%%%%%%%%%%%%%%%%
\begin{figure}[ht]
   \epsfxsize=4.0in 
\centerline{\epsffile{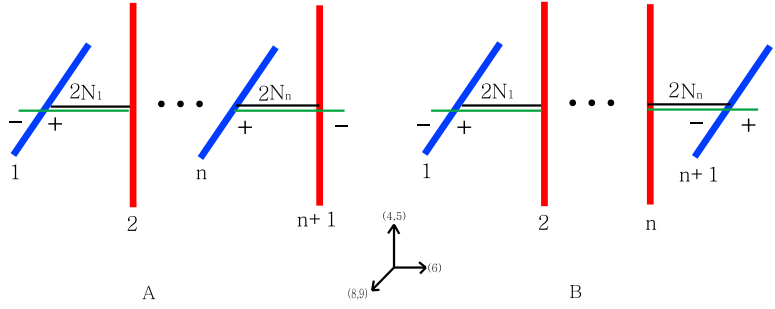}}
   \caption[FIG. \arabic{figure}.]{ 
The 
  ${\cal N}=1$ supersymmetric 
electric brane configuration for the gauge group 
$\left[\prod_{i=1}^{n-2} Sp(N_{c,i}) \times
SO(2N_{c,i+1})\right] \times Sp(N_{c,n})$ for 8A and 
 $ \prod_{i=1}^{n-1} Sp(N_{c,i}) \times
SO(2N_{c,i+1}) $ for 8B 
and bifundamentals $F_i$   with vanishing mass
for the bifundamental. 
%This Figure is obtained by adding
%  $O4^{+}$-plane for a symplectic gauge group and $O4^{-}$-plane for
%  an orthogonal gauge group
%  into the Figure 1 and the number of D4-branes is doubled.  
}
\end{figure}
%%%%%%%%%%%%%%%%%%%%%%%%%%%%%%%%%%%%%%%%%%%%%%%%%%%%%%%%%%%%%%%%%%%%%
%Figure 8A and 8B
%%%%%%%%%%%%%%%%%%%%%%%%%%%%%%%%%%%%%%%%%%%%%%%%%%%%%%%%%%%%%%%%%%%%%

Let us place the $NS5_1'$-brane at the origin $x^6=0$
and denote the $x^6$ 
coordinates for 
the  
$NS5_2$-brane, $\cdots$, the $NS5_{n+1}$-brane for odd $n$(or 
the $NS5_{n+1}'$-brane for even $n$)
are given by $x^6=y_1, y_1+y_2, \cdots, \sum_{j=1}^{n-1} y_j + y_n$
respectively.
The $2N_{c,1}$ D4-branes 
are suspended between the 
$NS5_1'$-brane and the $NS5_2$-brane, 
the $2N_{c,2}$ D4-branes 
are suspending between the 
$NS5_2$-brane and the $NS5_3'$-brane, $\cdots$ and 
the $2N_{c,n}$ D4-branes  
are suspended between the $NS5_n'$-brane and the $NS5_{n+1}$-brane for
odd $n$(or 
between the $NS5_n$-brane and the $NS5_{n+1}'$-brane for even $n$).
The fields $F_i$ correspond to 4-4 strings connecting 
the $2N_{c,i}$-color D4-branes with $2N_{c,i+1}$-color D4-branes.
We draw this ${\cal N}=1$ supersymmetric 
electric brane configuration in Figure 8A(8B) 
when $n$ is odd(even) for the vanishing mass
for the fields $F_i$. 

Let us deform the theory by Figure 8A.
Displacing the two NS5'-branes, $NS5_{i}'$-brane and
$NS5_{i+2}'$-brane, 
relative each other in the 
$v$
direction, characterized by $(\Delta x)_{i+1}$, 
corresponds to turning on a quadratic
mass-deformed superpotential
for the fields $F_i$  as follows:
\bea
W_{elec} = m_{i+1} F_i F_i (\equiv m_{i+1} \Phi_{i+1}), \qquad
\mbox{when $i$ is odd}
\label{Massodd}
\eea
where 
the $i$-th gauge group indices in $F_i$ and $F_i$ 
are contracted, each $(i+1)$-th gauge group index in them is encoded in 
$\Phi_{i+1}$ and the mass $m_{i+1}$ is given by
\bea
m_{i+1} 
%=\frac{(\Delta x)_{i+1}}{2\pi \alpha'} 
= 
\frac{(\Delta x)_{i+1}}{\ell_s^2}.
\label{M}
\eea

The gauge-singlet $\Phi_{i+1}$ for the $i$-th  gauge group is in the 
adjoint representation for the $(i+1)$-th  gauge group, 
i.e., 
\bea
{(\bf 1_1, \cdots, 1_i, 
(N_{c,i+1}-N_{c,i+2})(2N_{c,i+1}-2N_{c,i+2}-1), 1_{i+2}, \cdots, 1_n)  }
\nonu
\eea 
under the  gauge group
where the gauge group is broken from
$SO(2N_{c,i+1})$ 
to $SO(2N_{c,i+1}-2N_{c,i+2})$. 
Then the $\Phi_{i+1}$ is a $2(N_{c,i+1}-N_{c,i+2}) 
\times 2(N_{c,i+1}-N_{c,i+2})$ matrix.
The $NS5_{i+2}'$-brane together with $2(N_{c,i+1}-N_{c,i+2})$-color D4-branes 
is moving to the $\pm v$ direction  for
fixed other branes during this mass deformation. 
In other words, the $2N_{c, i+2}$ D4-branes among $2N_{c,i+1}$ D4-branes 
are not participating in 
the mass deformation.
Then the $x^5$ coordinate($\equiv x$) 
of $NS5_i'$-brane is equal to
zero
while the $x^5$ coordinate of $NS5_{i+2}'$-brane is given by 
$\pm(\Delta x)_{i+1}$.
Giving an expectation value to the meson field $\Phi_{i+1}$
corresponds to recombination of $2N_{c,i}$- and $2N_{c,i+1}$- color 
D4-branes, which will become $2N_{c,i}$- or $2N_{c,i+1}$-color D4-branes
in Figure 8A such that they are suspended between 
the $NS5_i'$-brane and the $NS5_{i+2}'$-brane 
and pushing them into the 
$w$
direction. We assume that the number of colors satisfies
$
N_{c,i+1} \geq N_{c,i}-N_{c,i-1}+2 \geq N_{c,i+2}$ where $i$ is odd.
Now 
we draw this brane configuration in Figure 9A for nonvanishing mass
for the fields $F_i$. 

%%%%%%%%%%%%%%%%%%%%%%%%%%%%%%%%%%%%%%%%%%%%%%%%%%%%%%%%%%%%%%%%%%%%%
%%%%%%%%%%%%%%%%%%%%%%%%%%%%%%%%%%%%%%%%%%%%%%%%%%%%%%%%%%%%%%%%%%%%%%
\begin{figure}[ht]
   \epsfxsize=4.0in 
\centerline{\epsffile{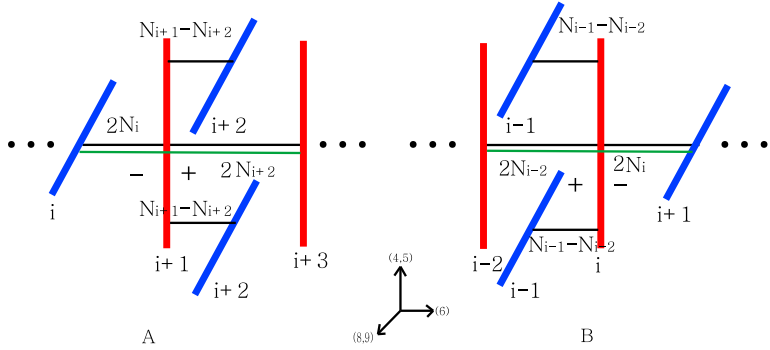}}
   \caption[FIG. \arabic{figure}.]{ 
The 
  ${\cal N}=1$ supersymmetric 
electric brane configuration for the gauge group 
$\left[\prod_{i=1}^{n-2} Sp(N_{c,i}) \times
SO(2N_{c,i+1})\right] \times Sp(N_{c,n})$ for 9A and 
 $\prod_{i=1}^{n-1} Sp(N_{c,i}) \times
SO(2N_{c,i+1})$ for 9B    with nonvanishing mass
for the bifundamental.  
The $2N_{c,i+1}$ D4-branes in 8A are decomposed into 
$2(N_{c,i+1}-N_{c,i+2})$ D4-branes which are moving to $\pm v$ direction in
  ${\bf Z}_2$ symmetric way in 9A 
and $2N_{c,i+2}$ D4-branes which are recombined with those D4-branes
connecting between $NS5_{i+2}'$-brane and $NS5_{i+3}$-brane in 9A.
The $2N_{c,i-1}$ D4-branes in 8B are decomposed into 
$2(N_{c,i-1}-N_{c,i-2})$ D4-branes which are moving to $\pm v$ direction in
  ${\bf Z}_2$ symmetric way in 9B 
and $2N_{c,i-2}$ D4-branes which are recombined with those D4-branes
connecting between $NS5_{i-2}$-brane and $NS5_{i-1}'$-brane in 9B. 
%The
%  $x$ coordinate for $NS5_{i+2}'$-brane is $ \pm (\Delta x)_{i+1}$ while
%  the one for $NS5_{i-1}'$-brane is $\pm (\Delta x)_{i-1}$.
}
\end{figure}
%%%%%%%%%%%%%%%%%%%%%%%%%%%%%%%%%%%%%%%%%%%%%%%%%%%%%%%%%%%%%%%%%%%%%
%Figure 9A and 9B
%%%%%%%%%%%%%%%%%%%%%%%%%%%%%%%%%%%%%%%%%%%%%%%%%%%%%%%%%%%%%%%%%%%%%%%

Let us deform the theory by Figure 8B.
Displacing the two NS5'-branes, the $NS_{i-1}'$-brane and the 
$NS_{i+1}'$-brane, 
relative each other in the 
$\pm v$ 
direction, characterized by $(\Delta x)_{i-1}$  
corresponds to turning on a quadratic
mass-deformed superpotential
for the fields $F_{i-1}$  as follows:
\bea
W = m_{i-1} F_{i-1} F_{i-1} (\equiv m_{i-1} \Phi_{i-1}), \qquad
\mbox{when $i$ is even}
\label{Mass1}
\eea
where 
the $i$-th gauge group indices in $F_{i-1}$ 
are contracted, each $(i-1)$-th gauge group index in them is encoded in 
$\Phi_{i-1}$ and the mass $m_{i-1}$ is given by
\bea
m_{i-1} 
%=\frac{(\Delta x)_{i+1}}{2\pi \alpha'} 
= 
\frac{(\Delta x)_{i-1}}{\ell_s^2}.
\label{M1}
\eea

The gauge-singlet $\Phi_{i-1}$ for the $i$-th  gauge group is in the 
adjoint representation for the $(i-1)$-th  gauge group, 
i.e., 
\bea
{(\bf 1_1, \cdots, 1_{i-2},
  (N_{c,i-1}-N_{c,i-2})(2N_{c,i-1}-2N_{c,i-2}+1), 
1_{i}, \cdots,
  1_n)}  
\nonu
\eea 
under the  gauge group
 where the gauge group is broken from
$Sp(N_{c,i-1})$ 
to $Sp(N_{c,i-1}-N_{c,i-2})$. 
Then the $\Phi_i$ is a $2(N_{c,i-1}-N_{c,i-2}) \times 
2(N_{c,i-1}-N_{c,i-2})$ matrix.
The $NS5_{i-1}'$-brane together with $2(N_{c,i-1}-N_{c,i-2})$-color D4-branes 
is moving to the $\pm v$ direction  for
fixed other branes during this mass deformation. 
In other words, the $2N_{c, i-2}$ D4-branes among $2N_{c,i-1}$ D4-branes 
are not participating in 
the mass deformation.
Then the $x^5$ coordinate($\equiv x$) 
of $NS5_{i+1}'$-brane is equal to
zero
while the $x^5$ coordinate of $NS5_{i-1}'$-brane is given by 
$\pm(\Delta x)_{i-1}$.
Giving an expectation value to the meson field $\Phi_{i-1}$
corresponds to recombination of $2N_{c,i-1}$- and $2N_{c,i}$- color 
D4-branes, which will become $2N_{c,i-1}$- or $2N_{c,i}$-color D4-branes
in Figure 8B such that they are suspended between 
the $NS5_{i-1}'$-brane and the $NS5_{i+1}'$-brane 
and pushing them into the 
$w$ direction. We assume that the number of colors satisfies
$
N_{c,i-1} \geq N_{c,i}-N_{c,i+1}-2 \geq N_{c,i-2}$ where $i$ is even.
Now 
we draw this brane configuration in Figure 9B for nonvanishing mass
for the fields $F_{i-1}$. 

%Next we describe five different magnetic dual theories by taking each
%corresponding mass deformation.
Basically the brane configurations are the same as the ones in
previous section if one includes the mirrors but the careful treatment
for the appropriate O4-plane charge is needed in order to obtain the
correct number of D4-branes \footnote{
When the O4-plane charges are reversed, then the gauge group will be 
either $\left[\prod_{i=1}^{n-2} SO(2N_{c,i}) \times
Sp(2N_{c,i+1})\right] \times SO(2N_{c,n})$ for odd $n$ or 
 $\prod_{i=1}^{n-1} SO(N_{c,i}) \times
Sp(2N_{c,i+1})$ for even $n$. One can analyze these cases also by
realizing that the adjoint of symplectic gauge group is symmetric
matrix while the adjoint of orthogonal gauge group is 
antisymmetric.
\label{footn}}. 

%%%%%%%%%%%%%%%%%%%%%%%%%%%%%%%%%%%%%%%%%%%%%%%%%%%%%%%%%%%%%%%%%%%%%%%%%
%%%%%%%%%%%%%%%%%%%%%%%%%%%%%%%%%%%%%%%%%%%%%%%%%%%%%%%%%%%%%%%%%%%%%%%%%
\subsection{${\cal N}=1$ 
$ Sp(N_{c,1}) \times \cdots 
\times SO(2\widetilde{N}_{c,i})[Sp(\widetilde{N}_{c,i})] 
\times \cdots $ magnetic theory}
%%%%%%%%%%%%%%%%%%%%%%%%%%%%%%%%%%%%%%%%%%%%%%%%%%%%%%%%%%%%%%%%%%%%%%%%
%%%%%%%%%%%%%%%%%%%%%%%%%%%%%%%%%%%%%%%%%%%%%%%%%%%%%%%%%%%%%%%%%%%%%%%%

%Let us first consider the Seiberg dual for the middle gauge group
%factor.
%There are two magnetic duals depending on whether the gauge group factor
%occurs at odd chain or even chain.

%%%%%%%%%%%%%%%%%%%%%%%%%%%%%%%%%%%%%%%%%%%%%%%%%%%%%%%%%%%%%%
%%%%%%%%%%%%%%%%%%%%%%%%%%%%%%%%%%%%%%%%%%%%%%%%%%%%%%%%%%%%%%
\subsubsection{ When the dual gauge group occurs at odd chain}
%%%%%%%%%%%%%%%%%%%%%%%%%%%%%%%%%%%%%%%%%%%%%%%%%%%%%%%%%%%%%%
%%%%%%%%%%%%%%%%%%%%%%%%%%%%%%%%%%%%%%%%%%%%%%%%%%%%%%%%%%%%%%

Starting from Figure 9A and interchanging the 
$NS5_i'$-brane and the $NS5_{i+1}$-brane,
one obtains the Figure 10A.
By introducing $-2N_{c,i+2}$ D4-branes and $-2N_{c,i+2}$ 
anti-D4-branes  between the  $NS5_{i+1}$-brane and   
$NS5_i'$-brane, reconnecting the former with  
the $N_{c,i+1}$ D4-branes connecting between  
$NS5_{i+1}$-brane
and the $NS5_i'$-brane  (therefore $2(N_{c,i+1}-N_{c,i+2})$ D4-branes)
and moving those combined
$(N_{c,i+1}-N_{c,i+2})$ 
D4-branes
to $+v$-direction(and their mirrors to $-v$ direction), 
one gets the final Figure 10A where we are left with 
$2(N_{c,i}-N_{c,i+2}-N_{c,i-1}+2)$ 
anti-D4-branes between the $NS5_{i+1}$-brane and   
$NS5_i'$-brane.

%%%%%%%%%%%%%%%%%%%%%%%%%%%%%%%%%%%%%%%%%%%%%%%%%%%%%%%%%%%%%%%%%%%%%
%%%%%%%%%%%%%%%%%%%%%%%%%%%%%%%%%%%%%%%%%%%%%%%%%%%%%%%%%%%%%%%%%%%%%%
\begin{figure}[ht]
   \epsfxsize=4.0in 
\centerline{\epsffile{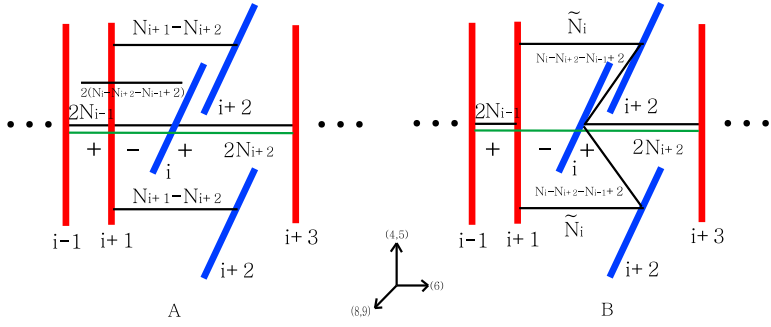}}
   \caption[FIG. \arabic{figure}.]{ 
The 
 ${\cal N}=1$ magnetic brane configuration for the gauge group 
containing $Sp(\widetilde{N}_{c,i}=N_{c,i-1}+N_{c,i+1}-N_{c,i}-2)$ 
with D4-
and $\overline{D4}$-branes(10A) and with 
a misalignment between D4-branes(10B) when the two NS5'-branes are close to
each other. 
The number of tilted D4-branes in 10B can be written as
$N_{c,i}-N_{c,i-1}-N_{c,i+2}+2
 =N_{c,i+1}-N_{c,i+2}-\widetilde{N}_{c,i}$.
The $x$ coordinate of $NS5_{i+2}'$-brane is given by $ \pm (\Delta x)_{i+1}$. 
%This Figure is obtained by adding
%  $O4^{+}$-plane for a symplectic gauge group and $O4^{-}$-plane for
%  an orthogonal gauge group
%  into the Figure 3 and the appropriate number of D4-branes is 
%considered.
}
\end{figure}
%%%%%%%%%%%%%%%%%%%%%%%%%%%%%%%%%%%%%%%%%%%%%%%%%%%%%%%%%%%%%%%%%%%%%
%Figure 10A and 10B
%%%%%%%%%%%%%%%%%%%%%%%%%%%%%%%%%%%%%%%%%%%%%%%%%%%%%%%%%%%%%%%%%%%%%

%Before arriving at the Figure 10A, there exists an intermediate 
%step where the $2(N_{c,i+1}+N_{c,i-1}-N_{c,i}-2)$ D4-branes are connecting between the 
%$NS5_{i+1}$-brane and the  $NS5_i'$-brane,  
%$(N_{c,i+1}-N_{c,i+2})$ D4-branes connecting between the  $NS5_i'$-brane and   
%$NS5_{i+2}'$-brane(and their mirrors), 
%and $2N_{c,i+2}$ D4-branes between the $NS5_i'$-brane and
%the $NS5_{i+3}$-brane. 

The dual gauge group is 
\bea
\cdots \times SO(2N_{c,i-1}) \times 
Sp(\widetilde{N}_{c,i} \equiv N_{c,i+1}+N_{c,i-1}-N_{c,i} -2) 
\times SO(2N_{c,i+1}) \times \cdots.
\label{Dual}
\eea
The matter contents are the field $f_i$ 
 charged under
$({\bf 1_1, \cdots, 1_{i-1}, 2\widetilde{N}_{c,i}, 2N_{c,i+1}, 1,
  \cdots, 1_n})$ under the dual gauge group
and  
the gauge-singlet $\Phi_{i+1}$ that is in the 
adjoint representation 
for the $(i+1)$-th dual gauge group, 
i.e., 
\bea
{(\bf 1_1, \cdots, 1_i,
   (N_{c,i+1}-N_{c,i+2})(2N_{c,i+1}-2N_{c,i+2}-1), 1_{i+2}, \cdots, 1_n) }
\nonu
\eea 
under the 
dual gauge group where the gauge group is broken from
$SO(2N_{c,i+1})$ 
to $SO(2N_{c,i+1}-2N_{c,i+2})$. That is, the $\Phi_{i+1}$ is  an 
 $2(N_{c,i+1}-N_{c,i+2}) \times 2(N_{c,i+1}-N_{c,i+2})$ antisymmetric matrix.

%When two NS5'-branes in Figure 10A are close to each other, then 
%it leads to Figure 10B
% by realizing that the number of $(N_{c,i+1}-N_{c,i+2})$
%D4-branes connecting between $NS5_{i+1}$-brane and $NS5_{i+2}'$-brane can
%be rewritten as $(N_{c,i}-N_{c,i+2}-N_{c,i-1}+2)$ plus $\widetilde{N}_{c,i}$.
%The Figure 7 of \cite{Ahn07-6} is contained in the Figure 10. In
%particular, the brane configuration from the $NS5_{i+1}$-brane to 
%the $NS5_{i+3}$-brane is exactly same as the one of \cite{Ahn07-6}.

The cubic superpotential with the mass term (\ref{Massodd}) and (\ref{M}) in the dual
theory 
\footnote{There are also 
the extra terms in the superpotential
$\Phi' f_{i-1} f_{i}  + \Phi_{i-1} f_{i-1} f_{i-1}$ where we define 
$\Phi' \equiv F_i F_{i-1}$, coming from 
different bifundamentals. However, the F- term condition,
$\Phi' f_i + 2 \Phi_{i-1} f_{i-1}=0$ leads to 
$<\Phi'>=<f_{i-1}>=0$. 
Therefore, these extra terms do not
contribute to the one loop computation up to quadratic order.} 
is given by
\bea
W_{dual} = \Phi_{i+1} f_i f_i + m_{i+1} \tr \Phi_{i+1}. 
\label{sup}
\eea
Here the magnetic field $f_i$  
corresponds to 4-4 strings connecting 
the $2\widetilde{N}_{c,i}$-color D4-branes(that are 
connecting between the $NS5_{i+1}$-brane
and the $NS5_{i+2}'$-brane including the mirrors) with $2N_{c,i+1}$-flavor 
D4-branes(that are 
a combination of three different D4-branes including the mirrors 
in Figure 10B).
Among these $2N_{c,i+1}$-flavor D4-branes, only the strings ending on
the upper and lower $2(N_{c,i+1}-N_{c,i}+N_{c,i-1}-2)$ D4-branes and 
on the tilted  $2(N_{c,i}-N_{c,i+2}-N_{c,i-1}+2)$ 
D4-branes including the mirrors 
in Figure 10B enter the cubic superpotential term (\ref{sup}). 

When the upper and lower half $NS5_{i+2}'$-branes 
are replaced by coincident $(N_{c,i+1}-N_{c,i+2})$ 
D6-branes 
and 
the $NS5_{i+3}$ is rotated by an angle $\frac{\pi}{2}$ in the $(v,w)$
plane in Figure 10B, 
this brane configuration reduces to the one 
found in \cite{Ahn07-8} where the gauge group was given by 
$ \cdots \times SO(2n_{c,i-1}) \times
Sp(n_{f,i}+n_{c,i+1}+n_{c,i-1}-n_{c,i}-2) 
\times SO(2n_{c,i+1}) \times \cdots$ 
with $2n_{f,i}$ multiplets, bifundamentals and gauge singlets. 
Then the present $(N_{c,i+1}-N_{c,i+2})$ corresponds to the $n_{f,i}$,
the number $N_{c,i-1}$ corresponds to $n_{c,i-1}$,
the number $N_{c,i+2}$ corresponds to $n_{c,i+1}$ and 
$N_{c,i}$ corresponds to the $n_{c,i}$. 
%In particular, the Figure 8B of \cite{Ahn07-8} with vanishing flavors
%$Q$ and $Q''$
%is contained in
%this modified Figure 10B running from the $NS5_{i-1}$-brane to 
%the $NS5_{i+3}$-brane. 
The dual gauge theory has  a meson field $\Phi_{i+1}$ and 
bifundamentals $f_i$  under the dual gauge
group (\ref{Dual}) and the superpotential (\ref{sup}) 
corresponding to Figures 10A and 10B is given by 
\bea
W_{dual} = h \Phi_{i+1} f_i f_i - h \mu_{i+1}^2 \tr
\Phi_{i+1}, \qquad h^2 = g_{i+1,
  mag}^2,
\qquad \mu_{i+1}^2 = -\frac{(\Delta x)_{i+1}}{ 2\pi g_s \ell_s^3}.
\nonu
\eea
Then $ f_i f_i$ is a $2\widetilde{N}_{c,i} \times 2\widetilde{N}_{c,i}$ 
matrix where the $(i+1)$-th gauge group indices for $f_i$  
are contracted with those
of $\Phi_{i+1}$ while the $\Phi_{i+1}$ is a 
$2(N_{c,i+1}-N_{c,i+2}) \times 2(N_{c,i+1}-N_{c,i+2})$ matrix.
%The product $f_i f_i$ has the same representation for the 
%product of quarks
%and moreover, 
%the $(i+1)$-th gauge group indices for the field $\Phi_{i+1}$ play the
%role of the flavor indices, as we observed above for the comparison
%with the brane configuration in the presence of D6-branes.
Therefore, the F-term equation, the derivative $W_{dual}$ with respect to the
meson field $\Phi_{i+1}$ cannot be satisfied if the $2(N_{c,i+1}-N_{c,i+2})$ exceeds
$2\widetilde{N}_{c,i}$.
So the supersymmetry is broken.   
That is, 
there exist two equations from F-term conditions:
$
f_i^a f_i^b -\mu_{i+1}^2 \delta^{a,b} =0$ and $ \Phi_{i+1} f_i =0$.
Then the solutions for these
are given by 
\bea
<f_i>   = 
\left(
\begin{array}{c}
\mu_{i+1}  {\bf 1}_{2\widetilde{N}_{c,i}}  \\
0
\end{array}
\right), 
\qquad
<\Phi_{i+1}> =
 \left(
\begin{array}{cc}
0  & 0  \\
0 & M_{i+1}  {\bf 1}_{(N_{c,i+1}-N_{c,i+2}-\widetilde{N}_{c,i}) 
\otimes i \sigma_2} 
\end{array}
\right). 
\nonu
%\label{poi1234}
\eea
%where the zero of $<f_i>$ is a $
%2(N_{c,i+1}-N_{c,i+2}-\widetilde{N}_{c,i}) \times 
%2\widetilde{N}_{c,i}$ 
%matrix and 
%the zeros of $<M_{i+1}>$ are $2\widetilde{N}_{c,i} \times 2\widetilde{N}_{c,i}$,
%$2\widetilde{N}_{c,i} \times 2(N_{c,i+1}-N_{c,i+2}-\widetilde{N}_{c,i})$, 
%and $2(N_{c,i+1}-N_{c,i+2}-\widetilde{N}_{c,i}) \times
%2\widetilde{N}_{c,i}$ matrices.
%Then one can expand these fields around on a point (\ref{poi1234})
%and one arrives at the relevant superpotential
%up to quadratic order in the fluctuation. 
%At one loop, the effective potential $V_{eff}^{(1)}$ for $M_{i+1}$
%leads to the positive value for $m_{M_{i+1}}^2$ implying that these
%vacua are stable.

%%%%%%%%%%%%%%%%%%%%%%%%%%%%%%%%%%%%%%%%%%%%%%%%%%%%%%%%%%%%%%
%%%%%%%%%%%%%%%%%%%%%%%%%%%%%%%%%%%%%%%%%%%%%%%%%%%%%%%%%%%%%%
\subsubsection{ When the dual gauge group occurs at even chain}
%%%%%%%%%%%%%%%%%%%%%%%%%%%%%%%%%%%%%%%%%%%%%%%%%%%%%%%%%%%%%%
%%%%%%%%%%%%%%%%%%%%%%%%%%%%%%%%%%%%%%%%%%%%%%%%%%%%%%%%%%%%%%

Let us discuss the other case for
the Seiberg dual of the middle gauge group
factor.
Let us consider other magnetic theory for the same electric theory in
previous subsection.
Starting from Figure 9B and interchanging the $NS5_i$-brane 
and the $NS5_{i+1}'$-brane,
one obtains the Figure 11A.

%%%%%%%%%%%%%%%%%%%%%%%%%%%%%%%%%%%%%%%%%%%%%%%%%%%%%%%%%%%%%%%%%%%%%%
%%%%%%%%%%%%%%%%%%%%%%%%%%%%%%%%%%%%%%%%%%%%%%%%%%%%%%%%%%%%%%%%%%%%%%
\begin{figure}[ht]
   \epsfxsize=4.0in 
\centerline{\epsffile{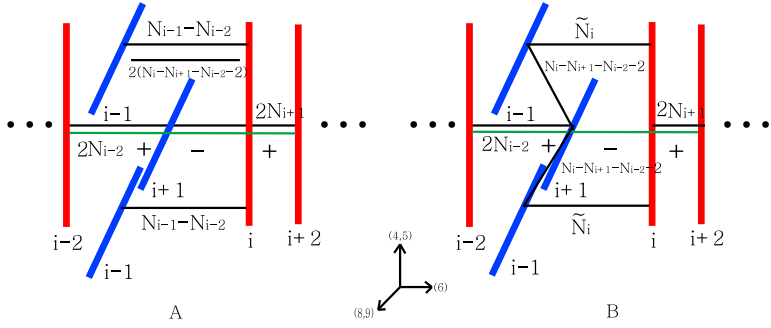}}
   \caption[FIG. \arabic{figure}.]{ 
The 
 ${\cal N}=1$ magnetic brane configuration for the gauge group 
containing $SO(2\widetilde{N}_{c,i}=2(N_{c,i-1}+N_{c,i+1}-N_{c,i}+2))$ 
with D4-
and $\overline{D4}$-branes(11A) and with 
a misalignment between D4-branes(11B) when the two NS5'-branes are close to
each other. 
The number of tilted D4-branes in 11B can be written as
$N_{c,i}-N_{c,i+1}-N_{c,i-2}-2
 =N_{c,i-1}-N_{c,i-2}-\widetilde{N}_{c,i}$.
The $x$ coordinate of $NS5_{i-1}'$-brane is given by $ \pm (\Delta
 x)_{i-1}$.
%This Figure is obtained by adding
%  $O4^{+}$-plane for a symplectic gauge group and $O4^{-}$-plane for
%  an orthogonal gauge group
%  into the Figure 4 and the appropriate number of D4-branes is 
%considered.
}
\end{figure}
%%%%%%%%%%%%%%%%%%%%%%%%%%%%%%%%%%%%%%%%%%%%%%%%%%%%%%%%%%%%%%%%%%%%%
%Figure 11A and 11B
%%%%%%%%%%%%%%%%%%%%%%%%%%%%%%%%%%%%%%%%%%%%%%%%%%%%%%%%%%%%%%%%%%%%%%%

%Before arriving at the Figure 11A, there exists an intermediate 
%step where the $(N_{c,i-1}-N_{c,i-2})$ D4-branes are connecting between the 
%$NS5_{i-1}'$-brane and the $NS5_{i+1}'$-brane(and their mirrors),  
%$2(N_{c,i-1}+N_{c,i+1}-N_{c,i}+2)$ D4-branes are connecting 
%between the $NS5_{i+1}'$-brane and   
%$NS5_i$-brane, and $2N_{c,i+1}$ D4-branes are suspended 
%between the $NS5_i$-brane and
%the $NS5_{i+2}$-brane. 
By introducing $-2N_{c,i-2}$ D4-branes and $-2N_{c,i-2}$ 
anti-D4-branes  between the  $NS5_{i+1}'$-brane and   
$NS5_i$-brane, reconnecting the former with  
the $2N_{c,i-1}$ D4-branes connecting between  
$NS5_{i+1}'$-brane
and the $NS5_i$-brane  (therefore $2(N_{c,i-1}-N_{c,i-2})$ D4-branes)
and moving those combined
$(N_{c,i-1}-N_{c,i-2})$ 
D4-branes
to $+v$-direction(and their mirrors to $-v$ direction), 
one gets the final Figure 11A where we are left with 
$2(N_{c,i}-N_{c,i+1}-N_{c,i-2}-2)$ 
anti-D4-branes between the $NS5_{i+1}'$-brane and   
$NS5_i$-brane.
%We assume  that the number of colors satisfies
%$
%N_{c,i-1}+N_{c,i+1} \geq N_{c,i}-2 \geq N_{c,i+1}+N_{c,i-2}$.
The dual gauge group is 
\bea
\cdots \times Sp(N_{c,i-1}) \times 
SO(2\widetilde{N}_{c,i} \equiv 2(N_{c,i+1}+N_{c,i-1}-N_{c,i} +2)) 
\times Sp(N_{c,i+1}) \times \cdots.
\label{11dualdual}
\eea
The matter contents are the field $f_{i-1}$ 
 charged under
$
({\bf 1, \cdots, 1_{i-2}, 2N_{c,i-1}, 2\widetilde{N}_{c,i}, 1_{i+1},
  \cdots, 1_n}) 
$
under the dual gauge group
and  
the gauge-singlet $\Phi_{i-1}$ for the $i$-th dual gauge group in the 
adjoint representation for the $(i-1)$-th dual gauge group, 
i.e., 
\bea
{(\bf 1_1, \cdots, 1_{i-2},  (N_{c,i-1}-N_{c,i-2})
(2N_{c,i-1}-2N_{c,i-2}+1),1_i,
  \cdots, 1_n) }
\nonu
\eea 
under the 
dual gauge group where the gauge group is broken from
$Sp(N_{c,i-1})$ 
to $Sp(N_{c,i-1}-N_{c,i-2})$.
Then the $\Phi_{i-1}$ is a $2(N_{c,i-1}-N_{c,i-2}) \times 2(
N_{c,i-1}-N_{c,i-2})$ matrix.

%When two NS5'-branes in Figure 11A are close to each other, then 
%it leads to Figure 11B
% by realizing that the number of $(N_{c,i-1}-N_{c,i-2})$
%D4-branes connecting between $NS5_{i-1}'$-brane and $NS5_i$-brane in
%Figure 11A can
%be rewritten as $(N_{c,i}-N_{c,i+1}-N_{c,i-2}-2)$ plus $\widetilde{N}_{c,i}$.
%The Figure 9 of \cite{Ahn07-6} is contained in the Figure 11. In
%particular, the brane configuration from the $NS5_{i-1}'$-brane to 
%the $NS5_{i+2}$-brane is exactly same as the one of \cite{Ahn07-6}.

The cubic superpotential with the mass term  (\ref{Mass1}) and (\ref{M1}) in the dual
theory 
\footnote{There are also 
the extra terms in the superpotential
$\Phi_{i+1} f_{i} f_{i} +  \Phi' f_{i-1} 
f_{i}$ where we define 
$\Phi' \equiv F_i F_{i-1}$, coming from 
different bifundamentals. However, the F- term condition,
$2\Phi_{i+1} f_i + \Phi' f_{i-1}=0$ leads to 
$<\Phi'>=<f_i>=0$. 
In this case also these extra terms do not
contribute to the one loop computation up to quadratic order.}
is given by
\bea
W_{dual} = \Phi_{i-1} f_{i-1} f_{i-1} + m_{i-1} \tr \Phi_{i-1} 
\label{11super}
\eea
where we define $\Phi_{i-1}$ as $\Phi_{i-1} \equiv F_{i-1} F_{i-1}$ and 
the $i$-th gauge group indices in $F_{i-1}$ 
are contracted, each $(i-1)$-th gauge group index in them is encoded in 
$\Phi_{i-1}$. 
Here the magnetic field $f_{i-1}$ 
corresponds to 4-4 strings connecting 
the $2\widetilde{N}_{c,i}$-color D4-branes(that are 
connecting between the $NS5_{i-1}'$-brane
and the $NS5_i$-brane in Figure 11B) with $2N_{c,i-1}$-flavor 
D4-branes including the mirrors(which  are realized 
as corresponding D4-branes in Figure 11A).

When the NS5'-brane(or $NS5_{i-1}'$-brane) 
is replaced by coincident $(N_{c,i-1}-N_{c,i-2})$ 
D6-branes 
and the $NS5_{i-2}$ is rotated by an angle 
$\frac{\pi}{2}$ in the $(v,w)$
plane in Figure 11B, 
this brane configuration leads to the one 
found in \cite{Ahn07-8} where the gauge group was given by 
$\cdots \times Sp(n_{c,i-1}) \times
SO(2n_{f,i}+2n_{c,i+1}+2n_{c,i-1}-2n_{c,i}+4) 
\times Sp(n_{c,i+1}) \times \cdots $ 
with $2n_{f,i}$ multiplets, bifundamentals and gauge singlets. 
Then the present $(N_{c,i-1}-N_{c,i-2})$ 
corresponds to the $n_{f,i}$, the number $N_{c,i}$
corresponds to $n_{c,i}$, 
the number $N_{c,i+1}$
corresponds to $n_{c,i+1}$ and
the number $N_{c,i-2}$
corresponds to $n_{c,i-1}$. 
%In particular, the Figure 8B of \cite{Ahn07-8} with vanishing flavors
%$Q$ and $Q''$
%is contained in
%this modified Figure 11B running from the $NS5_{i-2}$-brane to 
%the $NS5_{i+2}$-brane. 
The dual gauge theory has  a meson field $\Phi_{i-1}$  and 
bifundamentals $f_{i-1}$ under the dual gauge
group (\ref{11dualdual}) and the superpotential (\ref{11super}) 
corresponding to Figures 11A and 11B is given by 
\bea
W_{dual} = h \Phi_{i-1} f_{i-1} f_{i-1} - h \mu_{i-1}^2 \tr \Phi_{i-1}, 
\qquad h^2 = g_{i-1,
  mag}^2,
\qquad \mu_{i-1}^2 = -\frac{(\Delta x)_{i-1}}{ 2\pi g_s \ell_s^3}.
\nonu
\eea
Then $ f_{i-1} f_{i-1}$ is a 
$2\widetilde{N}_{c,i} \times 2\widetilde{N}_{c,i}$ 
matrix where the $(i-1)$-th gauge group indices for $f_{i-1}$  
are contracted with those
of $\Phi_{i-1}$ while $\Phi_{i-1}$ is a 
$2(N_{c,i-1}-N_{c,i-2}) \times 2(N_{c,i-1}-N_{c,i-2})$ matrix.
%The product $f_{i-1} f_{i-1}$ has the same representation for the 
%product of quarks
%and moreover, 
%the $(i-1)$-th gauge group indices for the field $\Phi_{i-1}$ play the
%role of the flavor indices as we observed above.
Therefore, the F-term equation, the derivative $W_{dual}$ with respect to the
meson field $\Phi_{i-1}$ cannot be satisfied if the $2(N_{c,i-1}-N_{c,i-2})$ exceeds
$2\widetilde{N}_{c,i}$.
So the supersymmetry is broken.   
That is, 
there exist two equations from F-term conditions:
$
f_{i-1}^a f_{i-1}^b -\mu_{i-1}^2 \delta^{a,b} =0$ and $ \Phi_{i-1} f_{i-1} =0$.
Then the solutions for these
are given by 
\bea
<f_{i-1}>   = 
\left(
\begin{array}{c}
\mu_{i-1}  {\bf 1}_{2\widetilde{N}_{c,i}}  \\
0
\end{array}
\right), 
\qquad
<\Phi_{i-1}> =
 \left(
\begin{array}{cc}
0  & 0  \\
0 & M_{i-1}  {\bf 1}_{2(N_{c,i-1}-N_{c,i-2}-\widetilde{N}_{c,i})} 
\end{array}
\right). 
\nonu
%\label{point234}
\eea
%where the zero of $<f_{i-1}>$ is a $
%2(N_{c,i-1}-N_{c,i-2}-\widetilde{N}_{c,i}) \times 2\widetilde{N}_{c,i}$ 
%matrix and 
%the zeros of $<\Phi_{i-1}>$ are $2\widetilde{N}_{c,i} \times 2\widetilde{N}_{c,i}$,
%$2\widetilde{N}_{c,i} \times 
%2(N_{c,i-1}-N_{c,i-2}-\widetilde{N}_{c,i})$ and
%$2(N_{c,i-1}-N_{c,i-2}-
%\widetilde{N}_{c,i}) \times
%2\widetilde{N}_{c,i}$ 
%matrices.
%Then one can expand these fields around on a point (\ref{point234})
%and one arrives at the relevant superpotential
%up to quadratic order in the fluctuation. 
%At one loop, the effective potential $V_{eff}^{(1)}$ for $M_{i-1}$
%leads to the positive value for $m_{M_{i-1}}^2$ implying that these
%vacua are stable.

%%%%%%%%%%%%%%%%%%%%%%%%%%%%%%%%%%%%%%%%%%%%%%%%%%%%%%%%%%%%%%%%%%%%%%
%%%%%%%%%%%%%%%%%%%%%%%%%%%%%%%%%%%%%%%%%%%%%%%%%%%%%%%%%%%%%%%%%%%%%%%
\subsection{ ${\cal N}=1$
$ Sp(N_{c,1}) \times \cdots  
\times Sp(\widetilde{N}_{c,n})[SO(2\widetilde{N}_{c,n})]$ magnetic theory}
%%%%%%%%%%%%%%%%%%%%%%%%%%%%%%%%%%%%%%%%%%%%%%%%%%%%%%%%%%%%%%%%%%%%%%%
%%%%%%%%%%%%%%%%%%%%%%%%%%%%%%%%%%%%%%%%%%%%%%%%%%%%%%%%%%%%%%%%%%%%%%%

%Let us next consider the Seiberg dual for the last gauge group
%factor.
%There are two magnetic duals depending on whether the gauge group factor
%occurs at odd chain or even chain.

%%%%%%%%%%%%%%%%%%%%%%%%%%%%%%%%%%%%%%%%%%%%%%%%%%%%%%%%%%%%%%
%%%%%%%%%%%%%%%%%%%%%%%%%%%%%%%%%%%%%%%%%%%%%%%%%%%%%%%%%%%%%%
\subsubsection{ When the dual gauge group occurs at odd chain}
%%%%%%%%%%%%%%%%%%%%%%%%%%%%%%%%%%%%%%%%%%%%%%%%%%%%%%%%%%%%%%
%%%%%%%%%%%%%%%%%%%%%%%%%%%%%%%%%%%%%%%%%%%%%%%%%%%%%%%%%%%%%%

Let us consider other magnetic theory for the same electric theory.
One can think of the following dual gauge group
\bea
Sp(N_c) \times \cdots \times SO(2N_{c,n-1}) \times 
Sp(\widetilde{N}_{c,n}=N_{c,n-1}-N_{c,n}-2)
\label{dualdualdual}
\eea
by performing the magnetic dual for the last gauge group.
By applying the Seiberg dual to the $Sp(N_{c,n})$ factor from Figure 9A and 
interchanging the $NS5_n'$-brane and the $NS5_{n+1}$-brane,
one obtains the Figure 12A.

%%%%%%%%%%%%%%%%%%%%%%%%%%%%%%%%%%%%%%%%%%%%%%%%%%%%%%%%%%%%%%%%%%%%%%
%%%%%%%%%%%%%%%%%%%%%%%%%%%%%%%%%%%%%%%%%%%%%%%%%%%%%%%%%%%%%%%%%%%%%%
\begin{figure}[ht]
   \epsfxsize=4.0in 
\centerline{\epsffile{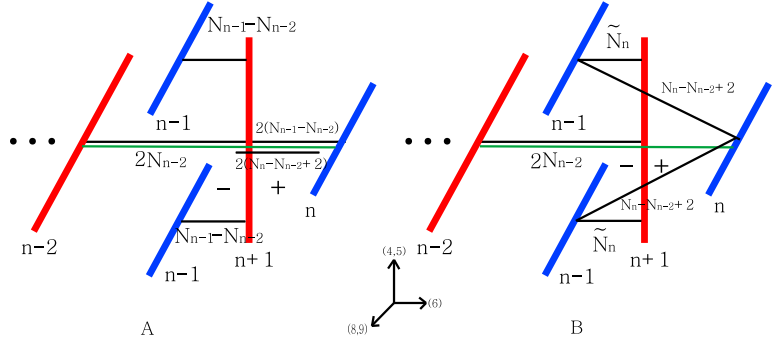}}
   \caption[FIG. \arabic{figure}.]{ 
The 
 ${\cal N}=1$ magnetic brane configuration for the gauge group 
containing $Sp(\widetilde{N}_{c,n}=N_{c,n-1}-N_{c,n}-2)$ 
with D4-
and $\overline{D4}$-branes(12A) and with 
a misalignment between D4-branes(12B) when the two NS5'-branes are close to
each other. 
The number of tilted D4-branes in 12B can be written as
$N_{c,n}-N_{c,n-2}+2 =(N_{c,n-1}-N_{c,n-2})-\widetilde{N}_{c,n}$.
The $x$ coordinate of $NS5_{n-1}'$-brane is given by $ \pm (\Delta
 x)_{n-1}$.
%This Figure is obtained by adding
%  $O4^{+}$-plane for a symplectic gauge group and $O4^{-}$-plane for
%  an orthogonal gauge group
%  into the Figure 5 and the appropriate number of D4-branes is 
%considered.
}
\end{figure}
%%%%%%%%%%%%%%%%%%%%%%%%%%%%%%%%%%%%%%%%%%%%%%%%%%%%%%%%%%%%%%%%%%%%%
%Figure 12A and 12B
%%%%%%%%%%%%%%%%%%%%%%%%%%%%%%%%%%%%%%%%%%%%%%%%%%%%%%%%%%%%%%%%%%%%%

%Before arriving at the Figure 12A, there exists an intermediate 
%step where 
%$2N_{c,n-2}$ D4-branes between
%$NS5_{n-2}'$-brane and the $NS5_{n-1}$-brane,
%the $2N_{c,n-1}$ D4-branes are connecting between the 
%$NS5_{n-1}$-brane and the  $NS5_{n+1}$-brane,  
%$(N_{c,n-1}-N_{c,n}-2)$ D4-branes are connecting between the  
%$NS5_{n+1}$-brane and   
%$NS5_n'$-brane(and their mirrors).
By rotating $NS5_{n-1}$-brane by an angle $\frac{\pi}{2}$, 
moving it with the $(N_{c,n-1}-N_{c,n-2})$ D4-branes 
to $+v$ direction where we introduce $2(N_{c,n}-N_{c,n-2})$ D4-branes and
$2(N_{c,n}-N_{c,n-2})$ anti D4-branes between the $NS5_{n+1}$-brane and the 
$NS5_{n}'$-brane, 
one gets the final Figure 12A where we are left with 
$2(N_{c,n}-N_{c,n-2}+2)$ 
anti-D4-branes between the $NS5_{n+1}$-brane and   
the $NS5_n'$-brane.
%We assume that $N_c' \geq N_c'' \geq 2N_c$.
When two NS5'-branes in Figure 12A are close to each other, then 
it leads to Figure 12B
 by realizing that the number of $(N_{c,n-1}-N_{c,n-2})$
D4-branes connecting between the 
$NS5_{n-1}'$-brane and the $NS5_{n+1}$-brane can
be rewritten as $(N_{c,n}-N_{c,n-2}+2)$ plus $\widetilde{N}_{c,n}$.
The Figure 10 of \cite{Ahn07-6} is contained in the Figure 12. In
particular, the brane configuration from the $NS5_{n-2}'$-brane to 
the $NS5_n'$-brane is exactly same as the one of \cite{Ahn07-6}.

%The brane configuration in Figure 12A is stable as long as the
%distance $(\Delta x)_{n-1}$ between the upper NS5'-brane and 
%the middle NS5'-brane(or $NS5_n'$-brane) 
%is large. If they are close to each other, then this brane
%configuration is unstable to decay to 
%the brane configuration in Figure
%12B.
%One can regard these brane configurations as particular states in the
%magnetic gauge theory with the gauge group and
%superpotential.
%The  upper $(N_{c,n-1}-N_{c,n-2}-\widetilde{N}_{c,n})$ flavor D4-branes of 
%straight brane configuration
%of
%Figure 12B  bend since there exists an attractive
%gravitational interaction
%between those flavor D4-branes and NS5-brane from the DBI action. 
%As mentioned in \cite{Ahn07-5},
%the two NS5'-branes are located at different side of $NS5_{n+1}$-brane in
%Figure 12B and the DBI action computation for this bending curve
%should be taken into account. 
%Of course, their mirrors, the lower 
%$(N_{c,n-1}-N_{c,n-2}-\widetilde{N}_{c,n})$ flavor D4-branes of 
%straight brane configuration
%of
%Figure 12B can bend and their trajectories connecting 
%two NS5'-branes should be preserved under the O4-plane.

The matter contents are the field $f_{n-1}$ 
 charged under
$({\bf 1_1, \cdots, 1_{n-2}, 2N_{c,n-1}, 2\widetilde{N}_{c,n}})$ 
under the dual gauge group
and  
the gauge-singlet $\Phi_{n-1}$ that is in the 
adjoint representation 
for the $(n-1)$-th dual gauge group,  
 ${(\bf 1_1, \cdots, 1_{n-2},
 (N_{c,n-1}-N_{c,n-2})(2N_{c,n-1}-2N_{c,n-2}-1), 1_n) }$ 
under the 
dual gauge group (\ref{dualdualdual}) where the gauge group is broken from
$SO(2N_{c,n-1})$ 
to $SO(2N_{c,n-1}-2N_{c,n-2})$. 
That is, the $\Phi_{n-1}$ is  an 
$2(N_{c,n-1}-N_{c,n-2}) \times 2(N_{c,n-1}-N_{c,n-2})$ antisymmetric matrix.

The cubic superpotential with the mass term  in the dual
theory is given by
\bea
W_{dual} = \Phi_{n-1} f_{n-1} f_{n-1} + m_{n-1} \tr \Phi_{n-1} 
\label{sup-11}
\eea
where we define $\Phi_{n-1}$ as $\Phi_{n-1} \equiv F_{n-1} F_{n-1}$ and 
the $n$-th gauge group indices in $F_{n-1}$  
are contracted, each $(n-1)$-th gauge group index in $F_{n-1}$ is encoded in 
$\Phi_{n-1}$. 
Here the magnetic field $f_{n-1}$  
correspond to 4-4 strings connecting 
the $2\widetilde{N}_{c,n}$-color D4-branes including the mirrors(that are 
connecting between the $NS5_{n-1}'$-brane
and the $NS5_{n+1}$-brane in Figure 12B) with $2N_{c,n-1}$-flavor 
D4-branes.
Among these $2N_{c,n-1}$-flavor D4-branes, only the strings ending on
the $2(N_{c,n-1}-N_{c,n}-2)$ D4-branes and 
on the tilted  $2(N_{c,n}-N_{c,n-2}+2)$ 
D4-branes in Figure 12B enter the cubic superpotential term (\ref{sup-11}). 
%Although the $(N_{c,n-1}-N_{c,n-2})$ D4-branes(and its mirrors) 
%for fixed other branes in Figure 12A cannot move any
%directions,
%the tilted $(N_{c,n}-N_{c,n-2}+2)$-flavor D4-branes(and its mirrors) 
%can move $w$ direction.
%The remaining upper and lower 
%$\widetilde{N}_{c,n}$ D4-branes are fixed also and cannot 
%move any direction. 
%Note that 
%there is a decomposition 
%\bea
%(N_{c,n-1}-N_{c,n-2})=(N_{c,n}-N_{c,n-2}+2)+\widetilde{N}_{c,n}.
%\nonu
%\eea 

%The brane configuration for zero mass for the bifundamental,
%which has only a cubic superpotential (\ref{sup-11}),
%can be obtained from Figure 12A by moving
%the upper and lower NS5'-branes together with $(N_{c,n-1}-N_{c,n-2})$ color D4-branes 
%into the origin $v=0$.
%Then the number of dual colors for D4-branes 
%becomes $2N_{c,n-2}$ between two NS5'-branes 
%and $2N_{c,n-1}$ between the $NS5_{n-1}'$-brane and the $NS5_{n+1}$-brane and 
%$2\widetilde{N}_{c,n}$ between the $NS5_{n+1}$-brane and the $NS5_n'$-brane.
%Or starting from Figure 8A and moving the $NS5_{n+1}$-brane to the left all the
%way past the $NS5_n'$-brane,
%one also obtains the corresponding magnetic brane configuration
%for massless case.

The dual gauge theory has  a meson field $\Phi_{n-1}$ and 
bifundamentals $f_{n-1}$ under the dual gauge
group (\ref{dualdualdual}) and the superpotential (\ref{sup-11}) 
corresponding to Figures 12A and 12B is given by 
\bea
W_{dual} = h \Phi_{n-1} f_{n-1} f_{n-1} - 
h \mu_{n-1}^2 \tr \Phi_{n-1}, \qquad h^2 = g_{n-1,
  mag}^2,
\qquad \mu_{n-1}^2 = -\frac{(\Delta x)_{n-1}}{ 2\pi g_s \ell_s^3}.
\nonu
\eea
Then $ f_{n-1} f_{n-1}$ is a $2\widetilde{N}_{c,n} \times 2\widetilde{N}_{c,n}$ 
matrix where the $(n-1)$-th gauge group indices for $f_{n-1}$  
are contracted with those
of $\Phi_{n-1}$ while $\Phi_{n-1}$ is a 
$2(N_{c,n-1}-N_{c,n-2}) \times 2(N_{c,n-1}-N_{c,n-2})$ matrix.
%The product $f_{n-1} f_{n-1}$ has the same representation for the 
%product of quarks
%and moreover, 
%the $(n-1)$-th gauge group indices for the field $\Phi_{n-1}$ play the
%role of the flavor indices.
Therefore, the F-term equation, the derivative $W_{dual}$ with respect to the
meson field $\Phi_{n-1}$ cannot be satisfied if the $2(N_{c,n-1}-N_{c,n-2})$ exceeds
$2\widetilde{N}_{c,n}$.
So the supersymmetry is broken.   
That is, 
there exist two equations from F-term conditions:
$
f_{n-1}^a f_{n-1}^b-\mu_{n-1}^2 \delta^{a,b} =0$ and $ \Phi_{n-1} f_{n-1} =0$.
Then the solutions for these
are given by 
\bea
<f_{n-1}>   = 
\left(
\begin{array}{c}
\mu_{n-1}  {\bf 1}_{2\widetilde{N}_{c,n}}  \\
0
\end{array}
\right), \qquad
<\Phi_{n-1}> =
 \left(
\begin{array}{cc}
0  & 0  \\
0 & M_{n-1}  {\bf 1}_{(N_{c,n-1}-N_{c,n-2}-\widetilde{N}_{c,n}) \otimes i \sigma_2} 
\end{array}
\right) 
\nonu
%\label{poi-123}
\eea
%where the zero of $<f_{n-1}>$ is a $
%2(N_{c,n-1}-N_{c,n-2}-\widetilde{N}_{c,n}) \times 2\widetilde{N}_{c,n}$ 
%matrix and 
%the zeros of $<\Phi_{n-1}>$ are $2\widetilde{N}_{c,n} \times 2\widetilde{N}_{c,n}$,
%$2\widetilde{N}_{c,n} \times 2(N_{c,n-1}-N_{c,n-2}-\widetilde{N}_{c,n})$, 
%and $2(N_{c,n-1}-N_{c,n-2}-\widetilde{N}_{c,n}) \times
%2\widetilde{N}_{c,n}$ matrices.
%Then one can expand these fields around on a point (\ref{poi-123})
%and one arrives at the relevant superpotential
%up to quadratic order in the fluctuation. 
%At one loop, the effective potential $V_{eff}^{(1)}$ for $M_{n-1}$
%leads to the positive value for $m_{M_{n-1}}^2$ implying that these
%vacua are stable.

%%%%%%%%%%%%%%%%%%%%%%%%%%%%%%%%%%%%%%%%%%%%%%%%%%%%%%%%%%%%%%
%%%%%%%%%%%%%%%%%%%%%%%%%%%%%%%%%%%%%%%%%%%%%%%%%%%%%%%%%%%%%%
\subsubsection{ When the dual gauge group occurs at even chain}
%%%%%%%%%%%%%%%%%%%%%%%%%%%%%%%%%%%%%%%%%%%%%%%%%%%%%%%%%%%%%%
%%%%%%%%%%%%%%%%%%%%%%%%%%%%%%%%%%%%%%%%%%%%%%%%%%%%%%%%%%%%%%

One can think of the following dual gauge group
\bea
Sp(N_{c,1}) \times \cdots \times 
Sp(N_{c,n-1}) \times SO(2\widetilde{N}_{c,n} 
\equiv 2(N_{c,n-1}-N_{c,n}+2))
\label{dualgauge}
\eea
by performing the magnetic dual for the last gauge group.
By applying the Seiberg dual to the $SO(2N_{c,n})$ factor from Figure 9B and 
interchanging the $NS5_n$-brane and the $NS5_{n+1}'$-brane,
one obtains the Figure 13A.

%%%%%%%%%%%%%%%%%%%%%%%%%%%%%%%%%%%%%%%%%%%%%%%%%%%%%%%%%%%%%%%%%%%%%%
%%%%%%%%%%%%%%%%%%%%%%%%%%%%%%%%%%%%%%%%%%%%%%%%%%%%%%%%%%%%%%%%%%%%%%
\begin{figure}[ht]
   \epsfxsize=4.0in 
\centerline{\epsffile{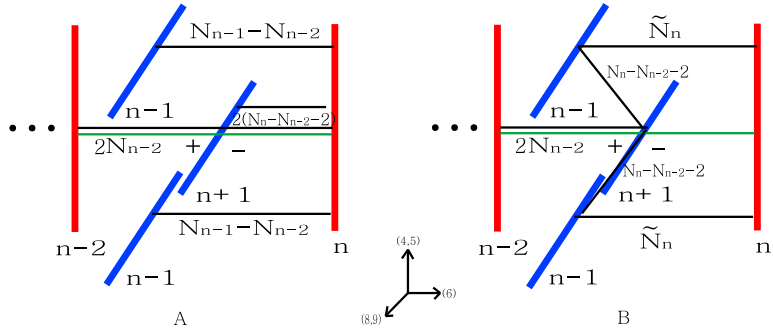}}
   \caption[FIG. \arabic{figure}.]{ 
The 
${\cal N}=1$ magnetic brane configuration for the gauge group 
containing $SO(2\widetilde{N}_{c,n}=2(N_{c,n-1}-N_{c,n}+2))$ 
with D4-
and $\overline{D4}$-branes(13A) and with 
a misalignment between D4-branes(13B) when the two NS5'-branes are close to
each other. 
The number of tilted D4-branes in 13B can be written as
$N_{c,n}-N_{c,n-2}-2 =(N_{c,n-1}-N_{c,n-2})-\widetilde{N}_{c,n}$.
The $x$ coordinate of $NS5_{n-1}'$-brane is given by $ \pm (\Delta
 x)_{n-1}$.
%This Figure is obtained by adding
%  $O4^{+}$-plane for a symplectic gauge group and $O4^{-}$-plane for
%  an orthogonal gauge group
%  into the Figure 6 and the appropriate number of D4-branes is 
%considered.
}
\end{figure}
%%%%%%%%%%%%%%%%%%%%%%%%%%%%%%%%%%%%%%%%%%%%%%%%%%%%%%%%%%%%%%%%%%%%%
%Figure 13A and 13B
%%%%%%%%%%%%%%%%%%%%%%%%%%%%%%%%%%%%%%%%%%%%%%%%%%%%%%%%%%%%%%%%%%%%%%%

%Before arriving at the Figure 13A, there exists an intermediate 
%step where 
%$2N_{c,n-2}$ D4-branes between
%$NS5_{n-2}$-brane and the $NS5_{n-1}'$-brane,
%the $2N_{c,n-1}$ D4-branes are connecting between the 
%$NS5_{n-1}'$-brane and the  $NS5_{n+1}'$-brane,  
%$(N_{c,n-1}-N_{c,n}+2)$ D4-branes are connecting between the  
%$NS5_{n+1}'$-brane and   
%$NS5_n$-brane(and their mirrors).
By  
moving the $NS5_{n-1}'$-brane with the $(N_{c,n-1}-N_{c,n-2})$ D4-branes 
to $+v$ direction where we introduce $-2N_{c,n-2}$ D4-branes and
$-2N_{c,n-2}$ anti D4-branes between the $NS5_{n+1}'$-brane and the 
$NS5_{n}$-brane, 
one gets the final Figure 13A where we are left with 
$2(N_{c,n}-N_{c,n-2}-2)$ 
anti-D4-branes between the $NS5_{n+1}$-brane and   
the $NS5_n'$-brane.
%We assume that $N_c' \geq N_c'' \geq 2N_c$.
When two NS5'-branes in Figure 13A are close to each other, then 
it leads to Figure 13B
 by realizing that the number of $(N_{c,n-1}-N_{c,n-2})$
D4-branes connecting between the 
$NS5_{n-1}'$-brane and the $NS5_{n}$-brane can
be rewritten as $(N_{c,n}-N_{c,n-2}-2)$ plus $\widetilde{N}_{c,n}$.

%The brane configuration in Figure 13A is stable as long as the
%distance $(\Delta x)_{n-1}$ between the upper NS5'-brane and 
%the middle NS5'-brane(or $NS5_{n+1}'$-brane) 
%is large. If they are close to each other, then this brane
%configuration is unstable to decay to 
%the brane configuration in Figure
%13B.
%One can regard these brane configurations as particular states in the
%magnetic gauge theory with the gauge group and
%superpotential.
%The  upper $(N_{c,n-1}-N_{c,n-2}-\widetilde{N}_{c,n})$ flavor D4-branes of 
%straight brane configuration
%of
%Figure 13B  bend since there exists an attractive
%gravitational interaction
%between those flavor D4-branes and NS5-brane from the DBI action. 
%Of course, their mirrors, the lower 
%$(N_{c,n-1}-N_{c,n-2}-\widetilde{N}_{c,n})$ flavor D4-branes of 
%straight brane configuration
%of
%Figure 13B can bend and their trajectories connecting 
%two NS5'-branes should be preserved under the O4-plane.

The matter contents are the field $f_{n-1}$ 
 charged under
$({\bf 1_1, \cdots, 1_{n-2}, 2N_{c,n-1}, 2\widetilde{N}_{c,n}})$ 
under the dual gauge group (\ref{dualgauge})
and  
the gauge-singlet $\Phi_{n-1}$ that is in the 
adjoint representation 
for the $(n-1)$-th dual gauge group, 
 ${(\bf 1_1, \cdots, 1_{n-2},
 (N_{c,n-1}-N_{c,n-2})(2N_{c,n-1}-2N_{c,n-2}+1), 1_n) }$ 
under the 
dual gauge group where the gauge group is broken from
$Sp(N_{c,n-1})$ 
to $Sp(N_{c,n-1}-N_{c,n-2})$. That is, the $\Phi_{n-1}$ is  an 
$2(N_{c,n-1}-N_{c,n-2}) \times 2(N_{c,n-1}-N_{c,n-2})$ symmetric matrix.
The cubic superpotential with the mass term  in the dual
theory is given by
\bea
W_{dual} = \Phi_{n-1} f_{n-1} f_{n-1} + m_{n-1} \tr \Phi_{n-1} 
\label{sup-1}
\eea
where we define $\Phi_{n-1}$ as $\Phi_{n-1} \equiv F_{n-1} F_{n-1}$ and 
the $n$-th gauge group indices in $F_{n-1}$  
are contracted, each $(n-1)$-th gauge group index in $F_{n-1}$ is encoded in 
$\Phi_{n-1}$. 
Here the magnetic field $f_{n-1}$  
correspond to 4-4 strings connecting 
the $2\widetilde{N}_{c,n}$-color D4-branes including the mirrors(that are 
connecting between the $NS5_{n-1}'$-brane
and the $NS5_{n}$-brane in Figure 13B) with $2N_{c,n-1}$-flavor 
D4-branes.
Among these $2N_{c,n-1}$-flavor D4-branes, only the strings ending on
the $2(N_{c,n-1}-N_{c,n}+2)$ D4-branes and 
on the tilted  $2(N_{c,n}-N_{c,n-2}-2)$ 
D4-branes in Figure 13B enter the cubic superpotential term (\ref{sup-1}). 
%Although the $(N_{c,n-1}-N_{c,n-2})$ D4-branes(and its mirrors) 
%in Figure 13A for fixed other branes cannot move any
%directions,
%the tilted $(N_{c,n}-N_{c,n-2}-2)$-flavor D4-branes(and its mirrors) 
%can move $w$ direction.
%The remaining upper and lower 
%$\widetilde{N}_{c,n}$ D4-branes are fixed also and cannot 
%move any direction. 
%Note that 
%there is a decomposition 
%\bea
%(N_{c,n-1}-N_{c,n-2})=(N_{c,n}-N_{c,n-2}-2)+\widetilde{N}_{c,n}.
%\nonu
%\eea 

%The brane configuration for zero mass for the bifundamental,
%which has only a cubic superpotential (\ref{sup-1}),
%can be obtained from Figure 13A by moving
%the upper and lower NS5'-branes together with $(N_{c,n-1}-N_{c,n-2})$ color D4-branes 
%into the origin $v=0$.
%Then the number of dual colors for D4-branes 
%becomes $2N_{c,n-2}$ between the $NS5_{n-2}$-brane and the $NS5_{n-1}'$-brane 
%and $2N_{c,n-1}$ between the $NS5_{n-1}'$-brane and the $NS5_{n+1}'$-brane and 
%$2\widetilde{N}_{c,n}$ between the $NS5_{n+1}'$-brane and the $NS5_n$-brane.
%Or starting from Figure 9B and moving the $NS5_{n+1}'$-brane to the left all the
%way past the $NS5_n$-brane,
%one also obtains the corresponding magnetic brane configuration
%for massless case.

When the upper NS5'-brane(or $NS5_{n-1}'$-brane) 
is replaced by coincident $(N_{c,n-1}-N_{c,n-2})$ 
D6-branes and 
the $NS5_{n-2}$ is rotated by an angle $\frac{\pi}{2}$ in the $(v,w)$
plane in Figure 13B, this brane configuration reduces to the one 
found in \cite{Ahn07-8} where the gauge group was given by 
$ \times \cdots \times SO(2n_{c,n-2}) 
\times Sp(n_{c,n-1}) \times SO(2n_{f,n}+2n_{c,n-1}-2n_{c,n}+4)  $ 
with $n_{f,n}$ multiplets, 
bifundamentals and gauge 
singlets. 
Then the present number $(N_{c,n-1}-N_{c,n-2})$ corresponds to the $n_{f,n}$, the
number $N_{c,n}$ corresponds to $n_{c,n}$ and 
the number $N_{c,n-2}$ corresponds to the $n_{c,n-1}$.
%Note that the number of D4-branes touching $NS5_{n-1}'$-brane in Figure 13B
%is equal to $(N_{c,n-1}-N_{c,n-2})$.
%In particular, the Figure 7B of \cite{Ahn07-8} with vanishing flavors
%$Q$ and $Q'$ and reversing the O4-plane charges
%is contained in
%this modified Figure 13B running from the $NS5_{n-2}$-brane to 
%the $NS5_{n}$-brane.
The dual gauge theory has  a meson field  $\Phi_{n-1}$  and 
bifundamentals $f_{n-1}$ under the dual gauge
group (\ref{dualgauge}) and the superpotential (\ref{sup-1}) 
corresponding to Figures 13A and 13B is given by 
\bea
W_{dual} = h \Phi_{n-1} f_{n-1} f_{n-1} - 
h \mu_{n-1}^2 \tr \Phi_{n-1}, \qquad h^2 = g_{n-1,
  mag}^2,
\qquad \mu_{n-1}^2 = -\frac{(\Delta x)_{n-1}}{ 2\pi g_s \ell_s^3}.
\nonu
\eea
Then $ f_{n-1} f_{n-1}$ is a $2\widetilde{N}_{c,n} \times 2\widetilde{N}_{c,n}$ 
matrix where the $(n-1)$-th gauge group indices for $f_{n-1}$  
are contracted with those
of $\Phi_{n-1}$ while $\Phi_{n-1}$ is a 
$2(N_{c,n-1}-N_{c,n-2}) \times 2(N_{c,n-1}-N_{c,n-2})$ matrix.
%The product $f_{n-1} f_{n-1}$ has the same representation for the 
%product of quarks
%and moreover, 
%the $(n-1)$-th gauge group indices for the field $\Phi_{n-1}$ play the
%role of the flavor indices.
Therefore, the F-term equation, the derivative $W_{dual}$ with respect to the
meson field $\Phi_{n-1}$ cannot be satisfied if the $2(N_{c,n-1}-N_{c,n-2})$ exceeds
$2\widetilde{N}_{c,n}$.
So the supersymmetry is broken.   
That is, 
there exist two equations from F-term conditions:
$
f_{n-1}^a f_{n-1}^b -\mu_{n-1}^2 \delta^{a,b} =0$ and $ \Phi_{n-1} f_{n-1} =0$.
Then the solutions for these
are given by
\bea
<f_{n-1}>   = 
\left(
\begin{array}{c}
\mu_{n-1}  {\bf 1}_{2\widetilde{N}_{c,n}}  \\
0
\end{array}
\right), \qquad
<\Phi_{n-1}> =
 \left(
\begin{array}{cc}
0  & 0  \\
0 & M_{n-1}  {\bf 1}_{2(N_{c,n-1}-N_{c,n-2}-\widetilde{N}_{c,n})} 
\end{array}
\right).
\nonu 
%\label{poi-1}
\eea
%where the zero of $<f_{n-1}>$ is a $
%2(N_{c,n-1}-N_{c,n-2}-\widetilde{N}_{c,n}) \times 2\widetilde{N}_{c,n}$ 
%matrix and 
%the zeros of $<\Phi_{n-1}>$ are $2\widetilde{N}_{c,n} \times 2\widetilde{N}_{c,n}$,
%$2\widetilde{N}_{c,n} \times 2(N_{c,n-1}-N_{c,n-2}-\widetilde{N}_{c,n})$, 
%and $2(N_{c,n-1}-N_{c,n-2}-\widetilde{N}_{c,n}) \times
%2\widetilde{N}_{c,n}$ matrices.
%Then one can expand these fields around on a point (\ref{poi-1}) 
%and one arrives at the relevant superpotential
%up to quadratic order in the fluctuation. 
%At one loop, the effective potential $V_{eff}^{(1)}$ for $M_{n-1}$
%leads to the positive value for $m_{M_{n-1}}^2$ implying that these
%vacua are stable.

%%%%%%%%%%%%%%%%%%%%%%%%%%%%%%%%%%%%%%%%%%%%%%%%%%%%%%%%%%%%%%%%%%%%%%%
%%%%%%%%%%%%%%%%%%%%%%%%%%%%%%%%%%%%%%%%%%%%%%%%%%%%%%%%%%%%%%%%%%%%%%%
\subsection{ ${\cal N}=1$
$Sp(\widetilde{N}_{c,1}) 
\times \cdots \times Sp(N_{c,n})[SO(2N_{c,n})]$ magnetic theory}
%%%%%%%%%%%%%%%%%%%%%%%%%%%%%%%%%%%%%%%%%%%%%%%%%%%%%%%%%%%%%%%%%%%%%%%%
%%%%%%%%%%%%%%%%%%%%%%%%%%%%%%%%%%%%%%%%%%%%%%%%%%%%%%%%%%%%%%%%%%%%%%%%

Let us consider the Seiberg dual for the first gauge group
factor.
Starting from Figure 8A, moving the $NS5_{3}'$-brane 
with $(N_{c,2}-N_{c,3})$
D4-branes 
to the $+v$ direction leading to Figure 9A, 
and interchanging the $NS5_{1}'$-brane and the $NS5_{2}$-brane,
one obtains the Figure 14A.

%%%%%%%%%%%%%%%%%%%%%%%%%%%%%%%%%%%%%%%%%%%%%%%%%%%%%%%%%%%%%%%%%%%
%%%%%%%%%%%%%%%%%%%%%%%%%%%%%%%%%%%%%%%%%%%%%%%%%%%%%%%%%%%%%%%%%%%%%%
\begin{figure}[ht]
   \epsfxsize=4.0in 
\centerline{\epsffile{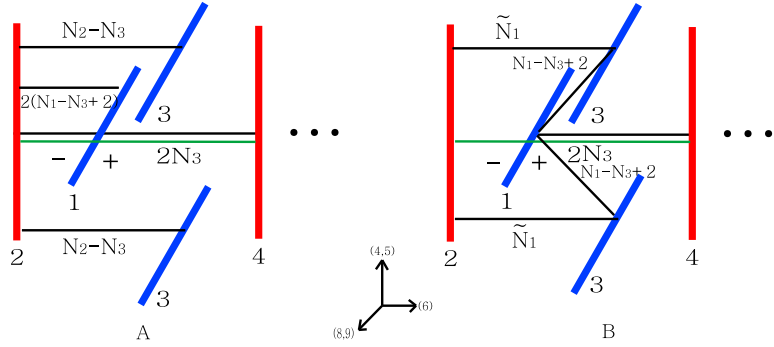}}
   \caption[FIG. \arabic{figure}.]{ 
The 
${\cal N}=1$ magnetic brane configuration for the gauge group 
containing $Sp(\widetilde{N}_{c,1}=N_{c,2}-N_{c,1}-2)$ 
with D4-
and $\overline{D4}$-branes(14A) and with 
a misalignment between D4-branes(14B) when the two NS5'-branes are close to
each other. 
The number of tilted D4-branes in 14B can be written as
$N_{c,1}-N_{c,3}+2 =(N_{c,2}-N_{c,3})-\widetilde{N}_{c,1}$.
The $x$ coordinate of $NS5_{3}'$-brane is given by $ \pm (\Delta
 x)_{2}$.
%This Figure is obtained by adding
%  $O4^{+}$-plane for a symplectic gauge group and $O4^{-}$-plane for
%  an orthogonal gauge group
%  into the Figure 7 and the appropriate number of D4-branes is 
%considered.
}
\end{figure}
%%%%%%%%%%%%%%%%%%%%%%%%%%%%%%%%%%%%%%%%%%%%%%%%%%%%%%%%%%%%%%%%%%%%%
%Figure 14A and 14B
%%%%%%%%%%%%%%%%%%%%%%%%%%%%%%%%%%%%%%%%%%%%%%%%%%%%%%%%%%%%%%%%%%%%%%%

%Before arriving at the Figure 14A, there exists an intermediate 
%step where the $2(N_{c,2}-N_{c,1}-2)$ D4-branes are 
%connecting between the 
%$NS5_{2}$-brane and the  $NS5_{1}'$-brane,  
%$(N_{c,2}-N_{c,3})$ D4-branes connecting between the  $NS5_{1}'$-brane and   
%$NS5_{3}'$-brane, and $N_{c,3}$ D4-branes between the $NS5_{1}'$-brane and
%the $NS5_{4}$-brane(and their mirrors). 
By introducing $-2N_{c,3}$ D4-branes and $-2N_{c,3}$ 
anti-D4-branes  between the  $NS5_{2}$-brane and   
$NS5_{1}'$-brane, reconnecting the former with  
the $2N_{c,2}$ D4-branes connecting between  
$NS5_{2}$-brane 
and the $NS5_{1}'$-brane (therefore $2(N_{c,2}-N_{c,3})$ D4-branes)
and moving those combined
$2(N_{c,2}-N_{c,3})$ 
D4-branes
to $\pm v$-direction, 
one gets the final Figure 14A where we are left with 
$2(N_{c,1}-N_{c,3}+2)$ 
anti-D4-branes between the $NS5_{2}$-brane and   
$NS5_{1}'$-brane.
The dual gauge group is given by 
\bea
Sp(\widetilde{N}_{c,1} \equiv N_{c,2}-N_{c,1}-2) \times 
SO(2N_{c,2}) \times \cdots \times Sp(N_{c,n})[SO(2N_{c,n})]
\label{magnetic}
\eea
where the matter contents are   
the bifundamentals $f_1$ in 
 $({\bf \Box_1, \Box_{2}, \cdots, 1_n})$,
in
addition to $(n-2)$ bifundamentals $F_j$, 
 $j= 2, 3,
\cdots, n$ and
the gauge singlet $\Phi_{2}$
for the first dual gauge group in the 
adjoint representation, 
i.e.,  
$
{(\bf 1_1, (N_{c,2}-N_{c,3})(2N_{c,2}-2N_{c,3}-1), 1_{3}, \cdots, 1_n)}
$
under the 
dual gauge group where the gauge group is broken from
$SO(2N_{c,2})$ 
to $SO(2N_{c,2}-2N_{c,3})$.

%When two NS5'-branes in Figure 14A are close to each other, then 
%it leads to Figure 14B by realizing that the number of $(N_{c,2}-N_{c,3})$
%D4-branes connecting between $NS5_{2}$-brane and $NS5_{3}'$-brane can
%be rewritten as $(N_{c,1}-N_{c,3}+2)$ plus $\widetilde{N}_{c,1}$.
%If we ignore all the D4-branes and NS-branes at the right hand side of 
%$NS_1'$-brane  
%from Figure 14, then
%the brane configuration becomes the one in \cite{GK}.

The cubic superpotential with the mass term  in the dual
theory is given by
\bea
W_{dual} = \Phi_{2} f_{1} f_{1}  + m_{2} \tr \Phi_{2}.
\label{Wdual33}
\eea
Here the magnetic fields $f_1$  
correspond to 4-4 strings connecting 
the $2\widetilde{N}_{c,1}$-color D4-branes(that are 
connecting between the $NS5_{2}$-brane
and the $NS5_{3}'$-brane in Figure 14B) with $2N_{c,2}$-flavor 
D4-branes(that are 
a combination of three different D4-branes in Figure 14B).
Among these $2N_{c,2}$-flavor D4-branes, only the strings ending on
the  $2(N_{c,2}-N_{c,1}-2)$ D4-branes and 
on the tilted middle $2(N_{c,1}-N_{c,3}+2)$ 
D4-branes in Figure 14B enter the cubic superpotential term (\ref{Wdual33}). 

When the upper NS5'-brane(or $NS5_{3}'$-brane) 
is replaced by coincident $(N_{c,2}-N_{c,3})$ 
D6-branes and 
the $NS5_{4}'$-brane is rotated by an angle $\frac{\pi}{2}$ in the $(v,w)$
plane in Figure 14B, this brane configuration reduces to the one 
found in \cite{Ahn07-8} where the gauge group was given by 
$ Sp(n_{f,1}+n_{c,2}-n_{c,1}-2) \times
SO(2n_{c,2}) 
\times \cdots $ 
with $n_{f,i}$ multiplets,  multiplets, bifundamentals and gauge 
singlets. 
Then the present number $(N_{c,2}-N_{c,3})$ corresponds to the $n_{f,1}$, the
number $N_{c,1}$ corresponds to $n_{c,1}$,
and
the number $N_{c,3}$ corresponds to the $n_{c,2}$.
%Note that the number of D4-branes touching $NS5_{3}'$-brane in Figure 3B
%is equal to $(N_{c,2}-N_{c,3})$.
%In particular, the Figure 7B of \cite{Ahn07-8} with vanishing flavors
%$Q$ and $Q'$
%is contained in
%this modified Figure 14B running from the $NS5_{2}$-brane to 
%the $NS5_{4}$-brane.
%The quantum corrections can be understood for small $(\Delta x)_{2}$ by 
%using the low energy field theory on the branes.
%The low energy dynamics of the magnetic brane configuration 
%can be described by the ${\cal N}=1$ supersymmetric gauge theory
%with gauge group (\ref{magnetic})
%and the gauge couplings for the three gauge group factors are
%given similarly.
The dual gauge theory has  a meson field $\Phi_{2}$ and 
bifundamentals $f_1$ and $F_j$ 
under the dual gauge (\ref{magnetic})
group and the superpotential (\ref{Wdual33}) 
corresponding to Figures 14A and 14B is given by 
\bea
W_{dual} = h \Phi_{2} f_1 f_1 - h \mu_2^2 \tr \Phi_{2}, 
\qquad h^2 = g_{2,
  mag}^2,
\qquad \mu_2^2 = -\frac{(\Delta x)_{2}}{ 2\pi g_s \ell_s^3}.
\nonu
\eea
Then $ f_1 f_1$ is a $2\widetilde{N}_{c,1} \times 2\widetilde{N}_{c,1}$ 
matrix where the second gauge group indices for $f_1$ 
are contracted with those
of $\Phi_{2}$ while $\Phi_2$ is a 
$2(N_{c,2}-N_{c,3}) \times 2(N_{c,2}-N_{c,3})$ matrix.
%Although the field $f_1$ itself is an antifundamental in the second gauge
%group
%which is a different  
%representation for the usual standard quark
%coming from D6-branes,
%the product $f_1 f_1$ has the same representation for the 
%product of quarks
%and moreover, 
%the second gauge group indices for the field $\Phi_{2}$ play the
%role of the flavor indices, as in comparison with the brane
%configuration in the presence of D6-branes before.
Therefore, the F-term equation, the derivative $W_{dual}$ with respect to the
meson field $\Phi_{2}$ cannot be satisfied if the $2(N_{c,2}-N_{c,3})$ exceeds
$2\widetilde{N}_{c,1}$.
So the supersymmetry is broken.   
That is, 
there exist three equations from F-term conditions:
$
f_1^a \widetilde{f}_1^b -\mu_2^2 \delta^{a,b} =0$, and 
$\Phi_{2} f_1 =0$.
Then the solutions for these
are given by 
\bea
<f_1>   = 
\left(
\begin{array}{c}
\mu_2  {\bf 1}_{2\widetilde{N}_{c,1}}  \\
0
\end{array}
\right), 
\qquad
<\Phi_{2}> =
 \left(
\begin{array}{cc}
0  & 0  \\
0 & M_{2}  {\bf 1}_{(N_{c,2}-N_{c,3}-\widetilde{N}_{c,1}) 
\otimes i \sigma_2} 
\end{array}
\right). 
\nonu
%\label{point3}
\eea
%where the zero of $<f_1>$ is a $
%2(N_{c,2}-N_{c,3}-\widetilde{N}_{c,1}) \times 2\widetilde{N}_{c,1}$ 
%matrix and 
%the zeros of $<\Phi_{2}>$ are $2\widetilde{N}_{c,1} \times 2\widetilde{N}_{c,1}$,
%$2\widetilde{N}_{c,1} \times 2(N_{c,2}-N_{c,3}-\widetilde{N}_{c,1})$, 
%and $2(N_{c,2}-N_{c,3}-\widetilde{N}_{c,1}) \times
%2\widetilde{N}_{c,1}$ matrices.
%Then one can expand these fields around on a point (\ref{point3})
%and one arrives at the relevant superpotential
%up to quadratic order in the fluctuation. 
%At one loop, the effective potential $V_{eff}^{(1)}$ for $M_{2}$
%leads to the positive value for $m_{M_{2}}^2$ implying that these
%vacua are stable.

As in footnote \ref{footn}, 
when the O4-plane charges are reversed, then the gauge group will be 
either $\left[\prod_{i=1}^{n-2} SO(2N_{c,i}) \times
Sp(2N_{c,i+1})\right] \times SO(2N_{c,n})$ for odd $n$ or 
 $\prod_{i=1}^{n-1} SO(N_{c,i}) \times
Sp(2N_{c,i+1})$ for even $n$. By following what we done so far,
one can analyze the magnetic duals for these cases also by
realizing that the adjoint of symplectic gauge group is symmetric
matrix while the adjoint of orthogonal gauge group is 
antisymmetric.

%%%%%%%%%%%%%%%%%%%%%%%%%%%%%%%%%%%%%%%%%%%%%%%%%%%%%%%%%%%%%%%%%%%%%%%%%%%%
%%%%%%%%%%%%%%%%%%%%%%%%%%%%%%%%%%%%%%%%%%%%%%%%%%%%%%%%%%%%%%%%%%%%%%%%%%%%
\section{Meta-stable brane configurations  with $2n$ NS-branes plus 
$O6^{-}$-plane}
%section4%%%%%%%%%%%%%%%%%%%%%%%%%%%%%%%%%%%%%%%%%%%%%%%%%%%%%%%%%%%%%%%%%%%
%%%%%%%%%%%%%%%%%%%%%%%%%%%%%%%%%%%%%%%%%%%%%%%%%%%%%%%%%%%%%%%%%%%%%%%%%%%%

The type IIA brane configuration, by generalizing the brane
configurations \cite{CSST,LO,Ahn07-3} to the case where there are more
NS-branes,  
corresponding to 
${\cal N}=1$ supersymmetric electric gauge theory(see also
\cite{Ahn07-9}) 
with
gauge group
\bea
Sp(N_{c,1}) \times SU(N_{c,2}) \cdots
\times SU(N_{c,n})
\nonu
\eea
and with 
the $(n-1)$ bifundametals $F_i$ charged under 
$({\bf 1_1, \cdots, 1, \Box_i, \overline{\Box}_{i+1}, 1, \cdots,  1_n})$
and their
complex conjugate fields $\widetilde{F}_i$ 
charged $({\bf 1_1, \cdots, 1, \overline{\Box}_i, \Box_{i+1}, 1, 
\cdots, 1_n})$ where $i=1, 2, \cdots, (n-1)$
can be described by 
the $NS5_1'$-brane, 
the  
$NS5_2$-brane, $\cdots$, the $NS5_{n}'$-brane for odd number
of gauge groups(or 
the $NS5_{n}$-brane for even number of gauge groups),
$2N_{c,1}$-, $N_{c,2}$-,  $\cdots$, and $N_{c,n}$-color D4-branes. 
See the Figure 15 for the details on the brane configuration. 
The $O6^{-}$-plane acts as $(x^4,x^5,x^6) \rightarrow
(-x^4,-x^5,-x^6)$ and has RR charge $-4$.

%%%%%%%%%%%%%%%%%%%%%%%%%%%%%%%%%%%%%%%%%%%%%%%%%%%%%%%%%%%%%%%%%%%%%
%%%%%%%%%%%%%%%%%%%%%%%%%%%%%%%%%%%%%%%%%%%%%%%%%%%%%%%%%%%%%%%%%%%%%%
\begin{figure}[ht]
   \epsfxsize=4.0in 
\centerline{\epsffile{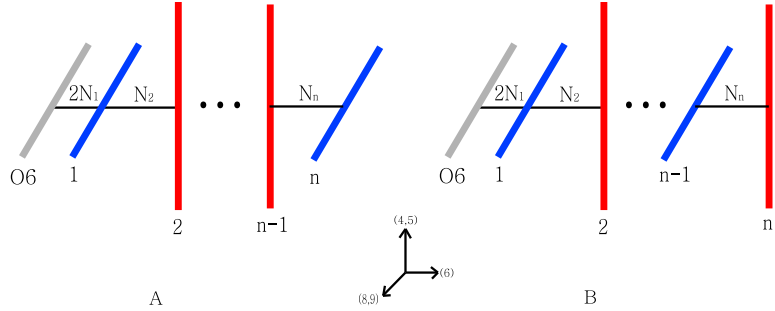}}
   \caption[FIG. \arabic{figure}.]{ 
The 
 ${\cal N}=1$ supersymmetric 
electric brane configuration for the gauge group $Sp(N_{c,1})
\times \prod_{i=2}^{n}
 SU(N_{c,i})$ 
and  bifundamentals $F_i$ and $\widetilde{F}_i$  with vanishing mass
for the bifundamentals when the number of gauge groups factor 
$n$ is odd(15A) and even(15B). 
We do not draw the mirrors of the branes appearing in the left hand
 side of $O6^{-}$-plane.  
%Note that the brane configuration 15A without 
%$O6^{+}$-plane looks similar to 1B while the one in 15B without 
%$O6^{+}$-plane looks similar to 1A. 
}
\end{figure}
%%%%%%%%%%%%%%%%%%%%%%%%%%%%%%%%%%%%%%%%%%%%%%%%%%%%%%%%%%%%%%%%%%%%%
%Figure 15A and 15B
%%%%%%%%%%%%%%%%%%%%%%%%%%%%%%%%%%%%%%%%%%%%%%%%%%%%%%%%%%%%%%%%%%%%%%%%

Let us place an O6-plane at the origin $x^6=0$
and denote the $x^6$ 
coordinates for 
the  
$NS5_1'$-brane, $\cdots$, the $NS5_{n}'$-brane for odd $n$(or 
the $NS5_{n}$-brane for even $n$)
are given by $x^6=y_1, y_1+y_2, \cdots, \sum_{j=1}^{n-1} y_j + y_{n}$
respectively.
The $2N_{c,1}$ D4-branes 
are suspended between the 
$NS5_1'$-brane and its mirror, 
the $N_{c,2}$ D4-branes 
are suspending between the 
$NS5_1'$-brane and the $NS5_2$-brane, $\cdots$ and 
the $N_{c,n}$ D4-branes  
are suspended between the $NS5_{n-1}$-brane and the $NS5_{n}'$-brane for
odd $n$(or between the $NS5_{n-1}'$-brane and
the $NS5_{n}$-brane for even $n$).
The fields $F_i$ and $\widetilde{F}_i$  correspond to 4-4 strings connecting 
the $N_{c,i}$-color D4-branes with $N_{c,i+1}$-color D4-branes.
We draw this ${\cal N}=1$ supersymmetric 
electric brane configuration in Figure 15A(15B) 
when $n$ is odd(even) for the vanishing mass
for the fields $F_i$ and $\widetilde{F}_i$. 

There is no superpotential in Figure 15A. Let us deform this theory.
Displacing the two NS5'-branes relative each other in the $+v$
direction, characterized by $(\Delta x)_{i-1}$,  
corresponds to turning on a quadratic
mass-deformed superpotential
for the field $F_{i-1}$ and $\widetilde{F}_{i-1}$ as follows:
\bea
W_{elec} = m_{i-1} F_{i-1} \widetilde{F}_{i-1} (\equiv m_{i-1}
\Phi_{i-1}), \qquad \mbox{when $i$ is odd}
\label{Mass11}
\eea
where 
the $i$-th gauge group indices in $F_{i-1}$ and $\widetilde{F}_{i-1}$ 
are contracted and the mass $m_{i-1}$ is given by .
\bea
m_{i-1} 
%=\frac{(\Delta x)_{i+1}}{2\pi \alpha'} 
= 
\frac{(\Delta x)_{i-1}}{\ell_s^2}.
\label{m22}
\eea
The gauge-singlet $\Phi_{i-1}$ for the $i$-th  gauge group is in the 
adjoint representation for the $(i-1)$-th  gauge group, 
i.e., ${\bf ( 1_1, \cdots, 1_{i-2}, (N_{c,i-1}-N_{c,i-2})^2-1, 
\cdots, 1_n)  \oplus (1_1, \cdots ,1_n) }$ 
under the  gauge group where the gauge group is broken from
$SU(N_{c,i-1})$ 
to $SU(N_{c,i-1}-N_{c,i-2})$. 
The $\Phi_{i-1}$ is a $(N_{c,i-1}-N_{c,i-2}) \times (N_{c,i-1}-N_{c,i-2})$ matrix.
The $NS5_{i-2}'$-brane together with $(N_{c,i-1}-N_{c,i-2})$-color D4-branes 
is moving to the $+v$ direction  for
fixed other branes during this mass deformation(and their mirrors to
$-v$ direction). 
Then the $x^5$ coordinate 
of $NS5_i'$-brane is equal to
zero
while the $x^5$ coordinate of $NS5_{i-2}'$-branes is given by 
$ (\Delta x)_{i-1}$.
Giving an expectation value to the meson field $\Phi_{i-1}$
corresponds to recombination of $N_{c,i-1}$- and $N_{c,i}$- color 
D4-branes, which will become $N_{c,i-1}$ or $N_{c,i}$-color D4-branes
in Figure 15A such that they are suspended between 
the $NS5_{i-2}'$-brane and the $NS5_i'$-brane 
and pushing them into the $w$
direction. 
We assume that the number of colors satisfies
$
N_{c,i-1} \geq N_{c,i}-N_{c,i+1} \geq N_{c,i-2}$.

Now 
we draw this brane configuration in Figure 16A for nonvanishing mass
for the fields $F_{i-1}$ and $\widetilde{F}_{i-1}$.

%%%%%%%%%%%%%%%%%%%%%%%%%%%%%%%%%%%%%%%%%%%%%%%%%%%%%%%%%%%%%%%%%%%%%%%
%%%%%%%%%%%%%%%%%%%%%%%%%%%%%%%%%%%%%%%%%%%%%%%%%%%%%%%%%%%%%%%%%%%%%%
\begin{figure}[ht]
   \epsfxsize=4.0in 
\centerline{\epsffile{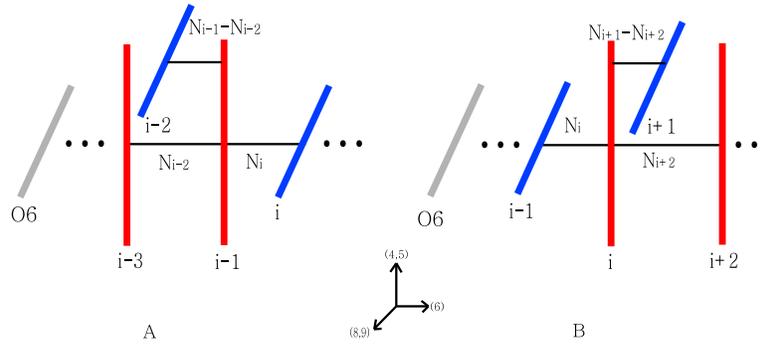}}
   \caption[FIG. \arabic{figure}.]{ 
The 
 ${\cal N}=1$ supersymmetric 
electric brane configuration for the gauge group $Sp(N_{c,1}) \times 
\prod_{i=2}^{n}
 SU(N_{c,i})$ 
and  bifundamentals $F_i$ and $\widetilde{F}_i$  with nonvanishing mass
for the bifundamentals when the number of gauge groups factor 
$n$ is odd(16A) and even(16B).  
The $N_{c,i-1}$ D4-branes in 15A are decomposed into 
$(N_{c,i-1}-N_{c,i-2})$ D4-branes which are moving to $+v$ direction
 in 16A 
and $N_{c,i-2}$ D4-branes which are recombined with those D4-branes
connecting between $NS5_{i-3}$-brane and $NS5_{i-2}'$-brane in 16A.
The $N_{c,i+1}$ D4-branes in 16B are decomposed into 
$(N_{c,i+1}-N_{c,i+2})$ D4-branes which are moving to $+v$ direction in 16B 
and $N_{c,i+2}$ D4-branes which are recombined with those D4-branes
connecting between $NS5_{i+1}'$-brane and $NS5_{i+2}$-brane in 16B.
%Note that the brane configuration 16A without 
%$O6^{+}$-plane looks similar to 2B while the one in 16B without 
%$O6^{+}$-plane looks similar to 2A.
}
\end{figure}
%%%%%%%%%%%%%%%%%%%%%%%%%%%%%%%%%%%%%%%%%%%%%%%%%%%%%%%%%%%%%%%%%%%%%
%Figure 16A and 16B
%%%%%%%%%%%%%%%%%%%%%%%%%%%%%%%%%%%%%%%%%%%%%%%%%%%%%%%%%%%%%%%%%%%%%%%

Let us deform the theory by Figure 15B.
Displacing the two NS5'-branes, the $NS_{i-1}'$-brane and the 
$NS_{i+1}'$-brane, 
relative each other in the 
$v$ 
direction, characterized by $(\Delta x)_{i+1}$, 
corresponds to turning on a quadratic
mass-deformed superpotential
for the fields $F_{i}$ and $\widetilde{F}_{i}$ as follows:
\bea
W = m_{i+1} F_{i} \widetilde{F}_{i} (\equiv m_{i+1} \Phi_{i+1}),
\qquad \mbox{when $i$ is even}
\label{mass111}
\eea
where 
the $i$-th gauge group indices in $F_{i}$ and $\widetilde{F}_{i}$ 
are contracted, each $(i+1)$-th gauge group index in them is encoded in 
$\Phi_{i+1}$ and the mass $m_{i+1}$ is given by
\bea
m_{i-1} 
%=\frac{(\Delta x)_{i+1}}{2\pi \alpha'} 
= 
\frac{(\Delta x)_{i+1}}{\ell_s^2}.
\label{m111}
\eea

The gauge-singlet $\Phi_{i+1}$ for the $i$-th  gauge group is in the 
adjoint representation for the $(i+1)$-th  gauge group, 
i.e., 
\bea
{(\bf 1_1, \cdots, 1_{i}, (N_{c,i+1}-N_{c,i+2})^2-1, 1_{i+2}, \cdots,
  1_n)  
\oplus (1_1, \cdots, 1_n)}
\nonu
\eea 
under the  gauge group where the gauge group is broken from
$SU(N_{c,i+1})$ 
to $SU(N_{c,i+1}-N_{c,i+2})$. 
Then the $\Phi_{i+1}$ is a $(N_{c,i+1}-N_{c,i+2}) \times 
(N_{c,i+1}-N_{c,i+2})$ matrix.
The $NS5_{i+1}'$-brane together with $(N_{c,i+1}-N_{c,i+2})$-color D4-branes 
is moving to the $+v$ direction  for
fixed other branes during this mass deformation. 
In other words, the $N_{c, i+2}$ D4-branes among $N_{c,i+1}$ D4-branes 
are not participating in 
the mass deformation.
Then the $x^5$ coordinate($\equiv x$) 
of $NS5_{i-1}'$-brane is equal to
zero
while the $x^5$ coordinate of $NS5_{i+1}'$-brane is given by 
$(\Delta x)_{i+1}$.
Giving an expectation value to the meson field $\Phi_{i+1}$
corresponds to recombination of $N_{c,i}$- and $N_{c,i+1}$- color 
D4-branes, which will become $N_{c,i}$- or $N_{c,i+1}$-color D4-branes
in Figure 15B such that they are suspended between 
the $NS5_{i-1}'$-brane and the $NS5_{i+1}'$-brane 
and pushing them into the 
$w$ direction. We assume that the number of colors satisfies
$
N_{c,i+1} \geq N_{c,i}-N_{c,i-1} \geq N_{c,i+2}$.
Now 
we draw this brane configuration in Figure 16B for nonvanishing mass
for the fields $F_{i}$ and $\widetilde{F}_{i}$. 

%Next we describe five different magnetic dual theories by taking each
%corresponding mass deformation.
%In principle, since the brane configurations in this section are similar to
%the ones in section 2, one can analyze those magnetic dual theories
%without any difficulty. The role of odd and even chains in section 2
%is reversed, as we noticed in Figures 15 and 16. 

%%%%%%%%%%%%%%%%%%%%%%%%%%%%%%%%%%%%%%%%%%%%%%%%%%%%%%%%%%%%%%%%%%%%%%%%%%%
%%%%%%%%%%%%%%%%%%%%%%%%%%%%%%%%%%%%%%%%%%%%%%%%%%%%%%%%%%%%%%%%%%%%%%%%%%%
\subsection{
${\cal N}=1$ 
$Sp(N_{c,1}) \times \cdots \times SU(\widetilde{N}_{c,i}) \times \cdots
\times
SU(N_{c,n})$ magnetic theory
}
%%%%%%%%%%%%%%%%%%%%%%%%%%%%%%%%%%%%%%%%%%%%%%%%%%%%%%%%%%%%%%%%%%%%%%%%%%%
%%%%%%%%%%%%%%%%%%%%%%%%%%%%%%%%%%%%%%%%%%%%%%%%%%%%%%%%%%%%%%%%%%%%%%%%%%%

%Let us first consider the Seiberg dual for the middle gauge group
%factor.
%There are two magnetic duals depending on whether the gauge group factor
%occurs at odd chain or even chain.

%%%%%%%%%%%%%%%%%%%%%%%%%%%%%%%%%%%%%%%%%%%%%%%%%%%%%%%%%%%%%%
%%%%%%%%%%%%%%%%%%%%%%%%%%%%%%%%%%%%%%%%%%%%%%%%%%%%%%%%%%%%%%
\subsubsection{When the dual gauge group occurs at odd chain}
%%%%%%%%%%%%%%%%%%%%%%%%%%%%%%%%%%%%%%%%%%%%%%%%%%%%%%%%%%%%%%
%%%%%%%%%%%%%%%%%%%%%%%%%%%%%%%%%%%%%%%%%%%%%%%%%%%%%%%%%%%%%%

Starting from Figure 16A and 
interchanging the $NS5_{i-1}$-brane and the $NS5_i'$-brane(and their mirrors),
one obtains the Figure 17A.

%%%%%%%%%%%%%%%%%%%%%%%%%%%%%%%%%%%%%%%%%%%%%%%%%%%%%%%%%%%%%%%%%%%%%
%%%%%%%%%%%%%%%%%%%%%%%%%%%%%%%%%%%%%%%%%%%%%%%%%%%%%%%%%%%%%%%%%%%%%%
\begin{figure}[ht]
   \epsfxsize=4.0in 
\centerline{\epsffile{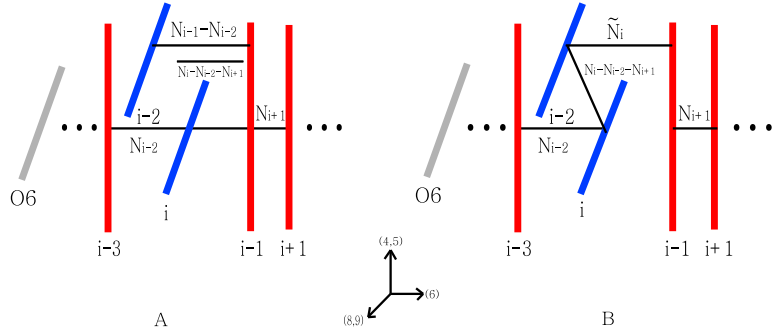}}
   \caption[FIG. \arabic{figure}.]{ 
The 
 ${\cal N}=1$ magnetic brane configuration for the gauge group 
containing $SU(\widetilde{N}_{c,i}=N_{c,i-1}+N_{c,i+1}-N_{c,i})$ 
with D4-
and $\overline{D4}$-branes(17A) and with 
a misalignment between D4-branes(18B) when the two NS5'-branes are close to
each other. 
The number of tilted D4-branes in 18B can be written as
$N_{c,i}-N_{c,i+1}-N_{c,i-2}
 =N_{c,i-1}-N_{c,i-2}-\widetilde{N}_{c,i}$.
%Note that the brane configuration Figure 17 without 
%$O6^{+}$-plane looks similar to Figure 4.
%The $x$ coordinate of $NS5_{i-2}'$-brane is given by $(\Delta
% x)_{i-1}$.
}
\end{figure}
%%%%%%%%%%%%%%%%%%%%%%%%%%%%%%%%%%%%%%%%%%%%%%%%%%%%%%%%%%%%%%%%%%%%%
%Figure 17A and 17B
%%%%%%%%%%%%%%%%%%%%%%%%%%%%%%%%%%%%%%%%%%%%%%%%%%%%%%%%%%%%%%%%%%%%%%

%Before arriving at the Figure 17A, there exists an intermediate 
%step where 
%the $(N_{c,i-1}-N_{c,i}+N_{c,i+1})$ D4-branes are connecting between the 
%$NS5_{i}'$-brane and the  $NS5_{i-1}$-brane,  
%$N_{c,i-1}$ D4-branes are connecting between the  $NS5_{i-2}'$-brane and   
%$NS5_i'$-brane(and their mirrors) as well as $N_{c,i-2}$ D4-branes between
%the
%$NS5_{i-3}$-brane and the $NS5_{i-2}'$-brane.
By introducing $-N_{c,i-2}$ D4-branes and $-N_{c,i-2}$ 
anti-D4-branes  between the  $NS5_{i}'$-brane and   
$NS5_{i-1}$-brane, reconnecting the former with  
the $N_{c,i-1}$ D4-branes connecting between  
$NS5_{i}'$-brane 
and the $NS5_{i-1}$-brane (therefore $(N_{c,i-1}-N_{c,i-2})$ D4-branes)
and moving those combined
$(N_{c,i-1}-N_{c,i-2})$ 
D4-branes
to $+v$-direction, 
one gets the final Figure 17A where we are left with 
$(N_{c,i}-N_{c,i+1}-N_{c,i-2})$ 
anti-D4-branes between the $NS5_i'$-brane and   
$NS5_{i-1}$-brane.

The dual gauge group is given by 
\bea
Sp(N_{c,1}) \times \cdots \times 
SU(\widetilde{N}_{c,i} \equiv N_{c,i+1}+N_{c,i-1}-N_{c,i}) 
\times \cdots \times SU(N_{c,n}).
\label{dualnew}
\eea
The matter contents are the field $f_{i-1}$ 
charged under
$({\bf 1_1, \cdots, 1_{i-2}, 
N_{c,i-1}, \overline{\widetilde{N}_{c,i}}, \cdots, 1_n})$, and its 
conjugate field  
$\widetilde{f}_{i-1}$  charged under 
$({\bf 1_1, \cdots, 1_{i-2}, 
\overline{N_{c,i-1}}, \widetilde{N}_{c,i}, \cdots, 1_n})$
under the dual gauge group (\ref{dualnew})
and  
the gauge-singlet $\Phi_{i-1}$ for the $i$-th dual gauge group in the 
adjoint representation for the $(i-1)$-th dual gauge group, 
i.e.,  ${(\bf 1_1, \cdots, 1_{i-2}, (N_{c,i-1}-N_{c,i-2})^2-1, 1,
  \cdots, 1_n)  }$ plus a singlet under the 
dual gauge group where the gauge group is broken from
$SU(N_{c,i-1})$ 
to $SU(N_{c,i-1}-N_{c,i-2})$.
Then the  $\Phi_{i-1}$ is a $(N_{c,i-1}-N_{c,i-2}) \times
(N_{c,i-1}-N_{c,i-2})$ 
matrix.

%When two NS5'-branes in Figure 17A are close to each other, it becomes 
%Figure 17B
%by realizing that the number of $(N_{c,i-1}-N_{c,i-2})$
%D4-branes connecting between $NS5_{i-2}'$-brane and $NS5_{i-1}$-brane in Figure
%17A can
%be rewritten as $(N_{c,i}-N_{c,i+1}-N_{c,i-2})$ plus $\widetilde{N}_{c,i}$. 

The cubic superpotential with the mass term  (\ref{Mass11}) and (\ref{m22})
is given by
\footnote{There exist also 
the extra terms in the superpotential
$\Phi_{i+1} f_{i} \widetilde{f}_{i} + 
\Phi'' \widetilde{f}_{i-1} \widetilde{f}_i + \Phi' f_{i-1} 
f_{i}$ where we define 
$\Phi' \equiv F_i F_{i-1}$ and 
$\Phi'' \equiv \widetilde{F}_{i} \widetilde{F}_{i-1}$, coming from 
different bifundamentals. However, the F- term conditions,
$\Phi_{i+1} \widetilde{f}_i + \Phi' f_{i-1}=0=\Phi_{i+1}
f_i + \Phi'' \widetilde{f}_{i-1}$ lead to 
$<\Phi'>=<\Phi''>=<f_i>=<\widetilde{f}_i>=0$. 
Also these extra terms do not
contribute to the one loop computation up to quadratic order.}
\bea
W_{dual} = \Phi_{i-1} f_{i-1} \widetilde{f}_{i-1} + m_{i-1} \tr \Phi_{i-1}. 
\label{superpo1new}
\eea
Here the magnetic fields $f_{i-1}$ and $\widetilde{f}_{i-1}$  
correspond to 4-4 strings connecting 
the $\widetilde{N}_{c,i}$-color D4-branes(that are 
connecting between the $NS5_{i-2}'$-brane
and the $NS5_{i-1}$-brane in Figure 17B) with $N_{c,i-1}$-flavor 
D4-branes(which  are realized 
as corresponding D4-branes in Figure 17A).

The low energy dynamics of the magnetic brane configuration 
can be described by the ${\cal N}=1$ supersymmetric gauge theory
with gauge group (\ref{dualnew})
and the gauge couplings for the three gauge group factors are
given by similarly.
The dual gauge theory has  a meson field $\Phi_{i-1}$  and 
bifundamentals $f_{i-1}$, and $\widetilde{f}_{i-1}$ under the dual gauge
group (\ref{dualnew}) and the superpotential (\ref{superpo1new}) 
corresponding to Figures 17A and 17B is given by 
\bea
W_{dual} = h \Phi_{i-1} f_{i-1} \widetilde{f}_{i-1} - 
h \mu_{i-1}^2 \tr \Phi_{i-1}, \qquad h^2 = g_{i-1,
  mag}^2,
\qquad \mu_{i-1}^2 = -\frac{(\Delta x)_{i-1}}{ 2\pi g_s \ell_s^3}.
\nonu
\eea
Then $ f_{i-1} \widetilde{f}_{i-1}$ is a 
$\widetilde{N}_{c,i} \times \widetilde{N}_{c,i}$ 
matrix where the $(i-1)$-th gauge group indices for $f_{i-1}$ and 
$\widetilde{f}_{i-1}$ 
are contracted with those
of $\Phi_{i-1}$ while $\Phi_{i-1}$ is a 
$(N_{c,i-1}-N_{c,i-2} )\times (N_{c,i-1}-N_{c,i-2})$ matrix.
%The product $f_{i-1} \widetilde{f}_{i-1}$ has the same representation for the 
%product of quarks
%and moreover, 
%the $(i-1)$-th gauge group indices for the field $\Phi_{i-1}$ play the
%role of the flavor indices.

When the upper NS5'-brane(or $NS5_{i-2}'$-brane) 
is replaced by coincident $(N_{c,i-1}-N_{c,i-2})$ 
D6-branes and the $NS5_{i-3}$ is rotated by $\frac{\i}{2}$ 
in Figure 17B, this brane configuration looks similar to the one 
found in \cite{Ahn07-9} where the gauge group was given by 
$Sp(n_{c,1}) 
\times \cdots \times SU(n_{f,i}+n_{c,i+1}+n_{c,i-1}-n_{c,i}) 
\times SU(n_{c,i+1}) \times
\cdots$ 
with $n_{f,i}$ multiplets, $\widetilde{n}_{f,i}$ multiplets, bifundamentals 
and singlets. 
Then the present $(N_{c,i-1}-N_{c,i-2})$ corresponds to the $n_{f,i}$,
$N_{c,i}$ corresponds to $n_{c,i}$,
$N_{c,i+1}$ corresponds to $n_{c,i+1}$
and 
$N_{c,i-2}$ corresponds to the $n_{c,i-1}$.
%In particular, the Figure 18B of \cite{Ahn07-9} with vanishing flavors
%$Q$ and $Q'$
%is contained in
%this modified Figure 17B running from the $NS5_{i-3}$-brane to 
%the $NS5_{i-1}$-brane.
 Therefore, the F-term equation, the derivative $W_{dual}$ with respect to the
meson field $\Phi_{i-1}$ cannot be satisfied if the $(N_{c,i-1}-N_{c,i-2})$ exceeds
$\widetilde{N}_{c,i}$.
So the supersymmetry is broken.   
That is, 
there exist three equations from F-term conditions:
$
f_{i-1}^a \widetilde{f}_{i-1,b} -\mu_{i-1}^2 \delta^a_b =0$ 
and $ \Phi_{i-1} f_{i-1} =0=
\widetilde{f}_{i-1} \Phi_{i-1}$.
Then the solutions for these
are given by 
\bea
<f_{i-1}>   & = & 
\left(
\begin{array}{c}
\mu_{i-1}  {\bf 1}_{\widetilde{N}_{c,i}}  \\
0
\end{array}
\right), 
\qquad
<\widetilde{f}_{i-1}>   = 
\left(
\begin{array}{cc}
\mu_{i-1}  {\bf 1}_{\widetilde{N}_{c,i}} & 0  \\
\end{array}
\right),
\nonu \\
<\Phi_{i-1}> & = &
 \left(
\begin{array}{cc}
0  & 0  \\
0 & M_{i-1}  {\bf 1}_{(N_{c,i-1}-N_{c,i-2}-\widetilde{N}_{c,i})} 
\end{array}
\right). 
\nonu
%\label{point20}
\eea
%where the zero of $<f_{i-1}>$ is a $
%(N_{c,i-1}-N_{c,i-2}-\widetilde{N}_{c,i}) \times \widetilde{N}_{c,i}$ 
%matrix, the zero of $<\widetilde{f}_{i-1}>$ is a
%$\widetilde{N}_{c,i} 
%\times (N_{c,i-1}-N_{c,i-2}-\widetilde{N}_{c,i}) $ matrix and 
%the zeros of $<\Phi_{i-1}>$ are $\widetilde{N}_{c,i} \times 
%\widetilde{N}_{c,i}$,
%$\widetilde{N}_{c,i} \times 
%(N_{c,i-1}-N_{c,i-2}-\widetilde{N}_{c,i})$ 
%and $(N_{c,i-1}-N_{c,i-2}-\widetilde{N}_{c,i}) \times
%\widetilde{N}_{c,i}$ 
%matrices.
%Then one can expand these fields around on a point (\ref{point20}) 
%and one arrives at the relevant superpotential
%up to quadratic order in the fluctuation. 
%At one loop, the effective potential $V_{eff}^{(1)}$ for $M_{i-1}$
%leads to the positive value for $m_{M_{i-1}}^2$ implying that these
%vacua are stable.

%%%%%%%%%%%%%%%%%%%%%%%%%%%%%%%%%%%%%%%%%%%%%%%%%%%%%%%%%%%%%%
%%%%%%%%%%%%%%%%%%%%%%%%%%%%%%%%%%%%%%%%%%%%%%%%%%%%%%%%%%%%%%
\subsubsection{When the dual gauge group occurs at even chain}
%%%%%%%%%%%%%%%%%%%%%%%%%%%%%%%%%%%%%%%%%%%%%%%%%%%%%%%%%%%%%%
%%%%%%%%%%%%%%%%%%%%%%%%%%%%%%%%%%%%%%%%%%%%%%%%%%%%%%%%%%%%%%

Let us consider other magnetic theory for the same electric theory.
Starting from Figure 16B and 
interchanging the $NS5_{i-1}'$-brane and the $NS5_i$-brane(and their mirrors),
one obtains the Figure 18A.

%%%%%%%%%%%%%%%%%%%%%%%%%%%%%%%%%%%%%%%%%%%%%%%%%%%%%%%%%%%%%%%%%%%%%
%%%%%%%%%%%%%%%%%%%%%%%%%%%%%%%%%%%%%%%%%%%%%%%%%%%%%%%%%%%%%%%%%%%%%%
\begin{figure}[ht]
   \epsfxsize=4.0in 
\centerline{\epsffile{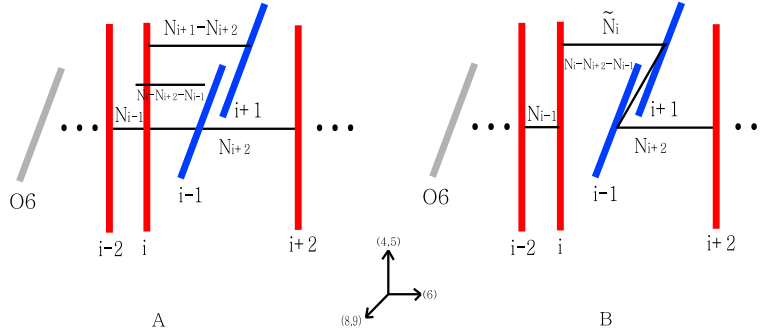}}
   \caption[FIG. \arabic{figure}.]{ 
The 
 ${\cal N}=1$ magnetic brane configuration for the gauge group 
containing $SU(\widetilde{N}_{c,i}=N_{c,i-1}+N_{c,i+1}-N_{c,i})$ 
with D4-
and $\overline{D4}$-branes(18A) and with 
a misalignment between D4-branes(18B) when the two NS5'-branes are close to
each other. 
The number of tilted D4-branes in 18B can be written as
$N_{c,i}-N_{c,i-1}-N_{c,i+2}
 =(N_{c,i+1}-N_{c,i+2})-\widetilde{N}_{c,i}$.
%Note that the brane configuration Figure 18 without 
%$O6^{+}$-plane looks similar to Figure 3.
%The $x$ coordinate of $NS5_{i+1}'$-brane is given by $(\Delta
% x)_{i+1}$.
}
\end{figure}
%%%%%%%%%%%%%%%%%%%%%%%%%%%%%%%%%%%%%%%%%%%%%%%%%%%%%%%%%%%%%%%%%%%%%
%Figure 18A and 18B 
%%%%%%%%%%%%%%%%%%%%%%%%%%%%%%%%%%%%%%%%%%%%%%%%%%%%%%%%%%%%%%%%%%%%%%%%%%%%

%Before arriving at the Figure 18A, there exists an intermediate 
%step where 
%the $(N_{c,i+1}+N_{c,i-1}-N_{c,i})$ D4-branes are connecting between the 
%$NS5_i$-brane and the  $NS5_{i-1}'$-brane,  
%$(N_{c,i+1}-N_{c,i+2})$ D4-branes are connecting between the  
%$NS5_{i-1}'$-brane and   
%$NS5_{i+1}'$-brane(and their mirrors) as well as $N_{c,i-1}$ D4-branes between
%$NS5_{i-2}$-brane and the $NS5_i$-brane.
By reconnecting the $(N_{c,i+1}-N_{c,i+2})$ D4-branes 
connecting between the 
$NS5_i$-brane and the  $NS5_{i-1}'$-brane
with  
those D4-branes connecting between  
$NS5_{i-1}'$-brane 
and the $NS5_{i+1}'$-brane where we introduce $-N_{c,i+2}$ D4-branes and
$-N_{c,i+2}$ anti D4-branes and moving those combined 
D4-branes
to $+v$-direction(and their mirrors to $-v$ direction), 
one gets the final Figure 18A where we are left with 
$(N_{c,i}-N_{c,i+2}-N_{c,i-1})$ 
anti-D4-branes between the $NS5_i$-brane and   
the $NS5_{i-1}'$-brane.
We assume  that the number of colors satisfies
$
N_{c,i+1} \geq N_{c,i}-N_{c,i-1} \geq N_{c,i+2}$.

The dual gauge group is given by
\bea
Sp(N_{c,1}) \times \cdots \times 
SU(\widetilde{N}_{c,i} \equiv N_{c,i+1}+N_{c,i-1}-N_{c,i}) 
\times \cdots \times SU(N_{c,n}).
\label{dualneww}
\eea
The matter contents are the field $f_{i}$ 
 charged under
$({\bf 1, \cdots, \Box_{i}, \overline{\Box}_{i+1}, \cdots, 1_n})$, 
 and their conjugates 
$\widetilde{f}_{i}$ charged $({\bf 1_1, \cdots, \overline{\Box}_{i},
  \Box_{i+1}, \cdots, 1_n })$ 
under the dual gauge group
(\ref{dualneww})
and  
the gauge-singlet $\Phi_{i+1}$ which is in the 
adjoint representation for the $i$-th dual gauge group, 
in other words,   
$ ({ \bf   1_1,  \cdots, 1_{i}, 
(N_{c,i+1}-N_{c,i+2})^2-1, \cdots, 1_n})  \oplus  ({\bf 1_1,
\cdots,1_n})$ 
under the 
dual gauge group (\ref{dualneww}) where the gauge group is broken from
$SU(N_{c,i+1})$ 
to $SU(N_{c,i+1}-N_{c,i+2})$.
Then the $\Phi_{i+1}$ is a $(N_{c,i+1}-N_{c,i+2}) \times
(N_{c,i+1}-N_{c,i+2})$ 
matrix.
Only $(N_{c,i+1}-N_{c,i+2})$ 
D4-branes can participate in the mass deformation.

%When two NS5'-branes in Figure 18A are close to each other, then 
%it leads to Figure 18B
%by realizing that the number of $(N_{c,i+1}-N_{c,i+2})$
%D4-branes connecting between $NS5_i$-brane and $NS5_{i+1}'$-brane in Figure
%18A can
%be rewritten as $(N_{c,i}-N_{c,i+2}-N_{c,i-1})$ plus $\widetilde{N}_{c,i}$.

The cubic superpotential with the mass term (\ref{mass111}) and (\ref{m111})
is given by
 \footnote{Of course, there are also 
the extra terms in the superpotential
$\Phi' f_{i-1} f_{i} + \Phi'' \widetilde{f}_{i-1} \widetilde{f}_i + \Phi_{i-1} f_{i-1} 
\widetilde{f}_{i-1}$ where we define 
$\Phi' \equiv F_i F_{i-1}$ and 
$\Phi'' \equiv \widetilde{F}_{i} \widetilde{F}_{i-1}$, coming from 
different bifundamentals. However, the F- term conditions,
$\Phi' f_i + \Phi_{i-1} \widetilde{f}_{i-1}=0=\Phi''
\widetilde{f}_i + \Phi_{i-1} f_{i-1}$ lead to 
$<\Phi'>=<\Phi''>=<f_{i-1}>=<\widetilde{f}_{i-1}>=0$. 
Then, these extra terms do not
contribute to the one loop computation up to quadratic order.}
\bea
W_{dual} = \Phi_{i+1} f_{i} \widetilde{f}_{i} + m_{i+1} \tr \Phi_{i+1}
\label{superpo11}
\eea
where we define $\Phi_{i+1}$ as $\Phi_{i+1} \equiv F_{i} \widetilde{F}_{i}$ and 
the $i$-th gauge group indices in $F_{i}$ and $\widetilde{F}_{i}$ 
are contracted, each $(i+1)$-th gauge group index in them is encoded in 
$\Phi_{i+1}$.  
Here the magnetic fields $f_{i}$ and $\widetilde{f}_{i}$  
correspond to 4-4 strings connecting 
the $\widetilde{N}_{c,i}$-color D4-branes(that are 
connecting between the $NS5_i$-brane
and the $NS5_{i+1}'$-brane in Figure 18B) with $N_{c,i+1}$-flavor 
D4-branes.
Among these $N_{c,i+1}$-flavor D4-branes, only the strings ending on
the upper $(N_{c,i-1}+N_{c,i+1}-N_{c,i})$ D4-branes and 
on the tilted  $(N_{c,i}-N_{c,i+2}-N_{c,i-1})$ 
D4-branes in Figure 18B enter the cubic superpotential term (\ref{superpo11}). 

When the upper NS5'-brane(or $NS5_{i+1}'$-brane) 
is replaced by coincident $(N_{c,i+1}-N_{c,i+2})$ 
D6-branes with a rotation of $NS5_{i+2}$-brane in Figure 18B, 
this brane configuration looks similar to the one 
found in \cite{Ahn07-9} where the gauge group was given by 
$Sp(n_{c,1}) 
\times \cdots \times SU(n_{f,i}+n_{c,i+1}+n_{c,i-1}-n_{c,i}) 
\times SU(n_{c,i+1}) \times
\cdots$ 
with $n_{f,i}$ multiplets, $\widetilde{n}_{f,i}$ multiplets,  
bifundamentals and gauge singlets. 
Then the present $(N_{c,i+1}-N_{c,i+2})$ corresponds to the $n_{f,i}$,
$N_{c,i}$ corresponds to $n_{c,i}$,
$N_{c,i-1}$ corresponds to $n_{c,i-1}$
and 
$N_{c,i+2}$ corresponds to the $n_{c,i+1}$. 
%In particular, the Figure 16B of \cite{Ahn07-9} with vanishing flavors
%$Q$ and $Q''$
%is contained in
%this modified Figure 18B running from the $NS5_{i}$-brane to 
%the $NS5_{i+2}$-brane.
The low energy dynamics of the magnetic brane configuration 
can be described by the ${\cal N}=1$ supersymmetric gauge theory
with gauge group (\ref{dualneww})
and the gauge couplings for the three gauge group factors are
given by similarly.
The dual gauge theory has  a meson field $\Phi_{i+1}$ and 
bifundamentals $f_{i}, \widetilde{f}_{i}$  under the dual gauge
group (\ref{dualneww}) and the superpotential (\ref{superpo11}) 
corresponding to Figures 18A and 18B is given by 
\bea
W_{dual} = h \Phi_{i+1} f_{i} \widetilde{f}_{i} - 
h \mu_{i+1}^2 \tr \Phi_{i+1}, \qquad h^2 = g_{i+1,
  mag}^2,
\qquad \mu_{i+1}^2 = -\frac{(\Delta x)_{i+1}}{ 2\pi g_s \ell_s^3}.
\nonu
\eea
Then $ f_{i} \widetilde{f}_{i}$ 
is a $\widetilde{N}_{c,i} \times \widetilde{N}_{c,i}$ 
matrix where the $(i+1)$-th gauge group indices for $f_{i+1}$ and 
$\widetilde{f}_{i+1}$ 
are contracted with those
of $\Phi_{i+1}$ while $\Phi_{i+1}$ is a 
$(N_{c,i+1}-N_{c,i+2}) \times (N_{c,i+1}-N_{c,i+2})$ matrix.
%The product $f_{i} \widetilde{f}_{i}$ has the same representation for the 
%product of quarks
%and moreover, 
%the $(i+1)$-th gauge group indices for the field $\Phi_{i+1}$ play the
%role of the flavor indices, as above.
Therefore, the F-term equation, the derivative $W_{dual}$ with respect to the
meson field $\Phi_{i+1}$ cannot be satisfied if the $(N_{c,i+1}-N_{c,i+2})$ exceeds
$\widetilde{N}_{c,i}$.
So the supersymmetry is broken.   
That is, 
there exist three equations from F-term conditions:
$
f_{i}^a \widetilde{f}_{i, b} -\mu_{i+1}^2 \delta^a_b =0$ and $ \Phi_{i+1} f_{i} =
0=\widetilde{f}_{i} \Phi_{i+1}$.
Then the solutions for these
are given by 
\bea
<f_{i}>  & = & 
\left(
\begin{array}{c}
\mu_{i+1}  {\bf 1}_{\widetilde{N}_{c,i}}  \\
0
\end{array}
\right), 
\qquad
<\widetilde{f}_{i}>   = 
\left(
\begin{array}{cc}
\mu_{i+1}  {\bf 1}_{\widetilde{N}_{c,i}} & 0  \\
\end{array}
\right), 
\nonu \\
<\Phi_{i+1}>  & = &
 \left(
\begin{array}{cc}
0  & 0  \\
0 & M_{i+1}  {\bf 1}_{(N_{c,i+1}-N_{c,i+2})-\widetilde{N}_{c,i}} 
\end{array}
\right).
\nonu 
%\label{point12-1}
\eea
%where the zero of $<f_{i}>$ is a $
%(N_{c,i+1}-N_{c,i+2}-\widetilde{N}_{c,i}) \times \widetilde{N}_{c,i}$ 
%matrix, the zero of $<\widetilde{f}_{i}>$ is a
%$\widetilde{N}_{c,i} \times (N_{c,i+1}-N_{c,i+2}-\widetilde{N}_{c,i}) $ matrix and 
%the zeros of $<\Phi_{i+1}>$ are $\widetilde{N}_{c,i} \times \widetilde{N}_{c,i}$,
%$\widetilde{N}_{c,i} \times 
%(N_{c,i+1}-N_{c,i+2}-\widetilde{N}_{c,i})$ and $(N_{c,i+1}-N_{c,i+2}-
%\widetilde{N}_{c,i}) \times
%\widetilde{N}_{c,i}$ 
%matrices.
%Then one can expand these fields around on a point (\ref{point12-1}) 
%and one arrives at the relevant superpotential
%up to quadratic order in the fluctuation. 
%At one loop, the effective potential $V_{eff}^{(1)}$ for $M_{i+1}$
%leads to the positive value for $m_{M_{i+1}}^2$ implying that these
%vacua are stable.

%%%%%%%%%%%%%%%%%%%%%%%%%%%%%%%%%%%%%%%%%%%%%%%%%%%%%%%%%%%%%%%%%%%%%%%%%%%%%%
%%%%%%%%%%%%%%%%%%%%%%%%%%%%%%%%%%%%%%%%%%%%%%%%%%%%%%%%%%%%%%%%%%%%%%%%%%%%%%
\subsection{${\cal N}=1$ 
$Sp(N_{c,1}) \times \cdots \times SU(\widetilde{N}_{c,n})$ magnetic theory}
%%%%%%%%%%%%%%%%%%%%%%%%%%%%%%%%%%%%%%%%%%%%%%%%%%%%%%%%%%%%%%%%%%%%%%%%%%%%%%
%%%%%%%%%%%%%%%%%%%%%%%%%%%%%%%%%%%%%%%%%%%%%%%%%%%%%%%%%%%%%%%%%%%%%%%%%%%%%%

%Let us first consider the Seiberg dual for the last gauge group
%factor.
%There are two magnetic duals depending on whether the gauge group factor
%occurs at odd chain or even chain.

%%%%%%%%%%%%%%%%%%%%%%%%%%%%%%%%%%%%%%%%%%%%%%%%%%%%%%%%%%%%%%
%%%%%%%%%%%%%%%%%%%%%%%%%%%%%%%%%%%%%%%%%%%%%%%%%%%%%%%%%%%%%%
\subsubsection{When the dual gauge group occurs at odd chain}
%%%%%%%%%%%%%%%%%%%%%%%%%%%%%%%%%%%%%%%%%%%%%%%%%%%%%%%%%%%%%%
%%%%%%%%%%%%%%%%%%%%%%%%%%%%%%%%%%%%%%%%%%%%%%%%%%%%%%%%%%%%%%

Starting from Figure 15A, moving the $NS5_{n-2}'$-brane 
with $(N_{c,n-1}-N_{c,n-2})$
D4-branes 
to the $+v$ direction leading to Figure 16A, 
and interchanging the $NS5_{n-1}$-brane and the $NS5_{n}'$-brane,
one obtains the Figure 19A.

%%%%%%%%%%%%%%%%%%%%%%%%%%%%%%%%%%%%%%%%%%%%%%%%%%%%%%%%%%%%%%%%%%%%%%
%%%%%%%%%%%%%%%%%%%%%%%%%%%%%%%%%%%%%%%%%%%%%%%%%%%%%%%%%%%%%%%%%%%%%%
\begin{figure}[ht]
   \epsfxsize=4.0in 
\centerline{\epsffile{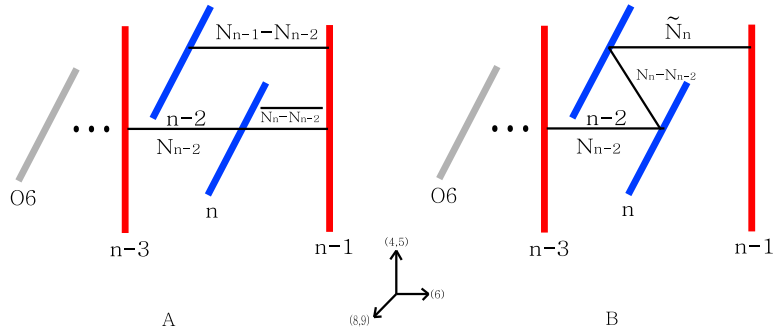}}
   \caption[FIG. \arabic{figure}.]{ 
The 
 ${\cal N}=1$ magnetic brane configuration for the gauge group 
containing $SU(\widetilde{N}_{c,n}=N_{c,n-1}-N_{c,n})$ 
with D4-
and $\overline{D4}$-branes(19A) and with 
a misalignment between D4-branes(19B) when the two NS5'-branes are close to
each other. 
The number of tilted D4-branes in 19B can be written as
$N_{c,n}-N_{c,n-2} =(N_{c,n-1}-N_{c,n-2})-\widetilde{N}_{c,n}$.
%Note that the brane configuration Figure 19 without 
%$O6^{+}$-plane looks similar to Figure 6.
%The $x$ coordinate of $NS5_{n-2}'$-brane is given by $(\Delta
% x)_{n-1}$.
}
\end{figure}
%%%%%%%%%%%%%%%%%%%%%%%%%%%%%%%%%%%%%%%%%%%%%%%%%%%%%%%%%%%%%%%%%%%%%
%Figure 19A and 19B
%%%%%%%%%%%%%%%%%%%%%%%%%%%%%%%%%%%%%%%%%%%%%%%%%%%%%%%%%%%%%%%%%%%%%

%Before arriving at the Figure 19A, there exists an intermediate 
%step where the $(N_{c,n-1}-N_{c,n})$ D4-branes are 
%connecting between the 
%$NS5_{n}'$-brane and the  $NS5_{n-1}$-brane,  
%$(N_{c,n-1}-N_{c,n-2})$ D4-branes connecting between the  $NS5_{n-2}'$-brane and   
%$NS5_{n}'$-brane, and $N_{c,n-2}$ D4-branes between the $NS5_{n-3}$-brane and
%the $NS5_{n}'$-brane. 
By introducing $-N_{c,n-2}$ D4-branes and $-N_{c,n-2}$ 
anti-D4-branes  between the  $NS5_{n}'$-brane and   
$NS5_{n-1}$-brane, reconnecting the former with  
the $N_{c,n-1}$ D4-branes connecting between  
$NS5_{n}'$-brane 
and the $NS5_{n-1}$-brane (therefore $(N_{c,n-1}-N_{c,n-2})$ D4-branes)
and moving those combined
$(N_{c,n-1}-N_{c,n-2})$ 
D4-branes
to $+v$-direction, 
one gets the final Figure 19A where we are left with 
$(N_{c,n}-N_{c,n-2})$ 
anti-D4-branes between the $NS5_{n}'$-brane and   
$NS5_{n-1}$-brane.

The dual gauge group is given by 
\bea
Sp(N_{c,1}) \times \cdots \times 
SU(N_{c,n-1}) \times SU(\widetilde{N}_{c,n} \equiv N_{c,n-1}-N_{c,n})
\label{dualgaugegroup}
\eea
and the matter contents are   
the bifundamentals $f_{n-1}$ in 
 $({\bf 1_1, \cdots, 1_{n-2}, \Box_{n-1}, \overline{\Box}_{n}, })$,
and $\widetilde{f}_{n-1}$ in the representation 
$({\bf 1_1, \cdots, 1_{n-2}, \overline{\Box}_{n-1}, \Box_{n}})$ in
addition to $(n-2)$ bifundamentals $F_j$ and $\widetilde{F}_j$, 
 $j=1,2,
\cdots, (n-2)$ and
the gauge singlet $\Phi_{n-1}$
for the $n$-th dual gauge group  is in the representation   
$
{(\bf 1_1, \cdots, 1_{n-2}, (N_{c,n-1}-N_{c,n-2})^2-1, 1_n)  
}
$ plus a singlet
under the 
dual gauge group (\ref{dualgaugegroup}) where the gauge group is broken from
$SU(N_{c,n-1})$ 
to $SU(N_{c,n-1}-N_{c,n-2})$.

%When two NS5'-branes in Figure 19A are close to each other, then 
%it leads to Figure 19B by realizing that the number of $(N_{c,n-1}-N_{c,n-2})$
%D4-branes connecting between $NS5_{n-2}'$-brane and $NS5_{n-1}$-brane can
%be rewritten as $(N_{c,n}-N_{c,n-2})$ plus $\widetilde{N}_{c,n}$.
%If we ignore all the D4-branes and NS-branes located at the left hand
%side of $NS_{n}'$-brane   
%from Figure 19, then
%the brane configuration becomes the one in \cite{GK}.
%The Figure 14 of \cite{Ahn07-6} is contained in the Figure 19. In
%particular, the brane configuration from the $NS5_{n-2}'$-brane to 
%the $NS5_{n-1}$-brane is exactly same as the one of \cite{Ahn07-6}.

The cubic superpotential with the mass term  in the dual
theory is given by
\bea
W_{dual} = \Phi_{n-1} f_{n-1} \widetilde{f}_{n-1}  + m_{n-1} \tr 
\Phi_{n-1}.
\label{spotential}
\eea
Here the magnetic fields $f_{n-1}$ and $\widetilde{f}_{n-1}$  
correspond to 4-4 strings connecting 
the $\widetilde{N}_{c,n}$-color D4-branes(that are 
connecting between the $NS5_{n-2}'$-brane
and the $NS5_{n-1}$-brane in Figure 19B) with $N_{c,n-1}$-flavor 
D4-branes(that are 
a combination of three different D4-branes in Figure 19B).
Among these $N_{c,n-1}$-flavor D4-branes, only the strings ending on
the upper $(N_{c,n-1}-N_{c,n})$ D4-branes and 
on the tilted $(N_{c,n}-N_{c,n-2})$ 
D4-branes in Figure 19B enter the cubic superpotential 
term (\ref{spotential}). 

When the upper NS5'-brane(or $NS5_{n-2}'$-brane) 
is replaced by coincident $(N_{c,n-1}-N_{c,n-2})$ 
D6-branes and the $NS5_{n-3}$ is rotated by $\frac{\pi}{2}$ 
in Figure 19B, this brane configuration looks similar to the one 
found in \cite{Ahn07-9} where the gauge group was given by 
$Sp(n_{c,1}) 
\times \cdots \times SU(n_{c,n-1}) \times SU(n_{f,n}+n_{c,n-1}-n_{c,n})$ 
with $n_{f,n}$ multiplets, $\widetilde{n}_{f,n}$ multiplets, bifundamentals 
and singlets. 
Then the present $(N_{c,n-1}-N_{c,n-2})$ corresponds to the $n_{f,n}$,
$N_{c,n}$ corresponds to $n_{c,n}$,
and 
$N_{c,n-2}$ corresponds to the $n_{c,n-1}$.
%In particular, the Figure 18B of \cite{Ahn07-9} with vanishing flavors
%$Q$ and $Q'$
%is contained in
%this modified Figure 19B running from the $NS5_{n-3}$-brane to 
%the $NS5_{n-1}$-brane.
%The quantum corrections can be understood for small $(\Delta x)_{n-1}$ by 
%using the low energy field theory on the branes.
%The low energy dynamics of the magnetic brane configuration 
%can be described by the ${\cal N}=1$ supersymmetric gauge theory
%with gauge group
%and the gauge couplings for the three gauge group factors are
%given by similarly.
The dual gauge theory has  a meson field  $\Phi_{n-1}$  and 
bifundamentals $f_{n-1}, \widetilde{f}_{n-1}, F_j$, and 
$\widetilde{F}_j$  and the superpotential 
corresponding to Figures 19A and 19B is given by 
\bea
W_{dual} = h \Phi_{n-1} f_{n-1} \widetilde{f}_{n-1} - 
h \mu_{n-1}^2 \tr \Phi_{n-1}, 
\qquad h^2 = g_{n-1,
  mag}^2,
\qquad \mu_{n-1}^2 = -\frac{(\Delta x)_{n-1}}{ 2\pi g_s \ell_s^3}.
\nonu
\eea
Then $ f_{n-1} \widetilde{f}_{n-1}$ is 
a $\widetilde{N}_{c,n} \times \widetilde{N}_{c,n}$ 
matrix where the $(n-1)$-th gauge group indices for $f_{n-1}$ and $\widetilde{f}_{n-1}$ 
are contracted with those
of $\Phi_{n-1}$ while $\Phi_{n-1}$ is a 
$(N_{c,n-1}-N_{c,n-2}) \times (N_{c,n-1}-N_{c,n-2})$ matrix.
%Although the field $f_{n-1}$ itself is an antifundamental in the $n$-th gauge
%group
%which is a different  
%representation for the usual standard quark
%coming from D6-branes,
%the product $f_{n-1} \widetilde{f}_{n-1}$ has the same representation for the 
%product of quarks
%and moreover, 
%the $(n-1)$-th gauge group indices for the field $\Phi_{n-1}$ play the
%role of the flavor indices, as in comparison with the brane
%configuration in the presence of D6-branes before.
Therefore, the F-term equation, the derivative $W_{dual}$ with respect to the
meson field $\Phi_{n-1}$ cannot be satisfied if the $(N_{c,n-1}-N_{c,n-2})$ exceeds
$\widetilde{N}_{c,n}$.
So the supersymmetry is broken.   
That is, 
there exist three equations from F-term conditions:
$
f_{n-1}^a \widetilde{f}_{n-1,b} -\mu_{n-1}^2 \delta^a_b =0$, and 
$\Phi_{n-1} f_{n-1} =0=\widetilde{f}_{n-1} \Phi_{n-1}$.
Then the solutions for these
are given by 
\bea
<f_{n-1}>   & = & 
\left(
\begin{array}{c}
\mu_{n-1}  {\bf 1}_{\widetilde{N}_{c,n}}  \\
0
\end{array}
\right), \qquad 
<\widetilde{f}_{n-1}>   = 
\left(
\begin{array}{cc}
\mu_{n-1}  {\bf 1}_{\widetilde{N}_{c,n}} & 0  \\
\end{array}
\right), \nonu \\
<\Phi_{n-1}> & = &
 \left(
\begin{array}{cc}
0  & 0  \\
0 & M_{n-1}  {\bf 1}_{(N_{c,n-1}-N_{c,n-2}-\widetilde{N}_{c,n})} 
\end{array}
\right).
\nonu 
%\label{point2}
\eea
%where the zero of $<f_{n-1}>$ is a $
%(N_{c,n-1}-N_{c,n-2}-\widetilde{N}_{c,n}) \times \widetilde{N}_{c,n}$ 
%matrix, the zero of $<\widetilde{f}_{n-1}>$ is a
%$\widetilde{N}_{c,n} \times (N_{c,n-1}-N_{c,n-2}-\widetilde{N}_{c,n}) $ matrix and 
%the zeros of $<\Phi_{n-1}>$ are $\widetilde{N}_{c,n} \times \widetilde{N}_{c,n}$,
%$\widetilde{N}_{c,n} \times (N_{c,n-1}-N_{c,n-2}-\widetilde{N}_{c,n})$, 
%and $(N_{c,n-1}-N_{c,n-2}-\widetilde{N}_{c,n}) \times
%\widetilde{N}_{c,n}$ matrices.
%Then one can expand these fields around on a point (\ref{point2})
%and one arrives at the relevant superpotential
%up to quadratic order in the fluctuation. 
%At one loop, the effective potential $V_{eff}^{(1)}$ for $M_{n-1}$
%leads to the positive value for $m_{M_{n-1}}^2$ implying that these
%vacua are stable.

%%%%%%%%%%%%%%%%%%%%%%%%%%%%%%%%%%%%%%%%%%%%%%%%%%%%%%%%%%%%%%
%%%%%%%%%%%%%%%%%%%%%%%%%%%%%%%%%%%%%%%%%%%%%%%%%%%%%%%%%%%%%%
\subsubsection{When the dual gauge group occurs at even chain}
%%%%%%%%%%%%%%%%%%%%%%%%%%%%%%%%%%%%%%%%%%%%%%%%%%%%%%%%%%%%%%
%%%%%%%%%%%%%%%%%%%%%%%%%%%%%%%%%%%%%%%%%%%%%%%%%%%%%%%%%%%%%%

Let us consider other magnetic theory for the same electric theory.
By applying the Seiberg dual to the $SU(N_{c,n})$ factor  and 
interchanging the $NS5_{n-1}'$-brane and the $NS5_{n}$-brane,
one obtains the Figure 20A.

%%%%%%%%%%%%%%%%%%%%%%%%%%%%%%%%%%%%%%%%%%%%%%%%%%%%%%%%%%%%%%%%%%%%%
%%%%%%%%%%%%%%%%%%%%%%%%%%%%%%%%%%%%%%%%%%%%%%%%%%%%%%%%%%%%%%%%%%%%%%
\begin{figure}[ht]
   \epsfxsize=4.0in 
\centerline{\epsffile{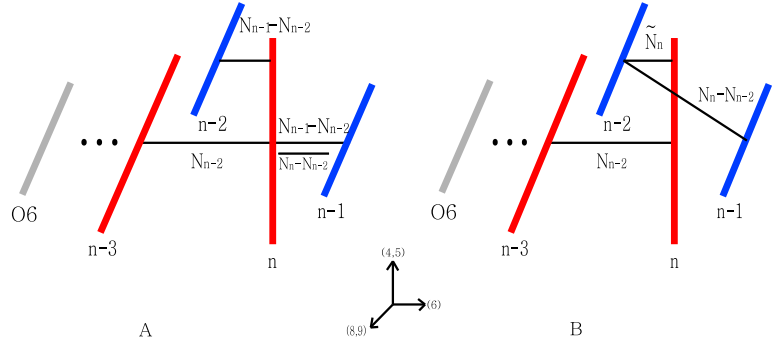}}
   \caption[FIG. \arabic{figure}.]{ 
The 
 ${\cal N}=1$ magnetic brane configuration for the gauge group 
containing $SU(\widetilde{N}_{c,n}=N_{c,n-1}-N_{c,n})$ 
with D4-
and $\overline{D4}$-branes(20A) and with 
a misalignment between D4-branes(20B) when the two NS5'-branes are close to
each other. 
The number of tilted D4-branes in 20B can be written as
$N_{c,n}-N_{c,n-2} =(N_{c,n-1}-N_{c,n-2})-\widetilde{N}_{c,n}$.
%Note that the brane configuration Figure 20 without 
%$O6^{+}$-plane looks similar to Figure 5.
%The $x$ coordinate of $NS5_{n-2}'$-brane is given by $(\Delta
% x)_{n-1}$.
}
\end{figure}
%%%%%%%%%%%%%%%%%%%%%%%%%%%%%%%%%%%%%%%%%%%%%%%%%%%%%%%%%%%%%%%%%%%%%
%Figure 20A and 20B
%%%%%%%%%%%%%%%%%%%%%%%%%%%%%%%%%%%%%%%%%%%%%%%%%%%%%%%%%%%%%%%%%%%%%

%Before arriving at the Figure 20A, there exists an intermediate 
%step where 
%$N_{c,n-1}$ D4-branes between
%$NS5_{n-2}'$-brane and the $NS5_{n}$-brane,
%the $N_{c,n-2}$ D4-branes are connecting between the 
%$NS5_{n-3}'$-brane and the  $NS5_{n-2}'$-brane, and  
%$(N_{c,n-1}-N_{c,n})$ D4-branes are connecting between the  $NS5_{n}$-brane and   
%$NS5_{n-1}'$-brane. 
By rotating $NS5_{n-2}$-brane by an angle $\frac{\pi}{2}$ which will
become $NS5_{n-2}'$-brane, 
moving it with the $(N_{c,n-1}-N_{c, n-2})$ D4-branes 
to $+v$ direction where we introduce $(N_{c,n}-N_{c,n-2})$ D4-branes and
$(N_{c,n}-N_{c,n-2})$ anti D4-branes between the $NS5_{n}$-brane and the 
$NS5_{n-1}'$-brane, 
one gets the final Figure 20A where we are left with 
$(N_{c,n}-N_{c,n-2})$ 
anti-D4-branes between the $NS5_{n}$-brane and   
the $NS5_{n-1}'$-brane.

The gauge group is given by
\bea
SU(N_{c,1}) \times \cdots \times 
SU(N_{c,n-1}) \times SU(\widetilde{N}_{c,n} \equiv N_{c,n-1}-N_{c,n})
\label{ggroup}
\eea
and the matter contents are the field $f_{n-1}$ 
 charged under
$( {\bf 1_1, \cdots, 1_{n-2}, \Box_{n-1}, \overline{\Box}_{n} })$ 
 and their conjugates 
$\widetilde{f}_{n-1}$ 
$( {\bf 1_1, \cdots, 1_{n-2}, \overline{\Box}_{n-1}, \Box_{n} })$ 
under the dual gauge group (\ref{ggroup})
and  
the gauge-singlet $\Phi_{n-1}$ which is in the 
adjoint representation for the $(n-1)$-th gauge group, 
in other words,   
$ ({ \bf   1_1, 
\cdots, 1_{n-2},  (N_{c,n-1}-N_{c,n-2})^2-1,1_n})  \oplus  ({\bf 1_1,
 \cdots, 1_n})$ under the
dual gauge group (\ref{ggroup})
where the gauge group is broken from
$SU(N_{c,n-1})$ 
to $SU(N_{c,n-1}-N_{c,n-2})$.
Then the $\Phi_{n-1}$ is a $(N_{c,n-1}-N_{c,n-2}) \times
 (N_{c,n-1}-N_{c,n-2})$ 
matrix.
Only $(N_{c,n-1}-N_{c,n-2})$ D4-branes can participate in the mass deformation.

%When two NS5'-branes in Figure 20A are close to each other, then 
%it leads to Figure 20B
% by realizing that the number of $(N_{c,n-1}-N_{c,n-2})$
%D4-branes connecting between $NS5_{n-2}'$-brane and $NS5_{n}$-brane can
%be rewritten as $(N_{c,n}-N_{c,n-2})$ plus $\widetilde{N}_{c,n}$.

%The brane configuration in Figure 20A is stable as long as the
%distance $(\Delta x)_{n-1}$ between the upper NS5'-brane and 
%the lower NS5'-brane(or $NS5_{n-1}'$-brane) 
%is large. If they are close to each other, then this brane
%configuration is unstable to decay to 
%the brane configuration in Figure
%20B.
%One can regard these brane configurations as particular states in the
%magnetic gauge theory with the gauge group and
%superpotential.
%The   $(N_{c,n-1}-N_{c,n-2}-\widetilde{N}_{c,n})$ flavor D4-branes of 
%straight brane configuration
%of
%Figure 20B  bend since there exists an attractive
%gravitational interaction
%between those flavor D4-branes and NS5-brane from the DBI action. 
%As mentioned in \cite{Ahn07-5},
%the two NS5'-branes are located at different side of $NS5_n$-brane in
%Figure 20B and the DBI action computation for this bending curve
%should be taken into account. 

The cubic superpotential with the mass term
is given by
\bea
W_{dual} = \Phi_{n-1} f_{n-1} \widetilde{f}_{n-1} + m_{n-1} \tr \Phi_{n-1}
\label{ssuper}
\eea
where we define $\Phi_{n-1}$ as $\Phi_{n-1} \equiv F_{n-1} \widetilde{F}_{n-1}$ and 
the $n$-th gauge group indices in $F_{n-1}$ and $\widetilde{F}_{n-1}$ 
are contracted, each $(n-1)$-th gauge group index in them is encoded in 
$\Phi_{n-1}$. 
Here the magnetic fields $f_{n-1}$ and $\widetilde{f}_{n-1}$  
correspond to 4-4 strings connecting 
the $\widetilde{N}_{c,n}$-color D4-branes(that are 
connecting between the $NS5_{n-2}'$-brane
and the $NS5_{n}$-brane in Figure 20B) with $N_{c,n-1}$-flavor 
D4-branes.
Among these $N_{c,n-1}$-flavor D4-branes, only the strings ending on
the upper $(N_{c,n-1}-N_{c,n})$ D4-branes and 
on the tilted $(N_{c,n}-N_{c,n-2})$ 
D4-branes in Figure 20B enter the cubic superpotential term (\ref{ssuper}). 
%Although the $(N_{c,n-1}-N_{c,n-2})$ D4-branes in Figure 20A for fixed
%other branes cannot move any
%directions,
%the tilted $(N_{c,n}-N_{c,n-2})$-flavor D4-branes can move $w$ direction.
%The remaining upper $\widetilde{N}_{c,n}$ D4-branes are fixed also and cannot 
%move any direction. 
%Note that 
%there is a decomposition 
%\bea
%(N_{c,n-1}-N_{c,n-2})=(N_{c,n}-N_{c,n-2})+\widetilde{N}_{c,n}.
%\nonu
%\eea 

%The brane configuration for zero mass for the bifundamental,
%which has only a cubic superpotential (\ref{ssuper}),
%can be obtained from Figure 20A by moving
%the upper  NS5'-brane together with $(N_{c,n-1}-N_{c,n-2})$ color D4-branes 
%into the origin $v=0$.
%Then the number of dual colors for D4-branes 
%becomes $N_{c,n-1}$ between the $NS5_{n-2}'$-brane and the $NS5_{n}$-brane, 
%$N_{c,n-2}$ between the $NS5_{n-3}'$-brane and the  $NS5_{n-2}'$-brane
%and $\widetilde{N}_{c,n}$ 
%between $NS5_{n}$-brane and $NS5_{n-1}'$-brane.
%Or starting from Figure 15B and moving the 
%$NS5_{n-1}'$-brane to the right all the
%way past the $NS5_{n}$-brane,
%one also obtains the corresponding magnetic brane configuration
%for massless case.

The low energy dynamics of the magnetic brane configuration 
can be described by the ${\cal N}=1$ supersymmetric gauge theory
with gauge group
and the gauge couplings for the three gauge group factors are
given by similarly.
The dual gauge theory has  a gauge singlet $\Phi_{n-1}$  and 
bifundamentals $f_{n-1}, \widetilde{f}_{n-1}, F_j$ and
$\widetilde{F}_j$ 
under the dual gauge
group and the superpotential 
corresponding to Figures 20A and 20B is given by 
\bea
W_{dual} = h \Phi_{n-1} f_{n-1} \widetilde{f}_{n-1} - 
h \mu_{n-1}^2 \tr \Phi_{n-1}, \qquad h^2 = g_{n-1,
  mag}^2,
\qquad \mu_{n-1}^2 = -\frac{(\Delta x)_{n-1}}{ 2\pi g_s \ell_s^3}.
\nonu
\eea
Then $ f_{n-1} \widetilde{f}_{n-1}$ is a $\widetilde{N}_{c,n} \times 
\widetilde{N}_{c,n}$ 
matrix where the $(n-1)$-th gauge group indices for $f_{n-1}$ and 
$\widetilde{f}_{n-1}$ 
are contracted with those
of $\Phi_{n-1}$ while the $\Phi_{n-1}$ is a 
$(N_{c,n-1}-N_{c,n-2}) \times (N_{c,n-1}-N_{c,n-2})$ matrix.
%The product $f_{n-1} \widetilde{f}_{n-1}$ has the same representation for the 
%product of quarks
%and moreover, 
%the $(n-1)$-th gauge group indices for the field $\Phi_{n-1}$ play the
%role of the flavor indices.
Therefore, the F-term equation, the derivative $W_{dual}$ with respect to the
meson field $\Phi_{n-1}$ cannot be satisfied if the $(N_{c,n-1}-N_{c,n-2})$ exceeds
$\widetilde{N}_{c,n}$.
So the supersymmetry is broken.   
That is, 
there exist three equations from F-term conditions:
$
f_{n-1}^a \widetilde{f}_{n-1,b} -\mu_{n-1}^2 \delta^a_b =0$ and $ \Phi_{n-1} f_{n-1} =0=
\widetilde{f}_{n-1} \Phi_{n-1}$.
Then the solutions for these
are given by 
\bea
<f_{n-1}>   & = & 
\left(
\begin{array}{c}
\mu_{n-1}  {\bf 1}_{\widetilde{N}_{c,n}}  \\
0
\end{array}
\right), \qquad
<\widetilde{f}_{n-1}>   = 
\left(
\begin{array}{cc}
\mu_{n-1}  {\bf 1}_{\widetilde{N}_{c,n}} & 0  \\
\end{array}
\right), \nonu \\
<\Phi_{n-1}>  & = &
 \left(
\begin{array}{cc}
0  & 0  \\
0 & M_{n-1}  {\bf 1}_{(N_{c,n-1}-N_{c,n-2})-\widetilde{N}_{c,n}} 
\end{array}
\right).
\nonu 
%\label{solutions}
\eea
%where the zero of $<f_{n-1}>$ is a $
%(N_{c,n-1}-N_{c,n-2}-\widetilde{N}_{c,n}) \times \widetilde{N}_{c,n}$ 
%matrix, the zero of $<\widetilde{f}_{n-1}>$ is a
%$\widetilde{N}_{c,n} \times (N_{c,n-1}-N_{c,n-2}-\widetilde{N}_{c,n}) $ matrix and 
%the zeros of $<\Phi_{n-1}>$ are $\widetilde{N}_{c,n} \times \widetilde{N}_{c,n}$,
%$\widetilde{N}_{c,n} \times 
%(N_{c,n-1}-N_{c,n-2}-\widetilde{N}_{c,n})$ and $(N_{c,n-1}-N_{c,n-2}-
%\widetilde{N}_{c,n}) \times
%\widetilde{N}_{c,n}$ 
%matrices.
%Then one can expand these fields around on a point (\ref{solutions})
%and one arrives at the relevant superpotential
%up to quadratic order in the fluctuation. 
%At one loop, the effective potential $V_{eff}^{(1)}$ for $M_{n-1}$
%leads to the positive value for $m_{M_{n-1}}^2$ implying that these
%vacua are stable.

%%%%%%%%%%%%%%%%%%%%%%%%%%%%%%%%%%%%%%%%%%%%%%%%%%%%%%%%%%%%%%%%%%%%%%%%%%%
%%%%%%%%%%%%%%%%%%%%%%%%%%%%%%%%%%%%%%%%%%%%%%%%%%%%%%%%%%%%%%%%%%%%%%%%%%%
\subsection{
${\cal N}=1$ 
$Sp(N_{c,1}) \times SU(\widetilde{N}_{c,2}) \times \cdots
\times
SU(N_{c,n})$ magnetic theory
}
%%%%%%%%%%%%%%%%%%%%%%%%%%%%%%%%%%%%%%%%%%%%%%%%%%%%%%%%%%%%%%%%%%%%%%%%%%%
%%%%%%%%%%%%%%%%%%%%%%%%%%%%%%%%%%%%%%%%%%%%%%%%%%%%%%%%%%%%%%%%%%%%%%%%%%%

Let us consider the Seiberg dual for the second gauge group
factor.
Starting from Figure 15A, moving the $NS5_{3}'$-brane 
with $(N_{c,3}-N_{c,4})$
D4-branes 
to the $+v$ direction leading to Figure 16A, 
and interchanging the $NS5_{1}'$-brane and the $NS5_{2}$-brane,
one obtains the Figure 21A.

%%%%%%%%%%%%%%%%%%%%%%%%%%%%%%%%%%%%%%%%%%%%%%%%%%%%%%%%%%%%%%%%%%%%%
%%%%%%%%%%%%%%%%%%%%%%%%%%%%%%%%%%%%%%%%%%%%%%%%%%%%%%%%%%%%%%%%%%%%%%
\begin{figure}[ht]
   \epsfxsize=4.0in 
\centerline{\epsffile{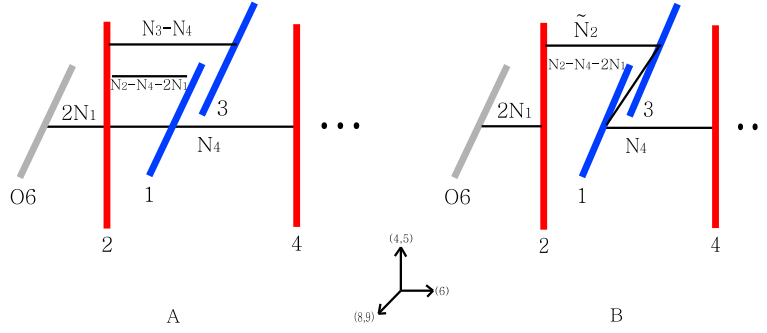}}
   \caption[FIG. \arabic{figure}.]{ 
The 
 ${\cal N}=1$ magnetic brane configuration for the gauge group 
containing $SU(\widetilde{N}_{c,2}=2N_{c,1}+N_{c,3}-N_{c,2})$ 
with D4-
and $\overline{D4}$-branes(21A) and with 
a misalignment between D4-branes(21B) when the two NS5'-branes are close to
each other. 
The number of tilted D4-branes in 21B can be written as
$N_{c,2}-N_{c,4}-2N_{c,1} =(N_{c,3}-N_{c,4})-\widetilde{N}_{c,2}$.
%Note that the brane configuration Figure 21 without 
%$O6^{+}$-plane looks similar to Figure 7.
%The $x$ coordinate of $NS5_{3}'$-brane is given by $(\Delta x)_{3}$.
}
\end{figure}
%%%%%%%%%%%%%%%%%%%%%%%%%%%%%%%%%%%%%%%%%%%%%%%%%%%%%%%%%%%%%%%%%%%%%
%Figure 21A and 21A 
%%%%%%%%%%%%%%%%%%%%%%%%%%%%%%%%%%%%%%%%%%%%%%%%%%%%%%%%%%%%%%%%%%%%%%%%%%%

%Before arriving at the Figure 21A, there exists an intermediate 
%step where the $(2N_{c,1}+N_{c,3}-N_{c,2})$ D4-branes are 
%connecting between the 
%$NS5_{2}$-brane and the  $NS5_{1}'$-brane,  
%$(N_{c,3}-N_{c,4})$ D4-branes connecting between the  $NS5_{1}'$-brane and   
%$NS5_{3}'$-brane, and $N_{c,4}$ D4-branes between the $NS5_{1}'$-brane and
%the $NS5_{4}$-brane. 
By introducing $-N_{c,4}$ D4-branes and $-N_{c,4}$ 
anti-D4-branes  between the  $NS5_{2}$-brane and   
$NS5_{1}'$-brane, reconnecting the former with  
the $N_{c,3}$ D4-branes connecting between  
$NS5_{2}$-brane 
and the $NS5_{1}'$-brane (therefore $(N_{c,3}-N_{c,4})$ D4-branes)
and moving those combined
$(N_{c,3}-N_{c,4})$ 
D4-branes
to $+v$-direction, 
one gets the final Figure 21A where we are left with 
$(N_{c,2}-N_{c,4}-2N_{c,1})$ 
anti-D4-branes between the $NS5_{2}$-brane and   
$NS5_{1}'$-brane.

The dual gauge group is given by 
\bea
Sp(N_{c,1}) \times 
SU(\widetilde{N}_{c,2} \equiv 2N_{c,1}+N_{c,3}-N_{c,2}) 
\times SU(N_{c,3}) \times \cdots \times SU(N_{c,n})
\label{dddual}
\eea
where the matter contents are   
the bifundamentals $f_2$ in 
 $({\bf 1_1, \Box_2, \overline{\Box}_{3}, 1, \cdots, 1_n})$,
and $\widetilde{f}_2$ in the representation 
$({\bf 1_1, \overline{\Box}_2, \Box_{3}, 1,
\cdots, 1_n})$ in
addition to $(n-2)$ bifundamentals $F_j$ and $\widetilde{F}_j$, 
 $j=1, 3, \cdots, (n-1)$ and
the gauge singlet $\Phi_{3}$
for the second dual gauge group in the 
adjoint representation for the third dual gauge group, 
i.e.,  
$
{(\bf 1_1, 1_2, (N_{c,3}-N_{c,4})^2-1, 1, \cdots, 1_n)  
}
$ plus a singlet
under the 
dual gauge group where the gauge group is broken from
$SU(N_{c,3})$ 
to $SU(N_{c,3}-N_{c,4})$.

%When two NS5'-branes in Figure 21A are close to each other, then 
%it leads to Figure 21B by realizing that the number of $(N_{c,3}-N_{c,4})$
%D4-branes connecting between $NS5_{2}$-brane and $NS5_{3}'$-brane can
%be rewritten as $(N_{c,2}-N_{c,4}-2N_{c,1})$ plus $\widetilde{N}_{c,2}$.
%The Figure 12 of \cite{Ahn07-6} is contained in the Figure 21. In
%particular, the brane configuration from the $NS5_2$-brane to 
%the $NS5_3'$-brane is exactly same as the one of \cite{Ahn07-6}.

The cubic superpotential with the mass term in the dual
theory is given by
\bea
W_{dual} = \Phi_{3} f_{2} \widetilde{f}_{2}  + m_{3} \tr \Phi_{3}.
\label{Wdual}
\eea
Here the magnetic fields $f_2$ and $\widetilde{f}_2$  
correspond to 4-4 strings connecting 
the $\widetilde{N}_{c,2}$-color D4-branes(that are 
connecting between the $NS5_{2}$-brane
and the $NS5_{3}'$-brane in Figure 21B) with $N_{c,3}$-flavor 
D4-branes(that are 
a combination of three different D4-branes in Figure 21B).
Among these $N_{c,3}$-flavor D4-branes, only the strings ending on
the upper $(2N_{c,1}+N_{c,3}-N_{c,2})$ D4-branes and 
on the tilted $(N_{c,2}-N_{c,4}-2N_{c,1})$ 
D4-branes in Figure 21B enter the cubic superpotential term (\ref{Wdual}). 

When the upper NS5'-brane(or $NS5_{3}'$-brane) 
is replaced by coincident $(N_{c,3}-N_{c,4})$ 
D6-branes and 
the $NS5_{4}$ is rotated by an angle $\frac{\pi}{2}$ in the $(v,w)$
plane in Figure 21B, this brane configuration reduces to the one 
found in \cite{Ahn07-9} where the gauge group was given by 
$ Sp(n_{c,1}) \times SU(n_{f,2}+n_{c,3}+2n_{c,1}-n_{c,2}) \times
SU(n_{c,3}) 
\times \cdots $ 
with $n_{f,i}$ multiplets,  $\widetilde{n}_{f,i}$ multiplets, 
bifundamentals and gauge 
singlets. 
Then the present number $(N_{c,3}-N_{c,4})$ corresponds to the $n_{f,2}$, the
number $N_{c,2}$ corresponds to $n_{c,2}$,
 the
number $N_{c,4}$ corresponds to $n_{c,3}$,
 and 
the number $N_{c,1}$ corresponds to the $n_{c,1}$.
%Note that the number of D4-branes touching $NS5_{3}'$-brane in Figure 21B
%is equal to $(N_{c,3}-N_{c,4})$.
%In particular, the Figure 16B of \cite{Ahn07-9} with vanishing flavors
%$Q$ and $Q''$
%is contained in
%this modified Figure 21B running from the $NS5_{2}$-brane to 
%the $NS5_{4}$-brane.
%The quantum corrections can be understood for small $(\Delta x)_{3}$ by 
%using the low energy field theory on the branes.
%The low energy dynamics of the magnetic brane configuration 
%can be described by the ${\cal N}=1$ supersymmetric gauge theory
%with gauge group 
%and the gauge couplings for the three gauge group factors are
%given.
The dual gauge theory has  a meson field $\Phi_{3}$  and 
bifundamentals $f_2, \widetilde{f}_2, F_j$, and $\widetilde{F}_j$ 
under the dual gauge
group  (\ref{dddual}) and the superpotential (\ref{Wdual}) 
corresponding to Figures 21A and 21B is given by 
\bea
W_{dual} = h \Phi_{3} f_2 \widetilde{f}_2 - h \mu_{3}^2 \tr \Phi_{3}, 
\qquad h^2 = g_{3,
  mag}^2,
\qquad \mu_{3}^2 = -\frac{(\Delta x)_{3}}{ 2\pi g_s \ell_s^3}.
\nonu
\eea
Then $ f_2 \widetilde{f}_2$ is a $\widetilde{N}_{c,2} \times \widetilde{N}_{c,2}$ 
matrix where the third gauge group indices for $f_2$ and $\widetilde{f}_2$ 
are contracted with those
of $\Phi_{3}$ while the $\Phi_{3}$ is a 
$(N_{c,3}-N_{c,4}) \times (N_{c,3}-N_{c,4})$ matrix.
%Although the field $f_2$ itself is an antifundamental in the third gauge
%group
%which is a different  
%representation for the usual standard quark
%coming from D6-branes,
%the product $f_2 \widetilde{f}_2$ has the same representation for the 
%product of quarks
%and moreover, 
%the third gauge group indices for the field $\Phi_{3}$ play the
%role of the flavor indices, as in comparison with the brane
%configuration in the presence of D6-branes before.
Therefore, the F-term equation, the derivative $W_{dual}$ with respect to the
meson field $\Phi_{3}$ cannot be satisfied if the $(N_{c,3}-N_{c,4})$ exceeds
$\widetilde{N}_{c,2}$.
So the supersymmetry is broken.   
That is, 
there exist three equations from F-term conditions:
$
f_2^a \widetilde{f}_{2,b} -\mu_{3}^2 \delta^a_b =0$,  and 
$\Phi_{3} f_2 =0=\widetilde{f}_2 \Phi_{3}$.
Then the solutions for these
are given by 
\bea
<f_2>   & = & 
\left(
\begin{array}{c}
\mu_{3}  {\bf 1}_{\widetilde{N}_{c,2}}  \\
0
\end{array}
\right), 
\qquad
<\widetilde{f}_2>   = 
\left(
\begin{array}{cc}
\mu_{3}  {\bf 1}_{\widetilde{N}_{c,2}} & 0  \\
\end{array}
\right), \nonu \\ 
<\Phi_{3}> & = &
 \left(
\begin{array}{cc}
0  & 0  \\
0 & M_{3}  {\bf 1}_{(N_{c,3}-N_{c,4}-\widetilde{N}_{c,2})} 
\end{array}
\right).
\nonu 
%\label{poi}
\eea
%where the zero of $<f_2>$ is a $
%(N_{c,3}-N_{c,4}-\widetilde{N}_{c,2}) \times \widetilde{N}_{c,2}$ 
%matrix, the zero of $<\widetilde{f}_2>$ is a
%$\widetilde{N}_{c,2} \times (N_{c,3}-N_{c,4}-\widetilde{N}_{c,2}) $ matrix and 
%the zeros of $<\Phi_{3}>$ are $\widetilde{N}_{c,2} \times \widetilde{N}_{c,2}$,
%$\widetilde{N}_{c,2} \times (N_{c,3}-N_{c,4}-\widetilde{N}_{c,2})$, 
%and $(N_{c,3}-N_{c,4}-\widetilde{N}_{c,2}) \times
%\widetilde{N}_{c,2}$ matrices.
%Then one can expand these fields around on a point (\ref{poi})
%and one arrives at the relevant superpotential
%up to quadratic order in the fluctuation. 
%At one loop, the effective potential $V_{eff}^{(1)}$ for $M_{3}$
%leads to the positive value for $m_{M_{3}}^2$ implying that these
%vacua are stable.

%%%%%%%%%%%%%%%%%%%%%%%%%%%%%%%%%%%%%%%%%%%%%%%%%%%%%%%%%%%%%%%%%%%%%%%%%%%%
%%%%%%%%%%%%%%%%%%%%%%%%%%%%%%%%%%%%%%%%%%%%%%%%%%%%%%%%%%%%%%%%%%%%%%%%%%%%
\section{Meta-stable brane configurations  with $2n$ NS-branes plus 
$O6^{+}$-plane}
%section5%%%%%%%%%%%%%%%%%%%%%%%%%%%%%%%%%%%%%%%%%%%%%%%%%%%%%%%%%%%%%%%%%%%
%%%%%%%%%%%%%%%%%%%%%%%%%%%%%%%%%%%%%%%%%%%%%%%%%%%%%%%%%%%%%%%%%%%%%%%%%%%%

The type IIA brane configuration,  
 by generalizing the brane
configurations \cite{LO,Ahn07-3} to the case where there are more
NS-branes,  
corresponding to 
${\cal N}=1$ supersymmetric electric gauge theory(see also
\cite{Ahn07-9})  with
gauge group
\bea
SO(N_{c,1}) \times SU(N_{c,2}) \cdots
\times SU(N_{c,n})
\nonu
\eea
and with 
the $(n-1)$ bifundametals $F_i$ charged under 
$({\bf 1_1, \cdots, 1, \Box_i, \overline{\Box}_{i+1}, 1, \cdots,  1_n})$
and their
complex conjugate fields $\widetilde{F}_i$ 
charged $({\bf 1_1, \cdots, 1, \overline{\Box}_i, \Box_{i+1}, 1, 
\cdots, 1_n})$ where $i=1, 2, \cdots, (n-1)$
can be described by 
the $NS5_1$-brane, 
the  
$NS5_2'$-brane, $\cdots$, the $NS5_{n}$-brane for odd number
of gauge groups(or 
the $NS5_{n}'$-brane for even number of gauge groups),
$N_{c,1}$-, $N_{c,2}$-,  $\cdots$, and $N_{c,n}$-color D4-branes. 
See the Figure 22 for the details on the brane configuration. 
The $O6^{+}$-plane acts as $(x^4,x^5,x^6) \rightarrow
(-x^4,-x^5,-x^6)$ and has RR charge $+4$.

%%%%%%%%%%%%%%%%%%%%%%%%%%%%%%%%%%%%%%%%%%%%%%%%%%%%%%%%%%%%%%%%%%%%%%%%
%%%%%%%%%%%%%%%%%%%%%%%%%%%%%%%%%%%%%%%%%%%%%%%%%%%%%%%%%%%%%%%%%%%%%%
\begin{figure}[ht]
   \epsfxsize=4.0in 
\centerline{\epsffile{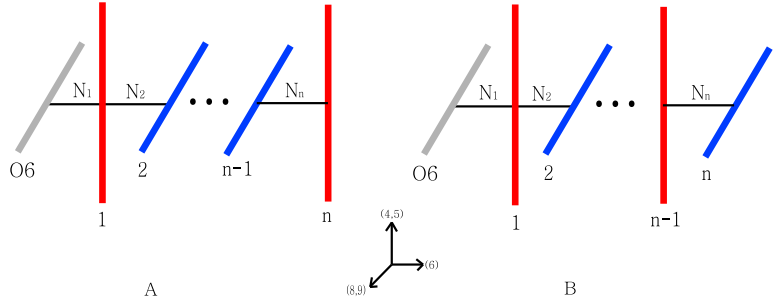}}
   \caption[FIG. \arabic{figure}.]{ 
The 
 ${\cal N}=1$ supersymmetric 
electric brane configuration for the gauge group $SO(N_{c,1})
\times \prod_{i=2}^{n}
 SU(N_{c,i})$ 
and  bifundamentals $F_i$ and $\widetilde{F}_i$  with vanishing mass
for the bifundamentals when the number of gauge groups factor 
$n$ is odd(22A) and even(22B). 
We do not draw the mirrors of the branes appearing in the left hand
 side of $O6^{+}$-plane. 
%This Figure can also be obtained from the
% Figure 15 by changing the O6-plane charge and exchanging $NS5$-branes
% and $NS5'$-branes.
}
\end{figure}
%%%%%%%%%%%%%%%%%%%%%%%%%%%%%%%%%%%%%%%%%%%%%%%%%%%%%%%%%%%%%%%%%%%%%
%Figure 22A and 22B
%%%%%%%%%%%%%%%%%%%%%%%%%%%%%%%%%%%%%%%%%%%%%%%%%%%%%%%%%%%%%%%%%%%%%%%

Let us place an O6-plane at the origin $x^6=0$
and denote the $x^6$ 
coordinates for 
the  
$NS5_1$-brane, $\cdots$, the $NS5_{n}$-brane for odd $n$(or 
the $NS5_{n}'$-brane for even $n$)
are given by $x^6=y_1, y_1+y_2, \cdots, \sum_{j=1}^{n-1} y_j + y_{n}$
respectively.
The $N_{c,1}$ D4-branes 
are suspended between the 
$NS5_1$-brane and its mirror, 
the $N_{c,2}$ D4-branes 
are suspending between the 
$NS5_1$-brane and the $NS5_2'$-brane, $\cdots$ and 
the $N_{c,n}$ D4-branes  
are suspended between the $NS5_{n-1}'$-brane and the $NS5_{n}$-brane for
odd $n$(or 
between the $NS5_{n-1}$-brane and  the $NS5_{n}'$-brane for even $n$).
The fields $F_i$ and $\widetilde{F}_i$  correspond to 4-4 strings connecting 
the $N_{c,i}$-color D4-branes with $N_{c,i+1}$-color D4-branes.
We draw this ${\cal N}=1$ supersymmetric 
electric brane configuration in Figure 22A(22B) 
when $n$ is odd(even) for the vanishing mass
for the fields $F_i$ and $\widetilde{F}_i$. 

There is no superpotential in Figure 22A. Let us deform this theory.
Displacing the two NS5'-branes relative each other in the $+v$
direction, characterized by $(\Delta x)_{i+1}$, 
corresponds to turning on a quadratic
mass-deformed superpotential
for the field $F_i$ and $\widetilde{F}_i$ as follows:
\bea
W = m_{i+1} F_i \widetilde{F}_i (\equiv m_{i+1} \Phi_{i+1})
\nonu
\eea
where 
the $i$-th gauge group indices in $F_i$ and $\widetilde{F}_i$ 
are contracted and
\bea
m_{i+1} 
%=\frac{(\Delta x)_{i+1}}{2\pi \alpha'} 
= 
\frac{(\Delta x)_{i+1}}{\ell_s^2}.
\nonu
\eea
The gauge-singlet $\Phi_{i+1}$ for the $i$-th  gauge group is in the 
adjoint representation for the $(i+1)$-th  gauge group, 
i.e., ${\bf ( 1_1, \cdots, 1_i, (N_{c,i+1}-N_{c,i+2})^2-1, 
\cdots, 1_n)  \oplus (1_1, \cdots ,1_n) }$ 
under the  gauge group
where the gauge group is broken from
$SU(N_{c,i+1})$ 
to $SU(N_{c,i+1}-N_{c,i+2})$. 
The $\Phi_{i+1}$ is a $(N_{c,i+1}-N_{c,i+2}) \times (N_{c,i+1}-N_{c,i+2})$ matrix.
The $NS5_{i+1}'$-brane together with $(N_{c,i+1}-N_{c,i+2})$-color D4-branes 
is moving to the $+v$ direction  for
fixed other branes during this mass deformation(and their mirrors to
$-v$ direction). 
Then the $x^5$ coordinate 
of $NS5_{i-1}'$-brane is equal to
zero
while the $x^5$ coordinate of $NS5_{i+1}'$-branes is given by 
$ (\Delta x)_{i+1}$.
Giving an expectation value to the meson field $\Phi_{i+1}$
corresponds to recombination of $N_{c,i}$- and $N_{c,i+1}$- color 
D4-branes, which will become $N_{c,i}$ or $N_{c,i+1}$-color D4-branes
in Figure 22A such that they are suspended between 
the $NS5_{i-1}'$-brane and the $NS5_{i+1}'$-brane 
and pushing them into the $w$
direction. 
We assume that the number of colors satisfies
$
N_{c,i+1} \geq N_{c,i}-N_{c,i-1} \geq N_{c,i+2}$.

Now 
we draw this brane configuration in Figure 23A for nonvanishing mass
for the fields $F_i$ and $\widetilde{F}_i$.

%%%%%%%%%%%%%%%%%%%%%%%%%%%%%%%%%%%%%%%%%%%%%%%%%%%%%%%%%%%%%%%%%%%%%%%
%%%%%%%%%%%%%%%%%%%%%%%%%%%%%%%%%%%%%%%%%%%%%%%%%%%%%%%%%%%%%%%%%%%%%%
\begin{figure}[ht]
   \epsfxsize=4.0in 
\centerline{\epsffile{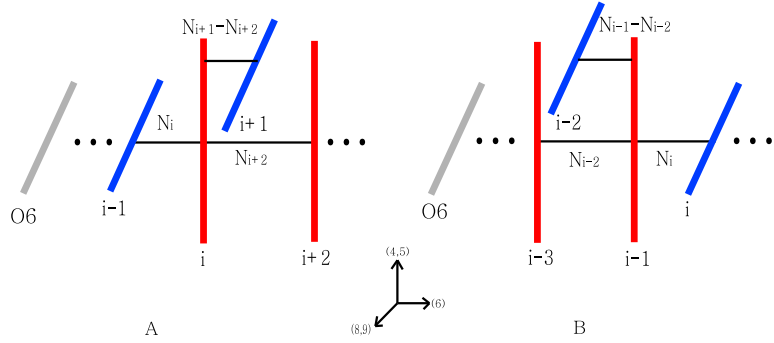}}
   \caption[FIG. \arabic{figure}.]{ 
The 
 ${\cal N}=1$ supersymmetric 
electric brane configuration for the gauge group $SO(N_{c,1}) \times 
\prod_{i=2}^{n}
 SU(N_{c,i})$ 
and  bifundamentals $F_i$ and $\widetilde{F}_i$  with nonvanishing mass
for the bifundamentals when the number of gauge groups factor 
$n$ is odd(23A) and even(23B).  
The $N_{c,i+1}$ D4-branes in 23A are decomposed into 
$(N_{c,i+1}-N_{c,i+2})$ D4-branes which are moving to $+v$ direction
 in 23A 
and $N_{c,i+2}$ D4-branes which are recombined with those D4-branes
connecting between $NS5_{i+1}'$-brane and $NS5_{i+2}$-brane in 23A.
The $N_{c,i-1}$ D4-branes in 23B are decomposed into 
$(N_{c,i-1}-N_{c,i-2})$ D4-branes which are moving to $+v$ direction in 23B 
and $N_{c,i-2}$ D4-branes which are recombined with those D4-branes
connecting between $NS5_{i-3}$-brane and $NS5_{i-2}'$-brane in 23B.
%Note that the brane configuration Figure 23A(23B) without 
%$O6^{-}$-plane looks similar to Figure 16B(16A).
%The $x$ coordinate of $NS5_{i+1}'$-brane is given by $(\Delta x)_{i+1}$
% while the 
%$x$ coordinate of $NS5_{i-2}'$-brane is given by $(\Delta x)_{i-1}$.
}
\end{figure}
%%%%%%%%%%%%%%%%%%%%%%%%%%%%%%%%%%%%%%%%%%%%%%%%%%%%%%%%%%%%%%%%%%%%%
%Figure 23A and 23B
%%%%%%%%%%%%%%%%%%%%%%%%%%%%%%%%%%%%%%%%%%%%%%%%%%%%%%%%%%%%%%%%%%%%%%%%%

Let us deform the theory by Figure 22B.
Displacing the two NS5'-branes, the $NS_{i-2}'$-brane and the 
$NS_{i}'$-brane, 
relative each other in the 
$v$ 
direction, characterized by $(\Delta x)_{i-1}$, 
corresponds to turning on a quadratic
mass-deformed superpotential
for the fields $F_{i-1}$ and $\widetilde{F}_{i-1}$ as follows:
\bea
W = m_{i-1} F_{i-1} \widetilde{F}_{i-1} (\equiv m_{i-1} \Phi_{i-1})
\nonu
\eea
where 
the $i$-th gauge group indices in $F_{i-1}$ and $\widetilde{F}_{i-1}$ 
are contracted, each $(i-1)$-th gauge group index in them is encoded in 
$\Phi_{i-1}$ and the mass $m_{i-1}$ is given by
\bea
m_{i-1} 
%=\frac{(\Delta x)_{i+1}}{2\pi \alpha'} 
= 
\frac{(\Delta x)_{i-1}}{\ell_s^2}.
\nonu
\eea

The gauge-singlet $\Phi_{i-1}$ for the $i$-th  gauge group is in the 
adjoint representation for the $(i-1)$-th  gauge group, 
i.e., 
\bea
{(\bf 1_1, \cdots, 1_{i-2}, (N_{c,i-1}-N_{c,i-2})^2-1, 1_{i}, \cdots,
  1_n)  
\oplus (1_1, \cdots, 1_n)}
\nonu
\eea 
under the  gauge group where the gauge group is broken from
$SU(N_{c,i-1})$ 
to $SU(N_{c,i-1}-N_{c,i-2})$.  
Then the $\Phi_{i-1}$ is a $(N_{c,i-1}-N_{c,i-2}) \times 
(N_{c,i-1}-N_{c,i-2})$ matrix.
The $NS5_{i-2}'$-brane together with $(N_{c,i-1}-N_{c,i-2})$-color D4-branes 
is moving to the $+v$ direction  for
fixed other branes during this mass deformation. 
In other words, the $N_{c, i-2}$ D4-branes among $N_{c,i-1}$ D4-branes 
are not participating in 
the mass deformation.
Then the $x^5$ coordinate($\equiv x$) 
of $NS5_{i}'$-brane is equal to
zero
while the $x^5$ coordinate of $NS5_{i-2}'$-brane is given by 
$(\Delta x)_{i-1}$.
Giving an expectation value to the meson field $\Phi_{i-1}$
corresponds to recombination of $N_{c,i-1}$- and $N_{c,i}$- color 
D4-branes, which will become $N_{c,i-1}$- or $N_{c,i}$-color D4-branes
in Figure 22B such that they are suspended between 
the $NS5_{i-2}'$-brane and the $NS5_{i}'$-brane 
and pushing them into the 
$w$ direction. 
%We assume that the number of colors satisfies
%$
%N_{c,i-1} \geq N_{c,i}-N_{c,i+1} \geq N_{c,i-2}$.
Now 
we draw this brane configuration in Figure 23B for nonvanishing mass
for the fields $F_{i-1}$ and $\widetilde{F}_{i-1}$. 

%Next we describe five different magnetic dual theories by taking each
%corresponding mass deformation.
%Many of the brane configurations in this section are coincident with the ones in
%previous section and we will mention the main results briefly. 

%%%%%%%%%%%%%%%%%%%%%%%%%%%%%%%%%%%%%%%%%%%%%%%%%%%%%%%%%%%%%%%%%%%%%%%%%%%
%%%%%%%%%%%%%%%%%%%%%%%%%%%%%%%%%%%%%%%%%%%%%%%%%%%%%%%%%%%%%%%%%%%%%%%%%%%
\subsection{
${\cal N}=1$ 
$SO(N_{c,1}) \times SU(\widetilde{N}_{c,2}) \times \cdots \times
SU(N_{c,n})$ magnetic theory
}
%%%%%%%%%%%%%%%%%%%%%%%%%%%%%%%%%%%%%%%%%%%%%%%%%%%%%%%%%%%%%%%%%%%%%%%%%%%
%%%%%%%%%%%%%%%%%%%%%%%%%%%%%%%%%%%%%%%%%%%%%%%%%%%%%%%%%%%%%%%%%%%%%%%%%%%

By applying the Seiberg dual to the $SU(N_{c,2})$ factor   and 
interchanging the $NS5_1$-brane and the $NS5_2'$-brane(and their mirrors),
one obtains the Figure 24A.

%%%%%%%%%%%%%%%%%%%%%%%%%%%%%%%%%%%%%%%%%%%%%%%%%%%%%%%%%%%%%%%%%%%%%
%%%%%%%%%%%%%%%%%%%%%%%%%%%%%%%%%%%%%%%%%%%%%%%%%%%%%%%%%%%%%%%%%%%%%%
\begin{figure}[ht]
   \epsfxsize=4.0in 
\centerline{\epsffile{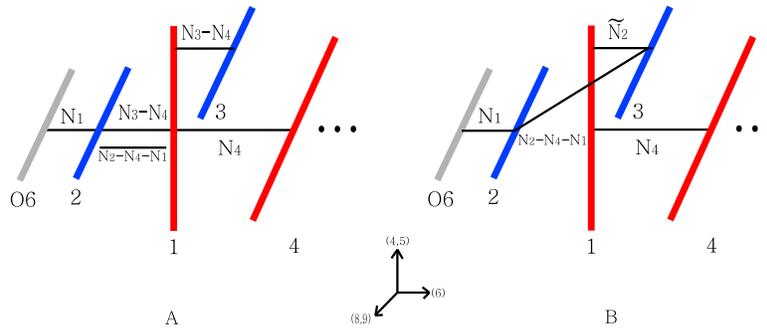}}
   \caption[FIG. \arabic{figure}.]{ 
The 
 ${\cal N}=1$ magnetic brane configuration for the gauge group 
containing $SU(\widetilde{N}_{c,2}=N_{c,1}+N_{c,3}-N_{c,2})$ 
with D4-
and $\overline{D4}$-branes(24A) and with 
a misalignment between D4-branes(24B) when the two NS5'-branes are close to
each other. 
The number of tilted D4-branes in 24B can be written as
$N_{c,2}-N_{c,4}-N_{c,1} =(N_{c,3}-N_{c,4})-\widetilde{N}_{c,2}$.
%The $x$ coordinate of $NS5_{3}'$-brane is given by $(\Delta x)_{3}$.
}
\end{figure}
%%%%%%%%%%%%%%%%%%%%%%%%%%%%%%%%%%%%%%%%%%%%%%%%%%%%%%%%%%%%%%%%%%%%%
%Figure 26A and 26B
%%%%%%%%%%%%%%%%%%%%%%%%%%%%%%%%%%%%%%%%%%%%%%%%%%%%%%%%%%%%%%%%%%%%%%%%%

%Before arriving at the Figure 24A, there exists an intermediate 
%step where 
%the $(N_{c,3}-N_{c,2}+N_{c,1})$ D4-branes are connecting between the 
%$NS5_2'$-brane and the  $NS5_1$-brane,  
%$N_{c,3}$ D4-branes are connecting between the  $NS5_1$-brane and   
%$NS5_3'$-brane(and their mirrors),
%$N_{c,4}$ D4-branes are connecting between the  $NS5_3'$-brane and   
%$NS5_4'$-brane(and their mirrors)
% as well as $N_{c,1}$ D4-branes between
%$NS5_2'$-brane and its mirror.
By rotating $NS5_3$-brane by an angle $\frac{\pi}{2}$, moving it with 
$(N_{c,3}-N_{c,4})$ 
D4-branes
to $+v$-direction(and their mirrors to $-v$ direction), 
one gets the final Figure 24A where we are left with 
$(N_{c,2}-N_{c,1}-N_{c,4})$ 
anti-D4-branes between the $NS5_2'$-brane and   
$NS5_1$-brane.

The dual gauge group is given by 
\bea
SO(N_{c,1}) \times 
SU(\widetilde{N}_{c,2} \equiv N_{c,1}+N_{c,3}-N_{c,2}) 
\times SU(N_{c,3}) \times \cdots \times SU(N_{c,n}).
\nonu
\eea
The matter contents are the field $f_2$ 
 charged under
$({\bf 1_1, \widetilde{N}_{c,2}, \overline{N_{c,3}}, \cdots, 1_n})$ 
and their conjugates 
$\widetilde{f}_2$ 
charged under $({\bf 1_1, \overline{\widetilde{N}_{c,2}}, N_{c,3}, 
\cdots, 1_n})$ 
under the dual gauge group
and  
the gauge-singlet $\Phi_3$ for the second dual gauge group in the 
adjoint representation for the third dual gauge group, 
i.e.,  ${(\bf 1_1, 1_2, (N_{c,3}-N_{c,4})^2-1, \cdots, 1_n)  \oplus
    (1_1,\cdots,1_n)}$ under the 
dual gauge group where the gauge group is broken from
$SU(N_{c,3})$ 
to $SU(N_{c,3}-N_{c,4})$. 
Then the  $\Phi_3$ is a $(N_{c,3}-N_{c,4}) \times (N_{c,3}-N_{c,4})$ 
matrix.

%When two NS5'-branes in Figure 24A are close to each other, it becomes 
%Figure 24B
% by realizing that the number of $(N_{c,3}-N_{c,4})$
%D4-branes connecting between $NS5_1$-brane and $NS5_3'$-brane can
%be rewritten as $(N_{c,2}-N_{c,1}-N_{c,4})$ plus $\widetilde{N}_{c,2}$. 
%The Figure 15 of \cite{Ahn07-6} is contained in the Figure 24. In
%particular, the brane configuration from the  $NS5_2'$-brane to 
%the $NS5_3$-brane is exactly same as the one of \cite{Ahn07-6}.

The cubic superpotential with the mass term  
is given by
(\ref{Wdual})
where we define $\Phi_3$ as $\Phi_3 \equiv F_2 \widetilde{F}_2$ and 
the second gauge group indices in $F_2$ and $\widetilde{F}_2$ 
are contracted, each third gauge group index in them is encoded in 
$\Phi_3$. 
Here the magnetic fields $f_2$ and $\widetilde{f}_2$  
correspond to 4-4 strings connecting 
the $\widetilde{N}_{c,2}$-color D4-branes(that are 
connecting between the $NS5_1$-brane
and the $NS5_3'$-brane in Figure 24B) with $N_{c,3}$-flavor 
D4-branes(which  are realized 
as corresponding D4-branes in Figure 24A).

The low energy dynamics of the magnetic brane configuration 
can be described by the ${\cal N}=1$ supersymmetric gauge theory
with gauge group
and the gauge couplings for the three gauge group factors are
given by the expressions in subsection 4.3.
The dual gauge theory has  a meson field  $\Phi_3$  and 
bifundamentals $f_2$ and $\widetilde{f}_2$ under the dual gauge
group  and the superpotential 
corresponding to Figures 24A and 24B is given by 
the expressions in subsection 4.3.
Then $ f_2 \widetilde{f}_2$ is a $\widetilde{N}_{c,2} \times \widetilde{N}_{c,2}$ 
matrix where the third gauge group indices for $f_2$ and 
$\widetilde{f}_2$ 
are contracted with those
of $\Phi_3$ while $\Phi_3$ is a 
$(N_{c,3}-N_{c,4}) \times (N_{c,3}-N_{c,4})$ matrix.
The product $f_2 \widetilde{f}_2$ 
has the same representation for the 
product of quarks
and moreover, 
the third gauge group indices for the field $\Phi_3$ play the
role of the flavor indices.

When the upper NS5'-brane(or $NS5_3'$-brane) 
is replaced by coincident $(N_{c,3}-N_{c,4})$ 
D6-branes and the $NS5_4'$-brane is rotated by $\frac{\pi}{2}$ in the 
$(v,w)$ plane in Figure 24B, this brane configuration looks similar to the one 
found in \cite{Ahn07-9} where the gauge group was given by 
$SO(n_{c,1}) \times SU(n_{f,2}+n_{c,1}+n_{c,3}-n_{c,2}) 
\times SU(n_{c,3}) \times \cdots$ 
with $n_{f,i}$ multiplets, $\widetilde{n}_{f,i}$ multiplets,
bifundamentals 
and gauge singlets. 
Then the present $N_{c,4}$ corresponds to the $n_{c,3}$,
$N_{c,2}$ corresponds to $n_{c,2}$,
$N_{c,1}$ corresponds to $n_{c,1}$,
and 
$(N_{c,3}-N_{c,4})$ corresponds to the $n_{f,2}$. 
%In particular, the Figure 12B of \cite{Ahn07-9} with vanishing flavors
%$Q$ and $Q''$
%is contained in
%this modified Figure 24B running from the $NS5_{2}'$-brane to 
%the $NS5_{4}'$-brane.
Therefore, the F-term equation, the derivative $W_{dual}$ with respect to the
meson field $\Phi_3$ cannot be satisfied if the $(N_{c,3}-N_{c,4})$ exceeds
$\widetilde{N}_{c,2}$.
So the supersymmetry is broken.   
That is, 
there exist three equations from F-term conditions:
$
f_2^a \widetilde{f}_{2,b} -\mu_3^2 \delta^a_b =0$ and $ \Phi_3 f_2 =0=\widetilde{f}_2 
\Phi_3$.
Then the solutions for these
are given by 
the expressions in subsection 4.3.
Then one can expand these fields around on a point 
and one arrives at the relevant superpotential
up to quadratic order in the fluctuation. 
At one loop, the effective potential $V_{eff}^{(1)}$ for $M_3$
leads to the positive value for $m_{M_3}^2$ implying that these
vacua are stable.

%%%%%%%%%%%%%%%%%%%%%%%%%%%%%%%%%%%%%%%%%%%%%%%%%%%%%%%%%%%%%%%%%%%%%%%%%%%
%%%%%%%%%%%%%%%%%%%%%%%%%%%%%%%%%%%%%%%%%%%%%%%%%%%%%%%%%%%%%%%%%%%%%%%%%%%
\subsection{
${\cal N}=1$ 
$SO(N_{c,1}) 
\times \cdots \times SU(\widetilde{N}_{c,i}) \times
\cdots \times SU(N_{c,n})$ magnetic theory
}
%%%%%%%%%%%%%%%%%%%%%%%%%%%%%%%%%%%%%%%%%%%%%%%%%%%%%%%%%%%%%%%%%%%%%%%%%%%
%%%%%%%%%%%%%%%%%%%%%%%%%%%%%%%%%%%%%%%%%%%%%%%%%%%%%%%%%%%%%%%%%%%%%%%%%%%

%Let us first consider the Seiberg dual for the middle gauge group
%factor.
%There are two magnetic duals depending on whether the gauge group factor
%occurs at odd chain or even chain.

%%%%%%%%%%%%%%%%%%%%%%%%%%%%%%%%%%%%%%%%%%%%%%%%%%%%%%%%%%%%%%
%%%%%%%%%%%%%%%%%%%%%%%%%%%%%%%%%%%%%%%%%%%%%%%%%%%%%%%%%%%%%%
\subsubsection{When the dual gauge group occurs at odd chain}
%%%%%%%%%%%%%%%%%%%%%%%%%%%%%%%%%%%%%%%%%%%%%%%%%%%%%%%%%%%%%%
%%%%%%%%%%%%%%%%%%%%%%%%%%%%%%%%%%%%%%%%%%%%%%%%%%%%%%%%%%%%%%

Let us consider other magnetic theory for the same electric theory.
Starting from Figure 23A and 
interchanging the $NS5_{i-1}'$-brane and the $NS5_i$-brane(and their mirrors),
one obtains the magnetic brane configuration which is exactly the same
as the Figure 18A, in subsection 4.1.2., 
with a replacement of $O6^{+}$-plane instead of $O6^{-}$-plane.
Let us call this as a ``modified'' Figure 18A.

The dual gauge group is given by
\bea
SO(N_{c,1}) \times \cdots  \times 
SU(\widetilde{N}_{c,i} \equiv N_{c,i+1}+N_{c,i-1}-N_{c,i}) 
 \times \cdots \times SU(N_{c,n}).
\nonu
\eea
The matter contents are the field $f_{i-1}$ 
 charged under
$({\bf 1, \cdots, \Box_{i-1}, \overline{\Box}_{i}, \cdots, 1_n})$, 
 and their conjugates 
$\widetilde{f}_{i-1}$ charged $({\bf 1_1, \cdots, \overline{\Box}_{i-1},
  \Box_{i}, \cdots, 1_n })$ 
under the dual gauge group
and  
the gauge-singlet $\Phi_{i+1}$ which is in the 
adjoint representation for the $i$-th dual gauge group, 
in other words,   
$ ({ \bf   1_1,  \cdots, 1_{i}, 
(N_{c,i+1}-N_{c,i+2})^2-1, \cdots, 1_n})  \oplus  ({\bf 1_1, \cdots,1_n})$ under the 
dual gauge group where the gauge group is broken from
$SU(N_{c,i+1})$ 
to $SU(N_{c,i+1}-N_{c,i+2})$. 
Then the $\Phi_{i+1}$ is a 
$(N_{c,i+1}-N_{c,i+2}) \times (N_{c,i+1}-N_{c,i+2})$ matrix.
Only $(N_{c,i+1}-N_{c,i+2})$ 
D4-branes can participate in the mass deformation.

%When two NS5'-branes in modified Figure 18A are close to each other, then 
%it leads to modified Figure 18B
%by realizing that the number of $(N_{c,i+1}-N_{c,i+2})$
%D4-branes connecting between $NS5_i$-brane and $NS5_{i+1}'$-brane in
%modified Figure
%18A can
%be rewritten as $(N_{c,i}-N_{c,i+2}-N_{c,i-1})$ plus $\widetilde{N}_{c,i}$.

The cubic superpotential with the mass term
is given by (\ref{superpo11}) in subsection 4.1.2
and we define $\Phi_{i+1}$ as $\Phi_{i+1} \equiv F_{i} \widetilde{F}_{i}$ and 
the $i$-th gauge group indices in $F_{i}$ and $\widetilde{F}_{i}$ 
are contracted, each $(i+1)$-th gauge group index in them is encoded in 
$\Phi_{i+1}$.  
The brane configuration for zero mass for the bifundamental,
which has only a cubic superpotential (\ref{superpo11}),
can be obtained from modified Figure 18A by moving
the upper  NS5'-brane together with $(N_{c,i+1}-N_{c,i+2})$ color D4-branes 
into the origin $v=0$(and their mirrors).

%The brane configuration in modified Figure 18A is stable as long as the
%distance $(\Delta x)_{i+1}$ between the upper NS5'-brane and 
%the lower NS5'-brane is large. 
%If they are close to each other, then this brane
%configuration is unstable to decay and leads to 
%the brane configuration in modified Figure
%18B.

%In particular, the Figure 14B of \cite{Ahn07-9} with vanishing flavors
%$Q$ and $Q'$
%is contained in
%the modified Figure 18B running from the $NS5_{i-2}$-brane to 
%the $NS5_{i+1}'$-brane.

The low energy dynamics of the magnetic brane configuration 
can be described by the ${\cal N}=1$ supersymmetric gauge theory
with gauge group 
and the gauge couplings for the three gauge group factors are
given by those in subsection 4.1.2
and 
the superpotential 
corresponding to modified Figures 18A and 18B is given by the one 
in subsection 4.1.2. 
Therefore, the F-term equation, the derivative $W_{dual}$ with respect to the
meson field $\Phi_{i+1}$ cannot be satisfied if the 
$(N_{c,i+1}-N_{c,i+2})$ exceeds
$\widetilde{N}_{c,i}$.
So the supersymmetry is broken.   
That is, 
there exist three equations from F-term conditions:
$
f_{i}^a \widetilde{f}_{i,b} -\mu_{i+1}^2 \delta^a_b =0$ and $ \Phi_{i+1} f_{i} =
0=\widetilde{f}_{i} \Phi_{i+1}$.
Then the solutions for these
are given by previous results. 
%At one loop, the effective potential $V_{eff}^{(1)}$ for $M_{i+1}$
%leads to the positive value for $m_{M_{i+1}}^2$ implying that these
%vacua are stable.

%%%%%%%%%%%%%%%%%%%%%%%%%%%%%%%%%%%%%%%%%%%%%%%%%%%%%%%%%%%%%%
%%%%%%%%%%%%%%%%%%%%%%%%%%%%%%%%%%%%%%%%%%%%%%%%%%%%%%%%%%%%%%
\subsubsection{When the dual gauge group occurs at even chain}
%%%%%%%%%%%%%%%%%%%%%%%%%%%%%%%%%%%%%%%%%%%%%%%%%%%%%%%%%%%%%%
%%%%%%%%%%%%%%%%%%%%%%%%%%%%%%%%%%%%%%%%%%%%%%%%%%%%%%%%%%%%%%

Starting from Figure 15B and 
interchanging the $NS5_{i-1}$-brane and the $NS5_i'$-brane(and their mirrors),
one obtains the Figure 17A, in subsection 4.1.1, 
with a replacement of $O6^{+}$-plane instead of $O6^{-}$-plane.
Let us call this as a ``modified'' Figure 17A.

The dual gauge group is given by 
\bea
SO(N_{c,1}) \times \cdots \times 
SU(\widetilde{N}_{c,i} \equiv N_{c,i+1}+N_{c,i-1}-N_{c,i}) 
\times \cdots \times SU(N_{c,n})
\nonu
\eea
and the matter contents are the field $f_{i-1}$ 
charged under
$({\bf 1_1, \cdots, 1_{i-2}, 
N_{c,i-1}, \overline{\widetilde{N}_{c,i}}, \cdots, 1_n})$, and its 
conjugate field  
$\widetilde{f}_{i-1}$  charged under 
$({\bf 1_1, \cdots, 1_{i-2}, 
\overline{N_{c,i-1}}, \widetilde{N}_{c,i}, \cdots, 1_n})$
under the dual gauge group
and  
the gauge-singlet $\Phi_{i-1}$ for the $i$-th dual gauge group in the 
adjoint representation for the $(i-1)$-th dual gauge group, 
i.e.,  ${(\bf 1_1, \cdots, 1_{i-2}, (N_{c,i-1}-N_{c,i-2})^2-1, 1,
  \cdots, 1_n)  }$ plus a singlet under the 
dual gauge group.
Then the  $\Phi_{i-1}$ is a $(N_{c,i-1}-N_{c,i-2}) \times
(N_{c,i-1}-N_{c,i-2})$ 
matrix.

%When two NS5'-branes in modified Figure 17A are close to each other, it becomes 
%modified Figure 17B
%by realizing that the number of $(N_{c,i-1}-N_{c,i-2})$
%D4-branes connecting between $NS5_{i-2}'$-brane and $NS5_{i-1}$-brane
%in modified
%Figure
%17A can
%be rewritten as $(N_{c,i}-N_{c,i+1}-N_{c,i-2})$ plus $\widetilde{N}_{c,i}$. 

The cubic superpotential with the mass term  
is given by (\ref{superpo1new}) in subsection 4.1.1
and 
the brane configuration for zero mass for the bifundamental,
which has only a cubic superpotential (\ref{superpo1new}),
can be obtained from modified Figure 17A by moving
the upper  NS5'-brane(or $NS5_{i-2}'$-brane) 
together with $(N_{c,i-1}-N_{c,i-2})$ color D4-branes 
into the origin $v=0$(and their mirrors).
The low energy dynamics of the magnetic brane configuration 
can be described by the ${\cal N}=1$ supersymmetric gauge theory
with gauge group 
and the gauge couplings for the three gauge group factors are
given by the ones in subsection 4.1.1.
The dual gauge theory has  a meson $\Phi_{i-1}$  and 
bifundamentals $f_{i-1}$, and $\widetilde{f}_{i-1}$ under the dual gauge
group and the superpotential 
corresponding to modified Figures 17A and 17B is the same as  
the one in subsection 4.1.1.
%
%In particular, the Figure 13B of \cite{Ahn07-9} with vanishing flavors
%$Q$ and $Q''$
%is contained in
%the modified Figure 17B running from the $NS5_{i-2}'$-brane to 
%the $NS5_{i+1}$-brane.
Therefore, the F-term equation, the derivative $W_{dual}$ with respect to the
meson field $\Phi_{i-1}$ cannot be satisfied if the 
$(N_{c,i-1}-N_{c,i-2})$ exceeds
$\widetilde{N}_{c,i}$.
So the supersymmetry is broken.   
That is, 
there exist three equations from F-term conditions:
$
f_{i-1}^a \widetilde{f}_{i-1,b} -\mu_{i-1}^2 \delta^a_b =0$ 
and $ \Phi_{i-1} f_{i-1} =0=
\widetilde{f}_{i-1} \Phi_{i-1}$.
Then the solutions for these
can be obtained. 
At one loop, the effective potential $V_{eff}^{(1)}$ for $M_{i-1}$
leads to the positive value for $m_{M_{i-1}}^2$ implying that these
vacua are stable.

%%%%%%%%%%%%%%%%%%%%%%%%%%%%%%%%%%%%%%%%%%%%%%%%%%%%%%%%%%%%%%%%%%%%%%%%%%%%%%
%%%%%%%%%%%%%%%%%%%%%%%%%%%%%%%%%%%%%%%%%%%%%%%%%%%%%%%%%%%%%%%%%%%%%%%%%%%%%%
\subsection{${\cal N}=1$ 
$SO(N_{c,1}) \times \cdots \times SU(\widetilde{N}_{c,n})$ magnetic theory}
%%%%%%%%%%%%%%%%%%%%%%%%%%%%%%%%%%%%%%%%%%%%%%%%%%%%%%%%%%%%%%%%%%%%%%%%%%%%%%
%%%%%%%%%%%%%%%%%%%%%%%%%%%%%%%%%%%%%%%%%%%%%%%%%%%%%%%%%%%%%%%%%%%%%%%%%%%%%%

%Let us consider the Seiberg dual for the last gauge group
%factor.
%There are two magnetic duals depending on whether the gauge group factor
%occurs at odd chain or even chain.

%%%%%%%%%%%%%%%%%%%%%%%%%%%%%%%%%%%%%%%%%%%%%%%%%%%%%%%%%%%%%%
%%%%%%%%%%%%%%%%%%%%%%%%%%%%%%%%%%%%%%%%%%%%%%%%%%%%%%%%%%%%%%
\subsubsection{When the dual gauge group occurs at odd chain}
%%%%%%%%%%%%%%%%%%%%%%%%%%%%%%%%%%%%%%%%%%%%%%%%%%%%%%%%%%%%%%
%%%%%%%%%%%%%%%%%%%%%%%%%%%%%%%%%%%%%%%%%%%%%%%%%%%%%%%%%%%%%%

Let us consider other magnetic theory for the same electric theory.
By applying the Seiberg dual to the $SU(N_{c,n})$ factor  and 
interchanging the $NS5_{n-1}'$-brane and the $NS5_{n}$-brane,
one obtains the Figure 20A which appears in subsection 4.2.2, 
with a replacement of $O6^{+}$-plane instead of $O6^{-}$-plane.
Let us call this as a ``modified'' Figure 20A.

The gauge group is given by
\bea
SO(N_{c,1}) \times \cdots \times 
SU(N_{c,n-1}) \times SU(\widetilde{N}_{c,n} \equiv N_{c,n-1}-N_{c,n})
\nonu
\eea
and the matter contents are the field $f_{n-1}$ 
 charged under
$( {\bf 1_1, \cdots, 1_{n-2}, \Box_{n-1}, \overline{\Box}_{n} })$ 
 and their conjugates 
$\widetilde{f}_{n-1}$ 
$( {\bf 1_1, \cdots, 1_{n-2}, \overline{\Box}_{n-1}, \Box_{n} })$ 
under the dual gauge group
and  
the gauge-singlet $\Phi_{n-1}$ which is in the 
adjoint representation for the $(n-1)$-th gauge group, 
in other words,   
$ ({ \bf   1_1, 
\cdots, 1_{n-2},  (N_{c,n-1}-N_{c,n-2})^2-1,1_n})  \oplus  ({\bf 1_1,
 \cdots, 1_n})$ under the
dual gauge group where the gauge group is broken from
$SU(N_{c,n-1})$ 
to $SU(N_{c,n-1}-N_{c,n-2})$.
Then the $\Phi_{n-1}$ is a $(N_{c,n-1}-N_{c,n-2}) \times
 (N_{c,n-1}-N_{c,n-2})$ 
matrix.
Only $(N_{c,n-1}-N_{c,n-2})$ D4-branes can participate in the mass deformation.

%When two NS5'-branes in modified Figure 20A are close to each other, then 
%it leads to Figure modified 20B
% by realizing that the number of $(N_{c,n-1}-N_{c,n-2})$
%D4-branes connecting between $NS5_{n-2}'$-brane and $NS5_{n}$-brane can
%be rewritten as $(N_{c,n}-N_{c,n-2})$ plus $\widetilde{N}_{c,n}$.
%The Figure 16 of \cite{Ahn07-6} is contained in the modified Figure 20. In
%particular, the brane configuration from the  $NS5_{n-2}'$-brane to 
%the $NS5_{n-1}'$-brane is exactly same as the one of \cite{Ahn07-6}.

%The brane configuration in modified Figure 20A is stable as long as the
%distance $(\Delta x)_{n-1}$ between the upper NS5'-brane and 
%the lower NS5'-brane(or $NS5_{n-1}'$-brane) 
%is large. If they are close to each other, then this brane
%configuration is unstable to decay to 
%the brane configuration in modified Figure
%20B.
%
%The brane configuration for zero mass for the bifundamental,
%which has only a cubic superpotential (\ref{ssuper}),
%can be obtained from modified Figure 20A by moving
%the upper  NS5'-brane together with $(N_{c,n-1}-N_{c,n-2})$ color D4-branes 
%into the origin $v=0$.

The low energy dynamics of the magnetic brane configuration 
can be described by the ${\cal N}=1$ supersymmetric gauge theory
with gauge group
and the gauge couplings for the three gauge group factors are
given by
similarly and 
the dual gauge theory has  a meson field $\Phi_{n-1}$  and 
bifundamentals $f_{n-1}, \widetilde{f}_{n-1}, F_j$ and
$\widetilde{F}_j$ 
under the dual gauge
group and the superpotential 
corresponding to modified Figures 20A and 20B is given by 
the one in subsection 4.2.2.
Therefore, the F-term equation, the derivative $W_{dual}$ with respect to the
meson field $\Phi_{n-1}$ cannot be satisfied if the $(N_{c,n-1}-N_{c,n-2})$ exceeds
$\widetilde{N}_{c,n}$.
So the supersymmetry is broken.   
That is, 
there exist three equations from F-term conditions:
$
f_{n-1}^a \widetilde{f}_{n-1,b} -\mu_{n-1}^2 \delta^a_b =0$ 
and $ \Phi_{n-1} f_{n-1} =0=
\widetilde{f}_{n-1} \Phi_{n-1}$.
Then the solutions for these
are given by similarly.
%At one loop, the effective potential $V_{eff}^{(1)}$ for $M_{n-1}$
%leads to the positive value for $m_{M_{n-1}}^2$ implying that these
%vacua are stable.

%%%%%%%%%%%%%%%%%%%%%%%%%%%%%%%%%%%%%%%%%%%%%%%%%%%%%%%%%%%%%%
%%%%%%%%%%%%%%%%%%%%%%%%%%%%%%%%%%%%%%%%%%%%%%%%%%%%%%%%%%%%%%
\subsubsection{When the dual gauge group occurs at even chain}
%%%%%%%%%%%%%%%%%%%%%%%%%%%%%%%%%%%%%%%%%%%%%%%%%%%%%%%%%%%%%%
%%%%%%%%%%%%%%%%%%%%%%%%%%%%%%%%%%%%%%%%%%%%%%%%%%%%%%%%%%%%%%

Starting from Figure 23B, moving the $NS5_{n-2}'$-brane 
with $(N_{c,n-1}-N_{c,n-2})$
D4-branes 
to the $+v$ direction leading to Figure 23B, 
and interchanging the $NS5_{n-1}$-brane and the $NS5_{n}'$-brane,
one obtains the Figure 19A in subsection 4.2.1,
with a replacement of $O6^{+}$-plane instead of $O6^{-}$-plane.
Let us call this as a ``modified'' Figure 19A.

The dual gauge group is given by 
\bea
SO(N_{c,1}) \times \cdots \times 
SU(N_{c,n-1}) \times SU(\widetilde{N}_{c,n} \equiv N_{c,n-1}-N_{c,n})
\nonu
\eea
the matter contents are   
the bifundamentals $f_{n-1}$ in 
 $({\bf 1_1, \cdots, 1, \Box_{n-1}, \overline{\Box}_{n}, })$,
and $\widetilde{f}_{n-1}$ in the representation 
$({\bf 1_1, \cdots, 1, \overline{\Box}_{n-1}, \Box_{n}})$ in
addition to $(n-2)$ bifundamentals $F_j$ and $\widetilde{F}_j$, 
 $j=1,2,
\cdots, (n-2)$ and
the gauge singlet $\Phi_{n-1}$
for the $n$-th dual gauge group in the 
adjoint representation for the $(n-1)$-th dual gauge group, 
i.e.,  
$
{(\bf 1_1, \cdots, 1_{n-2}, (N_{c,n-1}-N_{c,n-2})^2-1, 1_n)  
}
$ plus a singlet
under the 
dual gauge group where the gauge group is broken from
$SU(N_{c,n-1})$ 
to $SU(N_{c,n-1}-N_{c,n-2})$.

%When two NS5'-branes in modified Figure 19A are close to each other, then 
%it leads to modified Figure 19B by realizing that the number of $(N_{c,n-1}-N_{c,n-2})$
%D4-branes connecting between $NS5_{n-2}'$-brane and $NS5_{n-1}$-brane can
%be rewritten as $(N_{c,n}-N_{c,n-2})$ plus $\widetilde{N}_{c,n}$.

%The brane configuration for zero mass for the bifundamental $F_{n-1}$
%and 
%$\widetilde{F}_{n-1}$,
%which has only a cubic superpotential (\ref{spotential}),
%can be obtained from modified Figure 19A by moving
%the upper $NS5_{n-2}'$-brane together with 
%$(N_{c,n-1}-N_{c,n-2})$ color D4-branes 
%into the origin $v=0$.
%The brane configuration in modified Figure 19A is stable as long as the
%distance $(\Delta x)_{n-1}$ between the upper NS5'-brane and 
%the lower NS5'-brane is large, as
%in \cite{GK}. If they are close to each other, then this brane
%configuration is unstable to decay and leads to 
%the brane configuration in modified Figure
%19B.

%In particular, the Figure 13B of \cite{Ahn07-9} with vanishing flavors
%$Q$ and $Q''$
%is contained in
%the modified Figure 19B running from the $NS5_{n-3}'$-brane to 
%the $NS5_{n-1}$-brane.

The dual gauge theory has  a meson field $\Phi_{n-1}$  and 
bifundamentals $f_{n-1}, \widetilde{f}_{n-1}, F_j$, and 
$\widetilde{F}_j$  and the superpotential 
corresponding to modified Figures 19A and 19B is given by 
previous results.
Therefore, the F-term equation, the derivative $W_{dual}$ with respect to the
meson field $\Phi_{n-1}$ cannot be satisfied if the $(N_{c,n-1}-N_{c,n-2})$ exceeds
$\widetilde{N}_{c,n}$.
So the supersymmetry is broken.   
%Then the solutions for these
%are given by similarly.
%At one loop, the effective potential $V_{eff}^{(1)}$ for $M_{n-1}$
%leads to the positive value for $m_{M_{n-1}}^2$ implying that these
%vacua are stable.

%%%%%%%%%%%%%%%%%%%%%%%%%%%%%%%%%%%%%%%%%%%%%%%%%%%%%%%%%%%%%%%%%%%%%
%%%%%%%%%%%%%%%%%%%%%%%%%%%%%%%%%%%%%%%%%%%%%%%%%%%%%%%%%%%%%%%%%%%%%
\section{Meta-stable brane configurations with $(2n+1)$ NS-branes plus
$O6^{+}$-plane}
%section6%%%%%%%%%%%%%%%%%%%%%%%%%%%%%%%%%%%%%%%%%%%%%%%%%%%%%%%%%%%%%%
%%%%%%%%%%%%%%%%%%%%%%%%%%%%%%%%%%%%%%%%%%%%%%%%%%%%%%%%%%%%%%%%%%%%%

The type IIA brane configuration,  
 by generalizing the brane
configurations \cite{LL,LLL} to the case where there are more
NS-branes,  
corresponding to 
${\cal N}=1$ supersymmetric gauge theory(see also
\cite{Ahn07-9})  with
gauge group
\bea
SU(N_{c,1}) \times SU(N_{c,2}) \cdots
\times SU(N_{c,n})
\nonu
\eea
and with a symmetric tensor field $S$ charged under $({\bf \frac{1}{2}
N_{c,1}(N_{c,1}+1), 1_2, \cdots, 1_n})$, 
the $(n-1)$ bifundametals $F_i$ charged under 
$({\bf 1_1, \cdots, 1, \Box_i, \overline{\Box}_{i+1}, 1, \cdots,  1_n})$
and their
complex conjugate fields $\widetilde{F}_i$ 
charged $({\bf 1_1, \cdots, 1, \overline{\Box}_i, \Box_{i+1}, 1, 
\cdots, 1_n})$ where $i=1, 2, \cdots, (n-1)$ as well as
a conjugate symmetric 
field  $\widetilde{S}$ charged under $({\bf \overline{\frac{1}{2}
N_{c,1}(N_{c,1}+1)}, 1_2, \cdots, 1_n})$
can be described by the NS5-brane located at the origin, 
the $NS5_1'$-brane, 
the  
$NS5_2$-brane, $\cdots$, the $NS5_{n}'$-brane for odd number
of gauge groups(or 
the $NS5_{n}$-brane for even number of gauge groups),
$N_{c,1}$-, $N_{c,2}$-,  $\cdots$, and $N_{c,n}$-color D4-branes. 
See the Figure 25 for the details on the brane configuration. 
The $O6^{+}$-plane acts as $(x^4,x^5,x^6) \rightarrow
(-x^4,-x^5,-x^6)$ and has RR charge $+4$.

%%%%%%%%%%%%%%%%%%%%%%%%%%%%%%%%%%%%%%%%%%%%%%%%%%%%%%%%%%%%%%%%%%%%%
%%%%%%%%%%%%%%%%%%%%%%%%%%%%%%%%%%%%%%%%%%%%%%%%%%%%%%%%%%%%%%%%%%%%%%
\begin{figure}[ht]
   \epsfxsize=4.0in 
\centerline{\epsffile{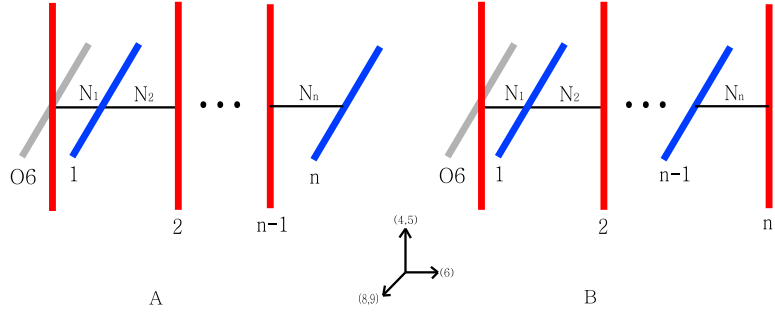}}
   \caption[FIG. \arabic{figure}.]{ 
The 
 ${\cal N}=1$ supersymmetric 
electric brane configuration for the gauge group $ \prod_{i=1}^{n}
 SU(N_{c,i})$ 
and  bifundamentals $F_i$ and $\widetilde{F}_i$  with vanishing mass
for the bifundamentals when the number of gauge groups factor 
$n$ is odd(25A) and even(25B). 
We do not draw the mirrors of the branes appearing in the left hand
 side of $O6^{+}$-plane. 
%This Figure looks similar to the Figure 15
% where there is no middle NS5-brane. 
}
\end{figure}
%%%%%%%%%%%%%%%%%%%%%%%%%%%%%%%%%%%%%%%%%%%%%%%%%%%%%%%%%%%%%%%%%%%%%
%Figure 29A and 29B
%%%%%%%%%%%%%%%%%%%%%%%%%%%%%%%%%%%%%%%%%%%%%%%%%%%%%%%%%%%%%%%%%%%%%%%

Let us place an O6-plane at the origin $x^6=0$
and denote the $x^6$ 
coordinates for 
the  
$NS5_1'$-brane, $\cdots$, the $NS5_{n}'$-brane for odd $n$(or 
the $NS5_{n}$-brane for even $n$)
are given by $x^6=y_1, y_1+y_2, \cdots, \sum_{j=1}^{n-1} y_j + y_{n}$
respectively.
The $N_{c,1}$ D4-branes 
are suspended between the 
$NS5_1'$-brane and its mirror, 
the $N_{c,2}$ D4-branes 
are suspending between the 
$NS5_1'$-brane and the $NS5_2$-brane, $\cdots$ and 
the $N_{c,n}$ D4-branes  
are suspended between the $NS5_{n-1}$-brane and the $NS5_{n}'$-brane for
odd $n$(or 
between the $NS5_{n-1}'$-brane and the $NS5_{n}$-brane for even $n$).
The fields $F_i$ and $\widetilde{F}_i$  correspond to 4-4 strings connecting 
the $N_{c,i}$-color D4-branes with $N_{c,i+1}$-color D4-branes.
We draw this ${\cal N}=1$ supersymmetric 
electric brane configuration in Figure 25A(25B) 
when $n$ is odd(even) for the vanishing mass
for the fields $F_i$ and $\widetilde{F}_i$.
The fields $S$ and $\widetilde{S}$  correspond to 4-4 strings connecting 
the $N_{c,1}$-color D4-branes with $x^6 < 0$ with $N_{c,1}$-color D4-branes
with $x^6 > 0$. 

There is no superpotential in Figure 25A. Let us deform this theory.
Displacing the two NS5'-branes relative each other in the $+v$
direction, characterized by $(\Delta x)_{i-1}$,  
corresponds to turning on a quadratic
mass-deformed superpotential
for the field $F_{i-1}$ and $\widetilde{F}_{i-1}$ as follows:
\bea
W = m_{i-1} F_{i-1} \widetilde{F}_{i-1} (\equiv m_{i-1} \Phi_{i-1})
\nonu
\eea
where 
the $i$-th gauge group indices in $F_{i-1}$ and $\widetilde{F}_{i-1}$ 
are contracted.
\bea
m_{i-1} 
%=\frac{(\Delta x)_{i+1}}{2\pi \alpha'} 
= 
\frac{(\Delta x)_{i-1}}{\ell_s^2}.
\nonu
\eea
The gauge-singlet $\Phi_{i-1}$ for the $i$-th  gauge group is in the 
adjoint representation for the $(i-1)$-th  gauge group, 
i.e., ${\bf ( 1_1, \cdots, 1_{i-2}, (N_{c,i-1}-N_{c,i-2})^2-1, 
\cdots, 1_n)  \oplus (1_1, \cdots ,1_n) }$ 
under the  gauge group
where the gauge group is broken from
$SU(N_{c,i-1})$ 
to $SU(N_{c,i-1}-N_{c,i-2})$. 
The $\Phi_{i-1}$ is a $(N_{c,i-1}-N_{c,i-2}) \times (N_{c,i-1}-N_{c,i-2})$ matrix.
The $NS5_{i-2}'$-brane together with $(N_{c,i-1}-N_{c,i-2})$-color D4-branes 
is moving to the $+v$ direction  for
fixed other branes during this mass deformation(and their mirrors to
$-v$ direction). 
Then the $x^5$ coordinate 
of $NS5_i'$-brane is equal to
zero
while the $x^5$ coordinate of $NS5_{i-2}'$-branes is given by 
$ (\Delta x)_{i-1}$.
Giving an expectation value to the meson field $\Phi_{i-1}$
corresponds to recombination of $N_{c,i-1}$- and $N_{c,i}$- color 
D4-branes, which will become $N_{c,i-1}$ or $N_{c,i}$-color D4-branes
in Figure 25A such that they are suspended between 
the $NS5_{i-2}'$-brane and the $NS5_i'$-brane 
and pushing them into the $w$
direction. 
%We assume that the number of colors satisfies
%$
%N_{c,i-1} \geq N_{c,i}-N_{c,i+1} \geq N_{c,i-2}$.

Now 
we obtain this brane configuration which is the same as the 
Figure 16A except the middle NS5-brane for nonvanishing mass
for the fields $F_i$ and $\widetilde{F}_i$.

Let us deform the theory by Figure 25B.
Displacing the two NS5'-branes, the $NS_{i-1}'$-brane and the 
$NS_{i+1}'$-brane, 
relative each other in the 
$v$ 
direction, characterized by $(\Delta x)_{i+1}$, 
corresponds to turning on a quadratic
mass-deformed superpotential
for the fields $F_{i}$ and $\widetilde{F}_{i}$ as follows:
\bea
W = m_{i+1} F_{i} \widetilde{F}_{i} (\equiv m_{i+1} \Phi_{i+1})
\nonu
\eea
where 
the $i$-th gauge group indices in $F_{i}$ and $\widetilde{F}_{i}$ 
are contracted, each $(i+1)$-th gauge group index in them is encoded in 
$\Phi_{i+1}$ and the mass $m_{i+1}$ is given by
\bea
m_{i+1} 
%=\frac{(\Delta x)_{i+1}}{2\pi \alpha'} 
= 
\frac{(\Delta x)_{i+1}}{\ell_s^2}.
\nonu
\eea

The gauge-singlet $\Phi_{i+1}$ for the $i$-th  gauge group is in the 
adjoint representation for the $(i+1)$-th  gauge group, 
i.e., 
\bea
{(\bf 1_1, \cdots, 1_{i}, (N_{c,i+1}-N_{c,i+2})^2-1, 1_{i}, \cdots,
  1_n)  
\oplus (1_1, \cdots, 1_n)}
\nonu
\eea 
under the  gauge group where the gauge group is broken from
$SU(N_{c,i+1})$ 
to $SU(N_{c,i+1}-N_{c,i+2})$.  
Then the $\Phi_{i+1}$ is a $(N_{c,i+1}-N_{c,i+2}) \times 
(N_{c,i+1}-N_{c,i+2})$ matrix.
The $NS5_{i+1}'$-brane together with $(N_{c,i+1}-N_{c,i+2})$-color D4-branes 
is moving to the $+v$ direction  for
fixed other branes during this mass deformation. 
In other words, the $N_{c, i+2}$ D4-branes among $N_{c,i+1}$ D4-branes 
are not participating in 
the mass deformation.
Then the $x^5$ coordinate($\equiv x$) 
of $NS5_{i-1}'$-brane is equal to
zero
while the $x^5$ coordinate of $NS5_{i+1}'$-brane is given by 
$(\Delta x)_{i+1}$.
Giving an expectation value to the meson field $\Phi_{i+1}$
corresponds to recombination of $N_{c,i}$- and $N_{c,i+1}$- color 
D4-branes, which will become $N_{c,i}$- or $N_{c,i+1}$-color D4-branes
in Figure 25B such that they are suspended between 
the $NS5_{i-1}'$-brane and the $NS5_{i+1}'$-brane 
and pushing them into the 
$w$ direction. We assume that the number of colors satisfies
$
N_{c,i+1} \geq N_{c,i}-N_{c,i-1} \geq N_{c,i+2}$.
Now 
we obtain this brane configuration that is the same as the 
Figure 16B except the middle NS5-brane for nonvanishing mass
for the fields $F_{i}$ and $\widetilde{F}_{i}$. 

%Next we describe five different magnetic dual theories by taking each
%corresponding mass deformation.
%Most of the brane configurations in this section are coincident with the ones in
%previous section if we neglect the presence of middle NS5-brane 
%and we will mention the main results briefly.

%%%%%%%%%%%%%%%%%%%%%%%%%%%%%%%%%%%%%%%%%%%%%%%%%%%%%%%%%%%%%%%%%%%%%%%%%
%%%%%%%%%%%%%%%%%%%%%%%%%%%%%%%%%%%%%%%%%%%%%%%%%%%%%%%%%%%%%%%%%%%%%%%%%
\subsection{${\cal N}=1$ 
$SU(N_{c,1}) \times \cdots \times SU(\widetilde{N}_{c,i}) \times \cdots \times
SU(N_{c,n})$ magnetic theory}
%%%%%%%%%%%%%%%%%%%%%%%%%%%%%%%%%%%%%%%%%%%%%%%%%%%%%%%%%%%%%%%%%%%%%%%%
%%%%%%%%%%%%%%%%%%%%%%%%%%%%%%%%%%%%%%%%%%%%%%%%%%%%%%%%%%%%%%%%%%%%%%%%

%Let us first consider the Seiberg dual for the middle gauge group
%factor.
%There are two magnetic duals depending on whether the gauge group factor
%occurs at odd chain or even chain.

%%%%%%%%%%%%%%%%%%%%%%%%%%%%%%%%%%%%%%%%%%%%%%%%%%%%%%%%%%%%%%
%%%%%%%%%%%%%%%%%%%%%%%%%%%%%%%%%%%%%%%%%%%%%%%%%%%%%%%%%%%%%%
\subsubsection{When the dual gauge group occurs at odd chain}
%%%%%%%%%%%%%%%%%%%%%%%%%%%%%%%%%%%%%%%%%%%%%%%%%%%%%%%%%%%%%%
%%%%%%%%%%%%%%%%%%%%%%%%%%%%%%%%%%%%%%%%%%%%%%%%%%%%%%%%%%%%%%

Starting from Figure 25A and 
interchanging the $NS5_{i-1}$-brane and the $NS5_i'$-brane(and their mirrors),
one obtains the Figure 17A, in subsection 4.1.1, 
with a replacement of a combination of $O6^{+}$-plane and a middle
NS5-brane,  
instead of $O6^{-}$-plane.
Let us denote this as the ``deformed'' Figure 17A.

The dual gauge group is given by 
\bea
SU(N_{c,1}) \times \cdots \times 
SU(\widetilde{N}_{c,i} \equiv N_{c,i+1}+N_{c,i-1}-N_{c,i}) 
\times \cdots \times SU(N_{c,n})
\nonu
\eea
The matter contents are the field $f_i$ 
charged under
$({\bf 1_1, \cdots, 1_{i-1}, 
\widetilde{N}_{c,i}, \overline{N_{c,i+1}}, \cdots, 1_n})$, and its 
conjugate field  
$\widetilde{f}_i$  charged under 
$({\bf 1_1, \cdots, 1_{i-1}, 
\overline{\widetilde{N}_{c,i}}, N_{c,i+1}, \cdots, 1_n})$
under the dual gauge group
and  
the gauge-singlet $\Phi_{i-1}$ for the $i$-th dual gauge group in the 
adjoint representation for the $(i-1)$-th dual gauge group, 
i.e.,  ${(\bf 1_1, \cdots, 1_{i-2}, (N_{c,i-1}-N_{c,i-2})^2-1, 1,
  \cdots, 1_n)  }$ plus a singlet under the 
dual gauge group
where the gauge group is broken from
$SU(N_{c,i-1})$ 
to $SU(N_{c,i-1}-N_{c,i-2})$. 
Then the  $\Phi_{i-1}$ is a $(N_{c,i-1}-N_{c,i-2}) \times
(N_{c,i-1}-N_{c,i-2})$ 
matrix.

%When two NS5'-branes in deformed Figure 17A are close to each other, it becomes 
%deformed Figure 17B
%by realizing that the number of $(N_{c,i-1}-N_{c,i-2})$
%D4-branes connecting between $NS5_{i-2}'$-brane and $NS5_{i-1}$-brane
%in 
%deformed Figure
%17A can
%be rewritten as $(N_{c,i}-N_{c,i+1}-N_{c,i-2})$ plus $\widetilde{N}_{c,i}$. 

The cubic superpotential with the mass term  
is given by the one in subsection 4.1.1
and 
the brane configuration for zero mass for the bifundamental,
which has only a cubic superpotential (\ref{superpo1new}),
can be obtained from deformed Figure 17A by moving
the upper  NS5'-brane(or $NS5_{i-2}'$-brane) 
together with $(N_{c,i-1}-N_{c,i-2})$ color D4-branes 
into the origin $v=0$(and their mirrors).
The low energy dynamics of the magnetic brane configuration 
can be described by the ${\cal N}=1$ supersymmetric gauge theory
with gauge group 
and the gauge couplings for the three gauge group factors are
given by the ones in subsection 4.1.1.
The dual gauge theory has  a meson $\Phi_{i-1}$  and 
bifundamentals $f_{i-1}$, and $\widetilde{f}_{i-1}$ under the dual gauge
group and the superpotential 
corresponding to deformed Figures 17A and 17B is the same as  
the one in subsection 4.1.1.
%In particular, the Figure 5B of \cite{Ahn07-9} with vanishing flavors
%$Q$ and $Q'$
%is contained in
%the deformed Figure 17B running from the $NS5_{i-3}'$-brane to 
%the $NS5_{i-1}$-brane.
Therefore, the F-term equation, the derivative $W_{dual}$ with respect to the
meson field $\Phi_{i-1}$ cannot be satisfied if the $(N_{c,i-1}-N_{c,i-2})$ exceeds
$\widetilde{N}_{c,i}$.
So the supersymmetry is broken.   
That is, 
there exist three equations from F-term conditions:
$
f_{i-1}^a \widetilde{f}_{i-1,b} -\mu_{i-1}^2 \delta^a_b =0$ 
and $ \Phi_{i-1} f_{i-1} =0=
\widetilde{f}_{i-1} \Phi_{i-1}$.
%Then the solutions for these
%can be obtained. 
%At one loop, the effective potential $V_{eff}^{(1)}$ for $M_{i-1}$
%leads to the positive value for $m_{M_{i-1}}^2$ implying that these
%vacua are stable.

%%%%%%%%%%%%%%%%%%%%%%%%%%%%%%%%%%%%%%%%%%%%%%%%%%%%%%%%%%%%%%
%%%%%%%%%%%%%%%%%%%%%%%%%%%%%%%%%%%%%%%%%%%%%%%%%%%%%%%%%%%%%%
\subsubsection{When the dual gauge group occurs at even chain}
%%%%%%%%%%%%%%%%%%%%%%%%%%%%%%%%%%%%%%%%%%%%%%%%%%%%%%%%%%%%%%
%%%%%%%%%%%%%%%%%%%%%%%%%%%%%%%%%%%%%%%%%%%%%%%%%%%%%%%%%%%%%%

Let us consider other magnetic theory for the same electric theory.
Starting from Figure 25B and 
interchanging the $NS5_{i-1}'$-brane and the $NS5_i$-brane(and their mirrors),
one obtains the magnetic brane configuration which is exactly the same
as the Figure 18A, in subsection 4.1.2., 
with a replacement of a combination of $O6^{+}$-plane and a middle
NS5-brane,  
instead of $O6^{-}$-plane.
Let us denote this as the ``deformed'' Figure 18A.

The dual gauge group is given by
\bea
SU(N_{c,1}) \times \cdots  \times 
SU(\widetilde{N}_{c,i} \equiv N_{c,i+1}+N_{c,i-1}-N_{c,i}) 
 \times \cdots \times SU(N_{c,n}).
\nonu
\eea
The matter contents are the field $f_{i}$ 
 charged under
$({\bf 1, \cdots, \Box_{i}, \overline{\Box}_{i+1}, \cdots, 1_n})$, 
 and their conjugates 
$\widetilde{f}_{i}$ charged $({\bf 1_1, \cdots, \overline{\Box}_{i},
  \Box_{i+1}, \cdots, 1_n })$ 
under the dual gauge group
and  
the gauge-singlet $\Phi_{i+1}$ which is in the 
adjoint representation for the $i$-th dual gauge group, 
in other words,   
$ ({ \bf   1_1,  \cdots, 1_{i}, 
(N_{c,i+1}-N_{c,i+2})^2-1, \cdots, 1_n})  \oplus  ({\bf 1_1, \cdots,1_n})$ under the 
dual gauge group
where the gauge group is broken from
$SU(N_{c,i+1})$ 
to $SU(N_{c,i+1}-N_{c,i+2})$.
Then the $\Phi_{i+1}$ is a $(N_{c,i+1}-N_{c,i+2}) 
\times (N_{c,i+1}-N_{c,i+2})$ matrix.
Only $(N_{c,i+1}-N_{c,i+2})$ 
D4-branes can participate in the mass deformation.

%When two NS5'-branes in deformed Figure 18A are close to each other, then 
%it leads to deformed Figure 18B
%by realizing that the number of $(N_{c,i+1}-N_{c,i+2})$
%D4-branes connecting between $NS5_i$-brane and $NS5_{i+1}'$-brane in
%deformed Figure
%18A can
%be rewritten as $(N_{c,i}-N_{c,i+2}-N_{c,i-1})$ plus $\widetilde{N}_{c,i}$.

The cubic superpotential with the mass term
is given by in subsection 4.1.2
and we define $\Phi_{i+1}$ as $\Phi_{i+1} \equiv F_{i} \widetilde{F}_{i}$ and 
the $i$-th gauge group indices in $F_{i}$ and $\widetilde{F}_{i}$ 
are contracted, each $(i+1)$-th gauge group index in them is encoded in 
$\Phi_{i+1}$.  
%The brane configuration for zero mass for the bifundamental,
%which has only a cubic superpotential (\ref{superpo11}),
%can be obtained from deformed Figure 18A by moving
%the upper  NS5'-brane together with $(N_{c,i+1}-N_{c,i+2})$ color D4-branes 
%into the origin $v=0$(and their mirrors).

%The brane configuration in deformed Figure 18A is stable as long as the
%distance $(\Delta x)_{i+1}$ between the upper NS5'-brane and 
%the lower NS5'-brane is large. 
%If they are close to each other, then this brane
%configuration is unstable to decay and leads to 
%the brane configuration in deformed Figure
%18B.

%In particular, the Figure 3B of \cite{Ahn07-9} with vanishing flavors
%$Q$ and $Q''$
%is contained in
%the deformed Figure 18B running from the $NS5_{i}$-brane to 
%the $NS5_{i+2}'$-brane.

The low energy dynamics of the magnetic brane configuration 
can be described by the ${\cal N}=1$ supersymmetric gauge theory
with gauge group 
and the gauge couplings for the three gauge group factors are
given by those in subsection 4.1.2
and 
the superpotential 
corresponding to deformed 
Figures 18A and 18B is given by the one in subsection 4.1.2. 
Therefore, the F-term equation, the derivative $W_{dual}$ with respect to the
meson field $\Phi_{i+1}$ cannot be satisfied if the $(N_{c,i+1}-N_{c,i+2})$ exceeds
$\widetilde{N}_{c,i}$.
So the supersymmetry is broken.   
That is, 
there exist three equations from F-term conditions:
$
f_{i}^a \widetilde{f}_{i,b} -\mu_{i+1}^2 \delta^a_b  =0$ 
and $ \Phi_{i+1} f_{i} =
0=\widetilde{f}_{i} \Phi_{i+1}$.
%Then the solutions for these
%are given by previous results. 
%At one loop, the effective potential $V_{eff}^{(1)}$ for $M_{i+1}$
%leads to the positive value for $m_{M_{i+1}}^2$ implying that these
%vacua are stable.

%%%%%%%%%%%%%%%%%%%%%%%%%%%%%%%%%%%%%%%%%%%%%%%%%%%%%%%%%%%%%%%%%%%%%%
%%%%%%%%%%%%%%%%%%%%%%%%%%%%%%%%%%%%%%%%%%%%%%%%%%%%%%%%%%%%%%%%%%%%%%%
\subsection{${\cal N}=1$ 
$SU(N_{c,1}) \times \cdots \times SU(\widetilde{N}_{c,n})$ magnetic theory}
%%%%%%%%%%%%%%%%%%%%%%%%%%%%%%%%%%%%%%%%%%%%%%%%%%%%%%%%%%%%%%%%%%%%%%%
%%%%%%%%%%%%%%%%%%%%%%%%%%%%%%%%%%%%%%%%%%%%%%%%%%%%%%%%%%%%%%%%%%%%%%%

%Let us first consider the Seiberg dual for the last gauge group
%factor.
%There are two magnetic duals depending on whether the gauge group factor
%occurs at odd chain or even chain.

%%%%%%%%%%%%%%%%%%%%%%%%%%%%%%%%%%%%%%%%%%%%%%%%%%%%%%%%%%%%%%
%%%%%%%%%%%%%%%%%%%%%%%%%%%%%%%%%%%%%%%%%%%%%%%%%%%%%%%%%%%%%%
\subsubsection{When the dual gauge group occurs at odd chain}
%%%%%%%%%%%%%%%%%%%%%%%%%%%%%%%%%%%%%%%%%%%%%%%%%%%%%%%%%%%%%%
%%%%%%%%%%%%%%%%%%%%%%%%%%%%%%%%%%%%%%%%%%%%%%%%%%%%%%%%%%%%%%

Starting from Figure 25A, moving the $NS5_{n-2}'$-brane 
with $(N_{c,n-1}-N_{c,n-2})$
D4-branes 
to the $+v$ direction leading to Figure 23B, 
and interchanging the $NS5_{n-1}$-brane and the $NS5_{n}'$-brane,
one obtains the Figure 19A in subsection 4.2.1,
with a replacement of a combination of $O6^{+}$-plane and a middle
NS5-brane,  
instead of $O6^{-}$-plane.
Let us denote this as the ``deformed'' Figure 19A.

The dual gauge group is given by 
\bea
SU(N_{c,1}) \times \cdots \times 
SU(N_{c,n-1}) \times SU(\widetilde{N}_{c,n} \equiv N_{c,n-1}-N_{c,n})
\nonu
\eea
the matter contents are   
the bifundamentals $f_{n-1}$ in 
 $({\bf 1_1, \cdots, 1, \Box_{n-1}, \overline{\Box}_{n}, })$,
and $\widetilde{f}_{n-1}$ in the representation 
$({\bf 1_1, \cdots, 1, \overline{\Box}_{n-1}, \Box_{n}})$ in
addition to $(n-2)$ bifundamentals $F_j$ and $\widetilde{F}_j$, 
 $j=1,2,
\cdots, (n-2)$ and
the gauge singlet $\Phi_{n-1}$
for the $n$-th dual gauge group in the 
adjoint representation for the $(n-1)$-th dual gauge group, 
i.e.,  
$
{(\bf 1_1, \cdots, 1_{n-2}, (N_{c,n-1}-N_{c,n-2})^2-1, 1_n)  
}
$ plus a singlet
under the 
dual gauge group
where the gauge group is broken from
$SU(N_{c,n-1})$ 
to $SU(N_{c,n-1}-N_{c,n-2})$.

%When two NS5'-branes in deformed Figure 19A are close to each other, then 
%it leads to deformed Figure 19B by realizing that the number of $(N_{c,n-1}-N_{c,n-2})$
%D4-branes connecting between $NS5_{n-2}'$-brane and $NS5_{n-1}$-brane can
%be rewritten as $(N_{c,n}-N_{c,n-2})$ plus $\widetilde{N}_{c,n}$.
%The Figure 4 of \cite{Ahn07-7} is contained in the modified Figure 19. In
%particular, the brane configuration from the  $NS5_{n-2}'$-brane to 
%the $NS5_{n-1}$-brane is exactly same as the one of \cite{Ahn07-7}.

%The brane configuration for zero mass for the bifundamental $F_{n-1}$
%and 
%$\widetilde{F}_{n-1}$,
%which has only a cubic superpotential (\ref{spotential}),
%can be obtained from deformed Figure 19A by moving
%the upper $NS5_{n-2}'$-brane together with 
%$(N_{c,n-1}-N_{c,n-2})$ color D4-branes 
%into the origin $v=0$.
%The brane configuration in deformed Figure 19A is stable as long as the
%distance $(\Delta x)_{n-1}$ between the upper NS5'-brane and 
%the lower NS5'-brane is large, as
%in \cite{GK}. If they are close to each other, then this brane
%configuration is unstable to decay and leads to 
%the brane configuration in deformed Figure
%19B.

%In particular, the Figure 5B of \cite{Ahn07-9} with vanishing flavors
%$Q$ and $Q'$
%is contained in
%the deformed Figure 19B running from the $NS5_{n-3}'$-brane to 
%the $NS5_{n-1}$-brane.

The dual gauge theory has  a $\Phi_{n-1}$  and 
bifundamentals $f_{n-1}, \widetilde{f}_{n-1}, F_j$, and 
$\widetilde{F}_j$  and the superpotential 
corresponding to deformed Figures 19A and 19B is given by 
previous results.
Therefore, the F-term equation, the derivative $W_{dual}$ with respect to the
meson field $\Phi_{n-1}$ cannot be satisfied if the $(N_{c,n-1}-N_{c,n-2})$ exceeds
$\widetilde{N}_{c,n}$.
So the supersymmetry is broken.   
%Then the solutions for these
%are given by similarly.
%At one loop, the effective potential $V_{eff}^{(1)}$ for $M_{n-1}$
%leads to the positive value for $m_{M_{n-1}}^2$ implying that these
%vacua are stable.

%%%%%%%%%%%%%%%%%%%%%%%%%%%%%%%%%%%%%%%%%%%%%%%%%%%%%%%%%%%%%%
%%%%%%%%%%%%%%%%%%%%%%%%%%%%%%%%%%%%%%%%%%%%%%%%%%%%%%%%%%%%%%
\subsubsection{When the dual gauge group occurs at even chain}
%%%%%%%%%%%%%%%%%%%%%%%%%%%%%%%%%%%%%%%%%%%%%%%%%%%%%%%%%%%%%%
%%%%%%%%%%%%%%%%%%%%%%%%%%%%%%%%%%%%%%%%%%%%%%%%%%%%%%%%%%%%%%

Let us consider other magnetic theory for the same electric theory.
By applying the Seiberg dual to the $SU(N_{c,n})$ factor  and 
interchanging the $NS5_{n-1}'$-brane and the $NS5_{n}$-brane,
one obtains the Figure 20A which appears in subsection 4.2.2, 
with a replacement of a combination of $O6^{+}$-plane and a middle
NS5-brane,  
instead of $O6^{-}$-plane.
Let us denote this as the ``deformed'' Figure 20A.

The gauge group is given by
\bea
SU(N_{c,1}) \times \cdots \times 
SU(N_{c,n-1}) \times SU(\widetilde{N}_{c,n} \equiv N_{c,n-1}-N_{c,n})
\nonu
\eea
and the matter contents are the field $f_{n-1}$ 
charged under
$( {\bf 1_1, \cdots, 1_{n-2}, \Box_{n-1}, \overline{\Box}_{n} })$ 
 and their conjugates 
$\widetilde{f}_{n-1}$ 
$( {\bf 1_1, \cdots, 1_{n-2}, \overline{\Box}_{n-1}, \Box_{n} })$ 
under the dual gauge group
 and  
the gauge-singlet $\Phi_{n-1}$ which is in the 
adjoint representation for the $(n-1)$-th gauge group, 
in other words,   
$ ({ \bf   1_1, 
\cdots, 1_{n-2},  (N_{c,n-1}-N_{c,n-2})^2-1,1_n})  \oplus  ({\bf 1_1,
 \cdots, 1_n})$ under the
dual gauge group
where the gauge group is broken from
$SU(N_{c,n-1})$ 
to $SU(N_{c,n-1}-N_{c,n-2})$.
Then the $\Phi_{n-1}$ is a $(N_{c,n-1}-N_{c,n-2}) \times
 (N_{c,n-1}-N_{c,n-2})$ 
matrix.
Only $(N_{c,n-1}-N_{c,n-2})$ D4-branes can participate in the mass deformation.

%When two NS5'-branes in deformed Figure 20A are close to each other, then 
%it leads to deformed Figure 20B
% by realizing that the number of $(N_{c,n-1}-N_{c,n-2})$
%D4-branes connecting between $NS5_{n-2}'$-brane and $NS5_{n}$-brane can
%be rewritten as $(N_{c,n}-N_{c,n-2})$ plus $\widetilde{N}_{c,n}$.

%The brane configuration in deformed Figure 20A is stable as long as the
%distance $(\Delta x)_{n-1}$ between the upper NS5'-brane and 
%the lower NS5'-brane(or $NS5_{n-1}'$-brane) 
%is large. If they are close to each other, then this brane
%configuration is unstable to decay to 
%the brane configuration in deformed Figure
%20B.
%
%The brane configuration for zero mass for the bifundamental,
%which has only a cubic superpotential (\ref{ssuper}),
%can be obtained from deformed Figure 20A by moving
%the upper  NS5'-brane together with $(N_{c,n-1}-N_{c,n-2})$ color D4-branes 
%into the origin $v=0$.

The low energy dynamics of the magnetic brane configuration 
can be described by the ${\cal N}=1$ supersymmetric gauge theory
with gauge group
and the gauge couplings for the three gauge group factors are
given by
similarly and 
the dual gauge theory has  a meson $\Phi_{n-1}$  and 
bifundamentals $f_{n-1}, \widetilde{f}_{n-1}, F_j$ and
$\widetilde{F}_j$ 
under the dual gauge
group and the superpotential 
corresponding to deformed Figures 20A and 20B is given by 
the one in subsection 4.2.2.
Therefore, the F-term equation, the derivative $W_{dual}$ with respect to the
meson field $\Phi_{n-1}$ cannot be satisfied if the $(N_{c,n-1}-N_{c,n-2})$ exceeds
$\widetilde{N}_{c,n}$.
So the supersymmetry is broken.   
That is, 
there exist three equations from F-term conditions:
$
f_{n-1}^a \widetilde{f}_{n-1,b} -\mu_{n-1}^2 \delta^a_b =0$ 
and $ \Phi_{n-1} f_{n-1} =0=
\widetilde{f}_{n-1} \Phi_{n-1}$.
%Then the solutions for these
%are given by similarly.
%At one loop, the effective potential $V_{eff}^{(1)}$ for $M_{n-1}$
%leads to the positive value for $m_{M_{n-1}}^2$ implying that these
%vacua are stable.

%%%%%%%%%%%%%%%%%%%%%%%%%%%%%%%%%%%%%%%%%%%%%%%%%%%%%%%%%%%%%%%%%%%%%%%
%%%%%%%%%%%%%%%%%%%%%%%%%%%%%%%%%%%%%%%%%%%%%%%%%%%%%%%%%%%%%%%%%%%%%%%
\subsection{
${\cal N}=1$ 
$SU(\widetilde{N}_{c,1}) \times \cdots \times SU(N_{c,n})$ magnetic theory}
%%%%%%%%%%%%%%%%%%%%%%%%%%%%%%%%%%%%%%%%%%%%%%%%%%%%%%%%%%%%%%%%%%%%%%%%
%%%%%%%%%%%%%%%%%%%%%%%%%%%%%%%%%%%%%%%%%%%%%%%%%%%%%%%%%%%%%%%%%%%%%%%%

Let us consider the Seiberg dual for the first gauge group and take 
the mass deformation by moving the $(N_{c,2}-N_{c,3})$ D4-branes between 
the middle NS5-brane and the $NS5_2'$-brane to $+v$ direction.
Starting from the Figure 25A, we apply the Seiberg dual to the first
gauge group 
$SU(N_{c,1})$ factor and the $NS5_1'$-brane and its mirror are 
interchanged each other. 
Then the brane configuration is the same as Figure 25A except that 
the number of color $\widetilde{N}_{c,1}$
is given by $\widetilde{N}_{c,1}=2N_{c,2}-N_{c,1}$ from \cite{Ahn07-4,Ahn07}.
By rotating the $NS5_2$-brane by $\frac{\pi}{2}$(leading to
$NS5_2'$-brane) and moving
it  together with $(N_{c,2}-N_{c,3})$ D4-branes
to $+ v$ direction, 
then the $(N_{c,2}-N_{c,3})$ 
D4-branes are connecting between the 
$NS5_1$-brane and the $NS5_2'$-brane and 
$\widetilde{N}_{c,1}$ D4-branes connecting between the middle NS5-brane and   
$NS5_1'$-brane as well as $N_{c,3}$ D4-branes between the $NS5_1'$-brane
and the $NS5_3'$-brane(and their mirrors). 

%%%%%%%%%%%%%%%%%%%%%%%%%%%%%%%%%%%%%%%%%%%%%%%%%%%%%%%%%%%%%%%%%%%%
%%%%%%%%%%%%%%%%%%%%%%%%%%%%%%%%%%%%%%%%%%%%%%%%%%%%%%%%%%%%%%%%%%%%%%
\begin{figure}[ht]
   \epsfxsize=4.0in 
\centerline{\epsffile{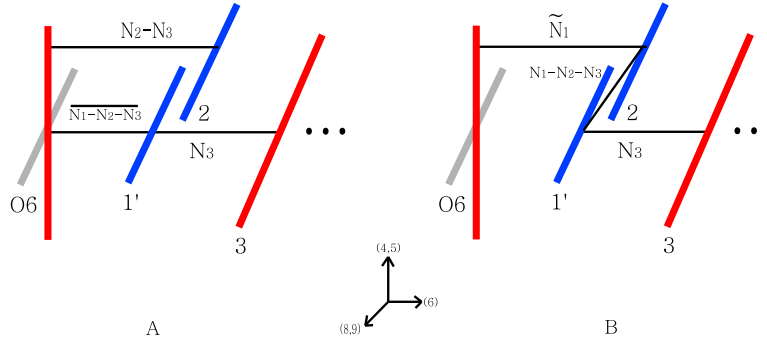}}
   \caption[FIG. \arabic{figure}.]{ 
The 
 ${\cal N}=1$ magnetic brane configuration for the gauge group 
containing $SU(\widetilde{N}_{c,1}=2N_{c,2}-N_{c,1})$ 
with D4-
and $\overline{D4}$-branes(26A) and with 
a misalignment between D4-branes(26B) when the two NS5'-branes are close to
each other. 
The number of tilted D4-branes in 26B can be written as
$N_{c,1}-N_{c,3}-N_{c,2} =(N_{c,2}-N_{c,3})-\widetilde{N}_{c,1}$.
%The $x$ coordinate of $NS5_{2}'$-brane is given by $(\Delta x)_{2}$.
}
\end{figure}
%%%%%%%%%%%%%%%%%%%%%%%%%%%%%%%%%%%%%%%%%%%%%%%%%%%%%%%%%%%%%%%%%%%%%
%Figure 30A and 30B
%%%%%%%%%%%%%%%%%%%%%%%%%%%%%%%%%%%%%%%%%%%%%%%%%%%%%%%%%%%%%%%%%%%%%%%

By introducing $(N_{c,2}-N_{c,3})$ D4-branes and $(N_{c,2}-N_{c,3})$ 
anti-D4-branes  between the middle NS5-brane and   
$NS5_1'$-brane, reconnecting the former with  
the $(N_{c,2}-N_{c,3})$ D4-branes connecting between the $NS5_1'$-brane and
the $NS5_2'$-brane 
and moving those combined D4-branes
to $+v$-direction(and their mirrors to $-v$ direction), 
one gets the final Figure 26A where we are left with 
$(N_{c,2}-N_{c,3}-\widetilde{N}_{c,1})$ 
anti-D4-branes between the middle NS5-brane and   
$NS5_1'$-brane.

Now 
we draw this ${\cal N}=1$ supersymmetric magnetic
brane configuration in Figure 26.
We assume, as before, that the number of colors satisfies
$
2N_{c,2} \geq N_{c,1} \geq N_{c,2} +N_{c,3}$.

%When two NS5'-branes in Figure 26A are close to each other, then 
%it leads to Figure 26B
%by realizing that the number of $(N_{c,2}-N_{c,3})$
%D4-branes connecting between the middle NS5-brane and the $NS5_2'$-brane in Figure
%26A can
%be rewritten as $(N_{c,1}-N_{c,2}-N_{c,3})$ plus $\widetilde{N}_{c,1}$.

The dual gauge group is given by
\bea
SU(\widetilde{N}_{c,1} \equiv 2N_{c,2}-N_{c,1}) \times 
SU(N_{c,2}) \times \cdots \times SU(N_{c,n})
\nonu
\eea
where the number of dual color can be obtained from the linking number 
counting, as done in \cite{Ahn07-4,Ahn07}. 
The matter contents are the flavor singlet $f_1$ in the bifundamental 
representation $({\bf \widetilde{N}_{c,1}, \overline{N_{c,2}}, 1_3, \cdots, 1_n})$
and its complex conjugate field $\widetilde{f}_1$ in the bifundamental 
representation  $({\bf \overline{\widetilde{N}_{c,1}}, N_{c,2}, 1_2,
  \cdots, 1_n})$,
under the dual gauge
group 
and  the gauge singlet $\Phi_2$ in the representation for 
$({\bf 1_1,(N_{c,2}-N_{c,3})^2-1, \cdots, 1_n}) 
\oplus ({\bf 1_1, \cdots, 1_n})$ under the dual gauge group
where the gauge group is broken from
$SU(N_{c,2})$ 
to $SU(N_{c,2}-N_{c,3})$. 
There are also
the symmetric flavor $s$ for $SU(\widetilde{N}_{c,1})$ and the conjugate 
symmetric flavor $\widetilde{s}$ for $SU(\widetilde{N}_{c,1})$ as well as
$F_j$ and $\widetilde{F}_j$.
Then the $\Phi_2$ is a $(N_{c,2}-N_{c,3}) \times (N_{c,2}-N_{c,3})$ matrix.
Only $(N_{c,2}-N_{c,3})$ D4-branes among $N_{c,2}$ D4-branes 
can participate in the mass deformation.
A cubic superpotential is an interaction between dual ``quarks''
and a meson. 

Then the dual magnetic superpotential, by adding the mass term
for the bifundamental $F_1$ which can be 
interpreted as a linear term in the meson $\Phi_2$ to this cubic
superpotential, is given by
\bea
W_{dual} =  \Phi_2 f_1 \widetilde{f}_1 + m_2 \tr \Phi_2, 
\nonu
\eea
where $\Phi_2$ was defined as $\Phi_2 \equiv F_1 \widetilde{F}_1$  and 
the first gauge group indices in $F_1$ and $\widetilde{F}_1$ 
are contracted and each second gauge group index in them is encoded in 
$\Phi_2$.

Here the magnetic fields $f_1$ and $\widetilde{f}_1$  
correspond to 4-4 strings connecting 
the $\widetilde{N}_{c,1}$-color D4-branes(that are 
connecting between the middle NS5-brane
and the $NS5_2'$-brane in Figure 26B) with $N_{c,2}$-flavor 
D4-branes.
Among these $N_{c,2}$-flavor D4-branes, only the strings ending on
the upper $(2N_{c,2}-N_{c,1})$ D4-branes and 
on the tilted  $(N_{c,1}-N_{c,2}-N_{c,3})$ 
D4-branes in Figure 26B enter the cubic superpotential term. 
Note that the summation of these D4-branes is equal to $(N_{c,2}-N_{c,3})$.

When the upper NS5'-brane(or $NS5_{2}'$-brane) 
is replaced by coincident $(N_{c,2}-N_{c,3})$ 
D6-branes and 
the $NS5_{3}'$ is rotated by an angle $\frac{\pi}{2}$ in the $(v,w)$
plane in Figure 26B, this brane configuration reduces to the one 
found in \cite{Ahn07-9} where the gauge group was given by 
$ SU(2n_{f,1}+2n_{c,2}-n_{c,1}) \times SU(n_{c,2}) \times   $ 
with $n_{f,1}$ multiplets,  $\widetilde{n}_{f,1}$ multiplets, 
 bifundamentals, a symmetric flavor, a
conjugate symmetric flavor, and various gauge singlets.  
Then the present number $(N_{c,2}-N_{c,3})$ 
corresponds to the $n_{f,1}$, the
number $N_{c,1}$ corresponds to $n_{c,1}$ and 
the number $N_{c,3}$ corresponds to the $n_{c,2}$.
%Note that the number of D4-branes touching $NS5_{2}'$-brane in Figure 26B
%is equal to $(N_{c,2}-N_{c,3})$.
%In particular, the Figure 2B of \cite{Ahn07-9} with vanishing flavors
%$Q'$ and $Q''$
%is contained in
%the modified Figure 26B running from the middle NS5-brane to 
%the $NS5_{3}'$-brane.

The low energy dynamics of the magnetic brane configuration 
can be described by the ${\cal N}=1$ supersymmetric gauge theory
with gauge group 
and the gauge couplings for the  gauge group factors are
given by similarly.
The dual gauge theory has  a meson field $\Phi_2$  and 
bifundamental $f_1$ in the representation 
 $({\bf \widetilde{N}_{c,1}, \overline{N_{c,2}}, \cdots, 1_n})$ 
under the dual gauge
group and the superpotential 
corresponding to the Figures 26A and 26B is given by
\bea
W_{dual} = h \Phi_2 f_1 \widetilde{f}_1 - h \mu_2^2 \tr \Phi_2, \qquad
h^2=  g_{2,mag}^2, \qquad 
\mu_{2}^2 = -\frac{(\Delta x)_{2}}{ 2\pi g_s \ell_s^3}. 
\nonu
\eea
Then $ f_1 \widetilde{f}_1$ is 
a $\widetilde{N}_{c,1} \times \widetilde{N}_{c,1}$ 
matrix where the second gauge group indices for $f_1$ and 
$\widetilde{f}_1$ 
are contracted with those
of $\Phi_2$ while $\Phi_2$ is a 
$(N_{c,2}-N_{c,3}) \times (N_{c,2}-N_{c,3})$ matrix.
%Although the field $f_1$ itself is an antifundamental in the second gauge
%group
%which is a different feature, compared with the singlet 
%representation for the usual quark
%coming from D6-branes \cite{Ahn07-4},
%the product $f_1 \widetilde{f}_1$ has the same representation with the 
%product of quarks.
%Moreover, the second gauge group indices for the field $\Phi_2$ play the
%role of the flavor indices.
Therefore, the F-term equation, the derivative of $W_{dual}$ with respect to the
meson field $\Phi_2$ cannot be satisfied if the $(N_{c,2}-N_{c,3})$ exceeds
$\widetilde{N}_{c,1}$.
So the supersymmetry is broken.   
That is, 
there are three equations from F-term conditions:
$
f_1^a \widetilde{f}_{1,b} -\mu_2^2 \delta^a_b =0, \Phi_2 f_1 =0$, and 
$\widetilde{f}_1 \Phi_2=0$.
Then the solutions for these
are given by 
\bea
<f_1>   = 
\left(
\begin{array}{c}
\mu_2   {\bf 1}_{\widetilde{N}_{c,1}}  \nonu \\
0
\end{array}
\right), \quad
<\widetilde{f}_1>   = 
\left(
\begin{array}{cc}
\mu_2   {\bf 1}_{\widetilde{N}_{c,1}} & 0 \nonu \\
\end{array}
\right), 
\quad
<\Phi_2> =
 \left(
\begin{array}{cc}
0  & 0
 \\
0 & M_2  {\bf 1}_{(N_{c,2}-N_{c,3}-\widetilde{N}_{c,1})} 
\end{array}
\right).
\nonu
\eea
%At one loop, the effective potential $V_{eff}^{(1)}$ for $M_2$
%leads to the positive value for $m_{M_2}^2$ implying that these
%vacua are stable.
%The gauge theory analysis where the theory will be strongly coupled in
%the IR region $N_{c,2} -N_{c,3}> 2\widetilde{N}_{c,1}-2$ 
%is only valid in the regime where 
%$(\Delta x)_2$ is smaller than $\exp(-\frac{C}{g_s})$ with some positive
%constant $C$, as done in \cite{Ahn07-5}. 

%%%%%%%%%%%%%%%%%%%%%%%%%%%%%%%%%%%%%%%%%%%%%%%%%%%%%%%%%%%%%%%%%%%%%
%%%%%%%%%%%%%%%%%%%%%%%%%%%%%%%%%%%%%%%%%%%%%%%%%%%%%%%%%%%%%%%%%%%%%
\section{Meta-stable brane configurations  with $(2n+1)$ NS-branes, 
$O6^{\pm}$-planes, and D6-branes }
%section7%%%%%%%%%%%%%%%%%%%%%%%%%%%%%%%%%%%%%%%%%%%%%%%%%%%%%%%%%%%%%%%%%%%%
%%%%%%%%%%%%%%%%%%%%%%%%%%%%%%%%%%%%%%%%%%%%%%%%%%%%%%%%%%%%%%%%%%%%%

The type IIA brane configuration,  
 by generalizing the brane
configurations \cite{LLL1,BHKL,EGKT} 
to the case where there are more
NS-branes,  
corresponding to 
${\cal N}=1$ supersymmetric electric gauge theory(see also
\cite{Ahn07-9}) with
gauge group
\bea
SU(N_{c,1}) \times SU(N_{c,2}) \cdots
\times SU(N_{c,n})
\nonu
\eea
and with 
 an antisymmetric tensor field $A$ charged under $({\bf \frac{1}{2}
N_{c,1}(N_{c,1}-1), 1_2, \cdots, 1_n})$, 
 a conjugate symmetric tensor field $\widetilde{S}$ 
charged under $({\bf \overline{\frac{1}{2}
N_{c,1}(N_{c,1}+1)}, 1_2, \cdots, 1_n})$,  
an eight fundamentals $\hat{Q}$ charged under
 $({\bf N_{c,1}, 1_2, \cdots, 1_n})$,
the $(n-1)$ bifundametals $F_i$ charged under 
$({\bf 1_1, \cdots, 1, \Box_i, \overline{\Box}_{i+1}, 1, \cdots,  1_n})$
and  $\widetilde{F}_i$ 
charged $({\bf 1_1, \cdots, 1, \overline{\Box}_i, \Box_{i+1}, 1, 
\cdots, 1_n})$ where $i=1, 2, \cdots, (n-1)$
can be described by 
the $NS5_1$-brane, 
the  
$NS5_2'$-brane, $\cdots$, the $NS5_{n}$-brane for odd number
of gauge groups(or 
the $NS5_{n}'$-brane for even number of gauge groups),
$N_{c,1}$-, $N_{c,2}$-,  $\cdots$, and $N_{c,n}$-color D4-branes. 
See the Figure 27 for the details on the brane configuration. 
The $O6^{\pm}$-planes act as $(x^4,x^5,x^6) \rightarrow
(-x^4,-x^5,-x^6)$ and has RR charge $ \pm 4$.

%%%%%%%%%%%%%%%%%%%%%%%%%%%%%%%%%%%%%%%%%%%%%%%%%%%%%%%%%%%%%%%%%%%%%%
%%%%%%%%%%%%%%%%%%%%%%%%%%%%%%%%%%%%%%%%%%%%%%%%%%%%%%%%%%%%%%%%%%%%%%
\begin{figure}[ht]
   \epsfxsize=4.0in 
\centerline{\epsffile{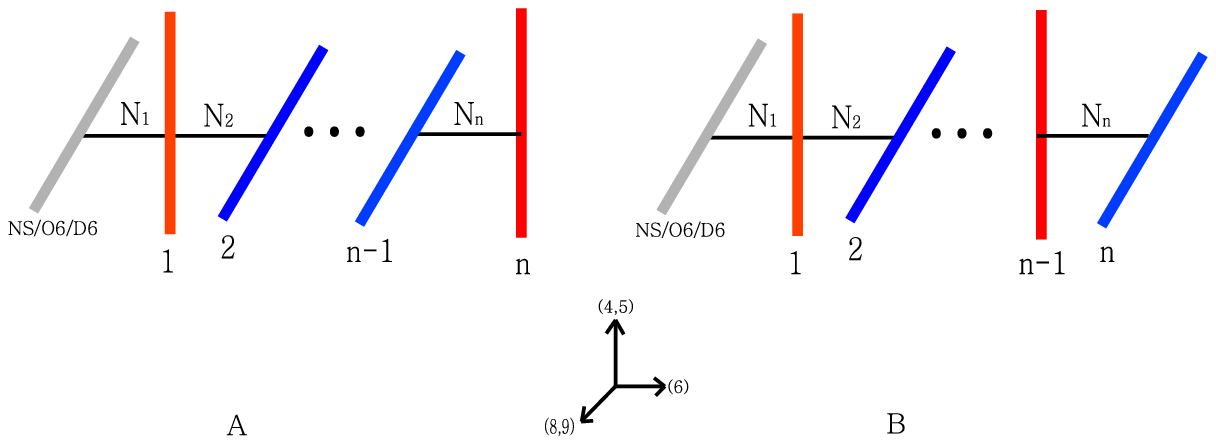}}
   \caption[FIG. \arabic{figure}.]{ 
The 
 ${\cal N}=1$ supersymmetric 
electric brane configuration for the gauge group $ \prod_{i=1}^{n}
 SU(N_{c,i})$ 
and  bifundamentals $F_i$ and $\widetilde{F}_i$  with vanishing mass
for the bifundamentals when the number of gauge groups factor 
$n$ is odd(27A) and even(27B). 
We do not draw the mirrors of the branes appearing in the left hand
 side of O6-plane. 
%This Figure looks similar to the Figure 22. 
We
 denote $NS/O6/D6$ as a combination of NS5'-brane, $O6^{\pm}$-planes
 and D6-branes.}
\end{figure}
%%%%%%%%%%%%%%%%%%%%%%%%%%%%%%%%%%%%%%%%%%%%%%%%%%%%%%%%%%%%%%%%%%%%%
%Figure 31A and 31B
%%%%%%%%%%%%%%%%%%%%%%%%%%%%%%%%%%%%%%%%%%%%%%%%%%%%%%%%%%%%%%%%%%%%%%%%

Let us place an O6-plane at the origin $x^6=0$
and denote the $x^6$ 
coordinates for 
the  
$NS5_1$-brane, $\cdots$, the $NS5_{n}$-brane for odd $n$(or 
the $NS5_{n}'$-brane for even $n$)
are given by $x^6=y_1, y_1+y_2, \cdots, \sum_{j=1}^{n-1} y_j + y_{n}$
respectively.
The $N_{c,1}$ D4-branes 
are suspended between the 
$NS5_1$-brane and its mirror, 
the $N_{c,2}$ D4-branes 
are suspending between the 
$NS5_1$-brane and the $NS5_2'$-brane, $\cdots$ and 
the $N_{c,n}$ D4-branes  
are suspended between the $NS5_{n-1}'$-brane and the $NS5_{n}$-brane for
odd $n$(or between the $NS5_{n-1}$-brane and 
the $NS5_{n}'$-brane for even $n$).
The fields $F_i$ and $\widetilde{F}_i$  correspond to 4-4 strings connecting 
the $N_{c,i}$-color D4-branes with $N_{c,i+1}$-color D4-branes.
The fields $A$ and $\widetilde{S}$  correspond to 4-4 strings connecting 
the $N_{c,1}$-color D4-branes with $x^6 < 0$ with $N_{c,1}$-color D4-branes
with $x^6 > 0$.
We draw this ${\cal N}=1$ supersymmetric 
electric brane configuration in Figure 27A(27B) 
when $n$ is odd(even) for the vanishing mass
for the fields $F_i$ and $\widetilde{F}_i$. 

There is no superpotential in Figure 27A. Let us deform this theory.
Displacing the two NS5'-branes relative each other in the $+v$
direction, characterized by $(\Delta x)_{i+1}$, 
corresponds to turning on a quadratic
mass-deformed superpotential
for the field $F_i$ and $\widetilde{F}_i$ as follows:
\bea
W = m_{i+1} F_i \widetilde{F}_i (\equiv m_{i+1} \Phi_{i+1})
\nonu
\eea
where 
the $i$-th gauge group indices in $F_i$ and $\widetilde{F}_i$ 
are contracted and the mass $m_{i+1}$ is given by 
\bea
m_{i+1} 
%=\frac{(\Delta x)_{i+1}}{2\pi \alpha'} 
= 
\frac{(\Delta x)_{i+1}}{\ell_s^2}.
\nonu
\eea
The gauge-singlet $\Phi_{i+1}$ for the $i$-th  gauge group is in the 
adjoint representation for the $(i+1)$-th  gauge group, 
i.e., ${\bf ( 1_1, \cdots, 1_i, (N_{c,i+1}-N_{c,i+2})^2-1, 
\cdots, 1_n)  \oplus (1_1, \cdots ,1_n) }$ 
under the  gauge group
where the gauge group is broken from
$SU(N_{c,i+1})$ 
to $SU(N_{c,i+1}-N_{c,i+2})$. 
The $\Phi_{i+1}$ is a $(N_{c,i+1}-N_{c,i+2}) \times (N_{c,i+1}-N_{c,i+2})$ matrix.
The $NS5_{i+1}'$-brane together with $(N_{c,i+1}-N_{c,i+2})$-color D4-branes 
is moving to the $+v$ direction  for
fixed other branes during this mass deformation(and their mirrors to
$-v$ direction). 
Then the $x^5$ coordinate 
of $NS5_{i-1}'$-brane is equal to
zero
while the $x^5$ coordinate of $NS5_{i+1}'$-branes is given by 
$ (\Delta x)_{i+1}$.
Giving an expectation value to the meson field $\Phi_{i+1}$
corresponds to recombination of $N_{c,i}$- and $N_{c,i+1}$- color 
D4-branes, which will become $N_{c,i}$ or $N_{c,i+1}$-color D4-branes
in Figure 27A such that they are suspended between 
the $NS5_{i-1}'$-brane and the $NS5_{i+1}'$-brane 
and pushing them into the $w$
direction. 
We assume that the number of colors satisfies
$
N_{c,i+1} \geq N_{c,i}-N_{c,i-1} \geq N_{c,i+2}$.

Now 
we obtain this brane configuration and this is the same as the 
Figure 23A except that there are NS5'-brane, $O6^{-}$-plane and eight
D6-branes at the origin $x^6=0$ for nonvanishing mass
for the fields $F_i$ and $\widetilde{F}_i$.

Let us deform the theory by Figure 27B.
Displacing the two NS5'-branes, the $NS_{i-2}'$-brane and the 
$NS_{i}'$-brane, 
relative each other in the 
$v$ 
direction, characterized by $(\Delta x)_{i-1}$, 
corresponds to turning on a quadratic
mass-deformed superpotential
for the fields $F_{i-1}$ and $\widetilde{F}_{i-1}$ as follows:
\bea
W = m_{i-1} F_{i-1} \widetilde{F}_{i-1} (\equiv m_{i-1} \Phi_{i-1})
\nonu
\eea
where 
the $i$-th gauge group indices in $F_{i-1}$ and $\widetilde{F}_{i-1}$ 
are contracted, each $(i-1)$-th gauge group index in them is encoded in 
$\Phi_{i-1}$ and the mass $m_{i-1}$ is given by
\bea
m_{i-1} 
%=\frac{(\Delta x)_{i+1}}{2\pi \alpha'} 
= 
\frac{(\Delta x)_{i-1}}{\ell_s^2}.
\nonu
\eea

The gauge-singlet $\Phi_{i-1}$ for the $i$-th  gauge group is in the 
adjoint representation for the $(i-1)$-th  gauge group, 
i.e., 
\bea
{(\bf 1_1, \cdots, 1_{i-2}, (N_{c,i-1}-N_{c,i-2})^2-1, 1_{i}, \cdots,
  1_n)  
\oplus (1_1, \cdots, 1_n)}
\nonu
\eea 
under the  gauge group
where the gauge group is broken from
$SU(N_{c,i-1})$ 
to $SU(N_{c,i-1}-N_{c,i-2})$. 
Then the $\Phi_{i-1}$ is a $(N_{c,i-1}-N_{c,i-2}) \times 
(N_{c,i-1}-N_{c,i-2})$ matrix.
The $NS5_{i-2}'$-brane together with $(N_{c,i-1}-N_{c,i-2})$-color D4-branes 
is moving to the $+v$ direction  for
fixed other branes during this mass deformation. 
In other words, the $N_{c, i-2}$ D4-branes among $N_{c,i-1}$ D4-branes 
are not participating in 
the mass deformation.
Then the $x^5$ coordinate($\equiv x$) 
of $NS5_{i}'$-brane is equal to
zero
while the $x^5$ coordinate of $NS5_{i-2}'$-brane is given by 
$(\Delta x)_{i-1}$.
Giving an expectation value to the meson field $\Phi_{i-1}$
corresponds to recombination of $N_{c,i-1}$- and $N_{c,i}$- color 
D4-branes, which will become $N_{c,i-1}$- or $N_{c,i}$-color D4-branes
in Figure 27B such that they are suspended between 
the $NS5_{i-2}'$-brane and the $NS5_{i}'$-brane 
and pushing them into the 
$w$ direction. We assume that the number of colors satisfies
$
N_{c,i-1} \geq N_{c,i}-N_{c,i+1} \geq N_{c,i-2}$.
Now 
we obtain this brane configuration and this is the same as the 
Figure 23B  
 except that there are NS5'-brane, $O6^{-}$-plane and eight
D6-branes at the origin $x^6=0$
for nonvanishing mass
for the fields $F_{i-1}$ and $\widetilde{F}_{i-1}$. 

%Next we describe five different magnetic dual theories by taking each
%corresponding mass deformation.
%Most of the brane configurations in this section are coincident with the ones in
%previous section if we neglect the presence of middle NS5'-brane, 
%$O6^{\pm}$-plane and D6-branes at the origin $x^6=0$ 
%and we will mention the main results briefly.

%%%%%%%%%%%%%%%%%%%%%%%%%%%%%%%%%%%%%%%%%%%%%%%%%%%%%%%%%%%%%%%%%%%%%%%%%
%%%%%%%%%%%%%%%%%%%%%%%%%%%%%%%%%%%%%%%%%%%%%%%%%%%%%%%%%%%%%%%%%%%%%%%%%
\subsection{${\cal N}=1$ 
$SU(N_{c,1}) \times \cdots \times 
SU(\widetilde{N}_{c,i}) \times \cdots \times SU(N_{c,n})$ magnetic theory}
%%%%%%%%%%%%%%%%%%%%%%%%%%%%%%%%%%%%%%%%%%%%%%%%%%%%%%%%%%%%%%%%%%%%%%%%
%%%%%%%%%%%%%%%%%%%%%%%%%%%%%%%%%%%%%%%%%%%%%%%%%%%%%%%%%%%%%%%%%%%%%%%%

%Let us first consider the Seiberg dual for the middle gauge group
%factor.
%There are two magnetic duals depending on whether the gauge group factor
%occurs at odd chain or even chain.

%%%%%%%%%%%%%%%%%%%%%%%%%%%%%%%%%%%%%%%%%%%%%%%%%%%%%%%%%%%%%%
%%%%%%%%%%%%%%%%%%%%%%%%%%%%%%%%%%%%%%%%%%%%%%%%%%%%%%%%%%%%%%
\subsubsection{When the dual gauge group occurs at odd chain}
%%%%%%%%%%%%%%%%%%%%%%%%%%%%%%%%%%%%%%%%%%%%%%%%%%%%%%%%%%%%%%
%%%%%%%%%%%%%%%%%%%%%%%%%%%%%%%%%%%%%%%%%%%%%%%%%%%%%%%%%%%%%%

Let us consider other magnetic theory for the same electric theory.
Starting from Figure 25B and 
interchanging the $NS5_{i-1}'$-brane and the $NS5_i$-brane(and their mirrors),
one obtains the magnetic brane configuration which is exactly the same
as the Figure 18A, in subsection 4.1.2., 
with a replacement of $NS/O6/D6$ instead of $O6^{-}$-plane.
Let us call this as ``modified'' Figure 18A.

The dual gauge group is given by
\bea
SU(N_{c,1}) \times \cdots  \times 
SU(\widetilde{N}_{c,i} \equiv N_{c,i+1}+N_{c,i-1}-N_{c,i}) 
 \times \cdots \times SU(N_{c,n}).
\nonu
\eea
The matter contents are the field $f_{i}$ 
 charged under
$({\bf 1, \cdots, \Box_{i}, \overline{\Box}_{i+1}, \cdots, 1_n})$, 
 and their conjugates 
$\widetilde{f}_{i}$ charged $({\bf 1_1, \cdots, \overline{\Box}_{i},
  \Box_{i+1}, \cdots, 1_n })$ 
under the dual gauge group
and  
the gauge-singlet $\Phi_{i+1}$ which is in the 
adjoint representation for the $i$-th dual gauge group, 
in other words,   
$ ({ \bf   1_1,  \cdots, 1_{i}, 
(N_{c,i+1}-N_{c,i+2})^2-1, \cdots, 1_n})  \oplus  ({\bf 1_1, \cdots,1_n})$ under the 
dual gauge group
where the gauge group is broken from
$SU(N_{c,i+1})$ 
to $SU(N_{c,i+1}-N_{c,i+2})$.
Then the $\Phi_{i+1}$ is a $(N_{c,i+1}-N_{c,i+2}) 
\times (N_{c,i+1}-N_{c,i+2})$ matrix.
Only $(N_{c,i+1}-N_{c,i+2})$ 
D4-branes can participate in the mass deformation.

%When two NS5'-branes in modified Figure 18A are close to each other, then 
%it leads to modified Figure 18B
%by realizing that the number of $(N_{c,i+1}-N_{c,i+2})$
%D4-branes connecting between $NS5_i$-brane and $NS5_{i+1}'$-brane in
%modified Figure
%18A can
%be rewritten as $(N_{c,i}-N_{c,i+2}-N_{c,i-1})$ plus $\widetilde{N}_{c,i}$.

The cubic superpotential with the mass term
is given by (\ref{superpo11}) in subsection 4.1.2
and we define $\Phi_{i+1}$ as $\Phi_{i+1} \equiv F_{i} \widetilde{F}_{i}$ and 
the $i$-th gauge group indices in $F_{i}$ and $\widetilde{F}_{i}$ 
are contracted, each $(i+1)$-th gauge group index in them is encoded in 
$\Phi_{i+1}$.  
The brane configuration for zero mass for the bifundamental,
which has only a cubic superpotential (\ref{superpo11}),
can be obtained from modified Figure 18A by moving
the upper  NS5'-brane together with $(N_{c,i+1}-N_{c,i+2})$ color D4-branes 
into the origin $v=0$(and their mirrors).

%The brane configuration in modified Figure 18A is stable as long as the
%distance $(\Delta x)_{i+1}$ between the upper NS5'-brane and 
%the lower NS5'-brane is large. 
%If they are close to each other, then this brane
%configuration is unstable to decay and leads to 
%the brane configuration in modified Figure
%18B.

%In particular, the Figure 10B of \cite{Ahn07-9} with vanishing flavors
%$Q$ and $Q'$
%is contained in
%the modified Figure 18B running from the $NS5_{i-2}$-brane to 
%the $NS5_{i+1}'$-brane. 

The low energy dynamics of the magnetic brane configuration 
can be described by the ${\cal N}=1$ supersymmetric gauge theory
with gauge group 
and the gauge couplings for the three gauge group factors are
given by those in subsection 4.1.2
and 
the superpotential 
corresponding to modified Figures 18A and 18B is given by the one in subsection 4.1.2. 
Therefore, the F-term equation, the derivative $W_{dual}$ with respect to the
meson field $\Phi_{i+1}$ cannot be satisfied if the $(N_{c,i+1}-N_{c,i+2})$ exceeds
$\widetilde{N}_{c,i}$.
So the supersymmetry is broken.   
That is, 
there exist three equations from F-term conditions:
$
f_{i}^a \widetilde{f}_{i,b} -\mu_{i+1}^2 \delta^a_b =0$ and $ \Phi_{i+1} f_{i} =
0=\widetilde{f}_{i} \Phi_{i+1}$.
%Then the solutions for these
%are given by previous results. 
%At one loop, the effective potential $V_{eff}^{(1)}$ for $M_{i+1}$
%leads to the positive value for $m_{M_{i+1}}^2$ implying that these
%vacua are stable.

%%%%%%%%%%%%%%%%%%%%%%%%%%%%%%%%%%%%%%%%%%%%%%%%%%%%%%%%%%%%%%
%%%%%%%%%%%%%%%%%%%%%%%%%%%%%%%%%%%%%%%%%%%%%%%%%%%%%%%%%%%%%%
\subsubsection{When the dual gauge group occurs at even chain}
%%%%%%%%%%%%%%%%%%%%%%%%%%%%%%%%%%%%%%%%%%%%%%%%%%%%%%%%%%%%%%
%%%%%%%%%%%%%%%%%%%%%%%%%%%%%%%%%%%%%%%%%%%%%%%%%%%%%%%%%%%%%%

Starting from Figure 25A and 
interchanging the $NS5_{i-1}$-brane and the $NS5_i'$-brane(and their mirrors),
one obtains the Figure 17A, in subsection 4.1.1, 
with a replacement of $NS/O6/D6$ instead of $O6^{-}$-plane.
Let us call this as ``modified'' Figure 17A.

The dual gauge group is given by 
\bea
SU(N_{c,1}) \times \cdots \times 
SU(\widetilde{N}_{c,i} \equiv N_{c,i+1}+N_{c,i-1}-N_{c,i}) 
\times \cdots \times SU(N_{c,n})
\nonu
\eea
The matter contents are the field $f_{i-1}$ 
charged under
$({\bf 1_1, \cdots, 1_{i-1}, 
N_{c,i-1}, \overline{\widetilde{N}_{c,i}}, \cdots, 1_n})$, and its 
conjugate field  
$\widetilde{f}_{i-1}$  charged under 
$({\bf 1_1, \cdots, 1_{i-2}, 
\overline{N_{c,i-1}}, \widetilde{N}_{c,i}, \cdots, 1_n})$
under the dual gauge group
and  
the gauge-singlet $\Phi_{i-1}$ for the $i$-th dual gauge group in the 
adjoint representation for the $(i-1)$-th dual gauge group, 
i.e.,  ${(\bf 1_1, \cdots, 1_{i-2}, (N_{c,i-1}-N_{c,i-2})^2-1, 1,
  \cdots, 1_n)  }$ plus a singlet under the 
dual gauge group
where the gauge group is broken from
$SU(N_{c,i-1})$ 
to $SU(N_{c,i-1}-N_{c,i-2})$.
Then the  $\Phi_{i-1}$ is a $(N_{c,i-1}-N_{c,i-2}) \times
(N_{c,i-1}-N_{c,i-2})$ 
matrix.

%When two NS5'-branes in modified Figure 17A are close to each other, it becomes 
%modified Figure 17B
%by realizing that the number of $(N_{c,i-1}-N_{c,i-2})$
%D4-branes connecting between $NS5_{i-2}'$-brane and $NS5_{i-1}$-brane
%in 
%modified Figure
%17A can
%be rewritten as $(N_{c,i}-N_{c,i+1}-N_{c,i-2})$ plus $\widetilde{N}_{c,i}$. 

The cubic superpotential with the mass term  
is given by the one (\ref{superpo1new}) in subsection 4.1.1
and 
the brane configuration for zero mass for the bifundamental,
which has only a cubic superpotential (\ref{superpo1new}),
can be obtained from modified Figure 17A by moving
the upper  NS5'-brane(or $NS5_{i-2}'$-brane) 
together with $(N_{c,i-1}-N_{c,i-2})$ color D4-branes 
into the origin $v=0$(and their mirrors).
The low energy dynamics of the magnetic brane configuration 
can be described by the ${\cal N}=1$ supersymmetric gauge theory
with gauge group 
and the gauge couplings for the three gauge group factors are
given by the ones in subsection 4.1.1.
The dual gauge theory has  a meson $\Phi_{i-1}$  and 
bifundamentals $f_{i-1}$, and $\widetilde{f}_{i-1}$ under the dual gauge
group and the superpotential 
corresponding to modified Figures 17A and 17B is the same as  
the one in subsection 4.1.1.
%In particular, the Figure 9B of \cite{Ahn07-9} with vanishing flavors
%$Q$ and $Q''$
%is contained in
%the modified Figure 17B running from the  $NS5_{i-2}'$-brane to 
%the $NS5_{i+1}$-brane.
Therefore, the F-term equation, the derivative $W_{dual}$ with respect to the
meson field $\Phi_{i-1}$ cannot be satisfied if the $(N_{c,i-1}-N_{c,i-2})$ exceeds
$\widetilde{N}_{c,i}$.
So the supersymmetry is broken.   
That is, 
there exist three equations from F-term conditions:
$
f_{i-1}^a \widetilde{f}_{i-1,b} -\mu_{i-1}^2 \delta^a_b =0$ and $ \Phi_{i-1} f_{i-1} =0=
\widetilde{f}_{i-1} \Phi_{i-1}$.
%Then the solutions for these
%can be obtained. 
%At one loop, the effective potential $V_{eff}^{(1)}$ for $M_{i-1}$
%leads to the positive value for $m_{M_{i-1}}^2$ implying that these
%vacua are stable.

%%%%%%%%%%%%%%%%%%%%%%%%%%%%%%%%%%%%%%%%%%%%%%%%%%%%%%%%%%%%%%%%%%%%%%
%%%%%%%%%%%%%%%%%%%%%%%%%%%%%%%%%%%%%%%%%%%%%%%%%%%%%%%%%%%%%%%%%%%%%%%
\subsection{${\cal N}=1$ 
$SU(N_{c,1}) \times \cdots \times SU(\widetilde{N}_{c,n})$ magnetic theory}
%%%%%%%%%%%%%%%%%%%%%%%%%%%%%%%%%%%%%%%%%%%%%%%%%%%%%%%%%%%%%%%%%%%%%%%
%%%%%%%%%%%%%%%%%%%%%%%%%%%%%%%%%%%%%%%%%%%%%%%%%%%%%%%%%%%%%%%%%%%%%%%

%Let us first consider the Seiberg dual for the last gauge group
%factor.
%There are two magnetic duals depending on whether the gauge group factor
%occurs at odd chain or even chain.

%%%%%%%%%%%%%%%%%%%%%%%%%%%%%%%%%%%%%%%%%%%%%%%%%%%%%%%%%%%%%%
%%%%%%%%%%%%%%%%%%%%%%%%%%%%%%%%%%%%%%%%%%%%%%%%%%%%%%%%%%%%%%
\subsubsection{When the dual gauge group occurs at odd chain}
%%%%%%%%%%%%%%%%%%%%%%%%%%%%%%%%%%%%%%%%%%%%%%%%%%%%%%%%%%%%%%
%%%%%%%%%%%%%%%%%%%%%%%%%%%%%%%%%%%%%%%%%%%%%%%%%%%%%%%%%%%%%%

Let us consider other magnetic theory for the same electric theory.
By applying the Seiberg dual to the $SU(N_{c,n})$ factor  and 
interchanging the $NS5_{n-1}'$-brane and the $NS5_{n}$-brane,
one obtains the Figure 20A which appears in subsection 4.2.2, 
with a replacement of $NS/O6/D6$ instead of $O6^{-}$-plane.
Let us call this as ``modified'' Figure 20A.

The gauge group is given by
\bea
SU(N_{c,1}) \times \cdots \times 
SU(N_{c,n-1}) \times SU(\widetilde{N}_{c,n} \equiv N_{c,n-1}-N_{c,n})
\nonu
\eea
and the matter contents are the field $f_{n-1}$ 
 charged under
$( {\bf 1_1, \cdots, 1_{n-2}, \Box_{n-1}, \overline{\Box}_{n} })$ 
 and their conjugates 
$\widetilde{f}_{n-1}$ 
$( {\bf 1_1, \cdots, 1_{n-2}, \overline{\Box}_{n-1}, \Box_{n} })$ 
under the dual gauge group
and  
the gauge-singlet $\Phi_{n-1}$ which is in the 
adjoint representation for the $(n-1)$-th gauge group, 
in other words,   
$ ({ \bf   1_1, 
\cdots, 1_{n-2},  (N_{c,n-1}-N_{c,n-2})^2-1,1_n})  \oplus  ({\bf 1_1,
 \cdots, 1_n})$ under the
dual gauge group
where the gauge group is broken from
$SU(N_{c,n-1})$ 
to $SU(N_{c,n-1}-N_{c,n-2})$.
Then the $\Phi_{n-1}$ is a $(N_{c,n-1}-N_{c,n-2}) \times
 (N_{c,n-1}-N_{c,n-2})$ 
matrix.
Only $(N_{c,n-1}-N_{c,n-2})$ D4-branes can participate in the mass deformation.

%When two NS5'-branes in modified Figure 20A are close to each other, then 
%it leads to modified Figure 20B
% by realizing that the number of $(N_{c,n-1}-N_{c,n-2})$
%D4-branes connecting between $NS5_{n-2}'$-brane and $NS5_{n}$-brane can
%be rewritten as $(N_{c,n}-N_{c,n-2})$ plus $\widetilde{N}_{c,n}$.
%The Figure 9 of \cite{Ahn07-7} is contained in the modified Figure 20. In
%particular, the brane configuration from the $NS5_{n-2}'$-brane to 
%the $NS5_{n-1}'$-brane is exactly same as the one of \cite{Ahn07-7}.

%The brane configuration in modified Figure 20A is stable as long as the
%distance $(\Delta x)_{n-1}$ between the upper NS5'-brane and 
%the lower NS5'-brane(or $NS5_{n-1}'$-brane) 
%is large. If they are close to each other, then this brane
%configuration is unstable to decay to 
%the brane configuration in modified Figure
%20B.
%
%The brane configuration for zero mass for the bifundamental,
%which has only a cubic superpotential (\ref{ssuper}),
%can be obtained from modified Figure 20A by moving
%the upper  NS5'-brane together with $(N_{c,n-1}-N_{c,n-2})$ color D4-branes 
%into the origin $v=0$.

The low energy dynamics of the magnetic brane configuration 
can be described by the ${\cal N}=1$ supersymmetric gauge theory
with gauge group
and the gauge couplings for the three gauge group factors are
given by
similarly and 
the dual gauge theory has  a meson $\Phi_{n-1}$  and 
bifundamentals $f_{n-1}, \widetilde{f}_{n-1}, F_j$ and
$\widetilde{F}_j$ 
under the dual gauge
group and the superpotential 
corresponding to modified Figures 20A and 20B is given by 
the one in subsection 4.2.2.
Therefore, the F-term equation, the derivative $W_{dual}$ with respect to the
meson field $\Phi_{n-1}$ cannot be satisfied if the $(N_{c,n-1}-N_{c,n-2})$ exceeds
$\widetilde{N}_{c,n}$.
So the supersymmetry is broken.   
That is, 
there exist three equations from F-term conditions:
$
f_{n-1}^a \widetilde{f}_{n-1,b} -\mu_{n-1}^2 \delta^a_b =0$ and $ \Phi_{n-1} f_{n-1} =0=
\widetilde{f}_{n-1} \Phi_{n-1}$.
%Then the solutions for these
%are given by similarly.
%At one loop, the effective potential $V_{eff}^{(1)}$ for $M_{n-1}$
%leads to the positive value for $m_{M_{n-1}}^2$ implying that these
%vacua are stable.

%%%%%%%%%%%%%%%%%%%%%%%%%%%%%%%%%%%%%%%%%%%%%%%%%%%%%%%%%%%%%%
%%%%%%%%%%%%%%%%%%%%%%%%%%%%%%%%%%%%%%%%%%%%%%%%%%%%%%%%%%%%%%
\subsubsection{When the dual gauge group occurs at even chain}
%%%%%%%%%%%%%%%%%%%%%%%%%%%%%%%%%%%%%%%%%%%%%%%%%%%%%%%%%%%%%%
%%%%%%%%%%%%%%%%%%%%%%%%%%%%%%%%%%%%%%%%%%%%%%%%%%%%%%%%%%%%%%

Starting from Figure 25A, moving the $NS5_{n-2}'$-brane 
with $(N_{c,n-1}-N_{c,n-2})$
D4-branes 
to the $+v$ direction leading to Figure 23B, 
and interchanging the $NS5_{n-1}$-brane and the $NS5_{n}'$-brane,
one obtains the Figure 19A,
with a replacement of $NS/O6/D6$ instead of $O6^{-}$-plane.
Let us call this as ``modified'' Figure 19A.

The dual gauge group is given by 
\bea
SU(N_{c,1}) \times \cdots \times 
SU(N_{c,n-1}) \times SU(\widetilde{N}_{c,n} \equiv N_{c,n-1}-N_{c,n})
\nonu
\eea
the matter contents are   
the bifundamentals $f_{n-1}$ in 
 $({\bf 1_1, \cdots, 1, \Box_{n-1}, \overline{\Box}_{n}, })$,
and $\widetilde{f}_{n-1}$ in the representation 
$({\bf 1_1, \cdots, 1, \overline{\Box}_{n-1}, \Box_{n}})$ in
addition to $(n-2)$ bifundamentals $F_j$ and $\widetilde{F}_j$, 
 $j=1,2,
\cdots, (n-2)$ and
the gauge singlet $\Phi_{n-1}$
for the $n$-th dual gauge group in the 
adjoint representation for the $(n-1)$-th dual gauge group, 
i.e.,  
$
{(\bf 1_1, \cdots, 1_{n-2}, (N_{c,n-1}-N_{c,n-2})^2-1, 1_n)  
}
$ plus a singlet
under the 
dual gauge group where the gauge group is broken from
$SU(N_{c,n-1})$ 
to $SU(N_{c,n-1}-N_{c,n-2})$.

%When two NS5'-branes in modified Figure 19A are close to each other, then 
%it leads to modified Figure 19B by realizing that the number of $(N_{c,n-1}-N_{c,n-2})$
%D4-branes connecting between $NS5_{n-2}'$-brane and $NS5_{n-1}$-brane can
%be rewritten as $(N_{c,n}-N_{c,n-2})$ plus $\widetilde{N}_{c,n}$.

%The brane configuration for zero mass for the bifundamental $F_{n-1}$
%and 
%$\widetilde{F}_{n-1}$,
%which has only a cubic superpotential (\ref{spotential}),
%can be obtained from modified Figure 19A by moving
%the upper $NS5_{n-2}'$-brane together with 
%$(N_{c,n-1}-N_{c,n-2})$ color D4-branes 
%into the origin $v=0$.
%The brane configuration in modified Figure 19A is stable as long as the
%distance $(\Delta x)_{n-1}$ between the upper NS5'-brane and 
%the lower NS5'-brane is large, as
%in \cite{GK}. If they are close to each other, then this brane
%configuration is unstable to decay and leads to 
%the brane configuration in modified Figure
%19B.

%In particular, the Figure 9B of \cite{Ahn07-9} with vanishing flavors
%$Q$ and $Q''$
%is contained in
%the modified Figure 19B running from the $NS5_{n-3}$-brane to 
%the $NS5_{n-1}$-brane.

The dual gauge theory has  a $\Phi_{n-1}$  and 
bifundamentals $f_{n-1}, \widetilde{f}_{n-1}, F_j$, and 
$\widetilde{F}_j$  and the superpotential 
corresponding to modified Figures 19A and 19B is given by 
previous results.
Therefore, the F-term equation, the derivative $W_{dual}$ with respect to the
meson field $\Phi_{n-1}$ cannot be satisfied if the $(N_{c,n-1}-N_{c,n-2})$ exceeds
$\widetilde{N}_{c,n}$.
So the supersymmetry is broken.   
%Then the solutions for these
%are given by similarly.
%At one loop, the effective potential $V_{eff}^{(1)}$ for $M_{n-1}$
%leads to the positive value for $m_{M_{n-1}}^2$ implying that these
%vacua are stable.

%%%%%%%%%%%%%%%%%%%%%%%%%%%%%%%%%%%%%%%%%%%%%%%%%%%%%%%%%%%%%%%%%%%%%%%
%%%%%%%%%%%%%%%%%%%%%%%%%%%%%%%%%%%%%%%%%%%%%%%%%%%%%%%%%%%%%%%%%%%%%%%
\subsection{
${\cal N}=1$ 
$SU(\widetilde{N}_{c,1}) \times \cdots \times SU(N_{c,n})$ magnetic theory}
%%%%%%%%%%%%%%%%%%%%%%%%%%%%%%%%%%%%%%%%%%%%%%%%%%%%%%%%%%%%%%%%%%%%%%%%
%%%%%%%%%%%%%%%%%%%%%%%%%%%%%%%%%%%%%%%%%%%%%%%%%%%%%%%%%%%%%%%%%%%%%%%%

Let us apply the Seiberg dual to the first gauge group $SU(N_{c,1})$ factor.
Starting from Figure 27A and moving the $NS5_1$-brane to the left all the
way past the middle NS5'-brane(and the mirror of $NS5_1$-brane to the
right of the middle NS5'-brane),
one obtains the Figure 28A.
%Before arriving at the Figure 28A, there exists an intermediate 
%step where 
%the $\widetilde{N}_{c,1}(= 2N_{c,2}-N_{c,1}+4)$ 
%D4-branes are connecting between the 
%$NS5_1$-brane and its mirror,  
%$(N_{c,2}-N_{c,3})$ D4-branes are connecting between the  $NS5_1$-brane and   
%$NS5_2'$-brane(and their mirrors) as well as $N_{c,3}$ D4-branes between
%$NS5_2'$-brane and the $NS5_3$-brane.
By introducing $(N_{c,2}-N_{c,3})$ D4-branes and $(N_{c,2}-N_{c,3})$ 
anti-D4-branes  between $NS5_1$-brane and   
its mirror, 
one gets the final Figure 28A where we are left with 
$(N_{c,2}-N_{c,3}-\widetilde{N}_{c,1})$ 
anti-D4-branes between the $NS5_1$-brane and   
its mirror.

%%%%%%%%%%%%%%%%%%%%%%%%%%%%%%%%%%%%%%%%%%%%%%%%%%%%%%%%%%%%%%%%%%%%%%
%%%%%%%%%%%%%%%%%%%%%%%%%%%%%%%%%%%%%%%%%%%%%%%%%%%%%%%%%%%%%%%%%%%%%%
\begin{figure}[ht]
   \epsfxsize=4.0in 
\centerline{\epsffile{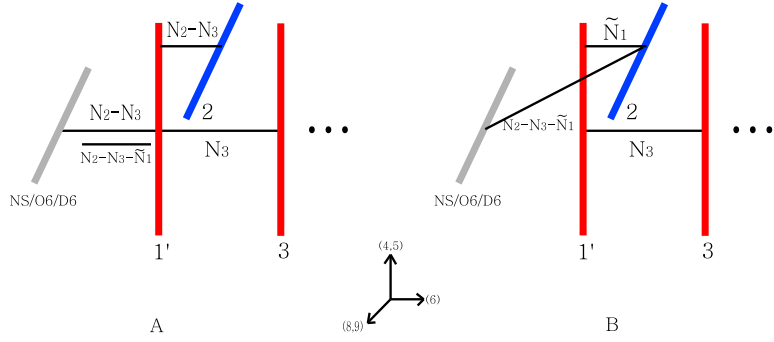}}
   \caption[FIG. \arabic{figure}.]{ 
The 
 ${\cal N}=1$ magnetic brane configuration for the gauge group 
containing $SU(\widetilde{N}_{c,1}=2N_{c,2}-N_{c,1}+4)$ 
with D4-
and $\overline{D4}$-branes(28A) and with 
a misalignment between D4-branes(28B) when the two NS5'-branes are close to
each other.
%The $x$ coordinate of $NS5_{2}'$-brane is given by $(\Delta x)_{2}$.
}
\end{figure}
%%%%%%%%%%%%%%%%%%%%%%%%%%%%%%%%%%%%%%%%%%%%%%%%%%%%%%%%%%%%%%%%%%%%%
%Figure 32A and 32B
%%%%%%%%%%%%%%%%%%%%%%%%%%%%%%%%%%%%%%%%%%%%%%%%%%%%%%%%%%%%%%%%%%%%%%%%

%When two NS5'-branes in Figure 28A are close to each other, it becomes 
%the meta-stable brane configuration Figure 28B
% by realizing that the number of $(N_{c,2}-N_{c,3})$
%D4-branes connecting between $NS5_1$-brane and $NS5_2'$-brane can
%be rewritten as $(N_{c,1}-N_{c,2}-N_{c,3}-4)$ 
%plus $\widetilde{N}_{c,1}$. 
%The Figure 7 of \cite{Ahn07-7} is contained in the Figure 28. In
%particular, the brane configuration from the middle NS5'-brane to 
%the $NS5_3$-brane is exactly same as the one of \cite{Ahn07-7}.

The gauge group is given by
\bea
SU(\widetilde{N}_{c,1} \equiv 2N_{c,2}-N_{c,1}+4) \times 
SU(N_{c,2}) \times \cdots \times SU(N_{c,n})
\nonu
\eea
where the number of dual color can be obtained from the linking number 
counting, as done in \cite{Ahn07-4,Ahn07-1}.
The matter contents are the flavor singlet $f_1$ in the bifundamental 
representation $({\bf \widetilde{N}_{c,1}, \overline{N_{c,2}}, \cdots,
  1_n})$
and its complex conjugate field $\widetilde{f}_1$ in the bifundamental 
representation  $({\bf \overline{\widetilde{N}_{c,1}}, N_{c,2}, \cdots, 1_n})$,
and  the gauge singlet $\Phi_2 \equiv F_1 \widetilde{F}_1$ 
in the representation for 
$({\bf 1_1, (N_{c,2}-N_{c,3})^2-1, 1_3, \cdots, 1_n}) 
\oplus ({\bf 1_1, \cdots, 1_n})$, under the dual gauge group
where the gauge group is broken from
$SU(N_{c,2})$ 
to $SU(N_{c,2}-N_{c,3})$.
There are also
the antisymmetric flavor $a$, the conjugate 
symmetric flavor $\widetilde{s}$ and eight fundamentals $\hat{q}$ for 
$SU(\widetilde{N}_{c,1})$ as well as $F_j$ and $\widetilde{F}_j$.

Then the dual magnetic superpotential, by adding the mass term
 for the bifundamental, is given by
\bea
W_{dual} =  \Phi_2 f_1 \widetilde{f}_1  + m_2 \tr \Phi_2
+ \hat{q}
 \widetilde{s} \hat{q}. 
\nonu
\eea
Here the magnetic fields $f_1$ and $\widetilde{f}_1$  
correspond to 4-4 strings connecting 
the $\widetilde{N}_{c,1}$-color D4-branes(that are 
connecting between the $NS5_1$-brane
and the $NS5_2'$-brane in Figure 28B) with $N_{c,2}$-flavor 
D4-branes(which  are realized 
as corresponding D4-branes in Figure 28A).
Among these $N_{c,2}$-flavor D4-branes, only the strings ending on
the upper $\widetilde{N}_{c,1}$ D4-branes and 
on the tilted  $(N_{c,2}-N_{c,3}-\widetilde{N}_{c,1})$ 
D4-branes in Figure 28B enter the above cubic superpotential term. 
Note that the summation of these D4-branes is equal to $(N_{c,2}-N_{c,3})$.
%Although the $(N_{c,2}-N_{c,3})$ D4-branes for fixed other branes 
%in Figure 28A cannot move any
%directions,
%the tilted $(N_{c,2}-N_{c,3}-\widetilde{N}_{c,1})$-flavor D4-branes can move 
%$w$ direction in Figure 28B(and its mirrors).
%The remaining upper $\widetilde{N}_{c,1}$ D4-branes are fixed also and cannot 
%move any direction. 
%Note that 
%there is a decomposition 
%\bea
%N_{c,2}-N_{c,3}=(N_{c,1}-N_{c,2}-N_{c,3}-4)+\widetilde{N}_{c,1}.
%\nonu
%\eea

%The brane configuration for zero mass for the bifundamental,
%which has only a cubic superpotential,
%can be obtained from Figure 28A by moving
%the upper  NS5'-brane together with $(N_{c,2}-N_{c,3})$ color D4-branes 
%into the origin $v=0$(and their mirrors).
%Then the number of dual colors for D4-branes 
%becomes $\widetilde{N}_{c,1}$ between the $NS5_1$-brane and its mirror, 
% $N_{c,2}$ between $NS5_1$-brane and the $NS5_2'$-brane
%and $N_{c,3}$ 
%between $NS5_2'$-brane and $NS5_3$-brane.
%
%The brane configuration in Figure 28A is stable as long as the
%distance $(\Delta x)_2$ between the upper $NS5_2'$-brane and 
%the middle NS5'-brane is large. If they are close to 
%each other then this brane
%configuration is unstable to decay and it becomes 
%the meta-stable brane configuration in Figure
%28B.
%Since the two NS5'-branes are located at different sides of $NS5_1$-brane
%in Figure 28B,
%for the DBI computation, this fact should be taken into account. 
%One can regard these brane configurations as particular states in the
%magnetic gauge theory with the gauge group  and
%superpotential.

When the $NS5_2'$-brane which is connected by $\widetilde{N}_{c,1}$
D4-branes 
is replaced by $(N_{c,2}-N_{c,3})$
D6-branes and the $NS5_3$-brane is rotated by $\frac{\pi}{2}$ in
Figure 28B,
the brane configuration reduces to the one 
in \cite{Ahn07-9}
where the gauge group is   
given by 
$SU(2n_{f,1}+2n_{c,2}-n_{c,1}+4) \times SU(n_{c,2}) \times SU(n_{c,3})
\times \cdots $ 
with $n_{f,1}$ fundamentals, bifundamentals, an antisymmetric flavor, a
conjugate symmetric flavor, eight fundamentals and various gauge singlets. 
Then the our $N_{c,1}$ corresponds to the $n_{c,1}$,
the number $(N_{c,2}-N_{c,3})$ corresponds to $n_{f,1}$,
and 
our $N_{c,3}$ corresponds to the $n_{c,2}$. 
%In particular, the Figure 7B of \cite{Ahn07-9} with vanishing flavors
%$Q'$ and $Q''$
%is contained in
%this modified Figure 28B running from the middle NS5-brane to 
%the $NS5_{3}$-brane.

The gauge couplings for the  gauge group factors are
given by similarly
and the superpotential 
corresponding to Figures 28A and 28B is given by
\bea
W_{dual} = h \Phi_2 f_1 \widetilde{f}_1 - h \mu_2^2 \tr \Phi_2 + \hat{q}
 \widetilde{s} \hat{q} , \qquad
h^2=  g_{2,mag}^2, \qquad
\mu_{2}^2 = -\frac{(\Delta x)_{2}}{ 2\pi g_s \ell_s^3}.
\nonu
\eea
Then the product $ f_1 \widetilde{f}_1$ is 
a $\widetilde{N}_{c,1} \times \widetilde{N}_{c,1}$ 
matrix where the second gauge group indices for $f_1$ and $\widetilde{f}_1$ 
are contracted with those
of $\Phi_2$ while $\mu_2^2$ is a 
$(N_{c,2}-N_{c,3}) \times (N_{c,2}-N_{c,3})$ matrix.
%Although the field $f_1$ itself is an antifundamental in the second gauge
%group,
%the product $f_1 \widetilde{f}_1$ has the same representation with the 
%product of dual quarks
%and the second gauge group indices for the field $\Phi_2$ play the
%role of the flavor indices.
Therefore, the F-term equation, the derivative of $W_{dual}$ with respect to the
meson field $\Phi_2$ cannot be satisfied if the $(N_{c,2}-N_{c,3})$ exceeds
$\widetilde{N}_{c,1}$.
So the supersymmetry is broken.   
The classical moduli space of vacua can be obtained from F-term
equations. 
That is, 
there are five equations from F-term conditions:
$
f_1^a \widetilde{f}_{1,b} -\mu_2^2 \delta^a_b =0,  \Phi_2 f_1 =0,  \widetilde{f}_1
\Phi_2=0, \hat{q} \widetilde{s} =0$, and $ \hat{q} \hat{q} =0$.
Then the solutions for these
are given by 
\bea
<f_1>   & = & 
\left(
\begin{array}{c}
\mu_2   {\bf 1}_{\widetilde{N}_{c,1}}  \nonu \\
0
\end{array}
\right), \quad
<\widetilde{f}_1>   = 
\left(
\begin{array}{cc}
\mu_2   {\bf 1}_{\widetilde{N}_{c,1}} & 0 \nonu \\
\end{array}
\right), 
\quad
<\Phi_2> =
 \left(
\begin{array}{cc}
0  & 0
 \\
0 & M_2  {\bf 1}_{(N_{c,2}-N_{c,3}-\widetilde{N}_{c,1})} 
\end{array}
\right),
\nonu \\
<\hat{q}> & = & 0,   \quad 
<\widetilde{s}> = 0.
\nonu
\eea
%One can expand around the solutions, as done in \cite{Ahn07-5}. Although there
%exists an extra last term  in the superpotential, this does not contribute to the
%one loop result.
%At one loop, the effective potential $V_{eff}^{(1)}$ for $M_2$
%leads to the positive value for $m_{M_2}^2$ implying that these
%vacua are stable.

%%%%%%%%%%%%%%%%%%%%%%%%%%%%%%%%%%%%%%%%%%%%%%%%%%%%%%%%%%%%%%%%
%%%%%%%%%%%%%%%%%%%%%%%%%%%%%%%%%%%%%%%%%%%%%%%%%%%%%%%%%%%%%%%%
\section{Conclusions}
%%%%%%%%%%%%%%%%%%%%%%%%%%%%%%%%%%%%%%%%%%%%%%%%%%%%%%%%%%%%%%%%%
%%%%%%%%%%%%%%%%%%%%%%%%%%%%%%%%%%%%%%%%%%%%%%%%%%%%%%%%%%%%%%%%

The meta-stable brane configurations we have found are summarized by
Figures 3-7 for the theory described in section
2, by  Figures 
10-14 for the theory given in section 3, by
Figures 17-21 for the theory in section 4. Moreover, those are
described  
by Figure 24 and ``modified'' Figures 17-20 for the theory described in
section 5, 
by Figure 26 and ``deformed'' Figures 17-20 for the theory described in
section 6, and
by Figure 28 and ``modified'' Figures 17-20 for the last theory described in
section 7.
If we replace the upper NS5'-brane in these Figures  
with the coincident D6-branes, 
the corresponding brane configurations become nonsupersymmetic
minimal energy brane configurations found in \cite{Ahn07-8,Ahn07-9} previously.

\vspace{.7cm}

%%%%%%%%%%%%%%%%%%%%%%%%%%%%%%%%%%%%%%%%%%%%%%%%%%%%%%%%%%%%%%%
%%%%%%%%%%%%%%%%%%%%%%%%%%%%%%%%%%%%%%%%%%%%%%%%%%%%%%%%%%%%%%%
\centerline{\bf Acknowledgments}
%%%%%%%%%%%%%%%%%%%%%%%%%%%%%%%%%%%%%%%%%%%%%%%%%%%%%%%%%%%%%%%
%%%%%%%%%%%%%%%%%%%%%%%%%%%%%%%%%%%%%%%%%%%%%%%%%%%%%%%%%%%%%%%

%I would like to thank KIAS(Korea Institute for Advanced Study)
%and Harvard High Energy Theory Group 
%for hospitality where this work was undertaken.
This work was supported by grant No.
R01-2006-000-10965-0 from the Basic Research Program of the Korea
Science \& Engineering Foundation.

\end{document}